\documentclass[a4paper,11pt]{article}
\usepackage{jheppub}
\usepackage{graphicx}
\usepackage{comment}
\usepackage{soul}
\usepackage{axodraw}
\usepackage{float}
\usepackage{slashed}
\usepackage{pbox}
\usepackage{multirow}
\usepackage{caption}
\usepackage{subcaption}
\usepackage{slashbox}
\allowdisplaybreaks
\def\lsim{\mathrel{\rlap{\lower4pt\hbox{\hskip1pt$\sim$}}
    \raise1pt\hbox{$<$}}}
\def\gsim{\mathrel{\rlap{\lower4pt\hbox{\hskip1pt$\sim$}}
    \raise1pt\hbox{$>$}}} 
\newcommand{\ktsq}{k_T^2}

\newcommand{\tktft}{\tilde{k}^4_T}

\newcommand{\ket}[1]{ | {#1} \rangle }
\newcommand{\minus}{\text{-}}

\newcommand{\mw}{m_{W}^{}}

\newcommand{\be}{\begin{eqnarray}}
\newcommand{\ee}{\end{eqnarray}}
\newcommand{\bray}{\begin{array}}
\newcommand{\eray}{\end{array}}
\def\order{\mathcal{O}}

\def\addresses#1#2{\hbox to \hsize{\@tablebox{#1}\hfil\@tablebox{#2}}}
\def\@tablebox#1{\vtop{\hsize=5in \begin{flushleft} #1 \end{flushleft}}}

\def\beq{\begin{equation}}
\def\eeq{\end{equation}}
\def\bit{\begin{itemize}}
\def\eit{\end{itemize}}
\def\beqa{\begin{eqnarray}}
\def\eeqa{\end{eqnarray}}

\def\MadGraph{{\tt MadGraph}}
\def\MadGraph5{{\tt MadGraph5}}

\def\ktsq{k_T^2}

\def\gv{g_V^{}}
\def\longit{L}
\def\zb{\bar z}
\def\figscale{0.7}
\def\calA{\mathcal{A}}
\def\calB{\mathcal{B}}
\def\calC{\mathcal{C}}
\def\calP{\mathcal{P}}
\newcommand{\Lm}{ \mathcal{L}}
\newcommand{\kt}{k_T^{}}

\newcommand{\met}{\displaystyle{\not}E_T}

\newcommand{\sm}{SU(2)_L\times U(1)_Y}
\definecolor{orange}{rgb}{1,0.5,0}
  
\title{Electroweak Splitting Functions and High Energy Showering}

\author{Junmou Chen,}
\author{Tao Han}
\author{and Brock Tweedie}

\affiliation{Pittsburgh Particle physics, Astrophysics, and Cosmology Center, \\
Department of Physics and Astronomy, University of Pittsburgh, \\
3941 O'Hara St., Pittsburgh, PA 15260, USA}

\emailAdd{juc44@pitt.edu}
\emailAdd{than@pitt.edu}
\emailAdd{bat42@pitt.edu}

\abstract{We derive the electroweak (EW) collinear splitting functions for the Standard Model, including the massive fermions, gauge bosons and the Higgs boson. We first present the splitting functions in the limit of unbroken $SU(2)_L\times U(1)_Y$ and discuss their general features in the collinear and soft-collinear regimes. These are the leading contributions at a splitting scale ($k_T$) far above the EW scale ($v$). We then systematically incorporate EW symmetry breaking (EWSB), which leads to the emergence of additional ``ultra-collinear'' splitting phenomena and naive violations of the Goldstone-boson Equivalence Theorem. We suggest a particularly convenient choice of non-covariant gauge (dubbed ``Goldstone Equivalence Gauge'') that disentangles the effects of Goldstone bosons and gauge fields in the presence of EWSB, and allows trivial book-keeping of leading power corrections in $v/k_T$. We implement a comprehensive, practical EW showering scheme based on these splitting functions using a Sudakov evolution formalism. Novel features in the implementation include a complete accounting of ultra-collinear effects, matching between shower and decay, kinematic back-reaction corrections in multi-stage showers, and mixed-state evolution of neutral bosons ($\gamma/Z/h$) using density-matrices. We employ the EW showering formalism to study a number of important physical processes at $\mathcal{O}$(1--10~TeV) energies. They include (a) electroweak partons in the initial state as the basis for vector-boson-fusion; (b) the emergence of ``weak jets'' such as those initiated by transverse gauge bosons, with individual splitting probabilities as large as $\mathcal{O}(35\%)$; (c) EW showers initiated by top quarks, including Higgs bosons in the final state; (d) the occurrence of $\mathcal{O}(1)$ interference effects within EW showers involving the neutral bosons; and (e) EW corrections to new physics processes, as illustrated by production of a heavy vector boson ($W'$) and the subsequent showering of its decay products.
}

\preprint{~~PITT-PACC 1602}

\begin{document}
\maketitle
\flushbottom
%\setcounter{page}{1}

%%%%%%%%%%%%%%%%%%%%%%
\section{Introduction}
\label{sec:intro}
%%%%%%%%%%%%%%%%%%%%%%

%-------------------------------------
\subsection{Electroweak parton showers}
%-------------------------------------

Process-independent parton showers in QED and QCD have long served as invaluable tools for particle physics in high energy collisions and decays.  By exploiting formal factorizations between hard/wide-angle physics and soft/collinear physics~\cite{Collins:1984kg,Collins:1989gx,Bengtsson:1986et}, the extremely complicated exclusive structure of high energy scattering events can be viewed in a modular fashion.  The dominant flows of energy and other quantum numbers are modeled with manageable, low-multiplicity matrix elements.  These are subsequently dressed with soft/collinear radiation, and hadronization applied to bare color charges.  Detailed implementations have varied significantly in specific approach, but showering programs such as {\tt PYTHIA}~\cite{Sjostrand:2007gs}, {\tt HERWIG}~\cite{Bahr:2008pv}, and {\tt SHERPA}~\cite{Gleisberg:2008ta} are now standard workhorses required for describing realistic collider events.  They have also found widespread use in modeling the interactions of high-energy cosmic rays \cite{Knapp:2002vs}, as well as the exclusive products of dark matter annihilation and decay \cite{Cirelli:2008pk,Cirelli:2010xx}.

Collinear parton showers become a ubiquitous phenomenon for processes at energies far above the mass scales of the relevant final-state particles, such as the electron mass in QED or the confinement scale in QCD. With the upgraded LHC and proposed future accelerators \cite{Arkani-Hamed:2015vfh,Mangano:2016jyj,Golling:2016gvc} and a growing suite of instruments sensitive to indirect signals of multi-TeV dark matter~\cite{Abramowski:2013ax,Lefranc:2015vza,Carr:2015hta}, we are now forced to confront processes at energies far above the next known mass threshold in Nature, the electroweak (EW) scale $v\approx 246$~GeV (the electroweak vacuum expectation value, ``VEV'' in short).  Consequently, we are entering a phase in particle physics where it becomes appropriate to consider electroweak parton showers, extending the usual $SU(3)_{\rm QCD} \times U(1)_{\rm EM}$ showers into the fully $SU(3)_{\rm QCD} \times SU(2)_L \times U(1)_Y$ symmetric framework of the Standard Model (SM).  In effect, we will start to see electroweak gauge bosons, Higgs bosons, and top quarks behaving like massless partons~\cite{Dawson:2014pea,Han:2014nja}, appearing both as constituents of jets \cite{Almeida:2008tp} as well as of initial-state beam particles. This is in stark contrast to the conventional perspective in which they are viewed as ``heavy'' particles that are only produced as part of the hard interaction.  

The concept of electroweak bosons as partons has a long history, beginning with the ``effective-$W$ approximation"~\cite{Kane:1984bb,Dawson:1984gx,Chanowitz:1985hj}. This picture of electroweak vector bosons radiating off of initial-state quarks is now strongly supported by the experimental observation of Higgs boson production via vector boson fusion (VBF) at the LHC~\cite{LHCVBF}. As we imagine probing VBF-initiated processes at even higher energies, with both the initial weak bosons and their associated tag jets becoming significantly more collinear to the beams, the idea of weak parton distribution functions (PDFs) within protons becomes progressively more appropriate.

Many calculations have further revealed large negative electroweak virtual corrections to a variety of exclusive high-energy processes, wherein real emission of additional weak bosons is not included. Such large ``non-emission'' rate penalties indicate the onset of the universal, logarithmically-enhanced Sudakov form-factors characteristic of massless gauge theories~\cite{Melles:2000gw,Beenakker:2000kb}. For example, exclusive di-jet production receives corrections from virtual $W/Z$ exchange that begin to exceed $-10\%$ for transverse momenta exceeding 3~TeV~\cite{Moretti:2006ea,Dittmaier:2012kx}, and grow to approximately $-$30\% at the 10's of TeV energies expected at future hadron colliders. For processes that include weak bosons at the hard event scale, such as $\gamma/Z/W$+jets or vector boson pair production, the corrections can quickly grow to $O(1)$~\cite{Kuhn:2004em,Kuhn:2005az,Kuhn:2005gv,Kuhn:2007qc,Kuhn:2007cv,Hollik:2007sq,Becher:2013zua}. A process-independent framework for extracting all such log-enhanced electroweak virtual corrections at fixed leading-order has been developed in~\cite{Denner:2000jv,Denner:2001gw}, and next-to-leading logarithmic resummation of the gauge corrections has been achieved using SCET formalism in~\cite{Chiu:2007yn,Chiu:2007dg,Chiu:2008vv,Chiu:2009mg,Chiu:2009ft}.

The total rates of real $W/Z$ emissions and other electroweak parton splittings have a direct correspondence with the ``lost'' event rates encoded in the negative electroweak virtual corrections, with matching logarithmic enhancements in accordance with the Kinoshita-Lee-Nauenberg theorem. Iterating this observation across all possible nested emissions and loops within a given process builds up the usual parton shower picture, allowing formal resummations of the logarithms that would otherwise still appear in well-defined exclusive rates. Many studies have addressed aspects of electroweak parton showering in the past several years~\cite{Ciafaloni:2000rp,Ciafaloni:2000gm,Ciafaloni:2005fm,Baur:2006sn,Bell:2010gi,Chiesa:2013yma,Christiansen:2014kba,Krauss:2014yaa,Bauer:2016kkv}. Parts of the complete shower are already available in public codes and are being tested at the LHC, with ATLAS recently making a first observation of collinear-enhanced $W/Z$ radiation within QCD jets~\cite{Aaboud:2016ylh}. A detailed listing of electroweak collinear splitting functions and PDF evolution equations, restricted to processes that survive in the unbroken limit, has been worked out in~\cite{Ciafaloni:2005fm}. There, the effects of electroweak symmetry breaking (EWSB) are addressed minimalistically by including a hard phase space cutoff and working in a preferred isospin basis. These results and more recent SCET-based calculations have also been adapted for the problem of TeV-scale dark matter annihilation in~\cite{Ciafaloni:2010ti,Cavasonza:2014xra,Bauer:2014ula,Ovanesyan:2014fwa,Baumgart:2014vma,Baumgart:2014saa,Baumgart:2015bpa}. For general-purpose applications, recent versions of {\tt PYTHIA} incorporate radiation of $W$ and $Z$ bosons off of light fermions~\cite{Christiansen:2014kba}, including a detailed model of how this component of the shower turns off due to $W/Z$ mass effects. A study using {\tt SHERPA}~\cite{Krauss:2014yaa} instead breaks down these emissions into separate transverse ($V_T$) and longitudinal ($V_{\longit}$) components, coupling in the latter strictly using Yukawa couplings by appealing to the Goldstone-boson Equivalence Theorem (GET) \cite{Lee:1977eg,Chanowitz:1985hj}. The problem has been approached in different way within {\tt ALPGEN}~\cite{Mangano:2002ea,Chiesa:2013yma}, by multiplying exclusive hard event rates with the fixed-order Sudakov factors of~\cite{Denner:2000jv,Denner:2001gw} and supplementing with exact fixed-order real emission processes. This approach, which is itself a first step towards electroweak shower matching, works well when the soft/collinear phase space enhancements are modest and the need for added accuracy of higher-multiplicity hard event generation balances the added computational complexity. However, a complete matching prescription will also ultimately involve a dedicated parton shower step, especially when convolved with QCD radiation. The simpler, process-independent parton shower approach will also become particularly useful in new physics applications~\cite{Hook:2014rka,Rizzo:2014xma}.

%---------------------------
\subsection{Our approach}
%---------------------------

Notably, no existing general-purpose parton showering algorithm that is capable of generating fully exclusive events has addressed the full scope of universal collinear electroweak physics. In particular, a complete treatment must include the high-rate of non-Abelian splittings amongst the weak bosons themselves, as well as showers that involve longitudinal/scalar states and many of the sometimes subtle effects of spontaneous symmetry breaking. The goal of the present paper is to outline such an algorithm, providing a comprehensive framework in which all collinear electroweak showering phenomena can be implemented, and including a systematic treatment of EWSB. Towards this end, we derive and tabulate the complete set of electroweak splitting functions in the broken phase, including the massive fermions, gauge bosons, and the Higgs boson. These generalize and unify both the unbroken-phase evolution equations of~\cite{Ciafaloni:2005fm} and the purely broken-phase effects already observed within the effective-$W$ approximation, namely the generation of longitudinal vector boson beams from massless fermions~\cite{Kane:1984bb,Dawson:1984gx,Chanowitz:1985hj}. We further investigate some of the physical consequences of these various electroweak showering phenomena.

Relative to QED and QCD showers, the complete electroweak parton shower exhibits many novel features. At the level of the unbroken theory at high energies, the shower becomes chiral and the particle content is extended to include an EW-charged scalar doublet. Most of the degrees of freedom contained in this scalar are to be identified with the longitudinal gauge bosons via the Goldstone-boson Equivalence Theorem. Including Yukawa couplings, the set of core splitting function topologies expands from the usual three to seven. EWSB also already makes a subtle imprint here due to the presence of a preferred isospin basis for asymptotic states, leading to interference and self-averaging effects between different exclusive isospin channels. The latter are intimately related to ``Bloch-Nordsieck violation'' when occurring in the initial state~\cite{Ciafaloni:2000rp,Bell:2010gi,Manohar:2014vxa}. As the shower evolves down through the weak scale, it becomes physically regulated by the appearance of gauge boson, scalar, and fermion masses. Unlike in QCD where the shower regulation occurs non-perturbatively due to confinement, or in QED where a small photon mass is sometimes used as an artificial regulator for soft emissions, the electroweak shower exhibits a perturbative transition with genuinely massive gauge bosons. It is possible to describe this transition rather accurately, but doing so requires a careful accounting of symmetry-violating effects beyond simple kinematic suppressions, and a consistent elimination of gauge artifacts. In particular, Goldstone-boson equivalence ceases to hold at relative transverse momenta of order the weak scale, allowing for an additional burst of many ``ultra-collinear'' radiation processes that do not exist in the unbroken theory, and are highly suppressed at energy scales $k_T \gg v$. To cleanly isolate these effects, we introduce a novel gauge dubbed ``Goldstone Equivalence Gauge'' (GEG). This is a particularly convenient choice of non-covariant gauge, allowing a completely transparent view of Goldstone-boson equivalence within the shower, as well as systematic corrections away from it in the splitting matrix elements, organized in a power series in VEV factors. The naively bad high energy behavior of the longitudinal gauge bosons is deleted, and the Goldstone fields allowed to interpolate physical states, at the cost of re-introducing explicit gauge-Goldstone boson mixing.

Our formalism developed here has deep implications and rich applications at TeV-scale energies and beyond. Some aspects include EW parton distribution functions associated with initial state radiation (ISR), multiple emissions in EW final state radiation (FSR), consistent merging of EW decays with EW showering, a quantum-coherent treatment of the Sudakov evolution of $\gamma/Z/h$ states, as well as modeling of general ultra-collinear processes including, e.g., $t_R \to h t_R$ and $h\to hh$. We also make some preliminary studies of the impact of EW showering on new physics searches in the context of a heavy $W'$ decay. Quite generally, we begin to see the emergence of the many nontrivial phenomena of ``weak jets'' across a broad range of SM and BSM phenomena.

Before proceeding, we also clarify what is {\it not} covered in our current treatment. 
We make exclusive use here of the collinear approximation, which, in physical gauges such as GEG, explicitly factorizes all soft and collinear divergences particle-by-particle, isolating them to $1\to 2$ real emission diagrams and self-energy loops. This furnishes a formally leading-log model of EW showering, capturing all double-log effects from the soft-collinear region of gauge emissions, as well as the single-logs associated to all hard-collinear splittings. The former are identical to the double-logs that would be inferred from the collinear limits of the eikonal approximation, whose particle-by-particle factorization can be seen upon application of Ward identities~\cite{Denner:2000jv,Denner:2001gw,Bell:2010gi}. However, there are additional single-log soft divergences within gauge emission interferences and virtual exchanges between different particles, which do not factorize so simply. For non-singlet EW ensembles, these contributions lead to global entanglements of isospin quantum numbers between different particles in the event, which are absent in our shower. These isospin entanglements are somewhat analogous to the global kinematic entanglements that occur due to soft gluon emissions/exchanges at NLL level in QCD. Nonetheless, the dominant effects of isospin rearrangements, in particular the Bloch-Nordsieck violation, arise already at the double-log level, and are modeled by our shower up to residual single-log ambiguities. We will address approaches to the NLL resummation of isospin entanglements in a future work~\cite{EWshower}.

The rest of the paper is organized as follows. We begin in Section~\ref{sec:split} with a generic discussion of splitting and evolution formalism with massive particles. We then outline some of the other nontrivial features such as PDFs for massive particles, interference between different mass eigenstates, showers interpolating onto resonances, and back-reaction effects from multiple emissions. 
In Section~\ref{sec:unbroken}, we introduce the splitting kernels for the unbroken electroweak theory, namely $SU(2)_L \times U(1)_Y$ gauge theory with massless fermions in SM representations, a single (massless) scalar doublet, and Yukawa interactions.  We then proceed in Section~\ref{sec:broken} to generalize these results to the broken phase. After a discussion of the violation of the Goldstone-boson Equivalence Theorem, we introduce the Goldstone Equivalence Gauge. We then discuss the EWSB modifications to the unbroken splitting functions and present a complete list of ultra-collinear processes that arise at leading-order in the VEV.
Section~\ref{sec:implementation} explores some key consequences of electroweak showering in final-state and initial-state splitting processes, including a discussion of EW parton distribution functions and multiple EW final state radiation.
We emphasize the novel features of the EW shower and illustrate some of the effects in the decay of a heavy vector boson $W'$.
We summarize and conclude in Section~\ref{sec:conclusions}. Appendices give supplementary details of Goldstone Equivalence Gauge, the corresponding Feynman rules and illustrative examples of practical calculations, more details on the density-matrix formalism for coherent Sudakov evolution, and a short description of our virtuality-ordered showering program used for obtaining numerical FSR results.

%%%%%%%%%%%%%%%%%%%%%%
\section{Showering Preliminaries and Novel Features with EWSB}
\label{sec:split}
%%%%%%%%%%%%%%%%%%%%%%

We first summarize the general formalism for the splitting functions and evolution equations with massive particles that forms the basis for the rest of the presentation. We then lay out some other novel features due to EWSB.

%----------------------------------
\subsection{Splitting formalism}
%----------------------------------

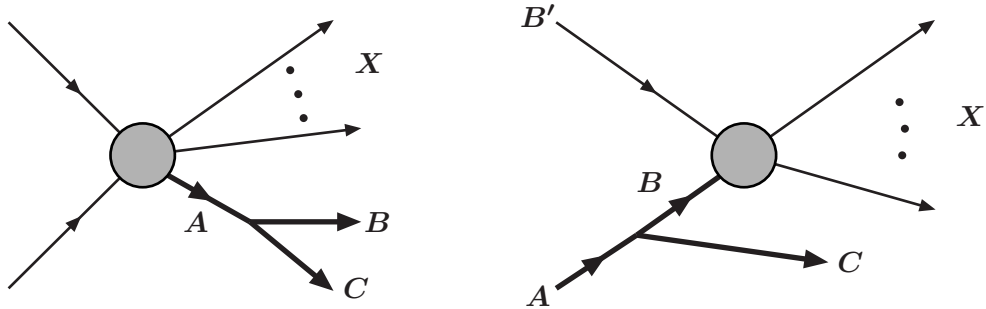
\begin{figure}[t]
\begin{center}
\begin{picture}(350,100)(0,0)
\SetColor{Black}
\SetWidth{2}
\ArrowLine(55,45)(90,25) % A
\LongArrow(90,25)(130,25) % B
\LongArrow(90,25)(120, 0) % C
\SetWidth{1}
\ArrowLine(0,0)(50,50)
\ArrowLine(0,100)(50,50)
\LongArrow(50,50)(120,100)
\LongArrow(50,50)(130, 60)
\Vertex(110,64){1.5}
\Vertex(108,73){1.5}
\Vertex(105,82){1.5}
\GCirc(50,50){12}{0.7}
\Text(135,85)[]{$\boldsymbol X$}
\Text(70,25)[]{$\boldsymbol A$}
\Text(138,25)[]{$\boldsymbol B$}
\Text(130,0)[]{$\boldsymbol C$}
\SetOffset(225,0)
\SetWidth{2}
\ArrowLine(-20,0)(10,20)  % A
\ArrowLine(10,20)(45,45)  % B
\LongArrow(10,20)(80,10)  % C
\SetWidth{1}
\ArrowLine(-20,100)(50,50)
\LongArrow(50,50)(120,100)
\LongArrow(50,50)(120, 30)
\Vertex(109,50){1.5}
\Vertex(110,60){1.5}
\Vertex(108,70){1.5}
\GCirc(50,50){12}{0.7}
\Text(135,65)[]{$\boldsymbol X$}
\Text(-27,-3)[]{$\boldsymbol A$}
\Text(15,40)[]{$\boldsymbol B$}
\Text(90,10)[]{$\boldsymbol C$}
\Text(-27,103)[]{$\boldsymbol{B'}$}
\end{picture}
\end{center}
\caption[]{Schematic processes involving a collinear splitting $A\to B+C$ in either the final state (left) or initial state (right).}
\label{fig:split}
\end{figure}

Consider a generic hard process nominally containing a particle $A$ in the final state, slightly off-shell and subsequently splitting to $B$ and $C$, as depicted in Fig.~\ref{fig:split} (left figure). In the limit where the daughters $B$ and $C$ are both approximately collinear to the parent particle $A$, the cross section can be expressed in a factorized form~\cite{Collins:1989gx}
\begin{equation}
d\sigma_{X,BC} \,\simeq\,  d\sigma_{X,A} \times d{\mathcal P}_{A\rightarrow B+C} \, , 
\label{eq:FSR}
\end{equation}
where $d{\mathcal P}$ is the {\it differential splitting function} (or probability distribution) for $A\to B+C$.
%\footnote{\Tao{It is the probability function for the splitting, and is normalized as the branching fraction for a physical decay.}} 
A given splitting can also act as the ``hard'' process for later splittings, building up jets. The factorization of collinear splittings applies similarly for initial-state particles, leading to the picture of parton distribution functions (PDFs) for an initial state parton $B$ or $C$, as in Fig.~\ref{fig:split} (right figure),
\begin{equation}
d\sigma_{AB'\to CX} \,\simeq\, d{\mathcal P}_{A\rightarrow B+C} \times d\sigma_{BB' \to X} \, .
\label{eq:ISR}
\end{equation}
We will discuss this situation in the next subsection. While our main focus here is on the leading-log resummation of these splitting effects in a parton shower/evolution framework, at a leading approximation Eqs.~(\ref{eq:FSR}) and~(\ref{eq:ISR}) can also be taken as-is, with a unique splitting in the event and no virtual/resummation effects, in order to quickly capture the tree-level collinear behavior of high energy processes. In our further analyses, we will refer to such a treatment as a ``fixed-order EW shower'' or ``fixed-order EW FSR (ISR).''

Integrating out the azimuthal orientation of the $B+C$ system, the splitting kinematics are parametrized with two variables: a dimensionful scale (usually chosen to be approximately collinear boost-invariant) and a dimensionless energy-sharing variable $z$. Common choices for the dimensionful variable are the daughter transverse momentum $k_T$ relative to the splitting axis, the virtuality $Q$ of the off-shell particle in the process, and variations proportional to the daughters' energy-weighted opening angle $\theta E_A$. Our descriptions here will mainly use $k_T$, as this makes more obvious the collinear phase space effects in the presence of masses. For our numerical results in Section~\ref{sec:implementation}, we switch to virtuality, which allows for a simpler matching onto $W/Z/t$ decays. Mapping between any of these different scale choices is however straightforward. The energy-sharing variable $z$~($\zb \equiv 1-z$) is commonly taken to be the energy fraction of $A$ taken up by $B$~($C$). The splitting kinematics takes the form
\beqa
E_B  \approx  z E_A,\quad
E_C \approx  \zb E_A, \quad
\kt \approx z\zb E_A \theta \ .
\eeqa
When considering splittings involving massive or highly off-shell particles, various possible definitions of $z$ exist which exhibit different non-relativistic limits. Besides strict energy fraction, a common choice is the light-cone momentum fraction, 
$z \equiv (E_B+\vec k_{B}\cdot\hat k_A)/(E_A+|\vec k_A|)$. Our specific implementation in Section~\ref{sec:implementation} uses the three-momentum fraction 
\be
z \equiv {|\vec k_B| \over |\vec k_B| + |\vec k_C| },
\ee
which makes phase space suppression in the non-relativistic limit more transparent. However, in the relativistic regime, where the collinear factorization is strictly valid, all of these definitions are equivalent, and we do not presently make a further distinction.\footnote{There is unavoidably some frame-dependence to this setup, as there is in all parton showers that are defined strictly using collinear approximations. A more complete treatment would exhibit manifest Lorentz-invariance and control of the low-momentum region, at the expense of more complicated book-keeping of the global event's kinematic and isospin structure, by using superpositions of different $2\to 3$ dipole splittings. Extending our treatment in this manner is in principle straightforward, but beyond the scope of the present work.} 

In the simplest cases, generalizing the collinear splitting function calculations to account for masses is straightforward. Up to the non-universal and convention-dependent factors that come into play in the non-relativistic/non-collinear limits, the splitting functions can be expressed as
\beq
\frac{d{\mathcal P}_{A\rightarrow B+C}}{dz\,d\ktsq} \,\simeq\,  {1\over 16\pi^2} \ { z \zb \: |{\cal M^{\rm (split)}}|^2 \over 
(\ktsq  + \zb m_B^2 + z m_C^2  - z \zb m_A^2)^2 } \ .
\label{eq:split}
\eeq
Here, ${\cal M^{\rm (split)}}$ is the $A\to B+C$ splitting matrix-element, which can be computed from the corresponding amputated $1\to 2$ Feynman diagrams with on-shell polarization vectors (modulo gauge ambiguities, which we discuss later). This may or may not be spin-averaged, depending on how much information is to be kept in the shower. Depending upon the kinematics, the mass-dependent factors in the denominator act to either effectively cut off collinear divergences at small $k_T$ or, in final-state showers, to possibly transition the system into a resonance region. In cases where interference between different mass eigenstates can be important, this basic framework must be further generalized. Resonance and interference effects are introduced in Section~\ref{sec:novel_features}.

On dimensional grounds, $|{\cal M^{\rm (split)}}|^2$ goes like either $k_T^2$ or some combination of the various $m^2$'s. Conventional splitting functions typically scale like $d\ktsq/k_T^2$, which is exhibited by all of the gauge and Yukawa splittings of the massless unbroken electroweak theory, as to be shown in Section~\ref{sec:unbroken}. There can also be mass-dependent splitting matrix elements that lead to $m^2 d\ktsq/k_T^4$ type scaling. These splittings are highly suppressed for $k_T \gsim m$. However, they are much more strongly power-enhanced at low $k_T$, a behavior which we call {\it ultra-collinear}.
% (borrowing a phrase from~\cite{Fleming:2007qr}). 
Upon integration over $k_T$, the total rate for an ultra-collinear splitting comes out proportional to dimensionless combinations of couplings and masses, with the vast majority of the rate concentrated near $k_T \sim m$. Such processes exist in familiar contexts like QED and QCD with massive fermions, for example the helicity-flipping splittings $e_L \to \gamma e_R$ and $g \to b_L \bar b_L$. They are usually not treated as distinct collinear physics with their own universal splitting functions, though they are crucial for systematically modeling shower thresholds. We choose to treat them on independent footing, since the threshold behaviors of the electroweak shower are highly nontrivial, including processes that are qualitatively different from the massless limit.

In both the conventional collinear and ultra-collinear cases, the remaining $z$ dependence after integrating over $k_T$ can be either $dz/z$ or $dz \times $(regular). The former yields additional soft logarithms (again, formally regulated by the particle masses), and appears only in splittings where $B$ or $C$ is a gauge boson.

%-------------------------------------------------
\subsection{Evolution equations}
\label{sec:evolv}
%-------------------------------------------------

When applied to the initial state, the splitting functions outlined in the previous section lead to both initial state radiation (ISR) as well as the dynamical generation of $B$ and $C$ parton distribution functions from a parent $A$. Considering a generic parton distribution function $f_i(z,\mu^2)$ with a factorization scale $\mu$ in $k_T$-space, the leading-order convolution relation is
\beq
f_B(z, \mu^2)  \,=\, f_B(z,\mu_0^2) \,+\, \sum_A \int^1_z {d\xi \over \xi} f_A(\xi,\mu_0^2) \int^{\mu^2}_{\mu_0^2} d\ktsq \, \frac{d{\mathcal P}_{A\rightarrow B+C}({z/ \xi}, \ktsq)}{dz \, d\ktsq} \, , \label{eq:convolution}
\eeq
where $\mu_0$ is an input factorization scale.
Differentiating with respect to $\mu^2$ and incorporating as well the evolution of the $f_A$ leads to the celebrated Dokshitzer-Gribov-Lipatov-Altarelli-Parisi (DGLAP) equation \cite{Dokshitzer:1977sg,Gribov:1972ri,Altarelli:1977zs}. 
\beq
{\partial f_B(z, \mu^2) \over \partial \mu^2} = \sum_A \int^1_z {d\xi \over \xi}\ 
{ d{\mathcal P}_{A\rightarrow B+C}({z/ \xi}, \mu^2) \over dz \, d\ktsq}  f_A(\xi, \mu^2) \, .
\eeq
Gauge theories such as QED and QCD predict that at high energies the splitting functions $d{\mathcal P}/d \ktsq$ go like $1 /\ktsq$, and thus that the PDFs evolve like $\ln (Q^2/\mu^2)$. This is the classic violation of the Bjorken scaling law~\cite{Bj}. In the broken electroweak theory, there are also the qualitatively different ultra-collinear splitting functions, which instead go as $m^2/k_T^4$. The PDFs arising from these splittings ``live'' only at the scale $k_T \sim m$. Instead of evolving logarithmically, they are cut off by a strong power-law suppression at $k_T \gsim m$. The corresponding PDFs preserve Bjorken scaling, up to contributions beyond leading order. In particular, longitudinal weak boson PDFs are practically entirely determined at splitting scales of ${\cal O}(m_W)$, even when used as inputs into processes at energies $E \gg m_W$.\footnote{This observation persists even in the presence of QCD corrections. We can imagine that a quark is first evolved to large $k_T$ (and hence large space-like virtuality $Q$) from multiple gluon emissions, and then splits into an on-shell quark and space-like longitudinal vector boson. The former emerges as an ISR jet and the latter participates in a hard interaction. We would find (e.g., using Goldstone Equivalence Gauge, introduced in Section~\ref{sec:GEG}) that the collinear-enhanced piece of the scattering amplitude carries a net suppression factor of ${\cal O}(m^2/Q^2)$, which cannot be compensated by integration over the collinear emission phase space.}

Numerical computation of electroweak PDFs with a proper scale evolution do not exist yet in the literature, though the complete unbroken-theory evolution equations appear in~\cite{Ciafaloni:2005fm}, and fixed-order results are straightforward to obtain with the simple convolution in Eq.~(\ref{eq:convolution}). In the resummed treatment, contributions from the region $k_T \sim m_W$ can perhaps most simply be incorporated as perturbative ``threshold'' effects, essentially adding in their integrated fixed-order contributions up to some scale (a~few)$\times m_W$ as $\delta$-functions in $k_T$-space. These would include the finite, mass-suppressed contributions from the turn-on of $f\to W_T f$ splittings, as well as the entire ultra-collinear $f \to W_{\longit} f$ contribution. Equivalently at leading-order, they may instead be folded continuously into the DGLAP evolution using the massive splitting functions defined as in Eq.~(\ref{eq:split}). This latter approach may also be simpler when alternative scaling variables are used, such as virtuality.

The other qualitatively new electroweak effects in the PDFs concern the treatment of weak isospin.
First, the chiral nature of the EW gauge interactions leads to more rapid evolution toward low-$x$ for left-handed fermions than for right-handed fermions. Furthermore, the isospin non-singlet nature of typical beam particles yields an additional interesting subtlety. In QED and color-averaged QCD evolution, the soft-singular limits of, e.g., $q\to gq$ at a given scale become indistinguishable from $q \to q$ with no splitting. Indeed, this allows for the balancing of real and virtual IR divergences as $z$ is formally taken to zero at fixed $k_T$, conventionally encoded in the plus-prescription. However, following this prescription for the electroweak evolution of fermion PDFs at $k_T \gg m_W$ leads to unregulated divergences in isospin-flipping transitions, such as $u_L \leftrightarrow d_L$ via arbitrarily soft $W^\pm$ emission. This is a manifestation of the so-called Bloch-Nordsieck violation effect~\cite{Ciafaloni:2000rp,Bell:2010gi,Manohar:2014vxa}. Regulation and resummation of this effect requires the introduction of some form of explicit cutoff $z \gsim k_T/E$ in the evolution equations when formulated in $(k_T,z)$ space, in order to avoid non-collinear emission regions~\cite{Ciafaloni:2005fm}.\footnote{In QED and QCD, these non-collinear emissions are implicitly and ``incorrectly'' integrated over in the plus-prescription. However, in the limit $E \gg k_T$, the numerical impact of doing so is of sub-leading importance.} The net effect is a gradual, controlled merging of the $u_L$ and $d_L$ PDFs (or $e_L$ and $\nu_L$ PDFs in the case of electron beams) into a common ``$q_L$'' (``$\ell_L$") PDF. Unlike conventional PDF evolution, implementing the $z$ cutoff in this way necessitates extending the arguments of the PDFs to explicitly include the (CM-frame) beam energy. While this is not a major complication, we do point out that different choices of scaling variables may yield the same non-collinear regulation without requiring the extra energy argument. A particularly simple choice would be the energy-weighted angle $\theta E_A$. We defer a detailed study of these issues to future work~\cite{EWshower}.

We caution that this treatment of the initial state using PDFs remains strictly valid only within the leading-log, collinear approximation. Soft $W^\pm$ virtual exchanges between the isospin non-singlet beams will induce single-log entanglements that do not factorize between the individual beams, and even more complicated entanglements emerge when we also consider isospin-exclusive final states. The proper generalization for the initial state is from running PDFs to running quantum-ensemble parton luminosities defined for {\it pairs} of beams. But it is also possible to define a scheme where these beam-entanglement effects are selectively treated at fixed-order, and PDF resummation still suffices~\cite{EWshower}. (The entanglement effects actually wash out as the scale is raised and the isospin ensembles become incoherent.) However, these PDFs will still likely reference the global beam setup via the aforementioned non-collinear cutoff.

Even applying the conventional factorization at leading-log, some of the PDFs must also still be treated as matrices~\cite{Ciafaloni:2005fm}. This is particularly relevant for the photon and transverse $Z$-boson PDFs, which develop sizable off-diagonal contributions. Indeed, the naive concept of independent ``photon PDF'' and ``$Z$ PDF'' at $k_T \gg m_Z$ is necessarily missing important physics, as $\gamma$ and $Z$ are not gauge eigenstates. We outline the appropriate treatment in Section~\ref{sec:interference} and Appendix~\ref{app:split}.

The same splitting functions that govern ISR and PDF generation also serve as the evolution kernels for final-state radiation (FSR). This integrates to the well-known Sudakov form factor $\Delta_A(t)$ characterizing the possible time-like branchings of parent $A$ at scales below $t \sim \log(k_T)$ or $\log(Q)$
\beqa
&& \Delta_A(t)  \,=\,  \exp \left[ -\sum_{BC} \int^t_{t_0} dt' \int dz \, \frac{d{\cal P}_{A\to B+C}(z,t')}{dz \, dt} \right] \, , %\\ 
%&& f_A(x, t)  \,=\,  \Delta_A(t) f_A(x, t_0) + \int^t_{t_0} {dt'\over t'} {\Delta(t)\over \Delta(t')} \int {dz\over z} P_{A\to BC}(z)\  f_A(x/z, t') \, .
\eeqa
where the allowed $z$ range is determined by kinematics. Practically, we perform the evolution starting at a high $k_T$ or virtuality scale characterized by the CM-frame energy of the hard partonic process, and running continuously down through the weak scale with the proper mass effects. The Sudakov factor, evaluated in small $t$ steps, functions as a survival probability for $A$, upon which the usual Markov chain monte carlo is constructed. (See, e.g.,~\cite{Sjostrand:2006za}.) If $A$ does not survive at some step, it is split into a state $B+C$. This splitting acts as the ``hard'' process that produced particles $B$ and $C$, and Sudakov evolution is continued on each of those particles. The ``resolution'' scale $t_0$ can be any scale well below $m_W$, at which conventional QED and QCD showers can take over.  Of course, the basic framework leaves many details unspecified, and allows for a great deal of freedom in specific implementation. For example, besides the choice of evolution variable, one must also specify a treatment of kinematic reshuffling. We elaborate on some additional aspects of our own implementation of final-state showers below and in Appendix~\ref{sec:FSR}. We will generally refer this treatment of Sudakov formalism as the ``full EW shower'' or ``full EW FSR'', in contrast to the fixed-order splitting calculations in Eqs.~(\ref{eq:FSR}) and (\ref{eq:ISR}).

%-------------------------------------------------
\subsection{Other novel features in EW showering}
\label{sec:novel_features}
%-------------------------------------------------

There are several additional novel features in EW showering beyond those encountered in the standard formalism. We outline a few relevant to our later discussions and also propose concrete schemes for their implementations.

%.............................
\subsubsection{Mass effects}
\label{sec:mass_effects}
%..................................

Besides the basic kinematic modifications and the emergence ultra-collinear splitting phenomena, the existence of a mass scale $m_{W,Z} \sim g v$ and $m_{f} \sim y_f v$ requires some special treatments as we approach kinematic thresholds and the boundaries of turnoff regions.

An immediate complication is that final-state weak showering smoothly connects onto the on-shell weak decays of top quarks, $W/Z$ bosons, and (to a much lesser extent) Higgs bosons. The shower describes the highly off-shell behavior of these particles, including resummed logarithmically-enhanced effects. But the effect of the pole is nonetheless visible, encoded in the last term in the denominator of Eq.~(\ref{eq:split}). Within the resonance region, the dominant behavior is more correctly captured by the standard Breit-Wigner line-shape governed by the physical width $\Gamma$, which involves a very different kind of resummation. However, a few $\Gamma$ above the peak, both descriptions can be expanded perturbatively and yield numerically similar predictions.\footnote{The agreement is further improved if $\Gamma$ is generalized to $\Gamma(Q)$. E.g., $\Gamma_Z \to \Gamma_Z(Q) \simeq (Q/m_Z)\Gamma_Z$.} It is therefore straightforward to define a well-behaved matching prescription. This is easiest to formulate within a virtuality-ordered shower: Halt the shower at some matching scale $Q_{\rm match} = m + $(a~few)$\Gamma$, and if the state has survived to this point, distribute its final mass according to a Breit-Wigner resonance below $Q_{\rm match}$. The exact choice of matching scale here is not crucial, as long as it is within the region where the Breit-Wigner and shower predictions are comparable. For other shower ordering variables, such as $k_T$, we can instead run the shower down to its nominal kinematic limit, but not integrating $z$ within the region that would yield $Q < Q_{\rm match}$. In either case, the parton shower may be restarted on the resonance's decay products.

Another place where mass effects can become important is in multiple emissions. In massless showers, sequential splittings are dominantly very strongly-ordered in scale, and as a consequence a given splitting rate can be computed without regard to the subsequent splittings while still capturing the leading behavior. However, in showers with massive particles, a large fraction of the available phase space for secondary splittings may require nontrivial kinematic rearrangements within the preceding splittings. For example, a $W$ boson might nominally be produced with a kinematic mass $m_W$ via emission off of a fermion. If the $W$ subsequently splits into a $W$ and a $Z$ boson at a virtuality $Q \gg m_W$, there is a chance that the off-shell $W$ now sits near a suppressed region (i.e., dead cone) for emission off of the mother fermion. In order to avoid badly mis-modeling such cases, secondary splittings can be weighted according to the relative rate modification that would be incurred on the previous splitting. This {\it back-reaction factor} depends in detail on how kinematic arrangements are done in the shower. Generally, a given $(z,Q)$ or $(z,k_T)$ parametrizing the mother splitting will be mapped onto a new $(z^*,Q^*)$ or $(z^*,k_T^*)$ for producing the off-shell daughter. The required back-reaction factor is the ratio of the new differential splitting function to the original one, multiplied by the Jacobian for the change of variables. For a final-state shower sequence $A^* \to B^*C \to (DE)C$, for the nested splitting we can use a splitting function multiplied by the back-reaction factor:
\beq
\frac{d{\cal P}(B^* \to DE)}{dz_{DE}\,dk_{T,DE}^2} \;\times\; \left( \frac{d{\cal P}(A^* \to B^*C)/dz^* dk_T^{2*}}{d{\cal P}(A^* \to BC)/dz\,d\ktsq}  \cdot \left| {\rm det} \left[\frac{dz^* dk_T^{2*}}{dz\,d\ktsq}  \right] \right| \right) \, .  \label{eq:weight}
\eeq
The simplest implementation would compute this factor independently for each daughter branch, assuming an on-shell sister and neglecting possible correlations in the potentially fully off-shell final configuration $A^* \to B^*C^*$. But a more thoroughly correlated weighting scheme could be pursued if deemed numerically relevant. The above prescription also generalizes beyond massive showers, wherein it has a sizable overlap with the effects of standard angular vetoing.
%\footnote{A strictly correct treatment of wide-angle charge/coherence for electroweak emissions would require a detailed accounting of hypercharge and $SU(2)$ dipole matrix factors for all pairs of particles at a given stage of the shower. One could also consider breaking down the event into classical weak isospin flows, analogous to the classical color flows often employed in QCD. But this is not obviously applicable since $SU(2)$ is hardly a ``large-$N$'' gauge theory. We defer detailed investigation of this issue, though we do investigate the effect of supplementing the above weight factor with an angular veto in Section~\ref{sec:implementation}.}
We further show below how back-reaction factors can be conveniently applied for a complete treatment of mixed neutral bosons, wherein an ``on-shell'' kinematic mass is not necessarily determined at their production.

The above back-reaction effects can be particularly important for ultra-collinear emissions, as these occur almost exclusively at the boundaries delineated by finite-mass effects. For example, the prototypical ultra-collinear emission is $f\to W_{\longit} f'$ with massless fermions~\cite{Kane:1984bb,Dawson:1984gx,Chanowitz:1985hj}. It proceeds only via a delicate balancing between a suppression factor $m_W^2/E^2$ in the squared splitting matrix element and a strong $1/k_T^4$ power enhancement from the fermion propagator that gets cut off at $k_T \sim m_W$, controlled by the form of the denominator in Eq.~(\ref{eq:split}). Within a final-state shower, if either the $W_{\longit}$ or its sister $f'$ is set far off-shell by a secondary splitting at some scale $Q$ (possibly a QCD splitting), that cutoff moves out to $k_T \sim Q$ but the original production matrix element stays approximately the same, and the total rate picks up an additional relative power suppression factor of $O(m_W^2/Q^2)$.\footnote{When the $W_L$ is off-shell, we would naively compensate by using an off-shell gauge polarization, yielding $Q^2/E^2$ instead of $m_W^2/E^2$. However, the appropriate treatment, discussed in more detail in Appendices~\ref{sec:gauge} and~\ref{sec:FeynmanRules}, uses on-shell polarization factors throughout. Additional non-collinear corrections might still be present, but are more appropriately viewed as contributions to $1\to 3$ splittings. New soft logarithms might also arise in these processes, but new {\it collinear} logs will not.} Roughly speaking, ultra-collinear processes can only occur near the ``end'' of the weak parton shower as it passes through the weak scale, or conversely near the ``beginning'' of weak PDF evolution. Such a feature is essentially built-into $k_T$-ordered parton evolution. The back-reaction correction ensures that it is also enforced in showers built on other ordering variables, such as virtuality, while still allowing further low-scale showering such as $q\to gq$ and $W_{\longit}\to \gamma W_{\longit}$.

%..................................
\subsubsection{Mixed-state evolution}
\label{sec:interference}
%..................................

Thus far, the shower formalism that we have presented neglects the possibility of interference between different off-shell intermediate particle states contributing to a specific splitting topology. Traditionally in QED and QCD showers, such interference leads to sub-leading effects associated with the unmeasured spin and color of intermediate particles~\cite{Nagy:2007ty}. 
However, the full electroweak theory at high energies presents us with cases where different mass and gauge eigenstates can also interfere at $O(1)$ level, most notably the neutral boson admixtures $\gamma/Z_T$ and $h/Z_L$~\cite{Ciafaloni:2005fm}. All other particles in the SM carry (approximately) conserved charge or flavor quantum numbers that can flow out into the asymptotic state, and therefore they do not tend to interfere in this manner. Interferences originating from CKM/PMNS flavor violations should be small and difficult to observe, and we neglect them for simplicity.

Showering involving superpositions of different particle species can be described using density matrix formalism. Let us consider the simpler case of final-state showers for illustration. The initial value of the density matrix is set proportional to the outer product of production amplitudes: $\rho_{ij} \propto {\mathcal M}_i^{{\rm (prod)}*} {\mathcal M}_j^{\rm (prod)}$, tracing out over other details of the rest of the event.\footnote{This treatment does not attempt to address quantum correlations between different branches of an event or shower.} Here, the indices run over the particle species. The probability for an initial mixed quantum state to subsequently split into a specific exclusive final state must be computed by generalizing the splitting functions to Hermitian splitting {\it matrices} $d{\mathcal P}_{ij}$. The exclusive splitting rates are then computed by tracing against the normalized density matrix,\footnote{In more complete generality, a mixed state can split into another mixed state, leading to an enlarged set of indices for the splitting matrices. However, in most cases, the final-state density matrices are fully determined by the initial-state density matrices, such that in practice a single pair of indices suffices.}
\beq
d{\cal P} \,=\, \frac{\rho_{ij}\ d{\mathcal P}_{ji}}{{\rm tr}[\rho]} \ .
\label{eq:rhoSplitting}
\eeq
Representing the propagator matrix as ${\cal D}_{ij}$, and the amputated splitting amplitudes as ${\cal M}^{\rm (split)}_i$, 
%the generalization from single-state evolution is
%\beq
%d{\cal P} \,\propto\, \frac{1}{q^4}|{\cal M^{\rm (split)}}|^2 \;\,\rightarrow\,\; d{\cal P}_{ij} \,\propto\, {\cal M}^{\rm (split)*}_k {\cal D}_{ki}^* {\cal D}_{jl} {\cal M}^{\rm (split)}_l. \label{eq:splittingMatrix}
%\eeq
this modifies Eq.~(\ref{eq:split}) to the more complete, yet more complicated form
\beq
\left[ \frac{d{\mathcal P}_{A\rightarrow B+C}}{dz\,d\ktsq} \right]_{ij} \,\simeq\,  {1\over 16\pi^2} \ \frac{1}{z\zb} \  {\cal M}^{\rm (split)*}_k {\cal D}_{ki}^* {\cal D}_{jl} {\cal M}^{\rm (split)}_l     \ .
\eeq
Note that large interference effects can persist even in the massless limit with unmixed propagators. A full treatment, including the Sudakov evolution for $\rho_{ij}$ and the explicit form of the propagators for $\gamma/Z_T$ and $h/Z_{\longit}$ systems, is given in Appendix~\ref{app:split}.

Handling the kinematics and decays of mixed states requires some additional steps. ``On-shell'' kinematics cannot be defined a priori, and we cannot collapse onto mass eigenstates or a showered final-state with well-defined mass until the coherent Sudakov evolution has run its course.  A simple prescription is to first produce a mixed boson with its minimum possible kinematic mass (zero for $\gamma/Z_T$, $m_Z$ for $h/Z_{\longit}$) in order to fully fill out the phase space. Splittings that occur before reaching the resonance are weighted by a back-reaction factor as per Eq.~(\ref{eq:weight}). If the state survives un-split down to the heavier resonance's matching threshold, we can decide to project onto a specific mass eigenstate according to the relative probabilities encoded in the surviving density matrix. The back-reaction factor may once again be employed here, implemented as a veto probability for the heavier resonance. (The factor will typically come out less than one for a sensibly-defined change of variables.) If the veto is thrown, the splitting that produced the mixed state is undone, and its mother's evolution continued. This prescription especially becomes relevant when evolving near kinematic thresholds or suppressed regions, for example where $Z$ boson emission would be suppressed but photon emission allowed.

For the mixed $\gamma/Z_T$ system, if a photon is projected out, we can restart a pure QED parton shower ($\gamma\to f\bar f$) with virtuality constrained below the $Z$ boson's $Q_{\rm match}$ scale at $\approx 100$~GeV. Interference effects below the matching scale can also be incorporated by coherently adding both the $\gamma$ and $Z$ contributions within the $Z$ resonance region. This requires delineating as well a lower virtuality boundary, ideally at a scale $O(1)$ smaller than $m_Z$. Depending on the integrated probability in this region (modulo the back-reaction veto), we would either create an $f\bar f$ state with an appropriately-distributed mass, or again set the state to a photon and continue running a pure QED shower, now constrained below the $Z$ resonance region.

We also comment that a fully consistent treatment here would require minor changes to the standard output formats of hard event generators. The standard practice of immediately collapsing onto mass eigenstates is equivalent to assuming trivial Sudakov evolution, and cannot formally be inverted such that a proper coherent parton shower can be applied. In particular, only one specific linear combination of $\gamma/Z_T$ states participates in the high-rate non-Abelian splittings to $W_T^\pm W_T^\mp$. While collapsing onto mass eigenstates is required to obtain well-defined hard event kinematics, a simple remedy here would be to supply for these particles their production density matrices, using some appropriately-mapped massless kinematics.

%%%%%%%%%%%%%%%%%%%%%%
\section{Splitting Functions in Unbroken $SU(2)_L\times U(1)_Y$}
\label{sec:unbroken}
%%%%%%%%%%%%%%%%%%%%%%

Before working out the complete set of electroweak splitting functions in the broken phase, it is important to first consider a conceptual limit with an unbroken $SU(2)_L \times U(1)_Y$ gauge symmetry with massless gauge bosons and fermions, supplemented by a massless complex scalar doublet field $H$ without a VEV. This last ingredient is the would-be Higgs doublet. This simplified treatment in the unbroken phase is not only useful to develop some intuition, but also captures the leading high-$k_T$ collinear splitting behavior of the broken SM electroweak sector. Some aspects of electroweak collinear splitting and evolution at this level have been discussed, e.g., in~\cite{Ciafaloni:2005fm}.

Anticipating electroweak symmetry breaking, we adopt the electric charge basis in weak isospin space. The corresponding $SU(2)_L$ bosons are $W^\pm$ and $W^0$, and the hyper-charge gauge boson we denote as $B^0$. Gauge boson helicities are purely transverse ($T$), and are averaged.\footnote{While the gauge helicity averaging is not strictly necessary, especially given that we will later make a distinction between transverse and longitudinal polarizations, it does simplify our presentation.  We also do not incorporate azimuthal interference effects, though this would be straightforward in analogy with QCD~\cite{Bahr:2008pv}.}  For the scalar doublet, we decompose as
\beq
H=\left({\begin{array}{c} 
 H^+ \\
 H^0  \end{array}}\right) = \left({\begin{array}{c} 
\phi^+ \\
 \frac{1}{\sqrt{2}}(h -  i\phi^0)    \end{array}}\right),  \label{eq:HiggsExpansion}
\eeq
where $\phi^\pm,\phi^0$ will later become the electroweak Goldstone bosons and $h$ the Higgs boson. However, at this stage, we will keep the neutral bosons $h$ and $\phi^0$ bundled into the complex scalar field $H^0$, as they are produced and showered together coherently. In the absence of the VEV, the doublet carries a perturbatively-conserved ``Higgs number,'' which may also be taken to flow through RH-chiral fermions in the Yukawa interactions.\footnote{We have expanded the neutral scalar field as $H^0 \propto h - i\phi^0$, adopting a phase convention such that $h$ and $\phi^0$ fields create/annihilate their respective one-particle states with trivial phases, and $H^0$ annihilates the one-particle state $\ket{H^0} \propto \ket{h} + i\ket{\phi^0}$. Treating $h$ and $\phi^0$ as independent showering particles would be analogous to adopting a Majorana basis instead of a Dirac basis for the fermions in QED or QCD. An incoherent parton shower set up in such a basis would not properly model the flow of fermion number and electric charge. Analogously, $H^0$ and $H^{0*}$ particles carry well-defined Higgs number that we choose to explicitly track through the shower. This leads to correlations between spins and electric charges within asymptotic states.} We denote a generic fermion of a given helicity by $f_s$ with $s=L,R$ (or equivalently $s=\mp$). We do not always specify the explicit isospin components of $f$ at this stage, but implicitly work in the usual $(u,d)$/$(\nu,e)$ basis. Isospin-flips (including RH-chiral isospin where appropriate) will be indicated by a prime, e.g.~$u' = d$. Effects of flavor mixing are ignored.

The $U(1)_Y$ and $SU(2)_L$ gauge couplings are respectively taken to be $g_1 \approx 0.36$ and $g_2 \approx 0.65$ (here evaluated near the weak scale, though in general run to a scale of $\order(k_T$)). For compactness we often represent a generic gauge coupling by $\gv$. We represent the gauge charge $Q$ of a particle $p$ coupling to gauge boson $V$ by $Q^V_p$, and we give the complete list of the gauge charges for the SM fermions and scalars in Table \ref{tab:charges} in Appendix~\ref{sec:conventions}. 
%e.g.: $Q^{B}_{u_R} = Y_{u_R} = 2/3$, $Q^{W^0}_{e_L} = T^3_{e_L} = -1/2$, $Q^{W^\pm}_{H^0} = Q^{W^\pm}_{\phi^\pm} = 1/\sqrt{2}$, etc. Yukawa couplings for fermions $f$ are defined as $y_f$, which would be related to the fermion masses by $\sqrt{2}m_f/v$ after the electroweak symmetry breaking (taking $v \approx 246$~GeV). A complete listing for our coupling conventions is provided in Appendix~\ref{sec:conventions}.

The splitting functions that involve only fermions and gauge bosons closely follow those of QED and QCD. Fermions with appropriate quantum numbers may emit transverse $SU(2)_L$ and $U(1)_Y$ gauge bosons with both soft and collinear enhancements, yielding total rates that grow double-logarithmically with energy. At this stage, fermion helicity coincides with the corresponding chirality, and is strictly conserved in these processes. The $SU(2)_L$ bosons also couple to one another via their non-Abelian gauge interactions, and similarly undergo double-logarithmic soft and collinear splittings $W^0\to W^+W^-$ and $W^\pm \to W^\pm W^0$. This is in direct analogy to $g\to gg$ in QCD, except that here we do not sum/average over gauge indices. All of the electroweak gauge bosons may also undergo single-log collinear splittings into fermion pairs, similar to $g\to q\bar q$ or $\gamma \to f \bar f$. 

The results can be cast into a familiar form. We write the probability function of finding a parton $B$ inside a parton $A$ with an energy or momentum fraction $z$ in terms of the collinear splitting kernels for $A\to B$ as $P_{BA}(z)$. Stripping the common $g^2/8\pi^2$ and $1/k_T^2$ factors, as well as group theory factors that depend on the gauge representations (hyper-charges or $SU(2)_L$ quadratic Casimirs and Dynkin indices), we are left with
\be
P_{Vf}(z) = {1+\bar z^2 \over z},\quad 
%P_{f'f}(z) = {1+ z^2 \over \bar z},\quad 
P_{V'V}(z) = {(1- z \bar z)^2\over z \bar z},\quad
P_{fV}(z) = {z^2+\bar z^2\over 2},~~
\label{eq:qcd}
\ee
with $\bar z \equiv 1-z$. Note that the other possible splitting $f\to f^{(\prime)}V$ is given by $P_{f^{(\prime)}f}(z) = {(1+ z^2)/ \bar z}$, but it is not independent and can be derived from $P_{Vf}$ with $z \leftrightarrow \bar z$. The factor of $1/2$ in $P_{fV}$, relative to the standard form in QED with the electric charge stripped (or in QCD with the $SU(3)$ Dynkin index stripped), is due to the fact that we treat each chiral fermion individually.

Interference between different gauge groups is a subtlety that is absent in the color-averaged $SU(3)_{\rm QCD} \times U(1)_{\rm EM}$ shower, and arises here from the fact that we have fixed a preferred gauge basis for asymptotic states instead of summing over gauge indices. Within different exclusive isospin channels in this basis, exchanges of $B^0$ and $W^0$ can exhibit $O(1)$ interference, and thus must be described using density matrices, which have briefly been discussed in Section~\ref{sec:interference}. In a truly massless theory, the physical preparation and identification of states in any preferred weak isospin basis is actually impossible, since arbitrarily soft $W^\pm$ can be radiated copiously at no energy cost and randomize the isospin.\footnote{Absent the quark chiral condensate at $O(100$~MeV), massless $SU(2)_L$ would also technically confine in the IR, so that asymptotic states would anyway be isospin-singlet bound states, making the situation even more analogous to QCD.} Our preferred basis here only becomes physical once we turn on the electroweak VEV and cut off the IR divergences. But the tendency for states to self-average in isospin space will persist at high energies.

%%%% Fermion splittings tables
\bgroup

\def\arraystretch{1.4}

\begin{table}[]
\centering
\vspace{-1cm}
\begin{tabular}{l|ccc}
\multicolumn{1}{l}{}
 & \multicolumn{2}{c}{
   \begin{picture}(50,40)(0,0)
     \SetColor{Black} \SetWidth{2} \SetScale{\figscale}
     \ArrowLine(  0,25)(50,25)
     \Photon(50,25)(95,45){3}{3}
     \ArrowLine( 50,25)(95,5)
     \Text(15,10)[]{$\boldsymbol \Leftarrow$}
     \Text(50,3)[]{\rotatebox{-26}{$\boldsymbol \Leftarrow$}}
   \end{picture} }
 & \begin{picture}(50,40)(0,0)
     \SetColor{Black} \SetWidth{2} \SetScale{\figscale}
     \ArrowLine(  0,25)(50,25)
     \DashLine(50,25)(95,45){5}
     \ArrowLine( 50,25)(95,5)
     \Text(15,10)[]{$\boldsymbol \Leftarrow$}
     \Text(50,3)[]{\rotatebox{-26}{$\boldsymbol \Rightarrow$}}
   \end{picture}
\\
%\multicolumn{1}{l}{}
    & \multicolumn{2}{c}{\rule[-3ex]{0pt}{0pt} \underline{\hspace{0.4cm}$\dfrac{1}{8\pi^2}\dfrac{1}{\ktsq}\left(\dfrac{1+\bar z^2}{z}\right)$\hspace{0.4cm}}}
    &  \underline{\hspace{0.5cm}$\dfrac{1}{8\pi^2}\dfrac{1}{\ktsq}\left(\dfrac{z}{2}\right)$\hspace{0.5cm}}
\\ 
%\backslashbox{$A$}{$\!\!\!\to BC$}
    &  $\to \ V_T \: f_{s}^{(\prime)}$  
    &  $[BW]_T^0 \: f_{s}$  
    &  $H^{0(*)} \: f_{\minus s}\ 
    %\overset{\rm or}
  {\rm or}  \ \ \phi^{\pm} \: f^{\prime}_{\minus s}$ 
\\  \hline%\hline
$f_{s=L,R}$\ 
    &  $\ \ g^2_V(Q^V_{f_s})^2$ 
    &  $g_1g_2Y_{f_s}T^3_{f_s}$ 
    &  $y^2_{f^{(\prime)}_R}$ 
\end{tabular}
\caption{Chiral fermion splitting functions $d{\mathcal P}/dz\,d\ktsq$ in the massless limit, with $z$ ($\bar z \equiv 1-z$) labeling the energy fraction of the first (second) produced particle. The fermion helicity is labelled by~$s$. Double-arrows in Feynman diagrams indicate example fermion helicity directions. Prime indicates isospin partner ($u_s' = d_s$, etc, independent of~$s$). Yukawa couplings are labelled by the participating RH-helicity fermion. The state $H^{0*}$ is the ``anti-$H^0$'', produced when the RH fermion is down-type and in the initial-state, or up-type in the final-state. Processes with $B^0$ and $W^0$ implicitly represent the respective diagonal terms in the neutral gauge boson's density matrix, whereas $[BW]^0$ indicates either of the off-diagonal terms (see text). Anti-fermion splittings are obtained by CP conjugation. The conventions for the couplings are given in \ref{sec:conventions}.
%Table \ref{tab:charges} in the appendix. 
}
\label{tab:massless_fermion_splittings}
%\end{table}
%\egroup

\vspace{0.5cm}

%%%% Transverse vector splittings
%\bgroup
%\def\arraystretch{1.5}
%\begin{table}[t]
%\centering
\begin{tabular}{l|cccc}
\multicolumn{1}{l}{}
 & \begin{picture}(50,40)(0,0)
     \SetColor{Black} \SetWidth{2} \SetScale{\figscale}
     \Photon(  0,25)(50,25){3}{3}
     \Photon(50,25)(95,45){3}{3}
     \Photon( 50,25)(95,5){3}{3}
   \end{picture}
 & \begin{picture}(50,40)(0,0)
     \SetColor{Black} \SetWidth{2} \SetScale{\figscale}
     \Photon(  0,25)(50,25){3}{3}
     \ArrowLine(50,25)(95,45)
     \ArrowLine(95,5)( 50,25)
     \Text(50,32)[]{\rotatebox{26}{$\boldsymbol \Leftarrow$}}
     \Text(50,3)[]{\rotatebox{-26}{$\boldsymbol \Rightarrow$}}
   \end{picture}
 & \multicolumn{2}{c}{
   \begin{picture}(50,40)(0,0)
     \SetColor{Black} \SetWidth{2} \SetScale{\figscale}
     \Photon(  0,25)(50,25){3}{3}
     \DashLine(50,25)(95,45){5}
     \DashLine(95,5)( 50,25){5}
   \end{picture} }
\\
%\multicolumn{1}{l}{}
 & \rule[-3ex]{0pt}{0pt} \underline{\hspace{0.0cm}$\dfrac{1}{8\pi^2}\dfrac{1}{\ktsq}\left(\dfrac{(1- z \bar z)^2}{z \bar z}\right)$\hspace{0.0cm}} 
 & \underline{\hspace{0.0cm}$\dfrac{1}{8\pi^2}\dfrac{1}{\ktsq}\left(\dfrac{z^2+\bar z^2}{2}\right)$\hspace{0.0cm}}
 & \multicolumn{2}{c}{\underline{\hspace{2.0cm}$\dfrac{1}{8\pi^2}\dfrac{1}{\ktsq}\left(z\zb\right)$\hspace{2.0cm}}}
\\ 
%\backslashbox{$A$}{$\!\!\!\to BC$}
    &  $\to \  W_T \: W_T$ \ 
    &  $f_s \: \bar f_{\minus s}^{(\prime)}$ 
    &  $\phi^+ \: \phi^- \ \overset{\rm or}\  \ H^0 \: H^{0*}$ 
    &  $\phi^+ \: H^{0*}\ \overset{\rm or}\  \ \phi^-\: H^0$  
\\  \hline%\hline
$V_T$\  
    &  $2 g_2^2\ (V\!\!=\!W^{0,\pm})$ 
    &  $N_f g_V^2(Q^V_{f_s})^2$
    &  $\frac14 g_V^2$ 
    &  $\frac12 g_2^2$ 
\\ 
$[BW]_T^0$\    
    &  $0$ 
    &  $N_f g_1g_2Y_{f_s}T^3_{f_s}$ 
    &  $\frac12 g_1g_2T^3_{\phi^+,H^0}$
    &  $0$ 
\end{tabular}
\caption{Transverse vector boson splitting functions $d{\mathcal P}/dz\,d\ktsq$ in the massless limit, where allowed by electric charge flow.  $N_f$ is a color multiplicity factor ($N_f=1$ for leptons, $N_f=3$ for quarks). Other conventions as in Table~\ref{tab:massless_fermion_splittings}.}
\label{tab:massless_vector_splittings}
%\end{table}
%\egroup

\vspace{0.5cm}

%%%% Scalar splittings
%\bgroup
%\def\arraystretch{1.3}
%\begin{table}[t]
%\centering
\begin{tabular}{l|ccccc}
\multicolumn{1}{l}{}
 & \multicolumn{3}{c}{
   \begin{picture}(50,40)(0,0)
     \SetColor{Black} \SetWidth{2} \SetScale{\figscale}
     \DashLine(  0,25)(50,25){6}
     \Photon(50,25)(95,45){3}{3}
     \DashLine( 50,25)(95,5){5}
   \end{picture} } 
 & \multicolumn{2}{c}{
   \begin{picture}(50,40)(0,0) \SetScale{\figscale}
     \SetColor{Black} \SetWidth{2}
     \DashLine(  0,25)(50,25){6}
     \ArrowLine(50,25)(95,45)
     \ArrowLine(95,5)( 50,25)
     \Text(50,32)[]{\rotatebox{26}{$\boldsymbol \Leftarrow$}}
     \Text(50,3)[]{\rotatebox{-26}{$\boldsymbol \Leftarrow$}}
   \end{picture}}
\\
%\multicolumn{1}{l}{}
    & \multicolumn{3}{c}{\rule[-3ex]{0pt}{0pt} \underline{\hspace{1.8cm}$\dfrac{1}{8\pi^2}\dfrac{1}{\ktsq}\left(\dfrac{2 \bar z}{z}\right)$\hspace{1.8cm}}}
    & \multicolumn{2}{c}{\underline{\hspace{1.0cm}$\dfrac{1}{8\pi^2}\dfrac{1}{\ktsq}\left(\dfrac12\right)$\hspace{1.0cm}}}   
\\ 
    &  \  $\rightarrow \ V_T^0 \, H$ \ 
    & \ $[BW]_T^0 \: H$ \ 
    & \ $W_T^\pm \: H'$ \ 
    & \ $u_R \: \bar u_R^{(\prime)}$ 
    &  $\bar d_L \: d_L^{(\prime)} \ \overset{\rm or}\  \ \bar e_L \: e_L^{(\prime)}$  
\\  \hline%\hline
$H=\phi^+,H^{0}$\   
    &  $\frac14 g_V^2$
    &  $\frac12 g_1g_2T^3_{\phi^+,H^0}$
    &  $\frac12 g_2^2$
    &  $3y^2_{u}$
    &  $N_{d,e}y^2_{d,e}$ 
\end{tabular}
\caption{Scalar splitting functions $d{\mathcal P}/dz\,d\ktsq$ in the massless limit via gauge couplings and Yukawa couplings. The symbol $H$ in the column headings represents the appropriate state $\phi^+,H^{0}$ for the given splitting, and $H'$ represents the $SU(2)_L$ isospin partner (e.g., $H^{0\prime} = \phi^+$). Anti-particle splittings are obtained by CP conjugation.  Other conventions as in Tables~\ref{tab:massless_fermion_splittings} and~\ref{tab:massless_vector_splittings}.} 
\label{tab:massless_scalar_splittings}
 \end{table}

\egroup

Beyond these, the major change is the introduction of the scalar doublet.\footnote{We neglect all $1\to 3$ splittings coming from either the scalar quartic or the scalar-gauge 4-point. These may feature single-logarithmic collinear divergences, but are expected to be rather highly numerically suppressed due to an additional $O(1/16\pi^2)$ phase space factor.}  First, the scalars may themselves radiate $SU(2)_L$ and $U(1)_Y$ gauge bosons. The soft-collinear behavior is identical to their fermionic counterparts, but the hard-collinear behavior is different.  Second, the electroweak gauge bosons can split into a pair of scalars, again in close analog with splittings to fermion pairs.  Third, fermions with appreciable Yukawa couplings to the scalar doublet can emit a scalar and undergo a helicity flip. Finally, the scalars can split into a pair of collinear, opposite-chirality (same-helicity) fermions. The corresponding splitting function kernels are found to be
\be
P_{H f}(z) = {z \over 2},\quad 
P_{H V}(z) = z \bar z,\quad 
P_{V H}(z) = {2 \bar z \over z},\quad
P_{f H}(z) = {1\over 2}.
%P_{\phi' \phi}(z) = {2 z \over \bar z}.
\label{eq:EW}
\ee
The other possible splittings $H \to H^{(\prime)} V$ and $f_s \to f_{\minus s}^{(\prime)} H$ are given by $P_{H^{(\prime)} H}(z) = 2z/ \bar z$ and $P_{f_{\minus s}^{(\prime)} f_s}(z) = \bar z/2$, derived from $P_{V H}$ and $P_{H f}$, respectively.\footnote{Note that transitions involving the scalars must conserve the Higgs number introduced earlier in this section. For example, we may have $H^0 \to W^-\phi^+$, but not $H^0 \to W^+\phi^-$. Similarly, $H^0 \to t_R \bar t_R$ is allowed but $H^0 \to t_L \bar t_L$ is not.} The splittings $W^0/B^0 \to H^0 H^{0(*)}$ can also be conveniently represented by the final-state $h\phi^0$, in what will ultimately become $hZ_{\longit}$ in mass/CP basis. Here the final-state bosons are entangled, but the effects of that entanglement are subtle and only become relevant if {\it both} bosons undergo secondary splittings and/or hard interactions. In practice, we will simply take the expedient of collapsing the final state to $h\phi^0$.

The complete set of splitting functions is summarized in Tables~\ref{tab:massless_fermion_splittings} through~\ref{tab:massless_scalar_splittings}.  The tables are organized according to the spin of the incoming particles: polarized fermions with helicity $s$, transverse gauge bosons ($V_T$), and scalars. Each table is further subdivided according to the spins of outgoing particles, all together corresponding to seven unique core splitting functions. The various table entries associated to a specific set of incoming and outgoing spins provide the remaining coupling and group theory factors.  All of the splitting functions have a conventional collinear logarithmic enhancement $d\ktsq/\ktsq$, and those involving emission of a massless gauge boson have an additional soft logarithmic enhancement $dz/z$. (The latter are the only emissions that preserve the leading particle's helicity in the soft emission limit.) To represent the off-diagonal terms for the neutral gauge bosons (either in production or splitting, where appropriate), we use the symbol $[BW]^0$. Otherwise, processes involving $B^0$ or $W^0$ alone implicitly represent the respective diagonal term in the density matrix.

%%%%%%%%%%%%%%%%%%%%%%%%%%%%%%%%%%%%%%%%%%%%%%%%%%%%%%%%%%%%%%%
\section{Splitting Functions in Spontaneously Broken $SU(2)_L\times U(1)_Y$}
\label{sec:broken}
%%%%%%%%%%%%%%%%%%%%%%%%%%%%%%%%%%%%%%%%%%%%%%%%%%%%%%%%%%%%%%%

While the parton shower formalism of the electroweak theory in the symmetric phase has much in common with that of $SU(3)_{\rm QCD}\times U(1)_{\rm EM}$, care needs to be taken when dealing with the broken phase and systematically accounting for the effects of the VEV ($v$). In a sense, we must extract the ``higher-twist'' effects of the broken electroweak theory in terms of powers of $v/E$. Although the regulating role of $v$ in the shower is somewhat analogous to that of $\Lambda_{QCD}$, the electroweak theory remains perturbative at $v$, and the unbroken QED shower continues into the deep infrared regime. The interplay between gauge and Goldstone degrees of freedom within the shower can also seem obscure, both technically and conceptually.

\begin{figure}[t]
\begin{center}
\begin{subfigure}[t]{0.495\textwidth}
\includegraphics[width=200pt]{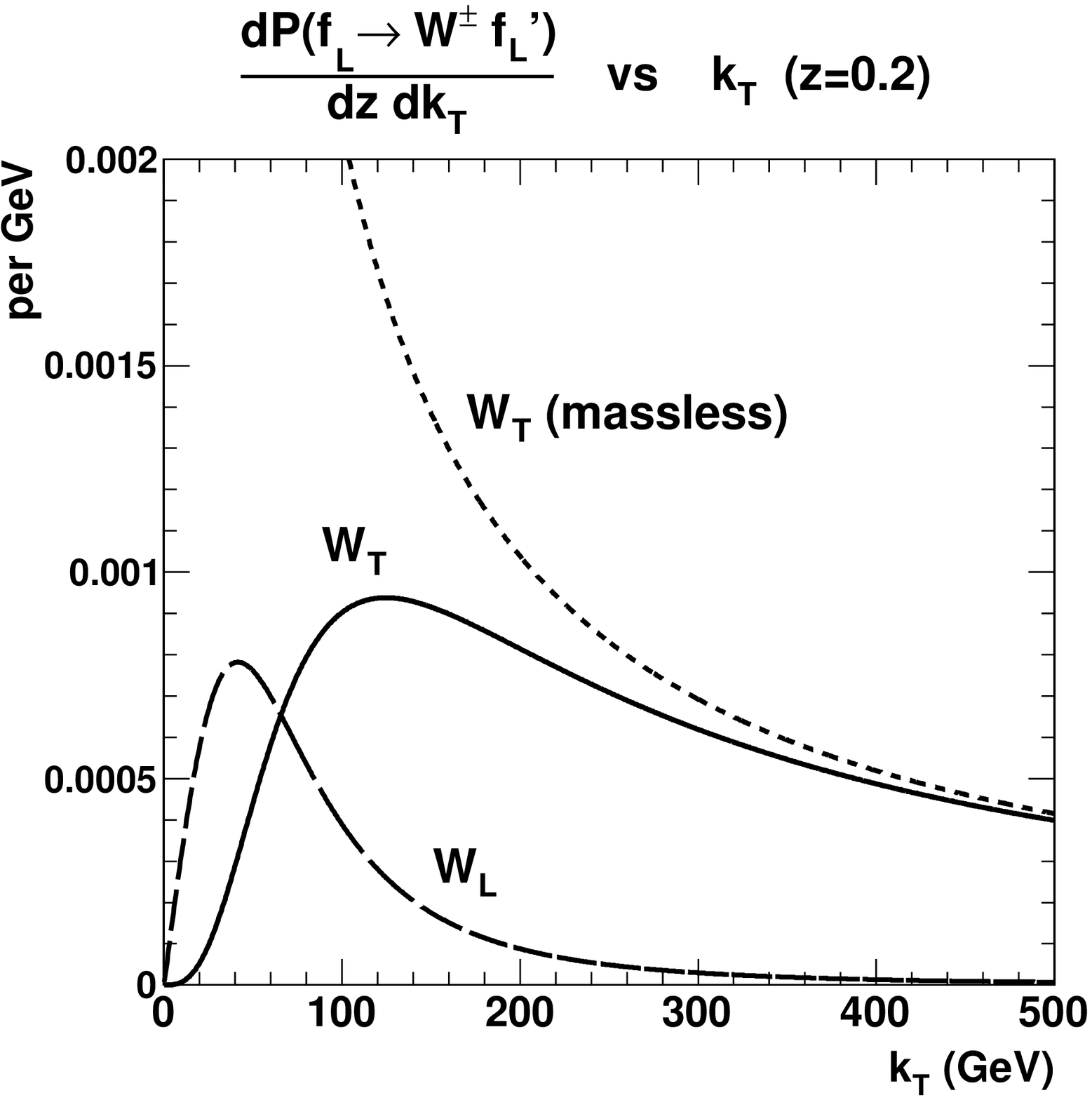}
\vspace{-0.4cm}
\caption{}
\end{subfigure}
\begin{subfigure}[t]{0.495\textwidth}
\includegraphics[width=200pt]{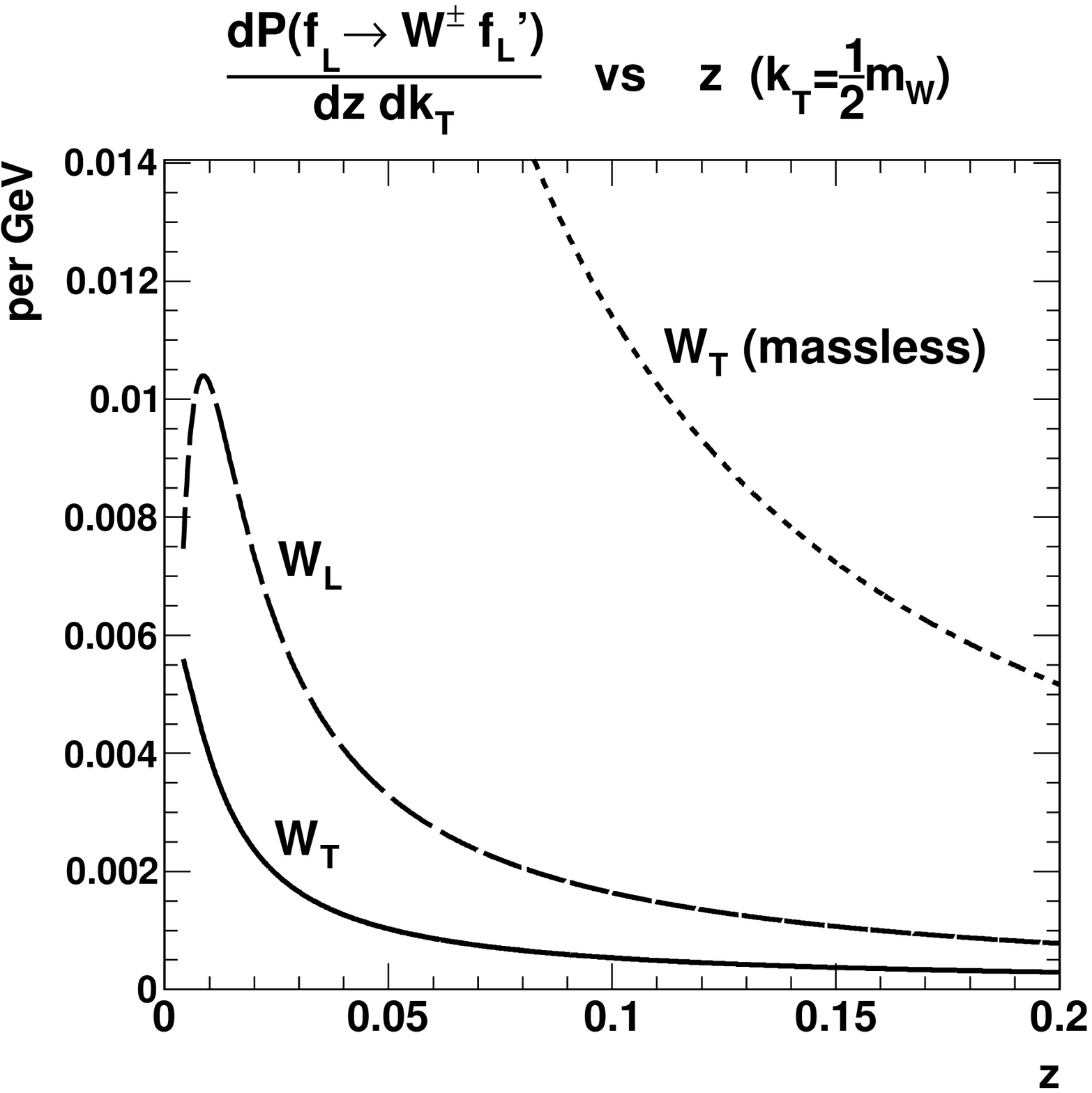}
\vspace{-0.4cm}
\caption{}
\end{subfigure}
\vspace{-0.8cm}
\end{center}
\caption[]{Fixed-order differential emission rate for $W^\pm$ bosons off a massless fermion at $E_f = 10$~TeV: (a) $k_T$ distribution at $z=0.2$, (b) $z$ distribution at $k_T = m_W/2$. The different curves correspond to massless transversely-polarized $W^\pm_T$ (dotted curves), massive transversely-polarized $W^\pm_T$ (solid curves), and massive longitudinally-polarized $W^\pm_L$ (dashed curves).
}
\label{fig:zkt}
\end{figure}

Most immediately, the splitting functions of the unbroken theory, already detailed in Section~\ref{sec:unbroken}, must be adjusted to account for the physical masses of the gauge bosons, Higgs boson, and top quark. To large extent, these constitute simple modifications, folding in the kinematic effects discussed in Section~\ref{sec:split}. As a straightforward example, in Fig.~\ref{fig:zkt} we illustrate the fixed-order emission rate for $W^\pm$ bosons off a massless fermion at $E_f = 10$~TeV.  Both the collinear and soft singularities of the massless theory (dotted curves) become regulated with $m_W \approx 80$~GeV (solid curves), as seen in the transversely-polarized boson $k_T$ distribution in Fig.~\ref{fig:zkt}(a) and the $z$ distribution in Fig.~\ref{fig:zkt}(b).\footnote{Note that in the region $z \lsim m_W/E$, the $W$s are non-relativistic, and collinear splitting function language ceases to be strictly appropriate or reliable. This region could more rigorously be matched onto universal soft Eikonal factors, e.g. as in~\cite{Denner:2000jv,Denner:2001gw}. But in practice, our treatment here still yields approximately correct rates for splitting angles $\lsim 1$ when the splitting is defined in the hard scatter frame.} Indeed, giving the gauge bosons a mass is a common trick for regulating QCD and QED calculations. In the electroweak theory, such regulated splitting functions become physically meaningful.

Figure~\ref{fig:zkt} also shows a contribution from longitudinal gauge boson radiation off of a massless fermion (dashed curves). This is a good example of an ``ultra-collinear'' process which emerges after EWSB at leading power in $v/E$. In this case it has a splitting probability of the form
\begin{equation}
d{\cal P} \sim {m^2_W\over k_T^2} {d\ktsq \over k_T^2} \ .
\label{eq:ultra}
\end{equation}
The rate is seen to be significant in the region $k_T \sim m_W$, and it can be larger than the conventional transverse emissions in the ultra-collinear region $k_T \lsim m_W$ as seen in Fig.~\ref{fig:zkt}(a). 
We further show in Fig.~\ref{fig:zkt}(b) the $z$ distribution at $k_T = m_W/2$, where we can see the dominance of the longitudinal polarization (dashed curve) over the transverse polarization (solid curve) for all values of $z$ at weak-scale values of $k_T$.
%The cut-off at low values generally depend on gauge and kinematic conventions in the $1\to 2$ splitting approximation. 
Here we have defined $z$ as three-momentum fraction, employed a strict kinematic cut-off $z > k_T/E$, and multiplied the splitting rate by the $W$ velocity to account for non-relativistic phase space suppression.

Considering emissions from light initial-state fermions, the ultra-collinear origins of these longitudinal weak bosons leads to quite distinctive PDFs~\cite{Kane:1984bb,Dawson:1984gx,Chanowitz:1985hj}. Due to the existence of an explicit mass scale $\mw \sim gv$, the resulting PDFs exhibit Bjorken scaling~\cite{Bj}. In other words, they do not run logarithmically and do not exhibit the usual scaling violations of conventional PDFs in massless gauge theories. Consequently, the ISR jets associated with their generation are constrained to the region $k_T \sim m_W$ even for arbitrarily-energetic hard processes. This observation has led to the concepts of ``forward-jet tagging''~\cite{Cahn:1986zv,Barger:1988mr,Kleiss:1987cj} for the $W_{\longit}W_{\longit}$ scattering signal and ``central-jet vetoing''~\cite{Barger:1990py} for separating the $f\to W_T f'$ backgrounds.

Such processes have no analogs in the unbroken theory. A naive application of the Goldstone-boson Equivalence Theorem (GET) \cite{Lee:1977eg,Chanowitz:1985hj} would have instructed us to identify longitudinal vector bosons with the eaten scalars from the Higgs doublet, and would have predicted zero rate because massless fermions have vanishing Yukawa couplings. More generally, we expect to see a variety of large effects of EWSB at $k_T \sim v$, beyond simple regulation of the unbroken-theory splitting functions. These will involve not only the broken-phase masses of the SM particles, but also broken-phase interactions such as scalar-vector-vector and the scalar cubics.

The more general role of Goldstone boson equivalence and its violations within the parton shower are rather subtle. We expect that the high-$k_T$ showering of longitudinal gauge bosons should closely follow the behavior of the scalars in the unbroken theory. But even this simple identification is obscured by longitudinal polarizations that diverge with energy and by the gauge/Goldstone boson propagators with gauge-dependent tensor and pole structure. For processes with multiple emissions, as well as with the introduction of the novel ultra-collinear emissions, complete isolation and removal of non-collinear gauge artifacts can appear rather complicated. We are thus compelled to seek out a more efficient treatment, such that the bad high energy behavior of the longitudinal gauge bosons is alleviated and the key features of EWSB are made more transparent.

%------------------------------------------------------------------------------
\subsection{Longitudinal gauge bosons and Goldstone Boson Equivalence}
\label{sec:GEG}
%------------------------------------------------------------------------------
%In the traditional $R_\xi$ gauges, 
The standard form for the polarization vector of an on-shell longitudinal gauge boson $W$ with a four-momentum $k_W^\mu = E_W(1,\beta_W \hat k_W)$ is 
\beq
\epsilon_{\longit}^\mu(W) \,=\, {E_W\over m_W} \left(\beta_W,\, \hat k_W\right) 
\,=\, {k^\mu_W\over m_W} - \frac{m_W} {E_W(1+\beta_W)}n^{\mu},
\label{eq:scalar}
\eeq
where we define the light-like four-vector
\beq
n^{\mu}\equiv(1,-\hat{k}_W) \, .
\label{eq:nmu}
\eeq
The second term in Eq.~(\ref{eq:scalar}) is of the order $m_W/E_W$, which could seemingly be ignored at very high energies in accordance with the GET.
%, when $E_W \gg m_W$ the longitudinal states should behave like the pseudo-scalars $\phi$ of the unbroken theory. 
However, there are caveats to this picture, and understanding how pseudo-scalars and longitudinal vector bosons behave as both external and intermediate states requires some care.

In the simplest approach, one would keep only the leading contribution, $k^\mu_W/\mw $. When contracted into scattering amplitudes, this piece effectively ``scalarizes'' the longitudinal vector boson, realizing the GET. This can often be seen at the level of individual Feynman diagrams. For example, in the decay of a heavy Higgs boson with $m_h \gg 2 \mw$, the vertex $g\, \mw hW^\mu W_\mu$ simply leads to a scalar interaction $(m_h^2 /v)h \phi^+ \phi^-$ after the substitution $\epsilon_{\longit}^\mu(W) \to {k^\mu_W/ \mw}$. In other cases, such as in couplings to fermion lines, the naively bad high-energy behavior $\propto E_W/\mw$ is fully cancelled thanks to Ward identities, up to possible chirality-flip effects that go like $m_f/E_W$. This reproduces the Yukawa couplings of the unbroken theory. When longitudinal and Goldstone bosons appear as off-shell intermediate states, it is also possible to show that neither the naively badly-behaved structure $k^\mu k^\nu/m_W^2$ (in unitarity gauge) nor spurious gauge/Goldstone poles (in more general gauges) can lead to new collinear behavior at zeroth-order in the VEV. The unbroken shower emerges as expected as long as $k_T \gg m_W$.

The major complication to the GET picture is that the naively sub-leading effects from EWSB can dominate in the relativistic ultra-collinear regime. Even if the $k^\mu_W/\mw $ piece of an emitted gauge boson is removed by Ward identities, the ${\cal O}(m_W/E_W)$ remainder of $\epsilon_{\longit}^\mu(W)$ can still receive a compensating ultra-collinear power-enhancement in the region $k_T \sim \mw$. There may also be comparable EWSB contributions lurking within off-shell propagators, including as well the propagators of Higgs bosons and massive fermions.

Disentangling all EWSB effects in an ultra-collinear parton splitting can be accomplished by isolating and removing all parts of a $1\to 2$ splitting amplitude that go like $(Q^2-m^2)/m_W^2$, where $Q^2$ and $m^2$ are respectively the squares of the four-momentum and pole mass of the off-shell particle in the splitting. Once multiplied by the propagators, such contributions are explicitly not collinear-enhanced, and would need to be combined with other non-collinear (and hence non-universal) diagrams from a hard process. Their extraction can generally be accomplished via manipulations between kinematic quantities, polarization vectors, and couplings. However, carrying out this extraction procedure process-by-process can be tedious, especially when multiple gauge bosons and/or nested collinear emissions are involved, and the effects of EWSB are often not immediately obvious. Within the gauge/Goldstone boson sector, we expect that the $k^\mu_W/\mw$ piece of the longitudinal polarization vector must generally reproduce the Goldstone scalar couplings, whereas the effects of EWSB are captured by the remainder term in Eq.~(\ref{eq:scalar}). A more convenient approach for tracking EWSB effects would be to keep the Goldstone scalar contributions manifest, and treat the remainder polarization as a separate entity.

We point out that such a division can be enforced by judicious gauge-fixing. We do so here via a novel gauge which we call {\it Goldstone Equivalence Gauge} (GEG). GEG is defined by generalizing off-shell the light-like four-vector $n^\mu$ that appears in Eq.~(\ref{eq:scalar}) and using it to perform the gauge-fixing in momentum-space. Taking $W_\mu$ to represent any specific real gauge adjoint, with contraction of gauge indices left implicit, we adopt the gauge-fixing term (dropping here and below the ``$W$'' subscript on energy/momentum variables)
\beq
{\cal L}_{\rm fix} \,=\,  -{1\over 2\xi} \big(n^\mu(k)\,W_\mu(k)\big)\big( n^\nu(k)\,W_\nu(-k)\big), \quad\quad (\xi \to 0) \ .
\label{eq:gauge}
\eeq
Taking the $\xi\to 0$ limit effectively introduces an infinite mass term for the gauge polarization associated with the collinear light-like direction $\bar n^\mu \equiv (1,\hat k)$, aligned with the large components of relativistic momentum modes. This reduces the naive number of dynamical gauge degrees of freedom from four to three. The transverse modes ($xy$ or helicity $\pm1$) are as usual, except that they gain a mass term after spontaneous symmetry breaking. The remaining gauge degree of freedom ``$W_n$" explicitly mixes into the Goldstone boson, and becomes associated with exactly the remainder polarization in Eq.~(\ref{eq:scalar}).

GEG is essentially a hybrid of Coulomb gauge \cite{Beenakker:2001kf} and light-cone gauge \cite{Srivastava:2002mw}, incorporating both the rotational-invariance of the former and the collinear boost-invariance of the latter, while isolating spurious gauge poles/discontinuities away from physical regions.\footnote{GEG falls into a more general class of non-covariant but physical gauges that exhibit many similar features in the broken phase. These include Coulomb~\cite{Beenakker:2001kf}, axial~\cite{Dams:2004vi}, and strict light-cone~\cite{Srivastava:2002mw} (as well as temporal, which has received little attention). In particular, splitting functions computed within GEG and Coulomb gauge should agree at high energies, but the latter can exhibit artificial singularities at zero three-momentum due to the residual gauge freedom.} This approach can be contrasted with the more commonly-used $R_\xi$ gauges, in which individual splitting diagrams often exhibit unphysical gauge artifacts scaling as $1/v$, Goldstone fields live purely off-shell, and Goldstone equivalence can become obscured. 

Canonically normalizing such that the gauge remainder field $W_n$ interpolates a longitudinal boson state with unit amplitude at tree level, its interaction vertices carry the polarization factor
\beq
\epsilon_{n}^\mu(k) \,\equiv\, \frac{-\sqrt{|k^2|}}{n(k)\cdot k}\ n^\mu(k) \,\,\overset{\overset{\text{\rm \footnotesize on-shell}}{}}{\to}\,\, \frac{m_W}{E+|\vec k|}\left(-1,\hat k\right).
\label{eq:L}
\eeq
The Goldstone field remains an integral part of the description here, but in a manner quite different from that in $R_\xi$ gauges. In particular, it interpolates onto the {\it same} external particle as the remainder gauge field. This particle, which may alternately be viewed as a ``longitudinal gauge boson'' or as a ``Goldstone boson'', takes on a kind of dual identity in interactions. Processes involving creation/annihilation of this particle are computed by coherently summing over Feynman diagrams interpolated by both remainder gauge fields and Goldstone fields.\footnote{For a different but related approach, see~\cite{Wulzer:2013mza}.}

More details and example calculations are presented in Appendices~\ref{sec:gauge} and~\ref{sec:FeynmanRules}. However, we can summarize here the key features of GEG that are relevant for parton shower physics:
\begin{itemize}
\item
Gauge artifacts proportional to $E/m_W$ are deleted from the description of the theory at the outset, and appear neither in external polarizations nor in propagators. Physical longitudinal gauge bosons are no longer interpolated by a gauge boson field $W_{\longit}$ and its associated ${\cal O}(E/m_W)$ polarization vector $\epsilon_{\longit}^\mu$, and no propagating component of the gauge field serves a proxy for the eaten Goldstone bosons in high-energy interactions via ``scalarization.'' Instead, only a remainder gauge field $W_n$ may still interpolate longitudinal gauge bosons. But it does so via the suppressed ${\cal O}(m_W/E)$ polarization vector $\epsilon_n^\mu$ in Eq.~(\ref{eq:L}).
\item
The high-energy equivalence between longitudinal gauge bosons and Goldstone bosons becomes trivially manifest at the level of individual Feynman diagrams. This is because the Goldstone fields behave almost identically as in the unbroken theory at high energies ($v/ E\to 0$). The equivalence extends off-shell, encountering neither the usual fake gauge nor Goldstone poles. All propagators exhibit the physical pole at $\mw$ or $m_Z^{}$ with positive residue. This greatly simplifies the interpretation of an ``almost on-shell'' boson as an intermediate state in a shower. 
\item 
Departures from Goldstone boson equivalence become organized in a systematic power expansion in $v/E$ factors. This allows general ultra-collinear splitting processes to be viewed as simple sums of well-behaved $1\to2$ Feynman diagrams. EWSB contributions in splitting matrix elements can come from remainder-longitudinal gauge insertions, fermion mass terms in spinor polarizations, and a small set of standard EWSB three-point vertices.
\end{itemize}

As a final remark of this section, we would like to point out that the GET has been shown to be valid including radiative corrections \cite{Yao:1988aj,Bagger:1989fc,He:1992nga}. Given the close relation between the GET and GEG, % seen in Eqs.~(\ref{eq:scalar}) and (\ref{eq:L}), 
we suspect that GEG should also be adequate in dealing with radiative corrections.

%------------------------------------------------------------------------------
\subsection{Splitting functions in the broken phase}
%------------------------------------------------------------------------------

%~~~~~~~~~~~~~~~~~~~~~~~~~~~~~~~~~~~~~~~~~
\subsubsection{Modifications to unbroken-phase splitting functions}
%~~~~~~~~~~~~~~~~~~~~~~~~~~~~~~~~~~~~~~~~~

The unbroken-phase splitting functions governed by the gauge and Yukawa couplings given in Tables~\ref{tab:massless_fermion_splittings} to~\ref{tab:massless_scalar_splittings} of Sec.~\ref{sec:unbroken} are still valid for $k_T$'s and virtualities far above the masses of all of the participating particles, provided we make the identification between pseudo-scalars and longitudinal gauge bosons in accordance with the GET. Indeed, in Goldstone Equivalence Gauge, this correspondence is completely transparent. The splitting matrix elements can be used largely unchanged as long as all of the particles are also relativistic, with corrections that typically scale as ${\cal O}(g^2v^2/E^2)$.

At $k_T$'s and virtualities approaching the physical masses, EWSB causes these splitting functions to either smoothly shut off or to transition into resonance decays. The modifications are captured by the propagator and kinematic effects outlined in Section~\ref{sec:split}. In particular, the propagator modifications effectively rescale the unbroken-phase splitting functions of Tables~\ref{tab:massless_fermion_splittings}--\ref{tab:massless_scalar_splittings} as
\beq
\frac{d{\cal P}}{dz\,d\ktsq} \,\to\, \frac{k_T^4}{{\tilde k}_T^4} \, \frac{d{\cal P}}{dz\,d\ktsq} \, \quad
{\rm where}\quad {\tilde k}_T^2 = k_T^2 + \bar z m_B^2 + z m_C^2 - z\bar z m_A^2 .
\label{eq:ktilde}
\eeq
Soft ($1/z$ type) singularities also generally become regulated, though in the $1\to 2$ collinear splitting function language this regulation is somewhat convention-dependent. For $k_T$'s far above the physical masses, soft singularities are anyway constrained by kinematics: $z,\zb \gsim k_T/E_A$. For lower $k_T$'s, such that non-relativistic splitting momenta can be approached, the $k_T$ suppression also sufficiently regulates any soft-singular behavior. But additional soft phase space factors can also be applied to reduce artificial spikes in the differential splitting rates. Minimalistically, this involves the product of velocities of the outgoing products in final-state showers, and for initial-state showers involves the product of the on-shell daughter's velocity and the space-like daughter's ``velocity''.
% $E/p$. 
We have seen a simple example in Fig.~\ref{fig:zkt}(b). 

For the neutral boson states, the propagator factors become matrices. These may be conveniently diagonalized by rotating from the interaction basis $B^0/W^0$ and $H^0/H^{0*}$ to the mass basis $\gamma/Z_T$ and $h/Z_{\longit}$. The former requires the usual rotation by $\theta_W$ in gauge space. The latter is accomplished by a $U(2)$ rotation into the standard CP-eigenstates. The showering must still be performed coherently in order to capture nontrivial effects such as the flow of weak isospin and Higgs number. The full treatment is detailed in Appendix~\ref{app:split}. One residual complication is that the off-diagonal terms in the splitting function matrices are proportional to products of different propagator factors. E.g., for a $\gamma/Z_T$ state, the appropriate modification factor for $d{\cal P}_{\gamma Z}$ would use instead
\beq
\tilde k_T^4 \,\to\, (k_T^2 + \bar z m_B^2 + z m_C^2)(k_T^2 + \bar z m_B^2 + z m_C^2 - z\bar z m_Z^2) \, . 
\label{eq:ktilde2}
\eeq
We also note that our convention here is to align the phases of external $Z_{\longit}$ states with those of the eaten scalar $\phi^0$. Consequently, terms like $d{\cal P}_{h Z_{\longit}}$ are pure imaginary.

The above modifications do not explicitly address possible running effects in the masses. Indeed, the numerical impact of the mass terms in the shower is anyway highly suppressed except at splitting scales of $\order(v)$. Still, some cases, such as kinematics with $k_T \sim v$ but $Q \gg v$, might require special care in the inclusion of higher-order radiative corrections. Similar considerations apply to the purely ultra-collinear splitting processes discussed below.

%~~~~~~~~~~~~~~~~~~~~~~~~~~~~~~~~~~~~~~~~~~~~~~~~
\subsubsection{Ultra-collinear broken-phase splitting functions}
%~~~~~~~~~~~~~~~~~~~~~~~~~~~~~~~~~~~~~~~~~~~~~~~~

%%%  Tables of initial fermions  %%%
\bgroup
\def\arraystretch{1.5}
\begin{table}
\centering
{\small 
\begin{tabular}{l|ccc}
\multicolumn{1}{l}{}
    & \begin{picture}(50,40)(0,0)
      \SetColor{Black} \SetWidth{2} \SetScale{\figscale}
      \ArrowLine(  0,25)(50,25)
      \DashLine(50,25)(95,45){6}
      \ArrowLine( 50,25)(95,5)
      \Text(15,10)[]{$\boldsymbol \Leftarrow$}
      \Text(50,3)[]{\rotatebox{-26}{$\boldsymbol \Leftarrow$}}
      \SetWidth{1}
      \GCirc(50,25){8}{0.7}
      \Text(70,35)[l]{$\phi/V_{\longit}$}
      \end{picture} %\includegraphics[width=100pt]{figs/fL-fL_Vlong}
    & \begin{picture}(50,40)(0,0)
      \SetColor{Black} \SetWidth{2} \SetScale{\figscale}
      \ArrowLine(  0,25)(50,25)
      \DashLine(50,25)(95,45){6}
      \ArrowLine( 50,25)(95,5)
      \Text(15,10)[]{$\boldsymbol \Leftarrow$}
      \Text(50,3)[]{\rotatebox{-26}{$\boldsymbol \Leftarrow$}}
      \SetWidth{1}
      \GCirc(50,25){8}{0.7}
      \Text(70,35)[l]{$h$} 
      \end{picture} %\includegraphics[width=100pt]{figs/fL-fL_h}       
    & \begin{picture}(50,40)(0,0)
      \SetColor{Black} \SetWidth{2} \SetScale{\figscale}
      \ArrowLine(  0,25)(50,25)
      \Photon(50,25)(95,45){3}{3}
      \ArrowLine( 50,25)(95,5)
      \Text(15,10)[]{$\boldsymbol \Leftarrow$}
      \Text(50,3)[]{\rotatebox{-26}{$\boldsymbol \Rightarrow$}}
      \SetWidth{1}
      \GCirc(50,25){8}{0.7}
      %\Text(65,40)[l]{$V_T$}
      \end{picture} % \includegraphics[width=100pt]{figs/fL-fR_VT}       
\\
%\multicolumn{1}{l}{}
    &  \underline{\hspace{0.7cm}$\dfrac{1}{16\pi^2}\dfrac{v^2}{\tilde{k}^4_T}\left(\dfrac{1}{z}\right)$\hspace{0.7cm}}
    &  \underline{\hspace{0.3cm}$\dfrac{1}{16\pi^2}\dfrac{v^2}{\tktft}$\hspace{0.3cm}} 
    &  \underline{\hspace{0.3cm}$\dfrac{1}{16\pi^2}\dfrac{v^2}{\tktft}$\hspace{0.3cm}} 
\\
%\backslashbox{$A$}{$\!\!\!\to BC$}
    &  $\to \ V_{\longit} \: f_{s}^{(\prime)} \ (V\!\ne\!\gamma)$
    &  $h \: f_{s}$        
    &  $V_T \: f_{\minus s}^{(\prime)}$        
\\
\hline%\hline
$f_{s=L}$
    &  $\big( I_f^V(y_f^2\zb - y_{f^{(\prime)}}^2)z - Q_{f_L}^Vg_V^2 \zb \big)^2$   
    &  $\frac14 y_f^4 z(1+\zb)^2$
    &  $g_V^2 z \big(Q_{f_R}^V y_{f}\zb - Q_{f_L}^V y_{f^{(\prime)}}\big)^2$
\\
$f_{s=R}$\ 
    &  $\big(  I_f^V y_f y_{f^{(\prime)}}z^2 - Q_{f_R}^Vg_V^2 \zb \big)^2  $   
    &  $\frac14 y_f^4 z(1+\zb)^2$
    &  $g_V^2 z \big(Q_{f_L}^V y_{f}\zb - Q_{f_R}^V y_{f^{(\prime)}}\big)^2$                                                                                                \end{tabular}
}
\caption{Ultra-collinear fermion splitting functions $d{\mathcal P}/dz\,d\ktsq$ in the broken phase. Wavy lines represent transverse gauge bosons, while the longitudinals/Goldstones and Higgs bosons are represented by dashed lines. 
The $\tilde k_T^4$ symbol is defined in Eq.~(\ref{eq:ktilde}).
The $I_f^V$ symbol is a shorthand for the ``charge'' of a fermion in its Yukawa coupling to the eaten Goldstone boson, or equivalently the fermion's axial charge under the vector $V$. These are normalized to approximately follow the weak isospin couplings, but are defined independently of the fermion's helicity: $I_u^Z = 1/2$, $I_{d/e}^Z = -1/2$, $I_u^{W^\pm} = I_{d/e}^{W^\pm} = 1/\sqrt{2}$. Other conventions are given in Appendix~\ref{sec:FeynmanRules}.}
\label{tab:broken_fermion_splittings}
\end{table}
\egroup

The remaining task is to compute all of the ultra-collinear splitting functions, proportional to the EWSB scale like in Eq.~(\ref{eq:ultra}). Generalizing the standard massless-fermion $f\to W_{\longit} f'$ calculation~\cite{Kane:1984bb,Dawson:1984gx,Chanowitz:1985hj}, we include the splittings involving arbitrary particles in the SM. 
The electroweak VEV ($v$), to which all of these splitting functions are proportionate, has been explicitly extracted, as well as universal numerical factors, the kinematic factor $\tilde k_T^4$
% \equiv (k_T^2 + \bar z m_B^2 + z m_C^2 - z\bar z m_A^2)^2$ (as per 
as in Eq.~(\ref{eq:ktilde}) or Eq.~(\ref{eq:ktilde2}), and the leading soft singularity structure ($1/z$, $1/\zb$, or $1/z\zb$). 
These are obtained quite straightforwardly in GEG, where individual $1\to 2$ ultra-collinear matrix elements all scale manifestly as $g^2v$, $y_f^2v$, or $gy_fv$. See Appendix~\ref{sec:FeynmanRules} for some explicit examples. 

We present these ``purely broken'' splitting functions in Tables~\ref{tab:broken_fermion_splittings}$-$\ref{tab:broken_scalar_splittings}, using similar logic as in Section~\ref{sec:unbroken}, though now working exclusively in mass basis for the neutral bosons.
Unlike conventional collinear splittings, ultra-collinear splittings do not lead to collinear logarithms. Instead, integrating the emissions at a fixed value of $z$ yields a rate that asymptotes to a fixed value as the input energy increases. However, they are also unlike ordinary finite perturbative corrections, in that they are highly collinear-beamed, and subject to maximally large Sudakov effects from the conventional parton showering that can occur at higher emission scales.

Ultra-collinear emissions of longitudinal gauge bosons, when formed by replacing a transverse boson in any conventional gauge emission by a longitudinal boson, retain soft-singular behavior $\sim 1/z$. (Within GEG, the $1/z$ factors within the splitting matrix elements become regulated to $2E_W/(E_W+k_W)$.) Fully integrating over emission phase space, these still lead to single-logarithmic divergences at high energy. This result might seem at odds with smoothly taking the unbroken limit. For $f\to W_{\longit}f'$, as we dial $v$ to zero at fixed fermion energy, the emission rate for longitudinal bosons grows unbounded. However, the spectrum of those bosons has a median energy fraction $z \sim \sqrt{m_W/E_f}$, and also tends to zero. Moreover, in theories where the fermion has a gauge-invariant mass, such as QED, the nominal ultra-collinear region $k_T \lsim m_W$ becomes subsumed by the usual emission dead cone at $k_T \lsim m_f$.

Many of the other (soft-regular) splitting functions are close analogs of the unbroken splittings, but with ``wrong'' helicities. For example, there are processes where a fermion emits a transverse gauge boson but undergoes a helicity flip, and also where a fermion emits a Higgs boson {\it without} flipping its helicity. There are also new processes such as $h\to h h$ where such an identification is not possible. Schematically, all of these processes can be viewed as arising from $1\to 3$ splittings in the unbroken theory, where one of the final-state particles is a Higgs boson set to its VEV.

%%%  Tables of initial transverse vector bosons  %%%
\bgroup
\def\arraystretch{1.5}
\begin{table}
\begin{subtable}[t]{{\textwidth}}
\centering
{\small 
\begin{tabular}{l|cccc}
\multicolumn{1}{l}{}
    &  \multicolumn{4}{c}{ \begin{picture}(50,40)(0,0)
      \SetColor{Black} \SetWidth{2} \SetScale{\figscale}
      \Photon(  0,25)(50,25){4}{3}
      \DashLine(50,25)(95,45){6}
      \Photon( 50,25)(95,5){4}{3}
      \SetWidth{1}
      \GCirc(50,25){8}{0.7}
      \Text(70,35)[l]{$\phi/V_{\longit}$}
      \end{picture}  }
\\
%\multicolumn{1}{l}{}
    & \multicolumn{4}{c}{\underline{\hspace{5.0cm}$\dfrac{1}{16\pi^2}\dfrac{v^2}{\tktft}\left(\dfrac{1}{z}\right)$\hspace{5.0cm}}}    
\\ 
    &  $\rightarrow W^{\pm}_{\longit} \: \gamma_T$
    &  $W^{\pm}_{\longit} \: Z_T$    
    &  $Z_{\longit} \: W^{\pm}_T$  
    &  $W_{\longit}^+ \: W_T^- \ \overset{\rm or}\ \ W_{\longit}^- \: W_T^+$ 
\\
\hline%\hline
$W_T^\pm$\    
    &  $e^2 g_2^2 \zb^3$  
%    &  $c_W^2 g_2^4 \zb \left(\zb + z/(2c_W^2)\right)^2$
    &  $\frac14 c_W^2 g_2^4 \zb \left((1+\zb) + t_W^2z\right)^2$
    &  $\frac14 g_2^4 \zb (1+\zb)^2$     
    &  $0$
\\
$\gamma_T$
    &  $0$
    &  $0$ 
    &  $0$
    &  $e^2 g_2^2 \zb$                        
\\
$Z_T$
    &  $0$
    &  $0$
    &  $0$
%    &  $c_W^2 g_2^4 \zb (1-z/(2c_W^2))^2$                        
    &  $\frac14 c_W^2 g_2^4 \zb \left((1+\zb) - t_W^2z\right)^2$                        
\\
$[\gamma Z]_T$ 
    &  $0$
    &  $0$
    &  $0$
%    &  $s_W c_W g_2^4 \zb (1-z/(2c_W^2))$             
    &  $\frac12 c_W e g_2^3 \zb \left((1+\zb) - t_W^2z\right)$             
\end{tabular}
}
\vspace{0.5cm}
\end{subtable}
%
%\vspace{0.5cm}
%
\begin{subtable}[t]{\textwidth}
\centering
{\small 
\begin{tabular}{c|cc}
\multicolumn{1}{l}{}
    &\begin{picture}(50,40)(0,0)
      \SetColor{Black} \SetWidth{2} \SetScale{\figscale}
      \Photon(  0,25)(50,25){4}{3}
      \DashLine(50,25)(95,45){6}
      \Photon( 50,25)(95,5){4}{3}
      \SetWidth{1}
      \GCirc(50,25){8}{0.7}
      \Text(70,35)[l]{$h$}
      \end{picture}
    & \begin{picture}(50,40)(0,0)
      \SetColor{Black} \SetWidth{2} \SetScale{\figscale}
      \Photon(  0,25)(50,25){4}{3}
      \ArrowLine(50,25)(95,45)
      \ArrowLine(95,5)( 50,25)
      \Text(50,32)[]{\rotatebox{26}{$\boldsymbol \Rightarrow$}}
      \Text(50,3)[]{\rotatebox{-26}{$\boldsymbol \Rightarrow$}}
      \SetWidth{1}
      \GCirc(50,25){8}{0.7}
      \end{picture}
\\
%\multicolumn{1}{l}{}
    &  \underline{\hspace{0.3cm}$\dfrac{1}{16\pi^2}\dfrac{v^2}{\tktft}$\hspace{0.3cm}}
    &  \underline{\hspace{0.3cm}$\dfrac{1}{16\pi^2}\dfrac{v^2}{\tktft}$\hspace{0.3cm}}
 \\
    &  \ $\to h \: V_T \ (V\!\ne\!\gamma)$ \ 
    &   $f_s \: \bar f^{(\prime)}_s$
\\
\hline%\hline
$V_T$   
    &  $\frac14 z\zb g_V^4$
    &  $\frac12 g_V^2 \left( Q^V_{f_s} y_{f^{(\prime)}} z + Q^V_{f_{\minus s}} y_{f} \zb \right)^2$
\\
$[\gamma Z]_T$   
    &  $0$
    &  $\frac12 e g_Z y_f^2 Q^\gamma_{f} \left( Q^Z_{f_s} z + Q^Z_{f_{\minus s}} \zb \right)$
\end{tabular}  
}
\end{subtable}
\caption{Ultra-collinear transverse vector splitting functions $d{\mathcal P}/dz\,d\ktsq$ in the broken phase. %Functions of the weak mixing angle are abbreviated as $c_W \equiv \cos\theta_W$ and $t_W \equiv \tan\theta_W$. 
For the off-diagonal incoming $[\gamma Z]_T$, 
%the $\tilde k_T^4$ symbol stands for $(k_T^2 + \zb m_B^2 + z m_C^2)\cdot(k_T^2 + \zb m_B^2 + z m_C^2 - z\zb m_Z^2)$. 
the $\tilde k_T^4$ symbol is defined in Eq.~(\ref{eq:ktilde2}).
Other conventions are as in Table~\ref{tab:broken_fermion_splittings} and in Appendix~\ref{sec:FeynmanRules}.}
\label{tab:broken_vector_splittings}
\end{table}
\egroup

To make Tables~\ref{tab:broken_fermion_splittings}$-$\ref{tab:broken_scalar_splittings} more compact, and to make closer contact with practical applications, we have made one additional simplification by neglecting neutral boson interference effects for outgoing particles. E.g., for an ultra-collinear process such as $t_{s} \to (h/Z_{\longit})t_{s}$ (helicity non-flipping scalar emission), we treat the outgoing Higgs and longitudinal $Z$ states incoherently. For final-state radiation, such a treatment is easily justified, since, as discussed in Section~\ref{sec:mass_effects}, the particles produced out of an ultra-collinear splitting have suppressed secondary showering. And for PDF evolution starting from an initial-state composed exclusively of light matter, there are simply no available ultra-collinear processes where such interference effects can occur (e.g., there is GET-violating $q_{s} \to Z_{\longit}q_{s}$, but not $q_{s} \to hq_{s}$). At higher scales, where heavier particles begin to populate the PDFs, further ultra-collinear splittings are again suppressed. Note, however, that we retain interference effects for {\it incoming} neutral bosons, which can remain important for final-state splittings like $\gamma/Z_T \to W^\pm_{\longit}W^\mp_T$. We also re-emphasize that interference effects for outgoing particles should still be retained for the conventional splitting functions, even in the broken phase. This is particularly important for the generation of the mixed $\gamma/Z_T$ PDF.

%%%  Tables of initial longitudinal and Higgs bosons  %%%
\bgroup
\def\arraystretch{1.5}
\begin{table}
\begin{subtable}[t]{\textwidth}
\centering
{\small 
\begin{tabular}{l|cc}
\multicolumn{1}{l}{}
    &\multicolumn{2}{c}{ \begin{picture}(50,40)(0,0)
      \SetColor{Black} \SetWidth{2} \SetScale{\figscale}
      \DashLine(  0,25)(50,25){6}
      \DashLine(50,25)(95,45){6}
      \DashLine( 50,25)(95,5){6}
      \SetWidth{1}
      \GCirc(50,25){8}{0.7}
      \Text(70,35)[l]{$\phi/V_{\longit}$}
      \Text(70, 5)[l]{$\phi/V_{\longit}$}
      \end{picture} }
\\
%\multicolumn{1}{l}{}
  & \multicolumn{2}{c}{\underline{\hspace{3.5cm}$\dfrac{1}{16\pi^2}\dfrac{v^2}{\tktft}\left(\dfrac{1}{z\zb}\right)$\hspace{3.5cm}}} \\
  &  $\rightarrow W^+_{\longit} \: W^-_{\longit}$
  &  $Z_{\longit} \: W^\pm_{\longit}/Z_{\longit}$
\\
\hline%\hline
$W^{\pm}_{\longit}$
    & $0$
    & $\frac{1}{16} g_2^4 \left( (\zb-z)(2+z\zb) - t_W^2\zb(1+\zb) \right)^2$
\\
$h$        
    &  $\frac14 \left( g_2^2(1-z\zb) - \lambda_h z\zb \right)^2$
    &  $\frac18 \left( g_Z^2(1-z\zb) - \lambda_h z\zb \right)^2$
\\
$Z_{\longit}$ 
    &  $\frac{1}{16} g_2^4 \left( (\zb-z) (2+z\zb - t_W^2z\zb) \right)^2$  
    &  $0$ 
\\
$[hZ_{\longit}]$
    &  $\frac{i}{8} g_2^2 \left( g_2^2(1-z\zb) - \lambda_h z\zb \right) \left(\zb-z\right) \left(2+z\zb - t_W^2z\zb  \right)$
    &  $0$
\end{tabular}
}
\vspace{0.5cm}
\end{subtable}
%
%\vspace{0.5cm}
%
\begin{subtable}[t]{\textwidth}
\centering
{\small
\begin{tabular}{l|cc}
\multicolumn{1}{l}{}
    & \ \ \ \ 
      \begin{picture}(50,40)(0,0)
      \SetColor{Black} \SetWidth{2} \SetScale{\figscale}
      \DashLine(  0,25)(50,25){6}
      \DashLine(50,25)(95,45){6}
      \DashLine( 50,25)(95,5){6}
      \SetWidth{1}
      \GCirc(50,25){8}{0.7}
      \Text(70,35)[l]{$h$}
      \Text(70, 5)[l]{$\phi/V_{\longit}$}
      \end{picture}
      \ \ \ \ 
    & \ \ \ \ 
      \begin{picture}(50,40)(0,0)
      \SetColor{Black} \SetWidth{2} \SetScale{\figscale}
      \DashLine(  0,25)(50,25){6}
      \DashLine(50,25)(95,45){6}
      \DashLine( 50,25)(95,5){6}
      \SetWidth{1}
      \GCirc(50,25){8}{0.7}
      \Text(70,35)[l]{$h$}
      \Text(70,5)[l]{$h$}
      \end{picture}
      \ \ \ \ 
\\
%\multicolumn{1}{l}{}
   &  \underline{\hspace{0.3cm}$\dfrac{1}{16\pi^2}\dfrac{v^2}{\tktft}\left(\dfrac{1}{\zb}\right)$\hspace{0.3cm}}
   &  \underline{\hspace{0.3cm}$\dfrac{1}{16\pi^2}\dfrac{v^2}{\tktft}$\hspace{0.3cm}}
\\
   &  $\rightarrow h \: W^\pm_{\longit}/Z_{\longit}$ 
   &  $h \: h$ 
\\
\hline%\hline
$W^{\pm}_{\longit}$
    &  $\frac14 z \left( g_2^2(1-z\zb) + \lambda_h \zb \right)^2$
    &  $0$
\\
$h$
    &  $0$
    &  $\frac98 \lambda_h^2 z\zb$
\\
$Z_{\longit}$
    &  $\frac14 z \left( g_Z^2(1-z\zb) + \lambda_h \zb \right)^2$
    &  $0$
\\
$[hZ_{\longit}]$
    &  $0$
    &  $0$
\end{tabular}
}
\vspace{0.5cm}
\end{subtable}
%
%\vspace{0.5cm}
%
\begin{subtable}[t]{\textwidth}
{\small
\begin{tabular}{l|cccc}
\multicolumn{1}{l}{}
    & \multicolumn{3}{c}{ \begin{picture}(50,40)(0,0)
      \SetColor{Black} \SetWidth{2} \SetScale{\figscale}
      \DashLine(  0,25)(50,25){6}
      \Photon(50,25)(95,45){4}{3}
      \Photon( 50,25)(95,5){4}{3}
      \SetWidth{1}
      \GCirc(50,25){8}{0.7}
      \end{picture}  }
    & \begin{picture}(50,40)(0,0)
      \SetColor{Black} \SetWidth{2} \SetScale{\figscale}
      \DashLine(  0,25)(50,25){6}
      \ArrowLine(50,25)(95,45)
      \ArrowLine(95,5)( 50,25)
      \Text(50,32)[]{\rotatebox{ 26}{$\boldsymbol \Leftarrow$}}
      \Text(50, 3)[]{\rotatebox{-26}{$\boldsymbol \Rightarrow$}}
      \SetWidth{1}
      \GCirc(50,25){8}{0.7}
      \end{picture}
      \\
%\multicolumn{1}{l}{}
    &   \multicolumn{3}{c}{\underline{\hspace{3.3cm}$\dfrac{1}{16\pi^2}\dfrac{v^2}{\tktft}$\hspace{3.3cm}}}
    &   \multicolumn{1}{c}{\underline{\hspace{0.3cm}$\dfrac{1}{16\pi^2}\dfrac{v^2}{\tktft}$\hspace{0.3cm}}}
    \\
    &  $\to \gamma_T \: W_T^\pm$
    &  $Z_T \: W_T^\pm/Z_T$
    &  $W_T^+ \: W_T^-$
    &  $f_s \: f^{(\prime)}_{\minus s}$ 
\\
\hline%\hline
\multirow{2}{*}{$W^{\pm}_{\longit}$}
    &  \multirow{2}{*}{$2 e^2 g_2^2 z^3\zb$} 
    &  \multirow{2}{*}{$\frac12 c_W^2 g_2^4 z\zb \left( (\zb-z) + t_W^2 \right)^2$}
    &  \multirow{2}{*}{$0$}
    &  \multicolumn{1}{l}{$s\!=\!L: \ \frac12 \left(y_f^2\zb + y_{f'}^2z - g_2^2z\zb \right)^2$}
\\
    &
    &
    &
    &  \multicolumn{1}{l}{$s\!=\!R: \qquad\qquad \frac12 y_f^2 y_{f'}^2$}
\\
$h$
    &  $0$ 
    &  $\frac14 g_Z^4 z\zb$
    &  $\frac12 g_2^4 z\zb$
    &  $\frac14 y_f^4 (\zb-z)^2$
\\
$Z_{\longit}$
    &  $0$
    &  $0$
    &  $\frac12 g_2^4 z\zb \left( \zb-z \right)^2$
    &  $\left(I_f^Z y_f^2 - Q_{f_s}^Z g_Z^2 z\zb \right)^2$ 
\\
$[hZ_{\longit}]$  
    &  $0$
    &  $0$
    &  $-\frac{i}{2} g_2^4 z\zb \left( \zb-z \right)$
    &  $(-1)^s\frac{i}{2} y_f^2 (\zb-z) \left(I_f^Z y_f^2 - Q_{f_s}^Z g_Z^2 z\zb \right)$
\end{tabular}
}
\end{subtable}
\caption{Ultra-collinear longitudinal vector boson and Higgs boson splitting functions $d{\mathcal P}/dz\,d\ktsq$. The Higgs quartic coupling $\lambda_h$ is normalized such that $m_h^2 = \lambda_h v^2/2$. For the off-diagonal incoming $[h Z_{\longit}]$, the $\tilde k_T^4$ symbol stands for $(k_T^2 + \zb m_B^2 + z m_C^2 - z\zb m_h^2)\cdot(k_T^2 + \zb m_B^2 + z m_C^2 - z\zb m_Z^2)$. Other conventions are as in Tables~\ref{tab:broken_fermion_splittings}, \ref{tab:broken_vector_splittings} 
and in Appendix~\ref{sec:FeynmanRules}. }
\label{tab:broken_scalar_splittings}
\end{table}
\egroup

%%%%%%%%%%%%%%%%%%%%%%%%%%%%%%%%%%%%%%%%
\section{Shower Implementation and Related New Phenomena}
%\section{Shower Implementation and Examples}
\label{sec:implementation}
%%%%%%%%%%%%%%%%%%%%%%%%%%%%%%%%%%%%%%%%

We are now in a position to implement the splitting formalism and to present some initial physics results. Our studies here involving PDFs have been generated using simple numerical integration techniques. Our studies involving final-state radiation, which provide much more exclusive event information, have been generated using a dedicated virtuality-ordered weak showering code. Some technical aspects of this code can be found in Appendix~\ref{sec:FSR}. We do not presently study the more technically-involved exclusive structure of weak ISR radiation. More detailed investigations of specific physics applications will appear in future work~\cite{EWshower}. 

We first show some representative integrated splitting rates for an illustrative set of electroweak splitting processes in Table~\ref{table:splitting_rates}, at incoming energies of 1 and 10~TeV, as well as the leading-log asymptotic behavior. We have mainly focused on examples from Sections~\ref{sec:unbroken} and~\ref{sec:broken} that exhibit single- or double-logarithmic scaling with energy. Unless otherwise noted, the rates are summed/averaged over spins and particle species. (For instance, $q=u_L,u_R,d_L,d_R$, and $f$ denotes all twelve fermion types of either spin.) The symbols in the parentheses denote the conventional collinear-enhanced (CL), infrared-enhanced (IR) and ultra-collinear (UC) behaviors, respectively. Radiation of a $V_T$ boson exhibits the usual CL+IR double-log behavior. Notably, the largest splitting rates occur for $V_T \to V_T V_T$, due to the large adjoint gauge charge. Splittings of this type occur with roughly $35\%$ probability at 10~TeV, a factor that is enormous for an ``EW correction'' and which clearly indicates the need for shower resummation. We also see the analogous UC+IR process $V_T\to V_L V_T$, which only grows single-logarithmically, but which still represents a sizable fraction of the total splitting rate (even more so if we focus on low-$k_T$ regions, similar to Fig.~\ref{fig:zkt}). Similarly, the other ultra-collinear channels are smaller but not negligible.

We next present our numerical results for various exclusive splitting phenomena, paying special attention to the novelties that arise in the EW shower.

\begin{table}
\begin{center}
\begin{tabular}{ l | c | c | c }
Process \ &  $\approx {\mathcal P}(E)$ (leading-log term)  & \ ${\mathcal P}(1~{\rm TeV})$ \  & \ ${\mathcal P}(10~{\rm TeV})$ \  \\   \hline 
$q \to V_Tq^{(\prime)}$   \ (CL+IR) &  $(3\times10^{-3})\left[\log\frac{E}{\mw} \right]^2$ &  1.6\% &  7\%    \\
$q \to V_{L}q^{(\prime)}$ \ (UC+IR) &  $ (2\times10^{-3})\log\frac{E}{\mw}$                &  0.4\% &  1.1\%   \\   \hline  
$t_R \to W_L^+ b_L$      \ (CL)    &  $(8\times10^{-3}) \log\frac{E}{\mw}$                &  2.5\%  &  4\%    \\
$t_R \to W_T^+ b_L$      \ (UC)    &  $(6\times10^{-3}) $                                 &  0.6\% &  0.6\%    \\ \hline
$V_T \to V_T V_T$        \ (CL+IR) & $(0.015)\left[\log\frac{E}{\mw} \right]^2$           &  7\% &  34\%   \\ 
$V_T \to V_{L}V_T$       \ (UC+IR)  & $(0.014)\log\frac{E}{\mw}$                          &  2.7\% &  7\%   \\
$V_T \to f\bar f$        \ (CL)    & $(0.02)\log\frac{E}{\mw}$                            &  5\% &  10\%   \\ \hline
$V_{L} \to V_T h$        \ (CL+IR)  & $(2\times10^{-3})\left[\log\frac{E}{\mw} \right]^2$  &  0.8\% &  4\%   \\
$V_{L} \to V_L h$        \ (UC+IR)  & $(2\times10^{-3})\log\frac{E}{\mw} $                 &  0.5\% &  1\%
\end{tabular}
\end{center}
\caption{Representative electroweak splitting behaviors and integrated fixed-order splitting probabilities for an illustrative set of processes at two parent energies $E=1,\ 10$ TeV. The symbols in the parentheses denote the collinear (CL), infrared (IR), and ultra-collinear (UC) behaviors, respectively.}
\label{table:splitting_rates}
\end{table}

%------------------------------------------
%\subsection{Electroweak effects in PDFs}
\subsection{Weak boson PDFs}
%------------------------------------------

We first revisit the classic calculation of weak boson PDFs within proton beams~\cite{Kane:1984bb,Dawson:1984gx}. The basic physical picture has been dramatically confirmed with the observation of the Higgs boson signal via vector boson fusion at the LHC~\cite{LHCVBF}. It is anticipated that at energies in the multi-TeV regime, the total production cross section for a vector boson fusion process $V_1 V_2\to X$ can be evaluated by convoluting the partonic production cross sections over the gauge boson PDFs, originated from the quark parton splittings $q\to W^\pm q',\ q \to \gamma/Z q$.\footnote{It should be noted that a formal factorization proof for electroweak processes in hadronic collisions is thus far lacking. For instance, it is not presently demonstrated whether contributions from gauge boson exchanges between the two incoming partons are factorizable. Nonetheless, we expect that the factorized PDF approach should furnish a reliable and useful calculation tool at very high energies at leading order, as indicated by simple scaling arguments~\cite{Kunszt:1987tk,Borel:2012by}.} A useful intermediate object in this calculation is the parton-parton luminosity, consisting of the convolutions of the PDFs from each proton. We write the cross section in terms of the parton luminosity of gauge boson collisions as
\beq
\sigma_{PP}(V_1 V_2 \to X)  \,=\, 
       \int_{\tau_{\rm low}}^{\tau_{\rm high}} d\tau \, \frac{d\mathcal{L}_{V_1 V_2}}{d\tau}\  \hat{\sigma}(V_1 V_2 \rightarrow \hat X_\tau) \, ,
\eeq
and can approximate this luminosity at fixed-order using the  concept of weak boson PDFs of individual quarks within the proton:
\beqa
\frac{d\mathcal{L}_{V_1 V_2}}{d\tau} & \,\simeq\, & 
       \frac{2}{(\delta_{V_1 V_2}+1)} \int^1_\tau\frac{d\xi}{\xi} ~\int^1_{\tau/\xi}\frac{dz_1}{z_1}~\int^1_{\tau/\xi/ z_1}\frac{dz_2}{z_2} \times \nonumber \\
 & & \sum_{q_1,q_2}  
      f_{V_1\in q_1}(z_1)f_{V_2\in q_2}(z_2)~f_{q_1\in P}(\xi)f_{q_2\in P}\left(\frac{\tau}{\xi z_1 z_2}\right) \, .  \label{eq:partonLumi}
\eeqa
Here, $\tau = s/S$ is the ratio of the partonic and hadronic energies squared, and $\tau_{\rm low}$ and $\tau_{\rm high}$ the kinematic boundaries (e.g., defining a bin in a histogram). We assume $\tau_{\rm low} \gg 4m_W^2/S$. The objects $f_{V\in q}$ are evaluated at fixed-order as
\beq
f_{V\in q}(z) \,\approx\, \int_{0}^{{\cal O}(s/4)}  d\ktsq  \, \frac{d{\cal P}_{q\to Vq^{(\prime)}}}{dz \, d\ktsq}(z,\ktsq) \, ,
\eeq
where the upper boundary of the $k_T$ integration is of order the partonic CM energy. For example~\cite{Kane:1984bb,Dawson:1984gx},
\beqa
f_{W_T^\pm \in u/d}(z)         \,\simeq\, \frac{\alpha_W}{8\pi} \frac{1+\zb^2}{z} \log\left(\frac{s}{4m_W^2}\right) , \quad
%\nonumber \\
f_{W_{\longit}^\pm \in u/d}(z) \,\simeq\, \frac{\alpha_W}{4\pi} \frac{\zb}{z} , \label{eq:fixed-order-PDFs}
\eeqa
where the PDFs have been integrated up to $k_T^2 = s/4$, assumed to be much larger than $m_W$.

We emphasize that in deriving these illustrative fixed-order weak boson PDFs, we have {\it not} resummed the logarithmic enhancement, which remains explicit in Eq.~(\ref{eq:fixed-order-PDFs}) for the transverse bosons. There are also corresponding double- and single-log EW enhancements in the virtual corrections for the sourcing quarks, arising from integrating over both $z$ and $k_T$, which we have not accounted for. While these are of formally higher-order concern in determining the weak boson PDFs, they would also be required for an all-orders resummation of the leading-order effects. (We comment on other novel EW effects on the quark PDFs at the end of this subsection.)

A related issue is that there are factorization scales implicit in the definition of the sourcing quark PDFs. Since the weak coupling and $\log(E/m_W)$ factors are together still below $\order(1)$ size at planned future machines, the choice of factorization scale might also seem to be of strictly higher-order concern. However, the interleaving of the much faster QCD evolution complicates the situation somewhat, especially at a large value of the energy fraction $z$. We have already noted above that the longitudinal $W/Z$ PDFs would not continue to be sourced above $m_W$, as their ultra-collinear generation is constrained to the region $k_T \sim m_W$. It is therefore important to fix a factorization scale of ${\cal O}(m_W)$ for the quark PDFs from which the fixed-order $W_{\longit}$ PDFs are derived, even for processes where $\sqrt{s} \gg m_W$~\cite{Han:1992hr}. However, the transverse $W/Z$ PDFs are sourced continuously at all scales. Higher-order calculations and/or full solution of the mixed QCD/EW DGLAP equations would be required to more fully resolve the issue of scale choices for the transverse bosons. Here we simply fix the scale for the sourcing quark PDFs to be the geometric mean of $\sqrt{s}$ and $m_W$ (e.g., ${\cal O}$(1~TeV) in a 10~TeV process).\footnote{This calculation uses only QCD evolution for the quark PDFs. The additional impact of electroweak evolution effects on the sourcing of the electroweak PDFs should indeed be small. Note also that mixed processes, such as $V_T V_{\longit} \to X$ would generally need a different factorization scale for each sourcing quark PDF.} 

\begin{figure}[t]
\begin{center}
\begin{subfigure}[t]{0.495\textwidth}
\includegraphics[width=210pt]{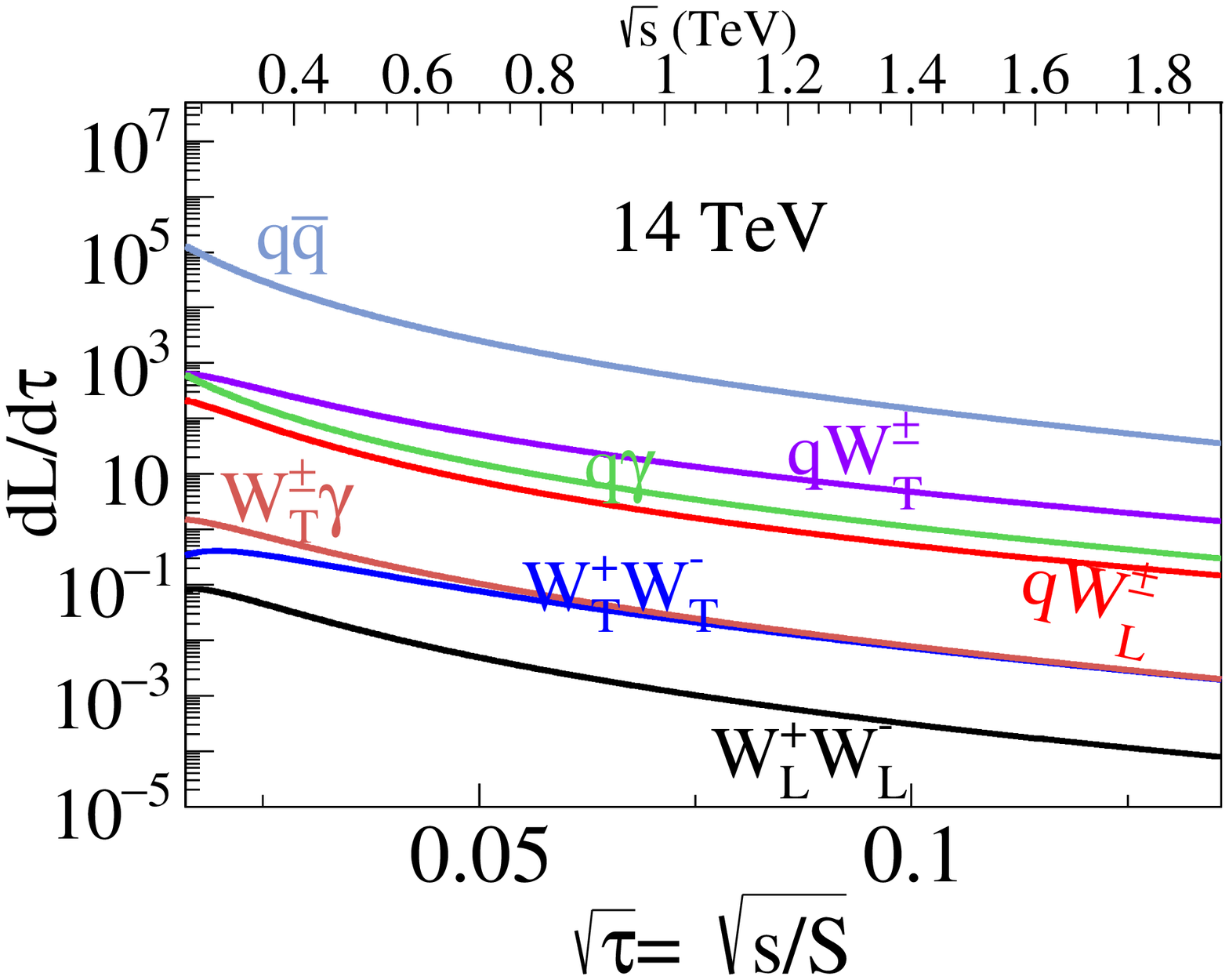}
\vspace{-0.1cm}\caption{}\end{subfigure} 
\begin{subfigure}[t]{0.495\textwidth}
\includegraphics[width=210pt]{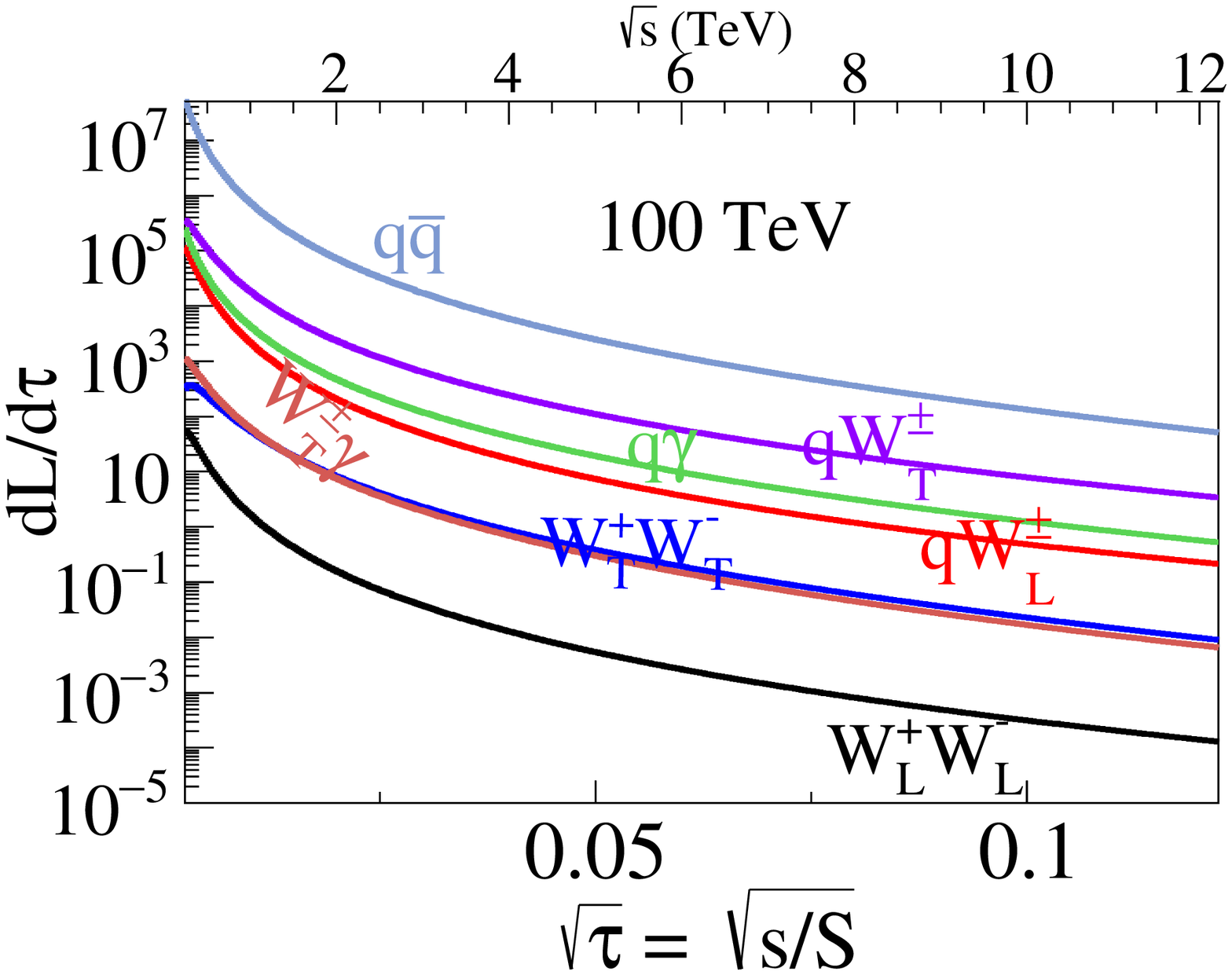} 
\vspace{-0.1cm}\caption{}\end{subfigure}
\\ \vspace{0.2cm}
\begin{subfigure}[t]{0.495\textwidth}
\includegraphics[width=210pt]{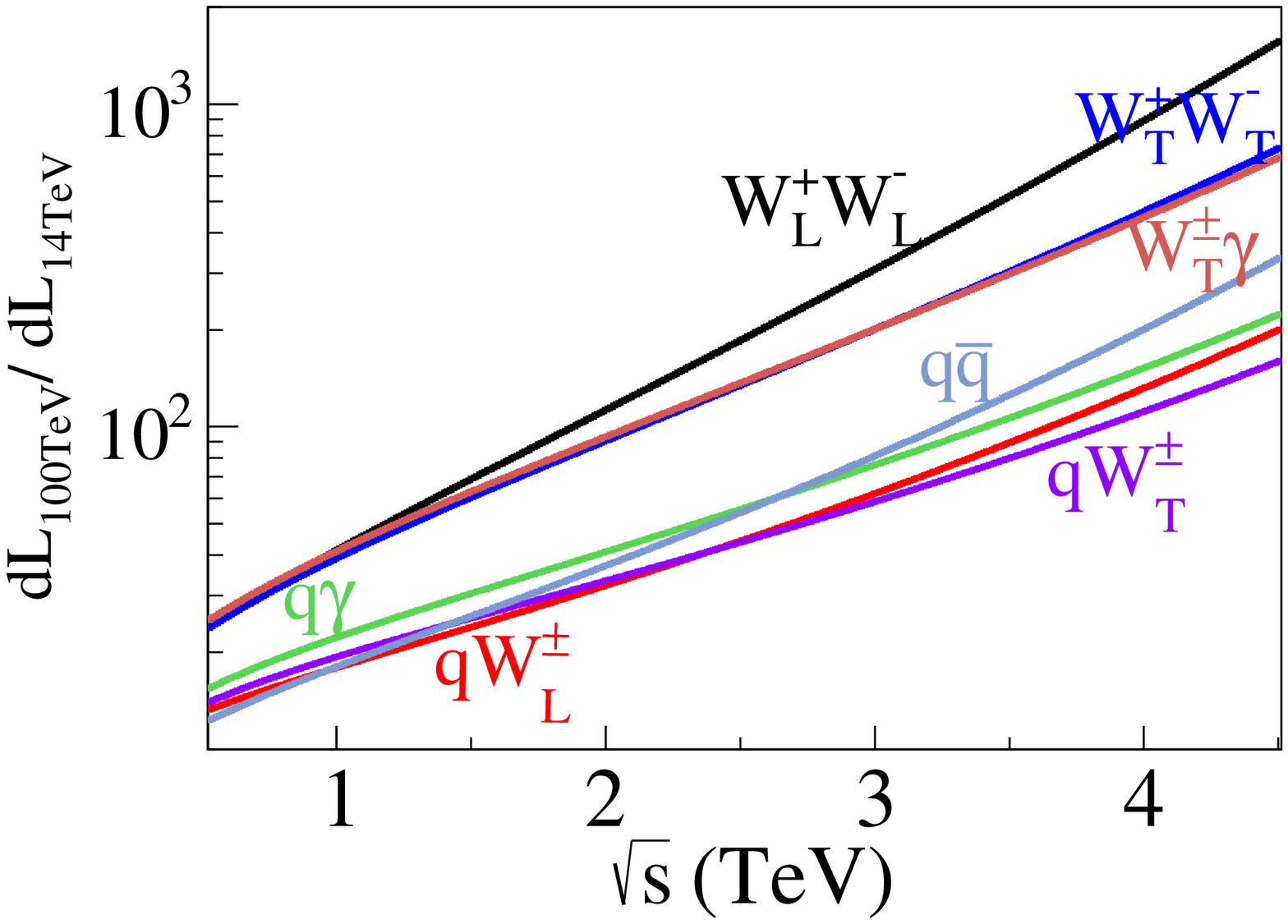}
\vspace{-0.1cm}\caption{}\end{subfigure}
\end{center}
\caption[]{Representative parton luminosities in $pp$ collisions at (a)~$\sqrt S=14$~TeV, (b)~$\sqrt S=100$~TeV, and (c) the ratio of luminosities between the two beam energies as a function of partonic CM energy $\sqrt{s}$.
}
\label{fig:PDF}
\end{figure}

Figures \ref{fig:PDF}(a) and \ref{fig:PDF}(b) show the predicted fixed-order luminosities for a variety of possible colliding partons, including quarks as well as polarized $W^\pm$ bosons and photons, at the 14 TeV LHC and a 100~TeV $pp$ collider. At low scales, the ``EW'' PDFs are of course wholly dominated by photons. However, at scales above $m_W$, the $W^\pm$ PDFs are of comparable size. This can be seen here by comparing the $q\gamma$ and $qW_T^\pm$ parton luminosities, as well as the $W_T^+\gamma$ and $W_T^+W_T^-$ luminosities. Note that in this comparison, we have also derived the photon PDF at fixed-order, sourced from quark PDFs.
Attempts at fitting the photon PDFs with LHC data have recently been made~\cite{Ababekri:2016kkj}. Some recent discussions regarding the factorization scale uncertainties can be found in Ref.~\cite{Alva:2014gxa}. 
More importantly, a complete description will ultimately require including as well the $Z_T$ and {\it mixed} $\gamma/Z_T$ PDFs~\cite{EWshower}.

The PDFs and corresponding parton luminosities for longitudinal gauge bosons can be seen to be significantly smaller than those of transverse bosons. Of course, these nonetheless remain uniquely important for probing the nature of the electroweak sector beyond the Standard Model \cite{Lee:1977eg,Chanowitz:1985hj,Barger:1990py,Bagger:1995mk,Agashe:2004rs,Giudice:2007fh}. In Fig.~\ref{fig:PDF}(c), we show the ratios of the partonic luminosities at the 100~TeV collider and the LHC $dL^{100}(s) /dL^{14}(s)$. The increase with energy is largest for $W_{\longit}W_{\longit}$, with an enhancement factor about two orders of magnitude for $\sqrt s = 1$--4~TeV.

As discussed in Sec.~\ref{sec:evolv}, some additional novel electroweak effects in the PDFs involve the different gauge interactions of left-handed and right-handed chiral fermions, and the isospin non-singlet nature of typical beam particles. The former leads to more rapid evolution to low-$x$ for left-handed fermions than for right-handed fermions. The latter leads to Bloch-Nordsieck violation \cite{Ciafaloni:2000rp,Bell:2010gi,Manohar:2014vxa}. In PDF language, this appears as a self-correcting instability wherein the two LH isospin components of the beam flip between one another at a progressively increasing double-logarithmic rate, via soft/collinear $W^\pm$ emissions. Both effects contribute to spontaneous beam polarization. In particular, in unpolarized proton beams the $u_L$ and $d_L$ PDFs will gradually split off from the $u_R$ and $d_R$ PDFs, and begin to asymptotically merge together into a common ``$q_L$'' PDF at high energies. We investigate these phenomena in future work~\cite{EWshower}.

%-----------------------------------------------------
\subsection{Final states with multiple gauge bosons}
%-----------------------------------------------------

The collinear showering approximation allows us to estimate the leading contributions for multiple EW gauge boson production at high energies. A major component is splittings amongst the gauge bosons themselves via their non-Abelian interactions, in analogy with $g \to gg$ splittings in QCD. These have so far received little dedicated study in the electroweak theory within a parton shower framework. For some earlier studies of the fixed-order Sudakov effects in high-$p_T$ gauge boson production, see for example~\cite{Kuhn:2005az,Kuhn:2005gv,Kuhn:2007cv}.\footnote{As a simple cross-check of our shower framework, we can make a comparison to the $p_T$-dependent EW radiative corrections in $Wj$ production, as computed to NLO and approximate NNLO~\cite{Kuhn:2007cv}. Since our shower is defined only for FSR, we study $Wq$ production and square the inferred Sudakov factor for the final-state quark. This approximately includes the Sudakov contribution of the initial-state quark. We select events without $W/Z$ emissions, but allow final-state photons. At $p_T = 1$~TeV, the EW correction to (NLO,NNLO) order is computed to be $-(27,24)\%$, whereas our resummed shower Sudakov also predicts $-24\%$. At $p_T = 2$~TeV, the EW correction to (NLO,NNLO) order is computed to be $-(42,34)\%$, whereas our resummed shower Sudakov predicts $-33\%$. (Following the exponentiation pattern of the corrections, the NNNLO contribution would be $\sim 1\%$.)}

\begin{figure}[t]
\begin{center}
%$\mathbf{WZj}$ \bf{at FCC, 10~ab}$\mathbf{^{-1}}$\bf{,} $\mathbf{p_T(j) > 3}$~\bf{TeV} \\ \strut \\
\begin{subfigure}[t]{0.495\textwidth}
\includegraphics[width=200pt]{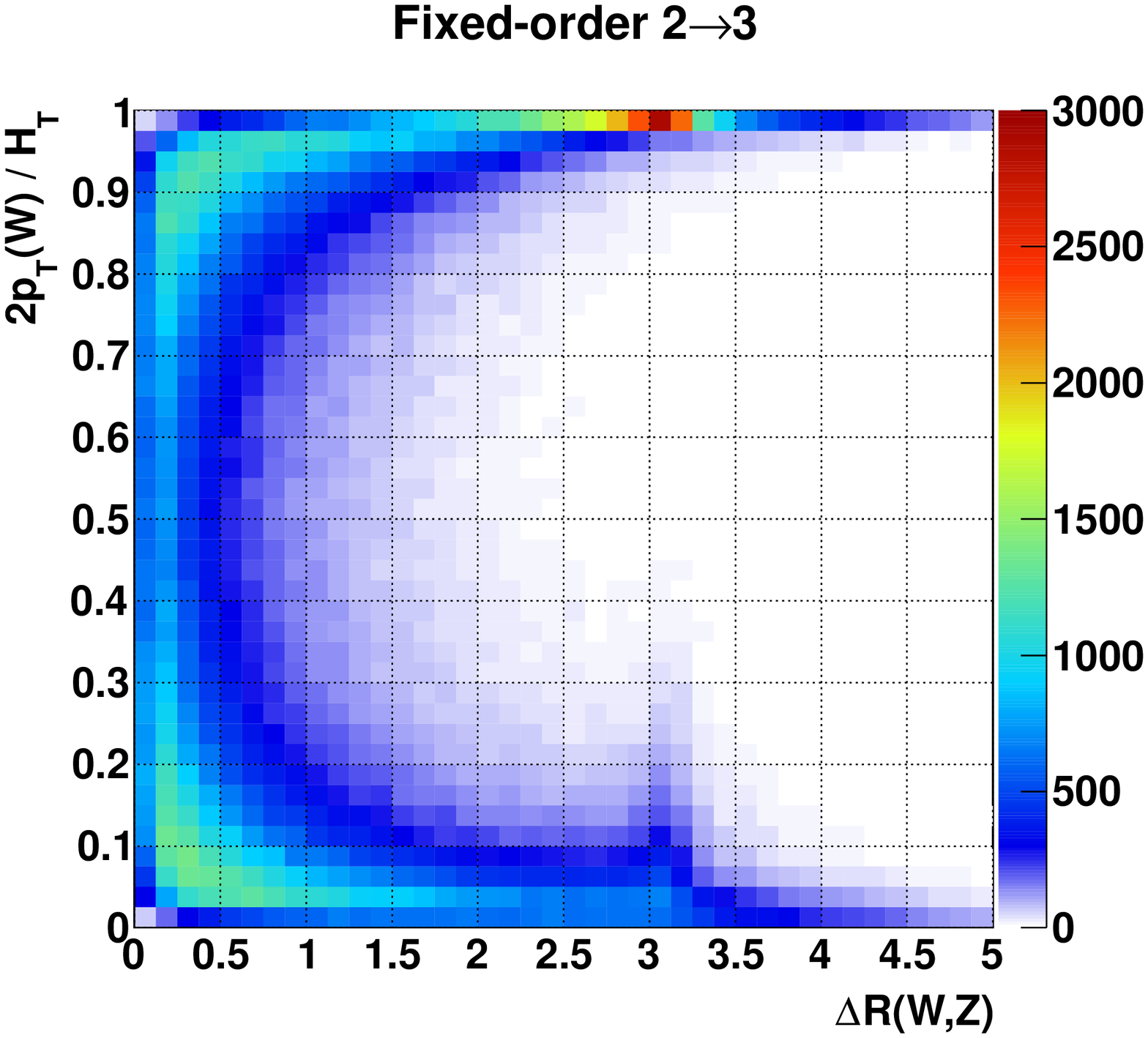}
\vspace{-0.5cm}\caption{}\end{subfigure}
\begin{subfigure}[t]{0.495\textwidth}
\includegraphics[width=200pt]{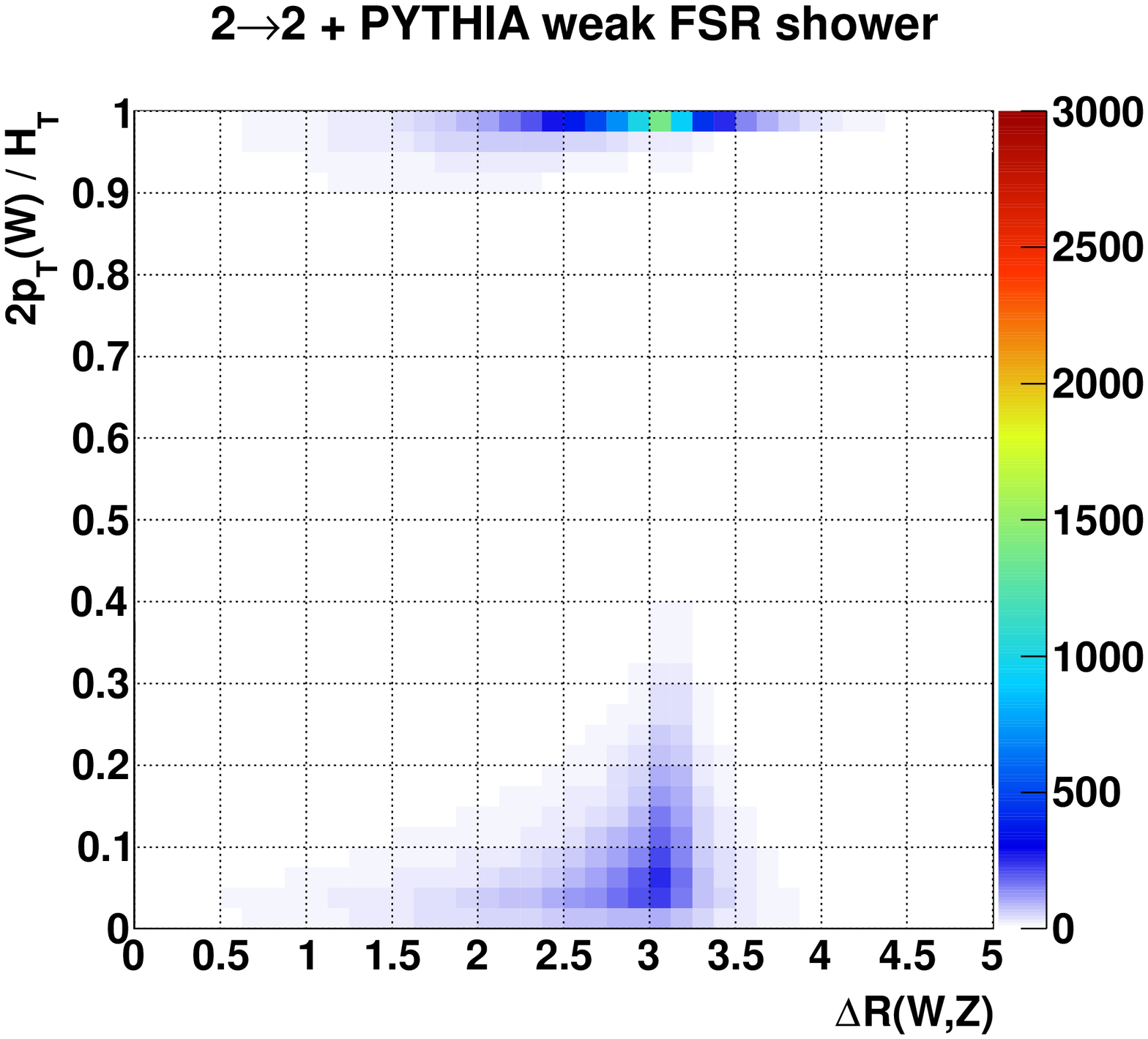} 
\vspace{-0.5cm}\caption{}\end{subfigure}
\\
\vspace{0.2cm}
\begin{subfigure}[t]{0.495\textwidth}
\includegraphics[width=200pt]{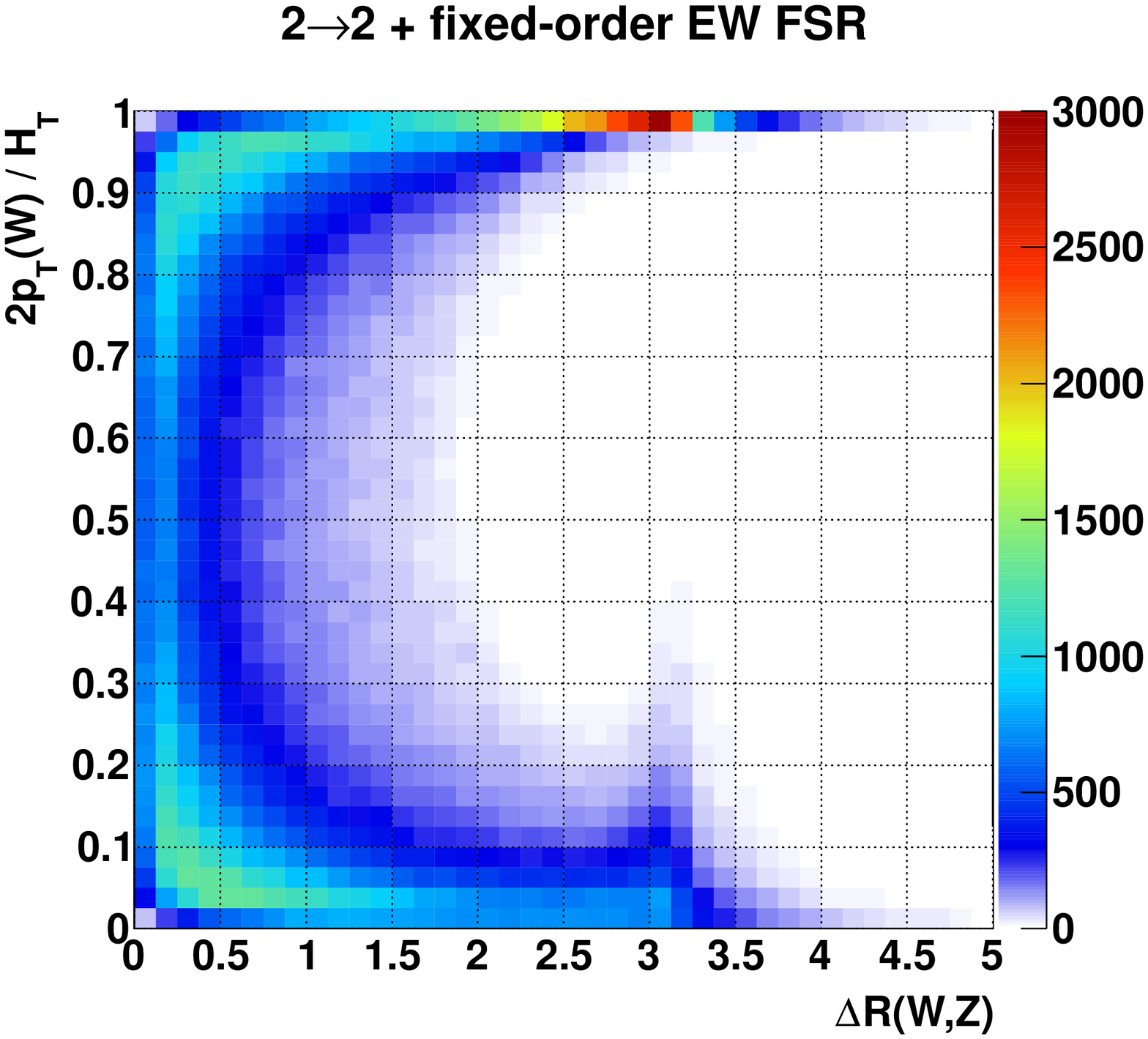}
\vspace{-0.5cm}\caption{}\end{subfigure}
\begin{subfigure}[t]{0.495\textwidth}
\includegraphics[width=200pt]{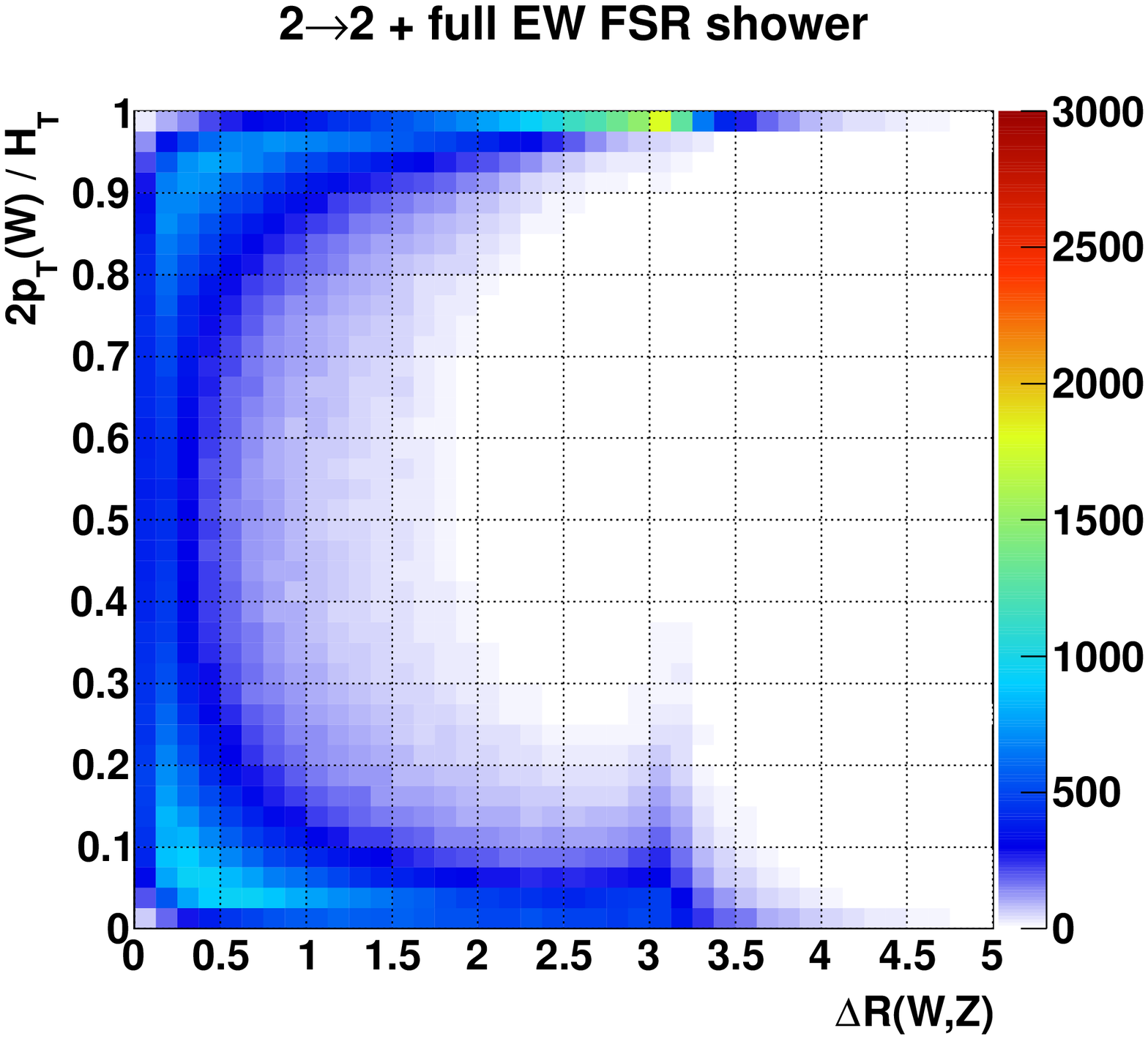}
\vspace{-0.5cm}\caption{}\end{subfigure}
\vspace{-0.6cm}
\end{center}
\caption[]{Event population for exclusive $WZ+j$ production in the plane of $2p_T(W)/H_T$ versus $\Delta R(W,Z)$ with $p_T(j) \ge 3$~TeV at a 100~TeV proton collider.  (a)~$2\to 3$ fixed-order $WZj$ production generated with {\tt MadGraph}; (b)~$2\to 2$ dressed with the {\tt PYTHIA} weak shower, which includes only $q\to Vq$ splittings; (c)~$2\to 2$ $Wj$ and $Zj$ production dressed with fixed-order FSR splitting functions; (d)~$2\to 2$ dressed with the full EW FSR shower, including all collinear final-state Sudakov effects. QCD showering is not incorporated. An integrated luminosity of 10~ab$^{-1}$ is used for illustration.}
\label{fig:WZjet}\end{figure}

As a simple illustration of the onset of shower-dominated behavior, we show in Fig.~\ref{fig:WZjet}(a) a 2D kinematic distribution in fixed-order $W^\pm Z + q/g$ production at a 100~TeV proton collider, generated with {\tt MadGraph5}~\cite{Alwall:2011uj}. A single kinematic cut $p_T(q/g) > 3$~TeV is applied. The horizontal axis is the $\Delta R$ separation between the $W$ and $Z$, and the vertical axis is the relative transverse momentum carried by the $W$: $2p_T(W)/H_T$ with $H_T$ defined as the scalar sum of all object $p_T$s. Several features are immediately apparent. Most of the rate is concentrated along a curved band at low $\Delta R(W,Z)$, indicating $W(q/g)$ production with a secondary collinear $W \to ZW$ splitting, and with enhancements at high (low) relative $p_T$ for $W\ (Z)$ events. A second clear concentration of events occurs at $\Delta R(W,Z) \simeq \pi$ and near-maximal relative $H_T$ indicating $Wq$ production with a secondary $q \to Zq$ splitting. A third, more subtle concentration is visible at $\Delta R(W,Z) \simeq \pi$ and low relative $H_T$, representing $Zq$ production with a secondary $q \to Wq'$ splitting.

We can show how portions of this distribution arise within an available showering framework by generating $Vj$ events within {\tt PYTHIA8}, and applying its native weak parton shower~\cite{Christiansen:2014kba}. This shower currently includes only $q\to Vq$ splittings, and does not model the $V\to VV$ splittings responsible for the dominant rate near $\Delta R(W,Z) \simeq 0$. The resulting incomplete distribution is shown in Fig.~\ref{fig:WZjet}(b).

As a step toward gaining a more complete picture, we show in Fig.~\ref{fig:WZjet}(c) the same distribution with hard $Vj$ events supplied by {\tt PYTHIA8} but dressed with our own EW FSR treatment (Appendix~\ref{sec:FSR}), for the moment using fixed-order splitting functions and without Sudakov evolution effects. Now including $V\to VV$ as well as $V\to Vq$, the agreement becomes quite good in all of the collinear-enhanced regions where we expect splitting functions to furnish a reliable description.\footnote{Physics parameters here and in the {\tt MadGraph} simulation are evaluated at a fixed scale of $m_Z$ for simplicity of comparison, using {\tt MadGraph}'s defaults. The PDF set is CTEQ6L1, evaluated at a factorization scale of 3~TeV. The {\tt PYTHIA} simulation does not track fermion chirality throughout the hard event, and directly collapses $\gamma/Z$ states into mass basis instead of providing a gauge-space wave function. We have explicitly corrected for both of these effects in this comparison and below.}  

Besides the simpler generation of high-multiplicity final-states in collinear regions, the advantage of the parton shower is the ability to automatically fold in Sudakov corrections, going beyond fixed-order predictions. We show the result of running the full parton shower evolution Fig.~\ref{fig:WZjet}(d), including as well important contributions such as $V \to f\bar f$. Exclusive $W^\pm Z(q/g)$ events are selected as including exactly one each of ``on-shell'' $W$ and $Z$, defined as lying within 10$\Gamma$ of their pole mass, and we allow for multiple photon emissions. While the distribution looks similar to that at fixed-order, the overall rates in the collinear regions are reduced by several tens of percent due to the Sudakov corrections.

\begin{figure}[t]
\begin{center}
\begin{subfigure}[t]{0.495\textwidth}
\includegraphics[width=210pt]{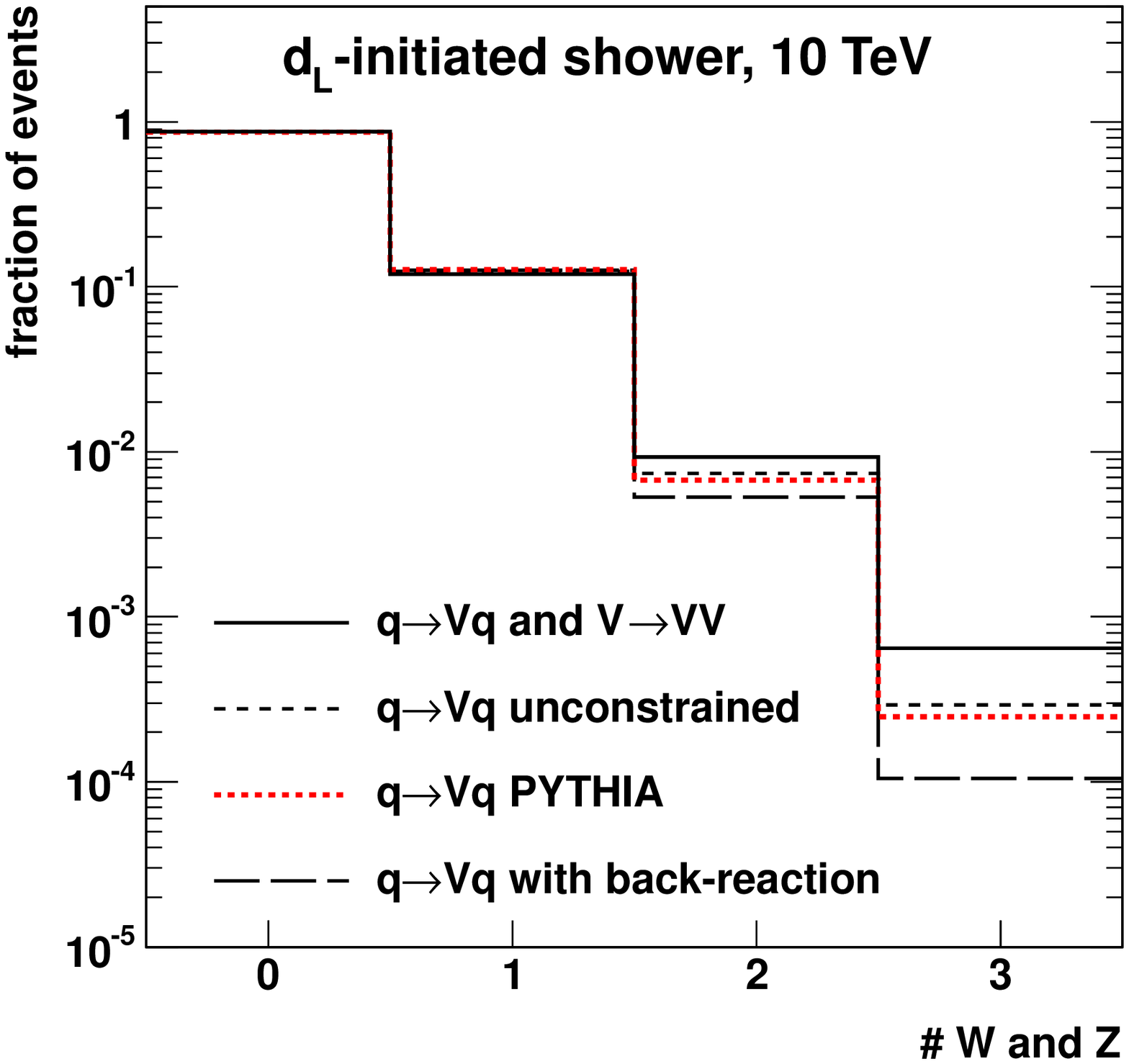}
\vspace{-0.5cm}\caption{}\end{subfigure}
\begin{subfigure}[t]{0.495\textwidth}
\includegraphics[width=210pt]{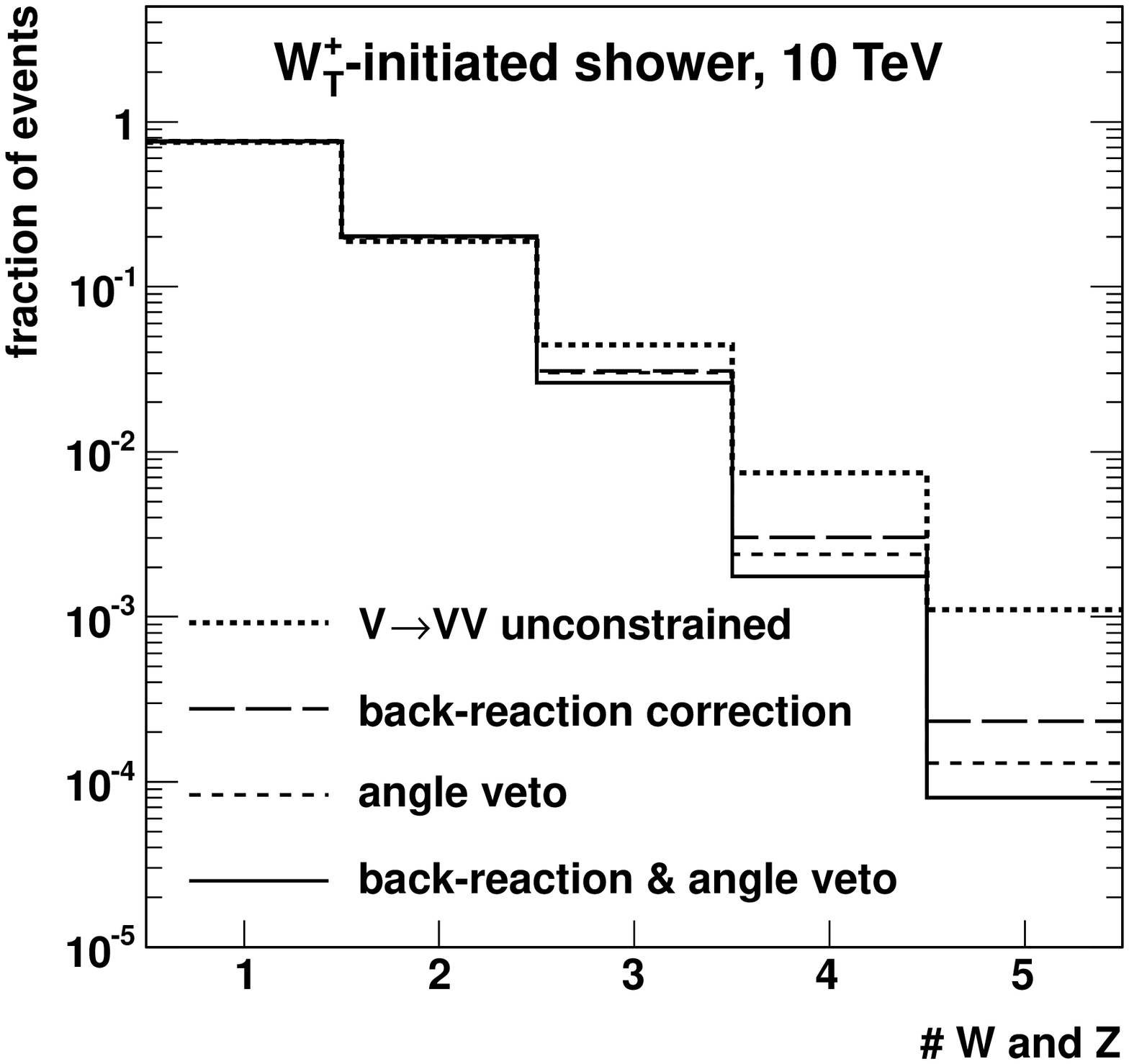}
\vspace{-0.5cm}\caption{}\end{subfigure}
\vspace{-0.6cm}
\end{center}
\caption[]{
Normalized rates versus the number of multiple final-state $W/Z$ emissions with a 10 TeV initial state particle, 
%clustering all particles into $R = \pi/2$ anti-$k_T$ jets, 
%initiated by (a)~$d_L$ and (b)~$W_T^+$. For 
(a) $d_L$-initiated showers for $q\to Vq$ and $V\to VV$ splittings with full EW FSR (solid histogram), $q\to Vq$ splitting only (long-dashed), and 
$q\to Vq$ without back-reaction correction (short-dashed). Output from {\tt PYTHIA} $q\to Vq$ weak shower is also included for comparison (dotted histogram). 
(b) $W_T$-initiated showers for fully constrained FSR (solid histogram), compared with various stages of approximations as labeled. 
}
\label{fig:multiple}
\end{figure}

While formally any secondary parton splittings involve rate penalties of $O(\alpha_W)$, they become progressively more log-enhanced at high energies. This is again in close analogy to QCD. However, unlike in QCD, individual weak splittings in arbitrarily soft/collinear limits are in principle both observable and subject to perturbative modeling.
Figure \ref{fig:multiple} shows the predicted number of $W/Z$ generated from showering off a highly energetic particle with $E = 10$~TeV. In this calculation, we keep the weak bosons stable and include only the splittings $f \to V f$ and $V \to VV$. QCD showering is also turned off. We construct ``weak jets'' by clustering particles with the anti-$k_T$ algorithm~\cite{Cacciari:2008gp} with $R = \pi/2$, and count the contained $W/Z$ bosons. 
In Fig.~\ref{fig:multiple}(a), we show the results for a left-handed chiral fermion $(d_L)$. Roughly speaking, we see that the emission of each additional gauge boson comes with an ${\cal O}$(10\%) suppression factor, which can be compared to the naive (not log-enhanced) ${\cal O}$(1\%) suppression typical of adding gauge bosons to lower-energy processes. The solid histogram shows the total rate and the long-dashed histogram indicates the rate with non-Abelian gauge splittings turned off. The difference indicates the large contribution from the gauge boson self-interaction beyond the first emission. 
%This behavior has been seen in QCD as well. 
As a cross-check, we include as well the prediction from the {\tt PYTHIA8} weak shower~\cite{Christiansen:2014kba}, as shown by the dotted histogram. Our own shower by default includes a back-reaction correction, discussed in Section~\ref{sec:mass_effects}, which approximates the expected suppression of multiple emissions due to dead cone-like effects for off-shell particles. To make a more direct comparison, we have also switched this off, and plotted the result as the short-dashed histogram. The two showers, both modeling unrestricted $q\to Vq$ emissions, are then seen to be in close agreement.

In Fig.~\ref{fig:multiple}(b), we show the predicted number of $W/Z$ contained in ``weak jets'' generated from showering off of a highly energetic transversely-polarized $W^\pm$ boson with $E_W = 10$~TeV. As already indicated in Table~\ref{table:splitting_rates}, the overall emission rates are much higher, close to 40\% for the first emission (including both photons and $Z$ bosons). Here we have again considered the effect of turning on/off back-reaction corrections. In addition, from experience with QCD showers, it is known that coherence effects in emission amplitudes lead to effective color-screening and approximate angular-ordering of nested emissions in non-Abelian splittings. To test this, we have also turned on/off a strict angular-ordering veto in our shower simulation. The results, visible in Fig.~\ref{fig:multiple}(b), are that both the back-reaction correction and the angular ordering can have an ${\cal O}(1)$ effect at high multiplicities, but that the two effects come with sizable overlap. Splittings with large opening angles tend to exhibit large back-reaction effects, and vice-versa. This observation provides some evidence that modeling of the high-multiplicity region might be made to quickly converge, though more study is required.

It should be noted that at higher energy scales, the production of multiple gauge bosons could be the characteristic signature in many scenarios for physics beyond the SM \cite{Agashe:2007ki,Dennis:2007tv}.

%----------------------------------------------
\subsection{EW Showers initiated by top quarks}
%----------------------------------------------
Top quarks are instrumental in searches for new physics related to the EWSB sector, and for exotica such as resonances with large couplings to the third generation, as well as third-generation squarks \cite{Agashe:2013hma}.
High-energy tops can be produced copiously at the LHC and at future accelerators, and 
multi-TeV top quarks offer a particularly rich laboratory to study the effects of weak showering. 

\begin{figure}[t]
\begin{center}
\begin{subfigure}[t]{0.495\textwidth}
\includegraphics[width=210pt]{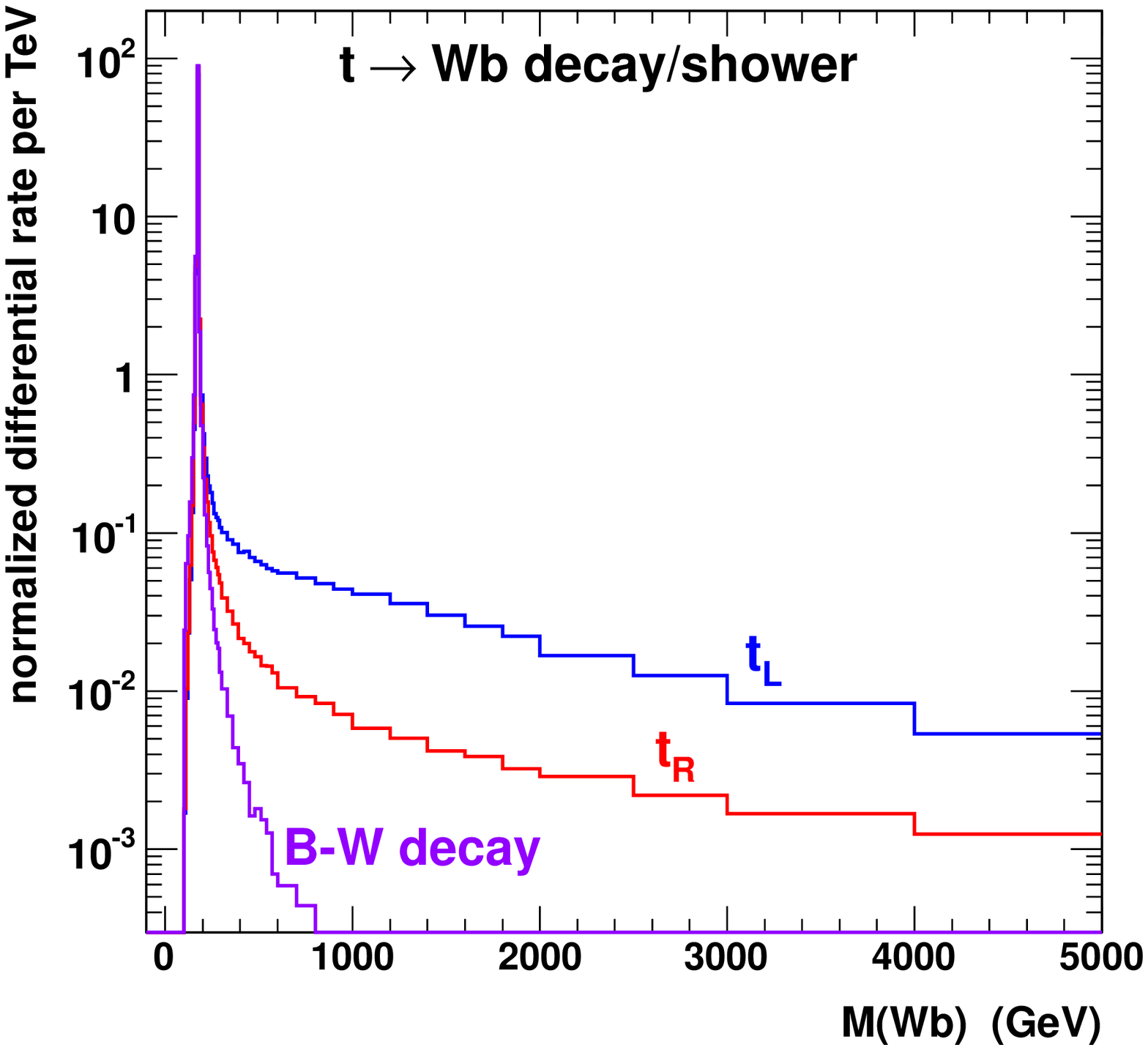}
\vspace{-0.5cm}\caption{}\end{subfigure}
\begin{subfigure}[t]{0.495\textwidth}
\includegraphics[width=210pt]{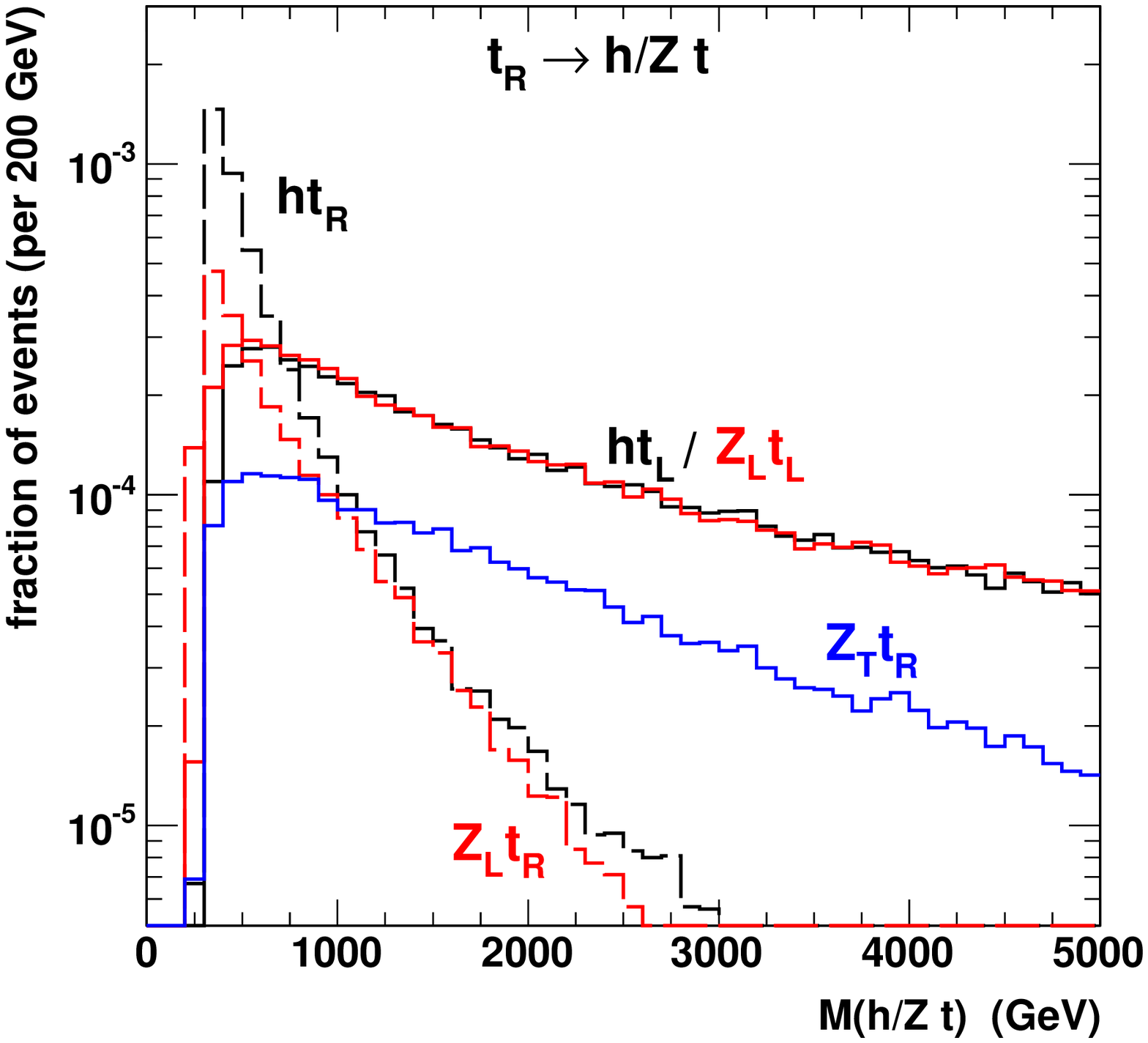}
\vspace{-0.5cm}\caption{}\end{subfigure}
\vspace{-0.6cm}
\end{center}
\caption[]{
Invariant mass distributions for EW splittings initiated by a 10~TeV polarized top quark (a) for $t_L\to Wb$ (top curve), $t_R\to Wb$ (middle curve) and a fixed-width Breit-Wigner for unpolarized top decay without shower (lower curve); 
(b) for $t_R \to ht_L/Z_{\longit} t_L,\ Z_T t_R$ (upper curves) and to $h t_R,\ Z_{\longit} t_R$ (lower curves), respectively. 
}
\label{fig:top}
\end{figure}

We start by considering splittings that follow the same structure as the top quark's weak decay, $t\to W^+b$. Figure \ref{fig:top}(a) shows the resulting $Wb$ mass spectrum from applying this splitting process to 10~TeV top quarks of left-handed or right-handed helicities. One immediate feature is the transition between shower and decay: the Breit-Wigner peak centered at $m_t$ continuously matches onto a high-virtuality shower dominated either by $W_T$ emission from left-handed top quarks, or $W_{\longit}$ emission from right-handed top quarks.\footnote{To improve the matching, we have distributed the ``decay'' events according to a Breit-Wigner distribution weighted by $\Gamma_t(Q)/\Gamma_t(m_t)$. This constitutes approximately a 30\% effect at the given matching scale of 187~GeV.} The former are simple manifestations of $SU(2)_L$ gauge showers with a larger rate (upper curve), whereas the latter are a due to the Goldstone-equivalent Yukawa showers with a smaller rate (middle curve). Ultra-collinear emissions are necessary for properly modeling the shower/decay transition, as shown in more detail in Appendix~\ref{sec:FeynmanRules} (see Fig.~\ref{app:top}).
We also show the unpolarized top decay with a fixed-width Breit-Wigner without shower (lower curve in Fig.~\ref{fig:top}(a)). The events are understandably much more constrained to the region $M(Wb)\simeq m_t$. It is very important to appreciate the difference, for example since one must properly model the properties of off-shell top quarks in searching for new physics~\cite{Agashe:2006hk,Lillie:2007yh,Frederix:2007gi,Han:2008xb,Degrande:2010kt,Agashe:2013hma} associated with the top quark as well as the Higgs sector.

Top quarks may also radiate Higgs bosons and, analogously, longitudinal $Z$ bosons. Both of these Yukawa-showering processes occur with similar rates off of left-handed and right-handed tops, and grow single-logarithmically with energy. In Fig.~\ref{fig:top}(b), we present a 10 TeV right-handed top quark splitting via the EW shower. The rates for $t_R \to ht_L$ and to $Z_L t_L$ are governed by the Yukawa coupling and essentially  the same, due to the GET. The channel $t_R \to Z_T t_R$, shown for reference, is via the gauge coupling of nearly pure $B^0$, which is rather small. The other two channels $t_R \to h t_R,\ Z_L t_R$ are helicity-conserving scalar emissions and are of the ultra-collinear nature. The integrated splitting rates for all the above channels are of similar size: 
$\calP(t_R \to h t_L) \simeq \calP(t_R \to Z_L t_L) \approx 7.2\times 10^{-3}$, 
$\calP(t_R \to h t_R)$ and $\calP(t_R \to Z_T t_R) \approx 4.5\times 10^{-3}$, and $\calP(t_R \to Z_L t_R) \approx  2.3\times 10^{-3}$.  Notably, the rates for the ultra-collinear processes are concentrated toward smaller virtualities (and correspondingly smaller $k_T$s). Though the total splitting rate represented in Fig.~\ref{fig:top}(b) is only a few percent, the fact that top quarks are produced through strong interactions can lead to significant numbers of showered events at a hadron collider. On the other hand, the splitting rates to a Higgs boson are in sharp contrast to the much smaller rate for an on-shell top quark decay to a Higgs boson in the Standard Model \cite{Han:2013sea}, of the order $10^{-9}$.

In considering determination of the top-quark Yukawa coupling in the processes $t\bar t h/t\bar t Z$ at high energies~\cite{Plehn:2015cta}, the qualitative features shown here should be informative.

%---------------------------------------------------
\subsection{EW Showers initiated by neutral bosons}
%---------------------------------------------------

The neutral bosons $\gamma$, $Z_T$, $h$, and $Z_{\longit}$ contain rich physics at high energies, but their showering requires special treatment due to the presence of sizable interference effects. 

%.....................................
\subsubsection{$\gamma/Z_T$ coherence}
%.....................................

For the $\gamma/Z_T$ system, these interference effects have two aspects: the mass basis is misaligned with the gauge interaction basis, and even when viewed within the $B^0/W^0$ interaction basis, the existence of a preferred physical isospin basis for asymptotic states leads to observable coherence between $B^0$ and $W^0$ exchanges. A rigorous final-state shower must address both of these aspects simultaneously by using Sudakov evolution based on density matrices, as outlined in Section~\ref{sec:interference}. More specific details can be found in Appendix~\ref{app:split}.

\begin{figure}[t]
\begin{center}
\begin{subfigure}[t]{0.495\textwidth}
\includegraphics[width=210pt]{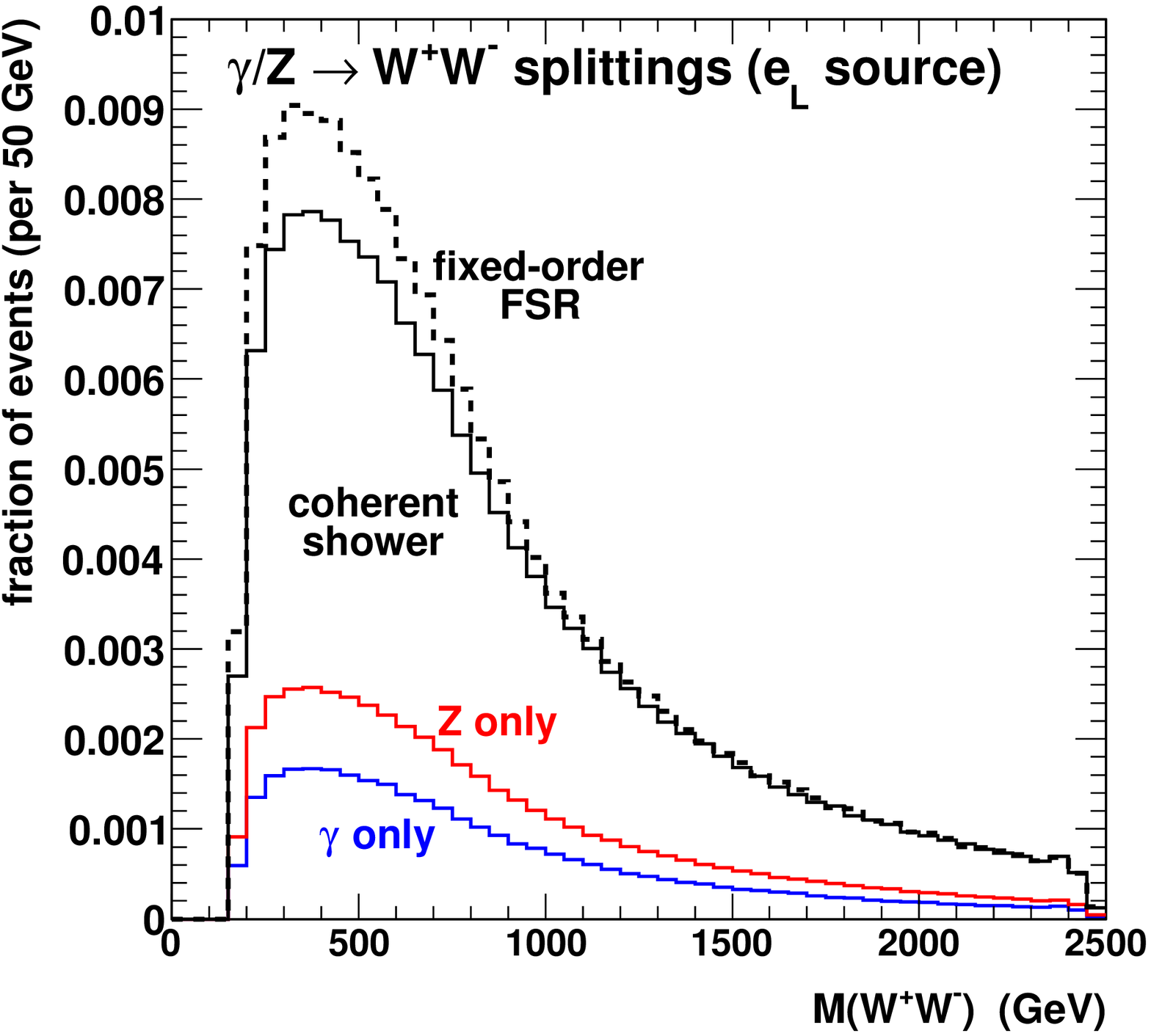}
\vspace{-0.5cm}\caption{}\end{subfigure}
\begin{subfigure}[t]{0.495\textwidth}
\includegraphics[width=210pt]{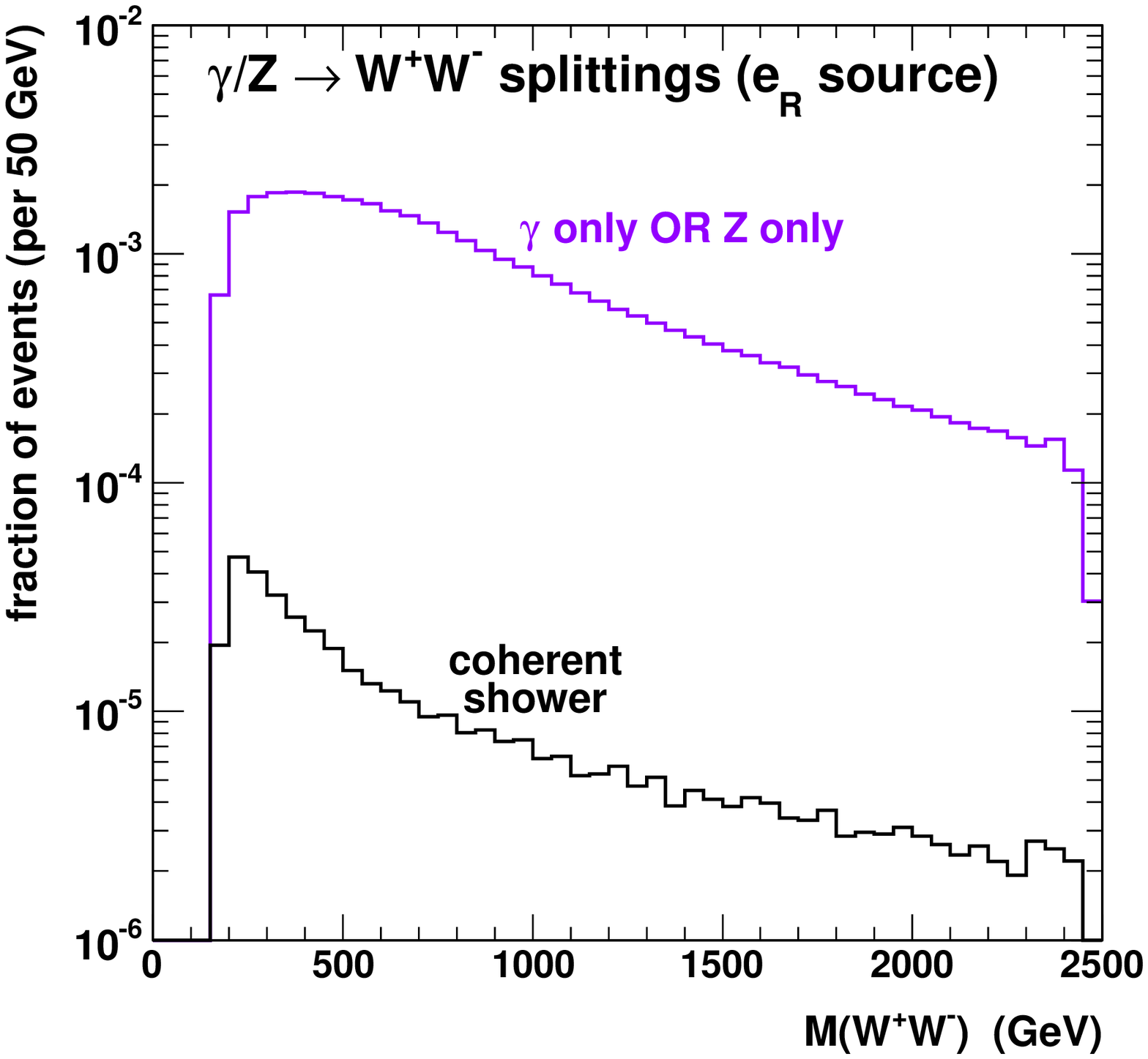}
\vspace{-0.5cm}\caption{}\end{subfigure}
\vspace{-0.6cm}
\end{center}
\caption[]{
Invariant mass distributions for $W^+W^-$ produced in the EW splitting of a 2.5 TeV $\gamma/Z$ neutral boson, initiated from 
(a) $e_L$ current with full coherent EW FSR (solid curve), fixed-order FSR (dashed curve), and the hypothetical incoherent $\gamma$ or $Z$ splittings (lower curves); 
(b) $e_R$ current with full coherent EW FSR (solid curve) and the hypothetical incoherent $\gamma$ or $Z$ splittings (upper curve). 
}
\label{fig:WW}
\end{figure}

As a simple example of the basis alignment issue, consider high energy showering of  neutral bosons $\gamma/Z \to W^+W^-$. 
%pair production of high-energy photons and/or $Z$ bosons at a polarized multi-TeV $e^+e^-$ collider. 
A naive treatment would shower the photon and $Z$ including the triple-vector processes $\gamma \to W^+ W^-$ and $Z \to W^+ W^-$.\footnote{Such a simplification has been made in~\cite{Ciafaloni:2010ti} for neutral bosons produced in dark matter annihilation.} However, depending on the gauge charges of the initial sources, the interference between these two mass-basis splitting channels can be ${\cal O}(1)$. In particular, for an energetic $\gamma/Z$ emitted from a right-handed chiral electron line, the $SU(2)_L$ content of the produced neutral gauge bosons is practically zero, suggesting a near absence of collinear $W^+W^-$ splittings in the final state. We explicitly compute these splittings assuming either an $e^-_L$ or $e^-_R$ source, which radiate off 2.5~TeV $\gamma/Z$ bosons (e.g., via neutral boson pair-production at a 5~TeV $e^-e^+$ collider). The results are displayed in Fig.~\ref{fig:WW}. Our full EW FSR treatment is labeled as ``coherent shower,'' contrasting with the hypothetical incoherent contributions from individual $\gamma$ or $Z$. For the $\gamma/Z$ produced by left-handed electrons in Fig.~\ref{fig:WW}(a), the $W^0$ fraction is prominent from the constructive interference between $\gamma/Z$, leading to a total splitting rate of roughly 15\% (black solid curve) and noticeable Sudakov distortions relative to a simple fixed-order splitting calculation (dashed curve). Fig.~\ref{fig:WW}(b) shows the result for a right-handed electron source, exhibiting the almost complete destructive interference between the $\gamma$ and $Z$ channels, due to the fact that the produced boson is nearly pure $B^0$ when viewed in gauge basis. The small residual rate at high virtualities is actually dominated by the unbroken-phase vector-to-scalar splitting $B^0 \to \phi^+\phi^- \sim W_{\longit}^+ W_{\longit}^-$. In our GEG approach, this is simply computed as a distinct process, rather than due to a delicate cancellation.

\begin{figure}[t]
\begin{center}
\begin{subfigure}[t]{0.495\textwidth}
\includegraphics[width=210pt]{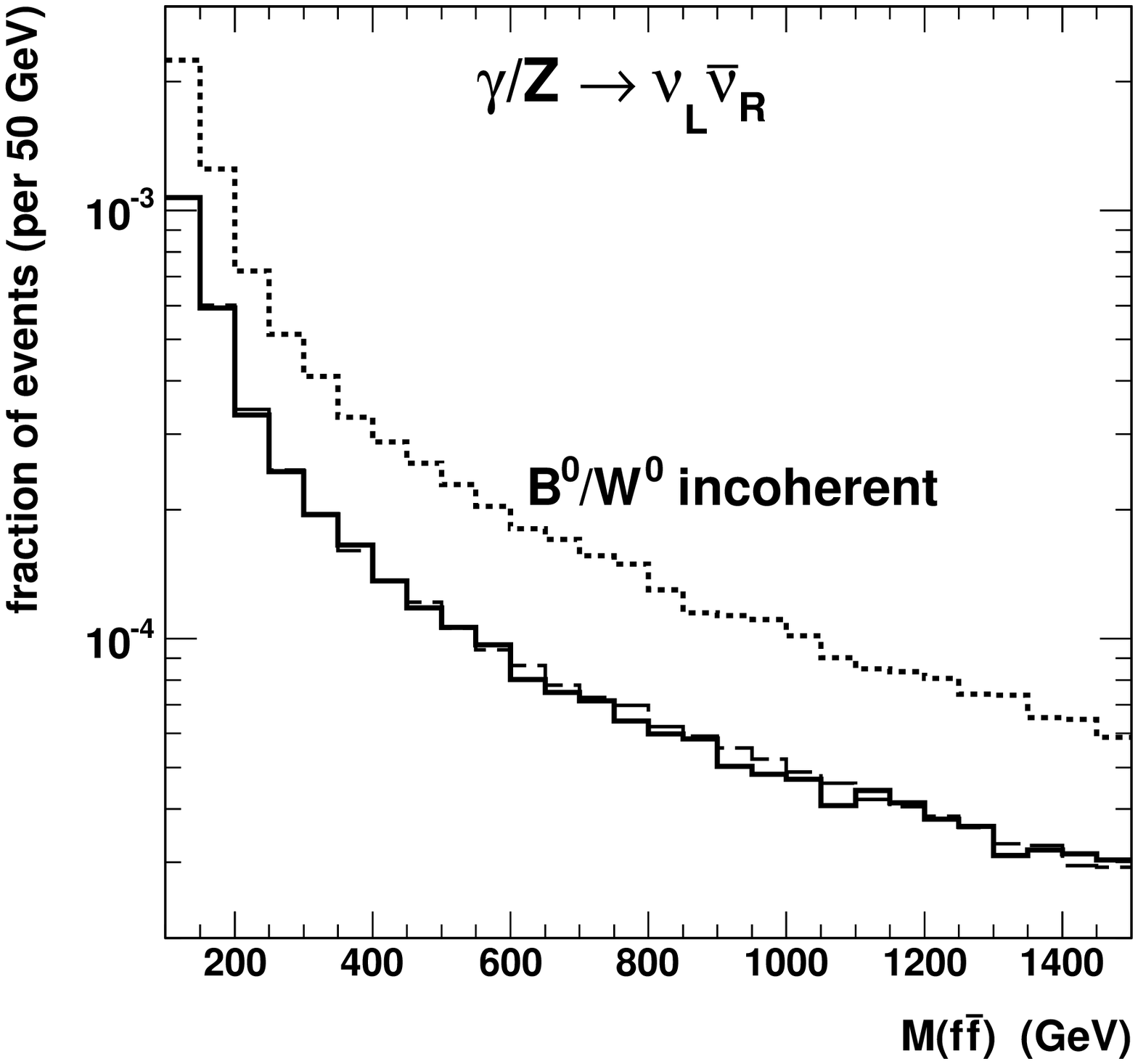}
\vspace{-0.5cm}\caption{}\end{subfigure}
\begin{subfigure}[t]{0.495\textwidth}
\includegraphics[width=210pt]{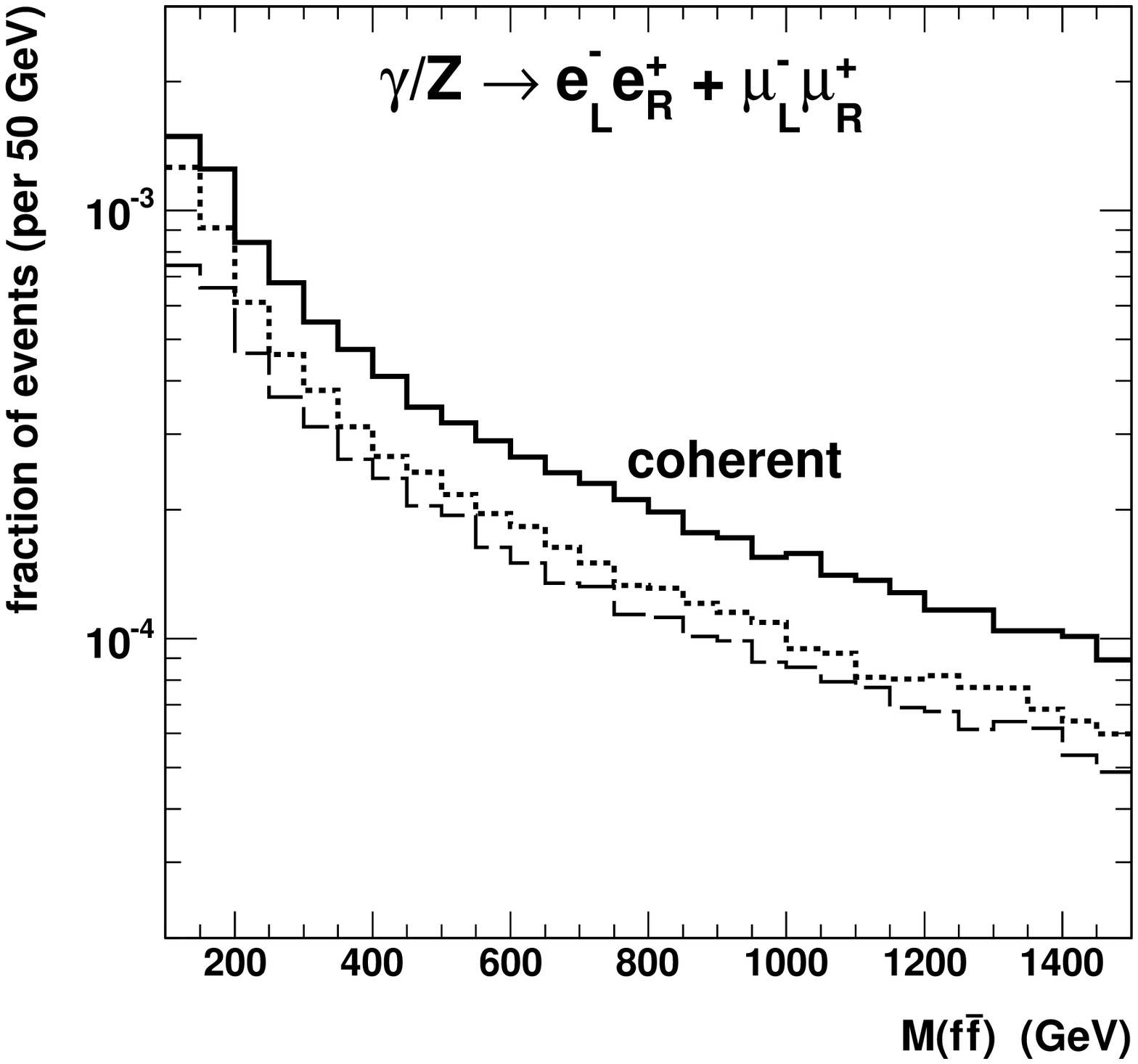}
\vspace{-0.5cm}\caption{}\end{subfigure}
\\
\vspace{0.2cm}
\begin{subfigure}[t]{0.495\textwidth}
\includegraphics[width=210pt]{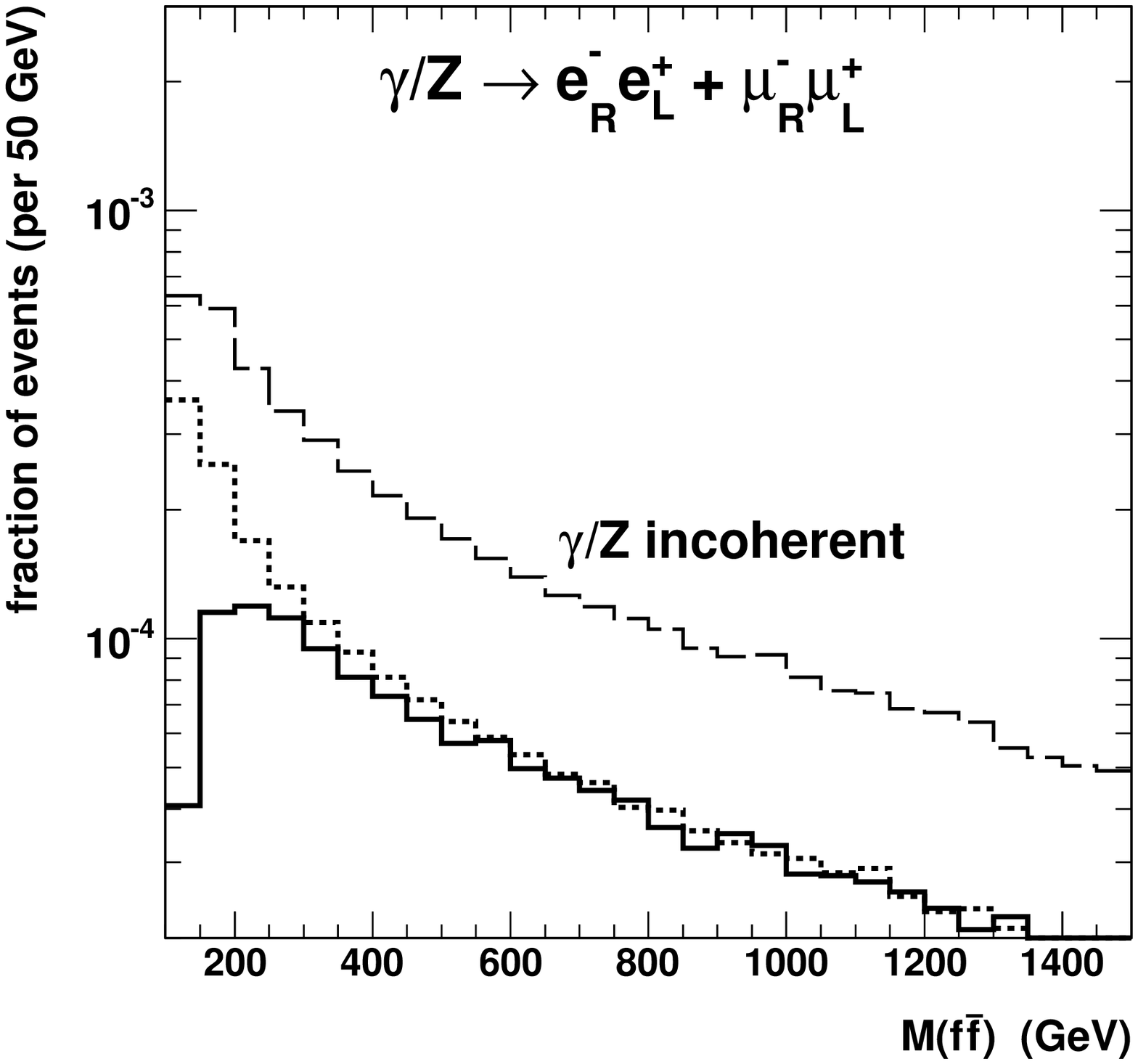}
\vspace{-0.5cm}\caption{}\end{subfigure}
\vspace{-0.6cm}
\end{center}
\caption[]{
Invariant mass distributions for fermion pairs produced in the EW splitting of a 2.5 TeV $\gamma/Z$ neutral boson, sourced by an $e_L$ current, for exclusive final states (a)  $\nu_L\bar \nu_R$, (b) $\ell_L^- \ell_R^+$, and (c) $\ell_R^- \ell_L^+$. Three treatments of the showering neutral bosons are: hypothetical incoherent $B^0/W^0$ (dotted), incoherent $\gamma/Z$ (dashed), and the full coherent EW evolution (solid).
 }
\label{fig:ff}
\end{figure}

Perhaps more subtle are the interference effects between different exclusive isospin channels. Naively, we might expect to be able to treat $SU(2)_L \times U(1)_Y$ in a manner analogous to $SU(3)_{\rm QCD} \times U(1)_{\rm EM}$, wherein the showers of the two gauge groups are simply run independently of one another. However, weak isospin quantum numbers are directly correlated with electric charge, and are therefore usually experimentally distinguishable. (Consider, e.g., the response of a detector to $e_L$ versus $\nu_L$.) Therefore, weak isospin cannot be summed/averaged like QCD color. As a consequence, observable rate asymmetries arise due to interference between the $SU(2)_L$ and $U(1)_Y$ gauge boson exchanges. Although a well-known effect, it has never been implemented in a parton shower framework. Again, we illustrate this by the splittings of 2.5 TeV $\gamma/Z$ neutral bosons, here produced off of a left-handed chiral electron line. This boson may subsequently split into a $\ell^- \ell^+$ or $\nu \bar\nu$ pair. The splitting rates with/without interference effects are shown in Fig.~\ref{fig:ff}.\footnote{For the incoherent sum over mass or gauge eigenstates, we have evolved separate samples starting from the individual pure-state density matrices, and recombined them according to their squared production amplitudes. Sudakov evolution of these density matrices has been switched off.} 
Besides the full coherent EW evolution (solid curves), two hypothetical incoherent treatments are shown using $\gamma$-$Z$ mass basis (dashed curves) and $B^0$-$W^0$ gauge basis (dotted curves). It is instructive to see that $Z\to \nu_L \bar \nu_R$ contribution alone gives the correct result as seen in Fig.~\ref{fig:ff}(a); $B^0 \to \ell^-_R \bar \ell^+_L$ alone also gives the correct result at high masses as seen in Fig.~\ref{fig:ff}(c), although it misses substantial destructive interference near $m_Z$ due to the unequal $\gamma$ and $Z$ masses; and $\ell^-_L \bar \ell^+_R$ would need coherent treatment in the whole kinematical regime as seen in Fig.~\ref{fig:ff}(b).
The same issues of course arise in hadron colliders, though the numerical impact is often smaller because of the healthy admixtures of $u/d$ flavors and LH/RH chiralities, as well as the charge-rearranging effects of hadronization. Nonetheless, we strongly advocate for a consistent treatment based on matrix-valued splitting functions and Sudakovs.

\begin{figure}[t]
\begin{center}
\begin{subfigure}[t]{0.495\textwidth}
\includegraphics[width=210pt]{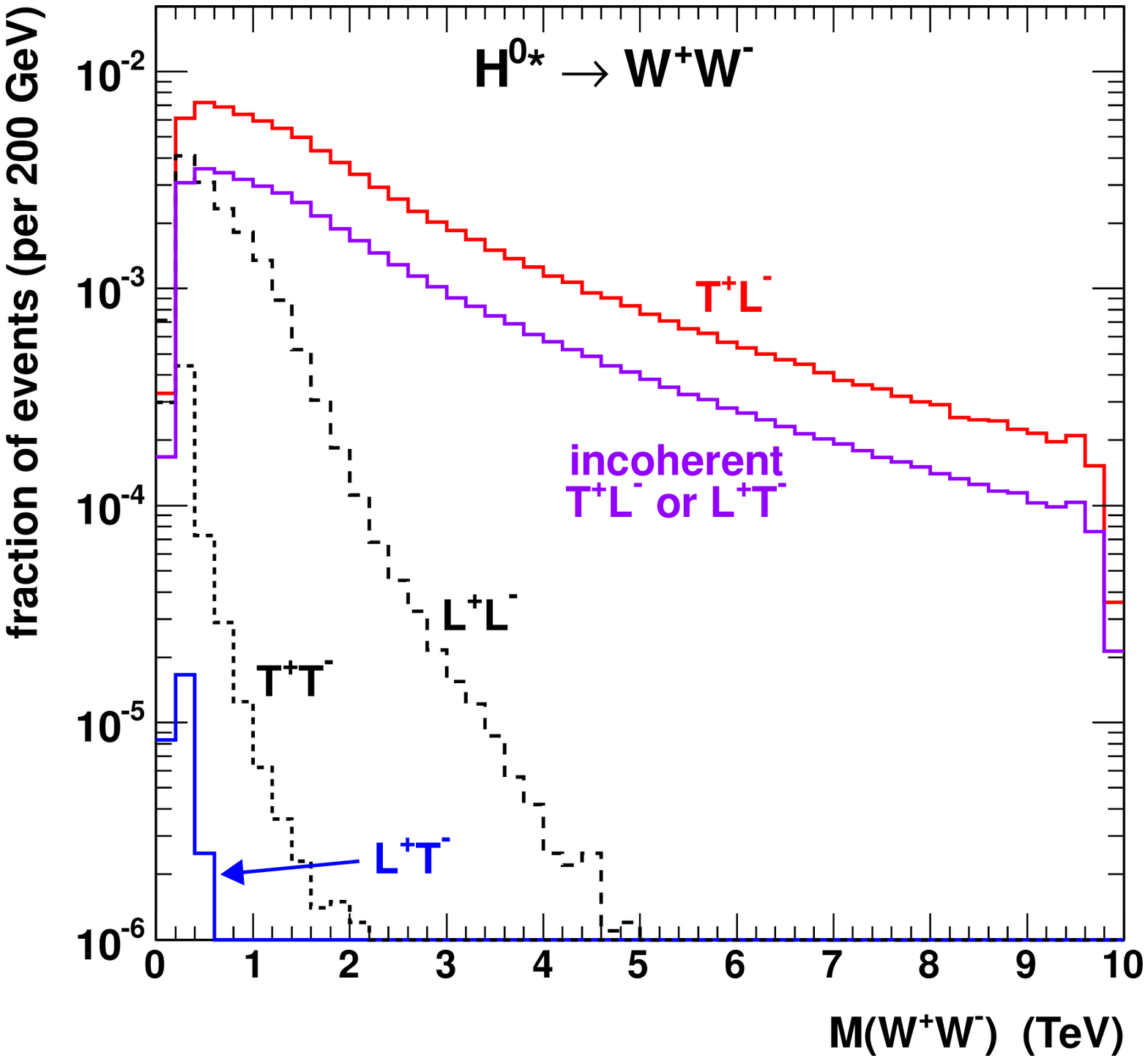}
\vspace{-0.5cm}\caption{}\end{subfigure}
\begin{subfigure}[t]{0.495\textwidth}
\includegraphics[width=210pt]{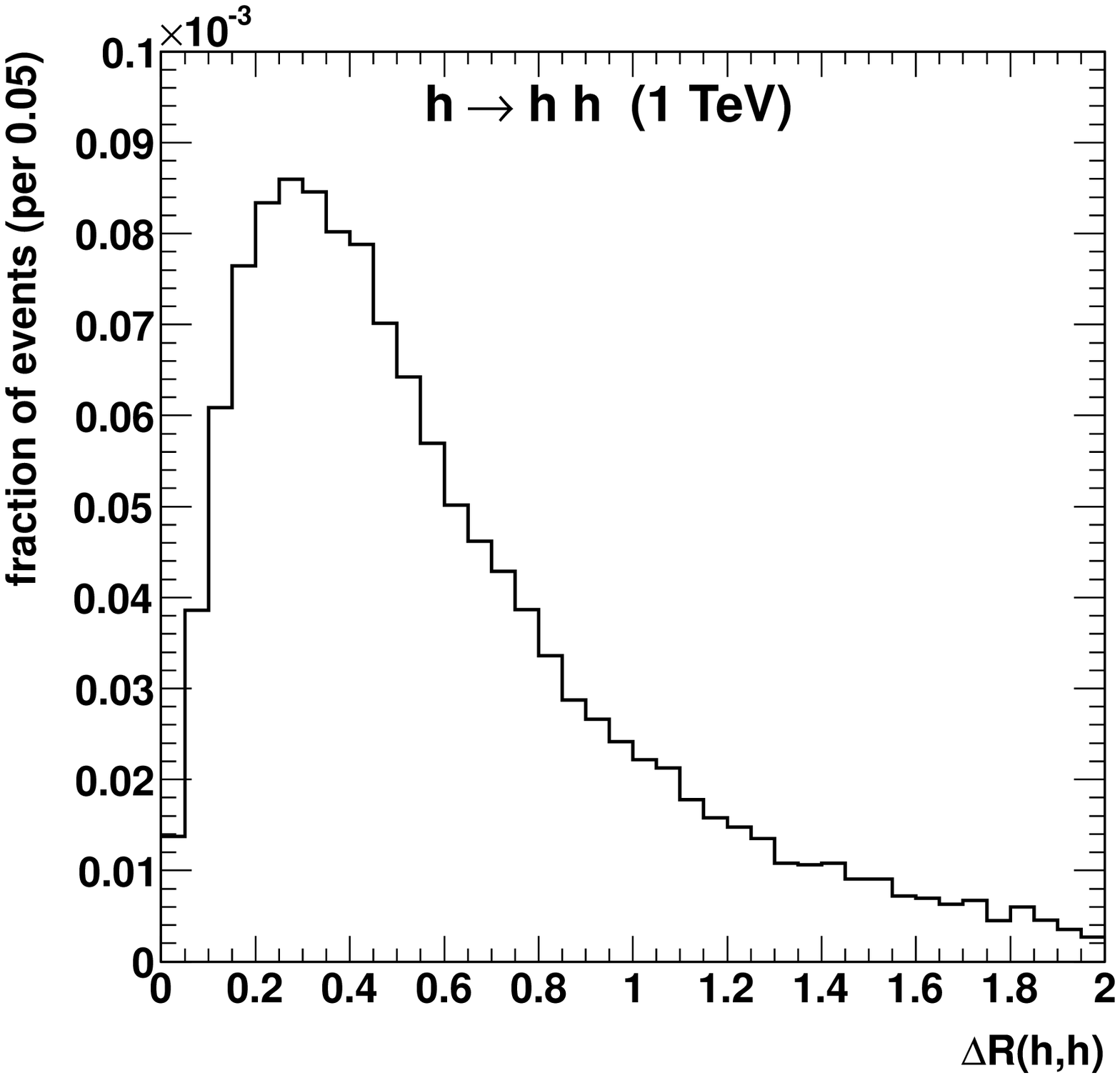}
\vspace{-0.5cm}\caption{}\end{subfigure}
\vspace{-0.6cm}
\end{center}
\caption[]{
(a) $W^+W^-$ invariant mass distributions from the EW splitting of a 10~TeV $h/Z_L\ (H^{0*}) \to W^+W^-$, labeled by the helicities and charges as $T^+L^-$, $L^+T^-$, $T^+T^-$, and $L^+L^-$. The ``incoherent $T^+L^-$ or $L^+T^-$'' curve shows the corresponding result from showering $h$ and $Z_{\longit}$ states independently. (b) Kinematic $\Delta R$ separation between the final state Higgs boson pair for the ultra-collinear splitting process $h \to hh$ from a 1~TeV Higgs boson. 
}
\label{fig:h_WW}
\end{figure}

%.....................................
\subsubsection{Higgs splitting and $h/Z_L$ coherence}
\label{sec:hzl}
%.....................................

Analogous interference effects also occur between the Higgs boson and longitudinal $Z$ boson. In the high-energy gauge theory, these appear as different components of the same complex scalar, and particular linear combinations carry a partially-conserved ``Higgs number'' that flows through the shower. As a simple illustration, consider high energy production of $W^{+}_T \to (h/Z_{\longit}) W^+_{\longit}$. 
The coherently mixed $h/Z_{\longit}$ carries Higgs number of $-1$, and corresponds to the ``anti-Higgs'' state $H^{0*}$. This state preferentially splits into $W^+_{T} W^-_{\longit}$ (or, equivalently, $W^+_{T} \phi^-$), as shown in the top curve of Fig.~\ref{fig:h_WW}(a), labeled by the $W$ helicities and charges as $T^+L^-$.
The charge conjugate state $W^+_{\longit} W^-_{T}$ (labeled $L^+T^-$) carries the opposite Higgs number and thus is highly suppressed. It arises only at low virtuality, mainly due to the Higgs-$Z$ mass difference. An incoherently-showered admixture of $h$ and $Z_{\longit}$ would instead distribute probability equally between these two different polarization channels, as shown in the figure with the middle curve. (A similar charge-polarization correlation also occurs in splittings to top quark pairs.)

The contributions from the other sub-leading ultra-collinear polarization channels are shown by curves labeled $L^+L^-$ and $T^+T^-$. Though not obvious from the virtuality distributions, we note that coherence effects also significantly influence these channels. In particular, the ultra-collinear splitting $H^{0*} \to W_{\longit}^+ W_{\longit}^-$ inherits the soft divergence from the regular gauge splitting $H^{0*} \to W_{T}^+ W_{\longit}^-$, but only in the limit as the $W_{\longit}^+$ becomes soft. Similarly for the CP-conjugate process. The individual $h$ and $Z_{\longit}$ incoherent showers, on the other hand, exhibit parts of the soft-singular behaviors of each of their $H^0$ and $H^{0*}$ components. See Table~\ref{tab:broken_scalar_splittings}.

As a final novel example of neutral boson showering, we consider the purely ultra-collinear splitting $h \to hh$. This proceeds through the Higgs cubic interaction that arises after EWSB, and it is the unique $1\to2$ splitting process in the SM that is strictly proportional to Higgs boson self-interaction $\lambda_h$. Isolating the $h$ component of a general energetic $h/Z_{\longit}$ state, the total splitting rate comes out to about $0.14\%$ for $E \gg m_h$. We illustrate in Fig.~\ref{fig:h_WW}(b) the kinematic distribution $\Delta R(h,h)$, for an example initial Higgs energy of 1~TeV. The distribution peaks at roughly $2m_h/E$, which in this example is close to $0.25$. Generally, the majority of the phase space for high-energy production $hhX$ for any $X$ becomes dominated by such collinear  configuration. While this ultra-collinear splitting process lacks any log-enhancements, integrating the splitting phase space yields a total rate relative to $hX$ that scales like $\lambda_h/16\pi^2$, whereas the non-collinear regions contribute a relative rate of order $\lambda_h^2/16\pi^2 \times v^2/E^2$. Therefore the ``collinear enhancement'' here is $E^2/\lambda_hv^2 \sim E^2/m_h^2$, rather than a conventional logarithm. Though the splitting rate is still quite small, for a 100~TeV $pp$ collider with 10's of~ab$^{-1}$ integrated luminosity, we expect thousands of such events arising from the (also novel) high-energy production process $qV_{\longit} \to q^{(\prime)}(h/Z_{\longit})$ at $p_T \sim 1$~TeV.
In future precision Higgs physics \cite{deFlorian:2016spz}, accurate description of such Higgs splittings could serve an interesting role.

%----------------------------------------------
\subsection{EW showers by a new heavy state: $W'$ example}  \label{sec:Wprime}
%----------------------------------------------

The possibility of multiple weak boson emissions in the same event, and indeed even from the same parent particle, leads us inevitably to start considering final-states in terms of ``weak jets'' rather than in terms of individual, well-separated EW-charged particles (possibly dressed with QCD and EM radiation). Besides altering the energy spectra of the particles emerging from a hard interaction, EW emissions can significantly alter the multiplicity and flavor structure of an event. In particular, this new feature could have major consequences for how a new physics signal would be detected and reconstructed. 

\begin{figure}[t]
\begin{center}
\begin{subfigure}[t]{0.495\textwidth}
\includegraphics[width=220pt]{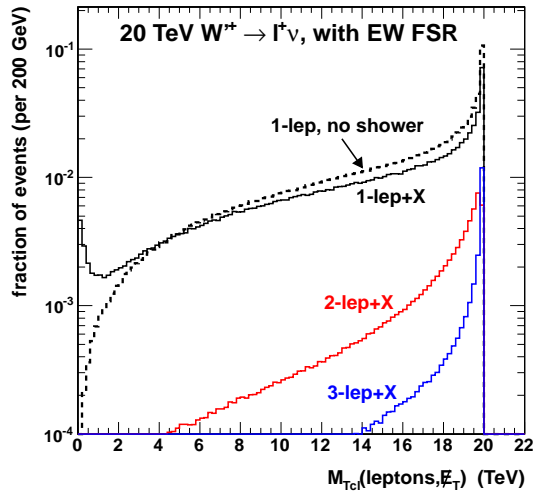}
\vspace{-0.5cm}\caption{}\end{subfigure}
\\
\vspace{0.2cm}
\begin{subfigure}[t]{0.495\textwidth}
\includegraphics[width=210pt]{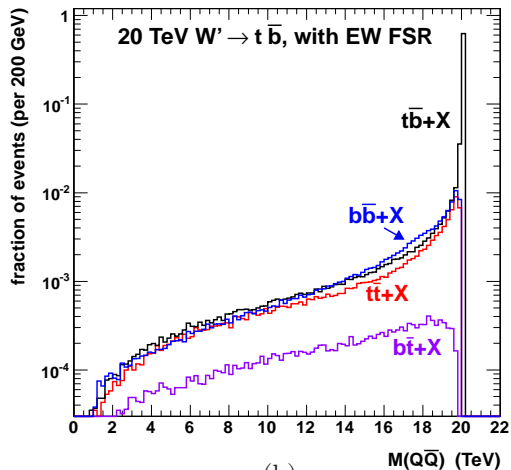}
\vspace{-0.5cm}\caption{}\end{subfigure}
\begin{subfigure}[t]{0.495\textwidth}
\includegraphics[width=210pt]{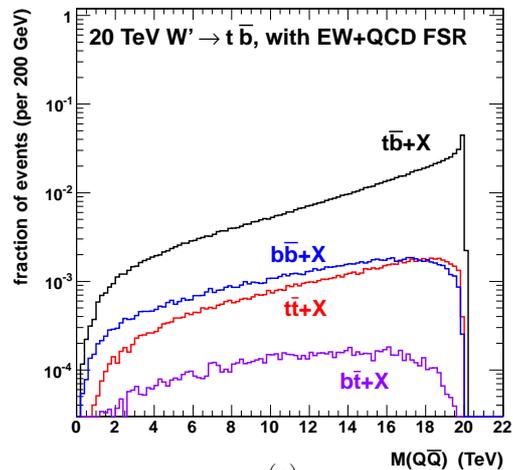} 
\vspace{-0.5cm}\caption{}\end{subfigure}
\vspace{-0.6cm}
\end{center}
\caption[]{
Showered events from 20 TeV $W'^+$ decays. 
(a)~$W'^+ \to \nu_L \ell_R^+$ cluster transverse mass distributions, running the full EW shower and breaking down the signal by inclusive lepton multiplicity (solid curves), as well as the uncorrected two-body decay result (dotted curve).
(b)~$W'^+ \to t_L \bar b_R$ quark-pair invariant mass distributions, running the full EW shower, and
(c)~combining EW and QCD showering. 
}
\label{fig:Wprime}
\end{figure}

While it is beyond the scope of this current paper to present detailed examples for physics beyond the SM in high energy collisions \cite{Morrissey:2009tf}, we study a simple case for illustration.
We consider the decay of a narrow heavy $W^{\prime +}$ resonance into $\nu_L\ell^+_R$, with a left-handed coupling and $M_{W'} \gg m_{W}$.
Nominally, the resonance is reconstructed from the charged lepton and the missing transverse momentum using the transverse mass variable $M_T(\ell,\met)$, which gives a Jacobian peak at $M_{W'}$. 
%(We ignore here continuum interference.) 
When multiple EW emissions are taken into account, various new flavor channels open up, as well as additional kinematic handles that can facilitate more accurate resonance reconstruction. For example, in~\cite{Hook:2014rka}, it was pointed out that collinear weak emissions $\nu \to Z\nu$ can effectively reveal the neutrino's direction-of-flight when the $Z$ decays visibly. For illustration here, we simply divide up the showered signal by inclusive lepton multiplicity, focusing on channels up to three charged leptons. Quarks and $\tau$-leptons may be present in the secondary $W/Z$ showering/decays, but are ignored here for simplicity. Within each lepton multiplicity channel, we approximately reconstruct the resonance using the ``cluster transverse mass'' variable $M_{T \rm cl}$, defined as \cite{Barger:1987re}
%the combined invariant mass of effective leptonic and invisible four-vectors. 
\begin{equation}
 M_{T \rm cl}^{2}
= \left(\sqrt{p_{T,\ell's}^2 + M_{\ell's}^2} + \met   \right)^2 
- (\vec p_{T,\ell's} + \vec{\met } )^2.
\label{eq:mc}
\end{equation}
%The former is simply the vector-sum of the leptons. The latter is defined by setting $p_x^{\rm inv} \equiv \mex$, $p_y^{\rm inv} \equiv \mey$, $m^{\rm inv} \equiv 0$, and $y^{\rm inv} \equiv y^{\rm lep}$ (equal rapidities). 
The result of this analysis is displayed in Fig~\ref{fig:Wprime}(a), taking $M_{W'} = 20$~TeV. 
%here is the legend. Black is 1-lep, red is 2-lep, blue is 3-lep. 
Solid curves are those from the nominal EW shower for 
$1\ell + X$, 
$2\ell + X$ and 
$3\ell + X$, where $X$ represents the rest of the particles in the event (mainly neutrinos and quarks).
The dotted line shows the result of the naive two-body decay calculation, without the parton shower. 
To focus on the weak-scale contributions, we have terminated the EW shower at a lower virtuality of 50~GeV.
The showering reduces the total visible rate within 10\% of the nominal peak by about $10\%$ due to the radiation. 
In this window, the relative contributions from 1-lepton, 2-lepton, and 3-lepton are respectively 0.81, 0.13 and 0.06. Although higher lepton multiplicities are rarer, their $M_{T \rm cl}$ distributions are also more sharply-peaked. It is also instructive to compare these predictions to those of a simple fixed-order splitting calculation, which captures the leading-log corrections but does not resum them. We find that this calculation predicts 9\% more 1-lepton events than the full EW shower in the near-peak region.

Like $e_L$ and $\nu_L$, left-handed top and bottom quarks live together in a weak isospin doublet, and can also convert into one another through soft/collinear $W^\pm$ emissions. Similar to the Bloch-Nordsieck violation effect discussed above for PDFs, the distinction between $t_L$- and $b_L$-jets therefore becomes somewhat blurred at high energy~\cite{Manohar:2014vxa}. This effect, which is double-log enhanced at fixed order, is automatically resummed in the parton shower. Consider again, as a simplified example, a narrow 20~TeV $W^{\prime +}$ resonance, this time decaying to $t_L \bar b_R$ of 10~TeV each in energy. The final flavor content of two heavy quarks should gradually average out.
%\footnote{A more realistic example would also include the $W^{\prime -}$ and $W^{\prime 0}$ production, which would already average-out the individual top and bottom production rates. While $W^{\prime +}$ production does tend to dominate at a proton collider, the full accounting is further complicated by the isospin-averaging effects within the initial-state.} 
We show in Fig.~\ref{fig:Wprime} the mass spectrum of the two-quark system resulting from the decay plus EW parton shower, individually in $t \bar b$, $b \bar b$, $t \bar t$, and $b \bar t$ channels. (For this purpose, the threshold between the ``shower'' and ``decay'' of a top quark is set to $m_t + 10\Gamma_t$.) Respectively, these are dominated by unshowered events, events with a single $t \to W^+ b$ splitting, events with a single $\bar b \to W^- \bar t$ splitting, and events with one of each such splitting. The relative rates of the four channels are about 0.77, 0.09, 0.12, and 0.015. Within 10\% of the $W'$ mass peak, the nominal $t\bar b$ signal would be reduced by almost 30\% from purely electroweak effects. Of course, this observation invites ``weak jet'' reconstructions that add back in the emitted gauge and scalar particles, though inferring the resonance's charge becomes somewhat more complicated.

Finally, we can consider the interplay of EW and QCD radiation, which is shown in Fig.~\ref{fig:Wprime}(c) for the mass spectra of the quarks when $t\to gt$ and $b\to gb$ emissions are also turned on. Again the shower is terminated at 50~GeV virtuality to focus on effects at and above the EW scale. The full Standard Model showering leads to dramatic distortions in both mass and flavor distributions. Now the $W'$ mass could be more accurately reconstructed by adding back in both-the EW {\it and} QCD radiation, which practically may overlap heavily since emitted weak bosons dominantly decay hadronically.

%%%%%%%%%%%%%%%%%%%%%%
\section{Summary and Conclusions}
\label{sec:conclusions}
%%%%%%%%%%%%%%%%%%%%%%

At very high energies, far above the electroweak scale, the full gauge and Yukawa structure of the Standard Model emerges, leading to an extremely rich set of parton showering phenomena. As this full SM parton shower evolves down in scale, it ultimately passes back through the electroweak scale. There it encounters additional showering phenomena that arise uniquely from EWSB, and then finally transitions back into the $SU(3)_{\rm QCD} \times U(1)_{\rm EM}$ gauge showers familiar from the past several decades of theoretical and experimental work.

With an eye towards experiments in the next decade and beyond, in this paper we have attempted lay out the above picture of electroweak showering in a more comprehensive manner. We have systematically presented the electroweak collinear splitting functions in the SM in the $SU(2)_L \times U(1)_Y$ symmetric phase as well as in the broken phase after electroweak symmetry breaking. We discussed their general features in the collinear and soft-collinear regimes and identified the general class of EWSB contributions that are uniquely ``ultra-collinear,'' namely localized at $k_T \sim v$ with appreciable rates, but otherwise absent in conventional showering regimes. Effects of the ultra-collinear part of the shower include counter-intuitive ``violations'' of the Goldstone-boson Equivalence Theorem. We have also identified a convenient way to isolate EWSB effects within the shower, especially by disentangling contributions from gauge bosons and Goldstone bosons at high energies, using a novel gauge choice which we call Goldstone Equivalence Gauge (GEG). We further implemented the full EW shower in a numerical monte carlo, and showed a number of new results regarding its subtleties and practical impact in SM processes and beyond.

Our main observations and results are as follows:\\
%\noindent
$\bullet$ 
The splitting functions of the unbroken $\sm$ theory, presented in Sec.~\ref{sec:unbroken}, typically act as the leading contributions to showering processes at energies far above the EW scale. \\
$\bullet$
At splitting scales $k_T \sim gv$ and $yv$, the unbroken splitting functions become regulated and the new ultra-collinear splitting functions arising from EWSB appear, as presented in Sec.~\ref{sec:broken}. The latter is the analogue of ``higher-twist'' terms in terms of the formal power counting. While they do not contribute to the leading logarithmic evolution, numerically they can be larger than the unbroken contributions at low $k_T$, and in some cases can also account for a sizable fraction of the integrated splitting rates. \\
$\bullet$ 
Goldstone-boson equivalence ceases to hold in the ultra-collinear regime, allowing, e.g., for emission of relativistic longitudinal bosons from massless fermions. This effect is generalized here to all splitting functions in the SM, often involving nontrivial interplays of EWSB effects in gauge, Yukawa, and scalar couplings. \\
$\bullet$ 
We introduced the Goldstone Equivalence Gauge (as detailed in Appendix~\ref{sec:gauge}) that practically as well as conceptually disentangles the effects from the Goldstone bosons and the gauge fields. Utilization of this gauge choice makes the GET transparent {\it and} organizes its leading violations in a straightforward diagrammatic expansion (see Appendix~\ref{sec:FeynmanRules}). The concept of a ``nearly on-shell'' gauge/Goldstone boson as an intermediate state in the shower also becomes unambiguous. \\
$\bullet$ 
We implemented a practical EW showering scheme based on the calculated collinear and ultra-collinear splitting kernels in a Sudakov formalism. As discussed in Sec.~\ref{sec:split}, some additional novel features in the implementation include matching between showering and resonance decay, kinematic back-reaction corrections for multiple emissions of massive particles, and a density matrix treatment for the mixed-state evolution of neutral bosons ($\gamma/Z/h$). Our treatment of EW showering is fully self-contained, and far beyond the currently existing monte carlo simulation packages. \\ 
$\bullet$
We applied the EW showering formalism to a number of important physical processes at high energies. They include: electroweak partons in PDFs as the basis for vector-boson-fusion; EW FSR as a leading source of multiple gauge boson production, with splitting probabilities at the level of 10s of percent; EW showers initiated by top quarks, including Higgs bosons in the final state; and showers initiated by neutral bosons $\gamma/Z/h$, for which care must be taken to obtain meaningful results. The emergence of ``weak jets'' from high-energy new physics processes was illustrated using a heavy $W'$ as an example.

%Our work leaves open a number of interesting directions for study, some of which will appear in forthcoming publications~\cite{EWshower}. One useful supplement to our collinear showering framework would be a more systematic inclusion of soft wide-angle corrections. Along the same lines, it would also be fruitful to consider how to construct a matching scheme into full matrix element calculations of hard wide-angle regions. Both tasks are made nontrivial by the presence of soft $W^\pm$ exchanges between different legs of an event and the coherence effects that occur in neutral boson exchanges. We also point out the limitation that our shower often cannot meaningfully be run on the output of present-day hard event generators, since these immediately collapse neutral bosons into mass basis (effectively assuming trivial Sudakov evolution at the outset). A more flexible output format that includes information on production density matrices for these states is therefore recommended.

In summary, we have derived the collinear splitting functions for the Standard Model electroweak sector, including the massive fermions, gauge bosons, and the Higgs boson, and implemented a collinear showering scheme in the Sudakov formalism for all SM particles at high energies. We have highlighted many novel features and the necessity to include them for physics explorations in and beyond the SM at high energies, including any physics at future colliders, as well as other processes in high energy regimes much above the electroweak scale. 

While our paper has explored collinear EW showering at a new level of detail compared to earlier works, it leaves open several interesting issues that we intend to address in future publications~\cite{EWshower}. One such issue is a more comprehensive picture of PDF evolution, folding together QCD and EW effects into a unified set of DGLAP equations that incorporate both quantum coherence effects and ultra-collinear effects, and allowing for a complete QCD+EW ISR showering scheme. Implications for the exclusive structure of multi-TeV VBF events would be particularly interesting to study. We also intend to address issues related to soft wide-angle EW exchanges, which lead to quantum entanglements between the isospins of the beams and the final state at NLL. These entanglements represent a formally subleading aspect of the notorious Bloch-Nordsieck violation, which naively implies double- and single-logarithmic divergences in inclusive cross sections sourced by isospin-exclusive initial states. The collinear formalism developed here would allow for simple LL resummation of the soft-collinear, double-logarithmic contributions. (See, e.g., Section~\ref{sec:Wprime} for simple examples in the final-state shower.) Capturing and resumming the remaining single-log, quantum-coherent contributions, as well as motivating factorization of the initial state at NLL, requires a more advanced formalism that uses the language of quantum ensembles.

%\Tao{We conclude with the following disclaimer: As troublesome as it seems to be due to the BN violation with bare gauge charges, the factorization and resummation which the current treatments have been based up on can be shown to work at LL with suitably-defined collinear evolution and shower scheme . The implementation can be extended to NLL in a natural way, allowing for a complete picture of evolution at leading order, fully exclusive in gauge charges and in kinematics. This is best illustrated by forming the entangled initial state/final state ensembles, which will be presented in a comprehensive manor in a separate publication \cite{EWshower}.}

\acknowledgments{We thank Matthew Baumgart, Andy Cohen, John Collins, Kaoru Hagiwara, Satyanarayan Mukhopadhyay, Juan Rojo, Ira Rothstein, Joshua Sayre, Varun Vaidya, and Susanne Westhoff for useful discussions.  Work supported in part by DoE grant No.~DE-FG02-95ER40896 and in part by PITT PACC.} 

\appendix

%%%%%%%%%%%%%%%%%%%%%%%%%%%%%%%%%%%%%%%%%%%%%
\section{Goldstone Equivalence}
\label{sec:gauge}
%%%%%%%%%%%%%%%%%%%%%%%%%%%%%%%%%%%%%%%%%%%%%

As discussed in Section~\ref{sec:broken}, there are considerable conceptual and technical complications in handling processes involving longitudinal gauge bosons at high energies. 
The behavior of longitudinal gauge bosons in high energy scattering and showering, both as off-shell intermediate states and as external particles participating in collinear splittings, becomes most transparent in ``physical'' non-covariant gauges where gauge-Goldstone mixing is left explicit, and the Goldstone fields remains capable of interpolating external particles~\cite{Beenakker:2001kf,Dams:2004vi,Srivastava:2002mw} (see also~\cite{Wulzer:2013mza}). 
We propose a particularly convenient physical gauge dubbed ``Goldstone Equivalence Gauge'' (GEG), wherein the emergence of Goldstone equivalence and its leading violations are manifest and easily calculable at tree-level, while maintaining some residual Lorentz symmetry and avoiding unphysical gauge poles. In this Appendix, we work out the details of this gauge.

GEG is essentially a hybrid of Coulomb and light-cone gauges. It employs a light-like gauge reference four-vector that rotates with momentum\footnote{$k^0$ can be negative for general off-shell modes. The given parametrization of $n^\mu$ is not unique. For example, (sign$(k^0),-\hat{k}$) and (sign$(k^0)|\vec{k}|,-\vec{k}$) also serve the same purpose.}
\beq 
n^\mu(k) \,=\, (n^0(k), \vec{n}(k)) \,\equiv\, (1, -\hat k \; {\rm sign}(k^0)), \qquad n^\mu n_\mu=0 . \label{eq:nmu2}
\eeq
Representing a generic gauge adjoint component of a vector field by $W^\mu$, we decompose the gauge degrees of freedom as the components of $W_n\ (W_{\bar n})$ aligned (anti-aligned) with $n^\mu$ and the two $\pm1$ helicity (or ``$xy$'') transverse modes, collectively $W_T$:
\beq
W^\mu(k) \,=\, W_T(k) \ \epsilon_T^\mu(k) \,+\, W_n(k) \ \epsilon_n^\mu(k) \,+\, W_{\bar n}(k) \ \epsilon_{\bar n}^\mu(k) \, , \label{eq:Wexpansion}
\eeq
with $\bar n^\mu \equiv (1,+\hat k \, {\rm sign}(k^0))$. 
Since $W^\mu$ is a real vector field here, we have chosen the above definition such that $n^\mu(k)^* = n^\mu(-k)$. 
Introducing the gauge-fixing Lagrangian in momentum space as
\beq\label{eq:gauge-fixing}
{\cal L}_{\rm fix} \,=\,  -{1\over 2\xi} \big(n(k)\cdot W(k)\big)\big( n(k)\cdot W(-k)\big), \quad\quad (\xi \to 0),
\eeq
the large light-like component of the on-shell longitudinal polarization, $W_{\bar n}$ field, ceases to propagate because of its infinite ``mass'' $1/\xi$. This is the key feature for GEG by design. We are left with three physical degrees of freedom that can propagate. 
It is interesting to note that GEG respects the rotational symmetry under $SO(3)$ by construction. The surviving polarization states are also invariant (up to a possible rescaling) under boosts collinear to $\vec k$. 

Incorporating EWSB, neither the gauge boson mass nor the would-be-Goldstone field $\phi$ are folded into the gauge-fixing procedure. The normalization of $W_n$ and its associated polarization vector $\epsilon_n^\mu \propto n^\mu$ can be chosen such that $W_n$ will interpolate external particles with unit amplitude:
\beq
\epsilon_n^\mu(k) \,\equiv\, \frac{-\sqrt{|k^2|}}{n(k)\cdot k}\ n^\mu(k) \,\,\overset{\overset{\text{\rm \footnotesize on-shell}}{}}{\to}\,\, \frac{m_W}{E+|\vec k|}\left(-1,\hat k\right).  \label{eq:epsilon_n}
\eeq
This polarization vector is what remains of the standard longitudinal polarization $\epsilon_{\longit}^\mu(k)$ upon subtraction of the Goldstone-equivalence term (scalarization term) $k^\mu/m_W$. Preserving Hermiticity of the $W_n$ field also necessitates introduction of a factor of $i$ into the polarization vector, such that $(i\epsilon_n^\mu(k))^* = i\epsilon_n^\mu(-k)$. This will also conveniently synchronize the phase of states created by the $W_n$ field and the $\phi$ field.\footnote{When working in a complex gauge basis, as for $W^\pm$, these polarization phase factors become simply $\pm 1$. In all cases, care must be taken to rigorously define the orientation of momentum flows when computing amplitudes, since $\epsilon_n^\mu(-k) = -\epsilon_n^\mu(k)$, and the sign is often needed to determine the relative phase between gauge-interpolated and Goldstone-interpolated diagrams.} 
Accounting for the gauge-Goldstone mixing term, the quadratic Lagrangian can then be expressed as
\beqa
\mathcal{L}_T(k)+{\rm h.c.} & \,=\, &   W_T(k) \big(k^2 - m^2\big) W_T(-k) \nonumber \\
\mathcal{L}_{n\phi}(k)+{\rm h.c.}  & \,=\, & 
\begin{bmatrix}
W_n(k) & \phi(k) )
\end{bmatrix} 
\begin{pmatrix}
 |k^2|  & -m_W\sqrt{|k^2|}  \\
 -m_W\sqrt{|k^2|}  & k^2 
\end{pmatrix}
\begin{bmatrix}
 W_n(-k) \\
 \phi(-k)
\end{bmatrix}
\eeqa
Inverting yields the propagators
\beqa
\big< W_T(k) W_T(-k) \big>  & \,=\, & \frac{i}{k^2-m_W^2},\qquad
\big< W_n(k) W_n(-k) \big>   \,=\,  \frac{i}{k^2-m_W^2}\,{\rm sign}(k^2) , \nonumber \\
\big< \phi(k) \phi(-k) \big> & \,=\, & \frac{i}{k^2-m_W^2},\quad 
\big< W_n(k) \phi(-k) \big>  \,=\,  \frac{i}{k^2-m_W^2}\frac{m_W}{\sqrt{|k^2|}} \, .
\label{app:prop}
\eeqa
These propagators are naively fully Lorentz-invariant, though choosing a polarization basis in the first place has anyway tied us to a specific frame. They share a unique, common pole at $k^2 = m_W^2$ with residue +1. The mixed $W_n$ and $\phi$ fields interpolate the same particle: the ``longitudinal gauge boson'' or ``Goldstone boson,'' depending on perspective.\footnote{This may be seen in various ways. Probably the most intuitive is to incorporate the $W$'s decay into massless fermions, as actually occurs in the SM. A $W_n$/$\phi$ created from some hard process would then coherently propagate and decay into the same final-state with the same amplitude.} Note that the apparent spurious pole at $k^2 = 0$ in the mixed propagator is purely an artifact of our momentum-dependent field normalization, and does not lead to light-like gauge poles in complete Feynman diagrams.\footnote{Such a pole arises in Lorenz-Landau gauge, where gauge-fixing on the light-cone is incomplete. Generally, gauge poles will cancel between gauge-exchange and Goldstone-exchange diagrams, but can lead to spurious singularities in individual diagrams. In GEG, the only such gauge pole occurs at the zero-mode, $k^\mu=0$, and only in the mixed gauge-Goldstone propagator. The loop-level and renormalization properties of this gauge could be interesting to study, assuming that there are no obvious analytic obstructions do doing so. However, as we here confine ourselves to tree-level, we save this topic for future work.}

Goldstone boson equivalence in the high-energy limit now emerges trivially, diagram-by-diagram. For a process where $|k^2| \gg m_W^2$ for all internal gauge/Goldstone lines and $E \gg m_W$ for all external bosons, the mixed propagators and $\epsilon_n$ factors scale away, leaving over only the Goldstone contributions. In addition, since there are no terms that go like $k/m_W$ or $E/m_W$, power-counting of corrections $\propto m_W$ becomes straightforward at the level of individual Feynman diagrams. Upon introduction of complete fermion and scalar sectors, we may generalize to counting VEV factors associated with arbitrary masses and interactions introduced by spontaneous symmetry breaking. Some simple examples for splitting calculations are given in Appendix~\ref{sec:FeynmanRules}.

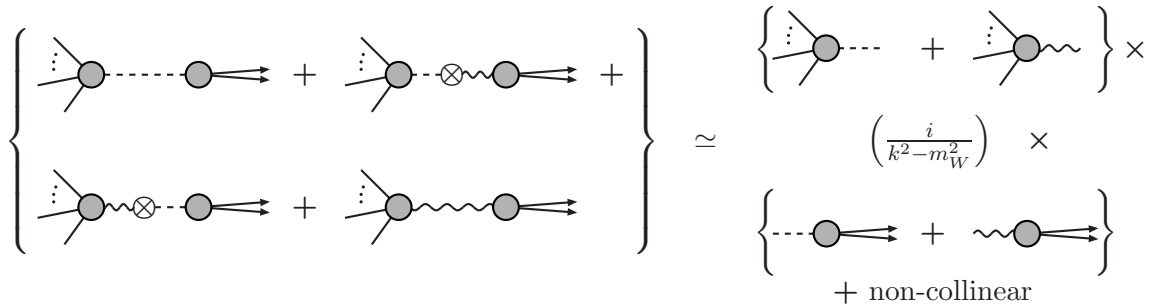
\begin{figure}[t]
\begin{center}
\begin{picture}(500,75)(0,0)
\SetColor{Black} \SetWidth{2}
\SetScale{0.4}
\Text(10,50)[]{$\left\{  \begin{array}{c} \\ \\  \\  \\  \\ \\ \end{array} \right. $}
%
%  gauge-Goldstone
\SetOffset(15,5)
\Line(25,15)(50,50)
\Line(0,40)(50,50)
\Line(15,85)(50,50)
\Vertex(15,55){1.5}
\Vertex(15,62){1.5}
\Vertex(17,69){1.5}
\Photon(50,50)(90,50){3}{3}
\Text(40,20)[]{$\boldsymbol \otimes$}
\DashLine(109,50)(150,50){6}
\LongArrow(150,50)(215,45)
\LongArrow(150,50)(215,55)
\GCirc(50,50){12}{0.7}
\GCirc(150,50){12}{0.7}
\Text(100,20)[]{\bf +}
%
%  gauge-gauge
\SetOffset(130,5)
\Line(25,15)(50,50)
\Line(0,40)(50,50)
\Line(15,85)(50,50)
\Vertex(15,55){1.5}
\Vertex(15,62){1.5}
\Vertex(17,69){1.5}
\Photon(50,50)(150,50){3}{5}
\LongArrow(150,50)(215,45)
\LongArrow(150,50)(215,55)
\GCirc(50,50){12}{0.7}
\GCirc(150,50){12}{0.7}
%
%  Goldstone-Goldstone
\SetOffset(15,55)
\Line(25,15)(50,50)
\Line(0,40)(50,50)
\Line(15,85)(50,50)
\Vertex(15,55){1.5}
\Vertex(15,62){1.5}
\Vertex(17,69){1.5}
\DashLine(50,50)(150,50){6}
\LongArrow(150,50)(215,45)
\LongArrow(150,50)(215,55)
\GCirc(50,50){12}{0.7}
\GCirc(150,50){12}{0.7}
\Text(100,20)[]{\bf +}
%
%  Goldstone-gauge
\SetOffset(130,55)
\Line(25,15)(50,50)
\Line(0,40)(50,50)
\Line(15,85)(50,50)
\Vertex(15,55){1.5}
\Vertex(15,62){1.5}
\Vertex(17,69){1.5}
\DashLine(50,50)(90,50){6}
\Text(40,20)[]{$\boldsymbol \otimes$}
\Photon(109,50)(150,50){3}{3}
\LongArrow(150,50)(215,45)
\LongArrow(150,50)(215,55)
\GCirc(50,50){12}{0.7}
\GCirc(150,50){12}{0.7}
\Text(100,20)[]{\bf +}
\SetOffset(-5,0)
\Text(245,50)[]{$\left.  \begin{array}{c} \\ \\  \\  \\  \\ \\ \end{array} \right\} $}
\Text(270,50)[]{$\simeq$}
%
%  Goldstone HARD
\SetOffset(290,65)
\Line(25,15)(50,50)
\Line(0,40)(50,50)
\Line(15,85)(50,50)
\Vertex(15,55){1.5}
\Vertex(15,62){1.5}
\Vertex(17,69){1.5}
\DashLine(50,50)(100,50){6}
\GCirc(50,50){12}{0.7}
\Text(60,20)[]{\bf +}
%
%  Gauge HARD
\SetOffset(365,65)
\Line(25,15)(50,50)
\Line(0,40)(50,50)
\Line(15,85)(50,50)
\Vertex(15,55){1.5}
\Vertex(15,62){1.5}
\Vertex(17,69){1.5}
\Photon(50,50)(100,50){3}{3}
\GCirc(50,50){12}{0.7}
%
%  Goldstone split
\SetOffset(250,-5)
\DashLine(100,50)(150,50){6}
\LongArrow(150,50)(215,45)
\LongArrow(150,50)(215,55)
\GCirc(150,50){12}{0.7}
\Text(100,20)[]{\bf +}
%
%  gauge split
\SetOffset(325,-5)
\Photon(100,50)(150,50){3}{3}
\LongArrow(150,50)(215,45)
\LongArrow(150,50)(215,55)
\GCirc(150,50){12}{0.7}
\SetOffset(290,84.5) \Text(0,0)[]{$\left\{  \begin{array}{c} \\ \\   \end{array} \right.  $}
\SetOffset(405,84.5) \Text(0,0)[l]{$\left.   \begin{array}{c} \\ \\   \end{array} \right\}  {\boldsymbol\times}$}
\SetOffset(360,50) \Text(0,0)[]{$\left(\frac{i}{k^2-m_W^2}\right) \quad {\boldsymbol\times}$}
\SetOffset(290,15.5) \Text(0,0)[]{$\left\{  \begin{array}{c} \\ \\   \end{array} \right.  $}
\SetOffset(405,15.5) \Text(0,0)[l]{$\left.   \begin{array}{c} \\ \\   \end{array} \right\} $}
\SetOffset(350,-7) \Text(0,0)[]{{\bf +} non-collinear}
\end{picture}
\end{center}
\caption[]{Schematic tree-level collinear factorization for an arbitrary process with a splitting Goldstone/longitudinal in the final state.}
\label{fig:factorizationEquation}
\end{figure}

We can also see how this gauge choice facilitates a factorized picture of longitudinal gauge/Goldstone boson production and splitting in the parton shower, beyond the simple Goldstone-equivalent picture at zeroth-order in the VEV. Fig.~\ref{fig:factorizationEquation} illustrates how this works schematically in a final-state shower. A generic hard process produces an off-shell gauge/Goldstone boson of virtuality $k^2$ with $m_W^2 \ll k^2 \ll E^2$, and this boson subsequently splits. There are four contributing classes of diagrams, corresponding to the four possible propagator exchanges between the production and splitting processes. We would like to approximate this as an on-shell production amplitude multiplied by a universal splitting amplitude. The decomposition is trivial for the leading pure Goldstone exchange diagram, but the other, subleading diagrams involve interplays between the propagators and the off-shell polarization vectors $\epsilon_n^\mu \propto (\sqrt{k^2}/E)n^\mu$. For the mixed diagrams, the propagator factor $m_W/\sqrt{k^2}$ can be combined with the polarization factor $\sqrt{k^2}/E$ to yield an approximate on-shell polarization proportional to $m_W/E$. Assuming that there is no large back-reaction in the hard production matrix element (at least to $O(m_W^2)$), contracting with the rescaled off-shell polarization approximately reproduces the on-shell hard process. For the mixed diagram where the gauge field contracts with the splitting process, this decomposition would simply instruct us to compute the splitting amplitude with an effective on-shell $\epsilon_n$. The pure gauge exchange does not immediately fit this pattern, but it can be separated into two pieces: $1/(k^2-m_W^2) = (m_W^2/k^2)/(k^2-m_W^2) + 1/k^2$. The former piece has the correct structure to provide $m_W/\sqrt{k^2}$ factors to each gauge polarization. The latter piece cancels the $\sqrt{k^2}$'s from each polarization vector, but leaves over no poles or mass factors. It therefore produces a non-collinear interaction that goes as $1/E^2$ instead of $1/(k^2-m_W^2)$, and can be grouped together with the neglected non-collinear diagrams. We can view all of the remaining collinear contributions as a simple product of on-shell gauge+Goldsone production and gauge+Goldstone splitting matrix elements, connected by the standard scalar propagator $i/(k^2-m_W^2)$.

Analogous results were obtained for the factorization of logarithmic virtual corrections to external gauge/Goldsone bosons in~\cite{Beenakker:2001kf} by working directly in Coulomb gauge, and in~\cite{Denner:2000jv,Denner:2001gw} by invoking the Goldstone Boson Equivalence Theorem in Feynman-'t~Hooft gauge. Our own approach directly exhibits the applicability of the Equivalence Theorem in the corresponding real emission processes at tree-level, and extends them beyond the strict Goldstone limit to $O(m_W/E)$.

%%%%%%%%%%%%%%%%%%%%%%%%%%%%%%%%%%%%%%%%%%%%%
\section{Couplings and Feynman Rules}
\label{sec:FeynmanRules}
%%%%%%%%%%%%%%%%%%%%%%%%%%%%%%%%%%%%%%%%%%%

%-----------------------------------------------------
\subsection{Lagrangian, couplings, and charge conventions}
\label{sec:conventions}
%-----------------------------------------------------

In Goldstone Equivalence Gauge, each physical longitudinal gauge boson state is interpolated by two fields: 
$V_n$ and $\phi_V$, where $V=W^\pm,Z$. Unlike, e.g., in $R_\xi$ gauges, the relative phases of $V_n$-mediated and $\phi_V$-mediated processes must be explicitly kept track of. Here, we first present the Lagrangian of the SM in GEG to set the conventions. Before electroweak symmetry breaking (EWSB), the Lagrangian with the gauge fixing is written as
\be\label{eq:lagrangian}
\cal{L}_{\rm Gauge} &\,=\,& -\frac{1}{4} W^{a\mu\nu}W^a_{\mu\nu} -\frac{1}{4} (B_{\mu\nu})^2-\frac{1}{2\xi}(n\cdot W)^2-\frac{1}{2\xi} (n\cdot B)^2, \nonumber \\
\cal{L}_{\rm fermion} &\,=\,& i\bar{\psi}\slashed{D}\psi, \\  %\,=\, i\bar{\psi}\slashed{\partial}\psi-(g_2\bar{\psi}\gamma^{\mu}t^a\psi W_{\mu}^a + g_Y Y \bar{\psi}\gamma^{\mu}\psi B_{\mu}), \\
\cal{L}_{\rm Yukawa} &\,=\,& -y_d\bar{Q}_LHd_R-y_u\epsilon_{ij}\bar{Q}_L^iH^{*j}u_R-y_e\bar{L}_L He_R + {\rm h.c.}\ , \nonumber \\
\cal{L}_{\rm Higgs} &\,=\,& (D^{\mu}H)^{\dagger}D_{\mu}H -\frac{\lambda_h}{4}\left(H^{\dagger}H-\frac{v^2}{2}\right)^2, \nonumber \\
%\quad \ 
\cal{L}_{\rm Ghost} &\,=\,& { \bar{c^a}n^{\mu}D_{\mu}^{ab}c^b. } 
\nonumber 
\ee
The flavor indices are suppressed since we do not consider the effects of flavor mixing. The covariant derivative $D_{\mu}$ and $SU(2)_L$ field strength component $W_{\mu\nu}^a$ are defined as 
\be
\label{eq:convent}
D_{\mu} = \partial_{\mu}-ig_2W_{\mu}^a T^a-ig_1YB_{\mu}, \qquad
W_{\mu\nu}^a = \partial_{\mu}W_{\nu}^a-\partial_{\nu}W_{\mu}^a+g_2f^{abc}W_{\mu}^bW_{\nu}^c .
\ee
The gauge-fixing vector $n^\mu$ of Eq.~(\ref{eq:nmu2}) would here be treated as a differential operator of schematic form $(1,-\partial_t\vec\nabla/\sqrt{\partial_t^2 \vec\nabla\cdot\vec\nabla})$. This becomes a well-defined operation in momentum-space. We take the formal $\xi\to 0$ limit in what follows.

After EWSB $\langle H^0 \rangle = v/\sqrt{2}$, and particles acquire masses. The neutral gauge fields $W^{\mu}_3$ and $B^{\mu}$ mixing to form mass eigenstates $Z^{\mu}$ and $A^{\mu}$. Gauge and fermion masses go as
\be
m_W = \frac12 g_2 v, \quad m_Z = \frac12 \sqrt{g_1^2 + g_2^2}\, v,\quad 
m_\gamma = 0,\quad 
m_f = \frac{1}{\sqrt 2} y_f  v,
\ee
with $g_1 \approx 0.36$ and $g_2\approx 0.65$ at the weak-scale, $y_t \approx 1$, and $v \approx 246$ GeV. 
The Higgs field self-coupling is normalized such that
\be
m_h^2 = \frac12\lambda_h v^2,
\ee
such that $\lambda_h \simeq 0.52$ for $m_h \simeq 125$~GeV.

%We are now in the position to define our coupling conventions. 
As for the gauge-fermion interactions in a general basis, we denote them using $g_V$ as the gauge coupling constant for a vector boson $V = B^0, W^0, W^\pm, \gamma, Z$,
\be
ig_V\gamma^{\mu}\sum_{\tau=L,R} g^V_{\tau} P_{\tau}  \,, 
\ee
where the chirality projection operators are $P_{R/L}=\frac{1}{2}(1\pm \gamma^5)$.
They are all built up from the underlying $U(1)_Y$ and $SU(2)_L$ gauge couplings. Specifically,
\beqa
g_{B^0} = g_1, \quad 
g_{W^0} = g_{W^\pm}  = g_2, \quad
g_{\gamma}  =  e = \frac{g_1 g_2}{\sqrt{g_1^2 + g_2^2}}, \quad 
g_Z  =  \sqrt{g_1^2 + g_2^2} .
\eeqa
As usual, the weak mixing angle is defined as  
\beqa
c_W \equiv \cos\theta_W = {g_2 \over g_Z} \quad {\rm or} \quad s_W \equiv \sin\theta_W = {g_1 \over g_Z} .
\eeqa
We denote the gauge charge $Q$ of a particle $p$ (chiral fermion or scalar) under a given gauge boson $V$ by $Q^V_p$.\footnote{For $V = W^\pm$, two different components of a left-handed doublet participate, but they can be assigned a common charge of $1/\sqrt{2}$, with either flavor plugged in.} We list the full set of charges in Table~\ref{tab:charges}.
%Chiral fermions couple to the scalars of the Higgs doublet via isospin-like charges $I_u^Z = 1/2$, $I_{d/e}^Z = -1/2$, $I_u^{W^\pm} = I_{d/e}^{W^\pm} = 1/\sqrt{2}$. These charges are defined independently of the fermion's helicity.

\begin{table}[]
\centering
\begin{tabular}{r|rrrrr}
          &  $Q^{B^0}_p = Y_p$ & $Q^{W^0}_p = T^3_p$ & \ $Q^{W^\pm}_p$ \ & $Q^{\gamma}_p = Q^{\rm EM}_p$  & $Q^Z_p = T^3_p-Q^{\rm EM}_p s^2_W$ \vspace{1mm} \\
\hline
$p = \qquad\ u_L$  &   1/6            &   1/2              &  $1/\sqrt{2}$ &   2/3  &  $1/2 - (2/3)s_W^2$ \\
    $u_R$  &   2/3            &   0                &   0           &   2/3  &  $- (2/3)s_W^2$     \\
    $d_L$  &   1/6            &  $-1/2$              &  $1/\sqrt{2}$ &  $-1/3$  &  $-1/2 + (1/3)s_W^2$ \\
    $d_R$  &  $-1/3$            &   0                &   0           &  $-1/3$  &  $(1/3)s_W^2$  \\
   $\nu_L$ &  $-1/2$            &   1/2              &  $1/\sqrt{2}$ &     0  &  $1/2$ \\
    $e_L$  &  $-1/2$            &  $-1/2$              &  $1/\sqrt{2}$ &    $-1$  &  $-1/2 + s_W^2$ \\
    $e_R$  &  $-1$              &   0                &   0           &    $-1$  &  $s_W^2$  \\
  $\phi^+$ &   1/2            &   1/2              &  $1/\sqrt{2}$ &     1  &  $1/2 - s_W^2$ \\
  $H^0 = \frac{h+i\phi^0}{\sqrt{2}}$ 
           &  1/2             &  $-1/2$              &  $1/\sqrt{2}$ &     0  &  $-1/2$ 
\end{tabular}
\caption{Gauge charges of chiral fermions and scalars in the Standard Model. For the fermions, first generation is used, but charges for second and third generations follow the same pattern.}
\label{tab:charges}
\end{table}
We now turn to the quadratic Lagrangian terms involving gauge fields and Goldstone fields. The quadratic terms of $Z$ and $\phi_Z$ Lagrangian are 
\be
\Lm_{Z^2}&=&-\frac{1}{2}\partial^{\mu}Z^{\nu}\partial_{\mu}Z_{\nu}+\frac{1}{2}\partial^{\mu}Z_{\mu}\partial^{\nu}Z_{\nu}+\frac{1}{2}m_Z^2Z_{\mu}Z^{\mu}
-\frac{1}{2\xi}(n^{\mu}Z_{\mu})^2 . \\
\Lm_{\phi_ZZ} &=& -m_ZZ^{\mu}\partial_{\mu}\phi_Z,\qquad
\Lm_{\phi_Z^2} = \frac{1}{2}(\partial^{\mu}\phi_Z)^2 .
\ee
Note that the minus sign in $\Lm_{\phi_ZZ}$ follows from the sign convention of the covariant derivative, Eq.~(\ref{eq:convent}), as well as our expansion of the Higgs doublet in Eq.~(\ref{eq:HiggsExpansion}), namely $H^0 \to (v+h - i\phi^0)/\sqrt{2}$. This in turn determines the phase factor of the polarization vector $Z_n$. (Though of course our convention choices ultimately have no effect on physical rates.)
For $W^{\pm}_{\mu}/\phi^{\pm}$, the unmixed kinetic and mass terms are analogous, and the quadratic mixing term is given by 
\be
\Lm_{W\phi}=-im_W W^+_{\mu}\partial^{\mu}\phi^- + {\rm h.c.}
\ee

%-----------------------------------------------------
\subsection{External polarizations and propagators}
\label{sec:polarizations}
%-----------------------------------------------------

We decompose all fermions and gauge bosons into helicity basis within the hard process CM frame, including off-shell particles. We emphasize that in computing leading-order $1\to2$ splitting functions, {\it all} particle polarization states should be set on-shell, since the off-shell corrections are strictly non-collinear. An on-shell polarization can be associated with an off-shell momentum, for example, by adjusting the three-momentum at fixed energy.

The fermion external spinors are as usual, though to facilitate extraction of $O(v)$ effects we Taylor expand in $m_f/E = (y_f/\sqrt{2})(v/E)$. Explicitly, for fermions moving approximately along the $z$-axis, possibly offset toward the $x$-axis by a small angle $\theta$,
\beq
\label{eq:ext_state}
u_{s=L} \,\simeq\, \sqrt{2E} \left(\begin{array}{r}  \left(\begin{array}{c} -\theta/2 \\ 1 \end{array}\right) \vspace{1mm} \\ \frac{m_f}{2E} \left(\begin{array}{c} -\theta/2 \\ 1 \end{array}\right) \end{array}\right)\,,  \qquad  u_{s=R} \,\simeq\, \sqrt{2E} \left(\begin{array}{r}   \frac{m_f}{2E} \left(\begin{array}{c} 1 \\ \theta/2 \end{array}\right) \vspace{1mm} \\ \left(\begin{array}{c} 1 \\ \theta/2 \end{array}\right) \end{array}\right) \,.
\eeq
Propagators are also as usual, but given our approximate decomposition into on-shell spin states, they fall into a factorizable form.
For a generic off-shell $k^\mu$, we can build an effective on-shell $\tilde k^\mu$ by keeping $k^0 \equiv E$ fixed but changing 
\be
\vec k = \hat k \sqrt{E^2-k^2} \;\to\; \hat k \sqrt{E^2-m_f^2} = 
\vec k + {\cal O}((k^2-m_f^2)/E).
\ee 
We may then rewrite the propagator as 
\beqa
\frac{\slashed{k}+m_f}{k^2-m_f^2} &\,=\,& \frac{(\slashed{\tilde k}+m_f) \,+\, \order((k^2-m_f^2)/E)}{k^2-m_f^2} \nonumber \\
                                  &\,=\,& \frac{\sum_{s=L,R}u_s(\tilde k)\,\bar u_s(\tilde k)}{k^2-m_f^2} \;+\; {\rm non \minus collinear\ terms},
\eeqa
exploiting the fact that the leading correction away from a factorized numerator is set up to cancel the propagator's denominator. We ignore possible coherence effects between different spin channels.

Transverse gauge bosons are also assigned their standard polarization vectors
\beq
\epsilon_{\pm}^\mu \,\simeq\, \frac{1}{\sqrt{2}}\big(0; 1,\pm i,-\theta  \big) \,,
\eeq
with the complex-conjugate $\epsilon^{\mu*}_{\pm}$ used for outgoing bosons. However, the longitudinal gauge/Goldstone sector is treated somewhat unconventionally. Longitudinal gauge bosons can be created by Goldstone/pseudo-scalar boson fields. We set our phase conventions so that these creation and annihilation amplitudes are unity, maintaining continuity with the unbroken theory. However, longitudinal bosons may also still be created by gauge fields, in association with the ``remainder'' field component $V_n$ expanded out in Eq.~(\ref{eq:Wexpansion}). Synchronizing these component fields such that they also create/annihilate external bosons with unit amplitude, their associated polarization vectors then carry nontrivial phases:
\beqa
\label{eq:phase}
&& {\rm incoming}\ Z: \ i\epsilon_n^{\mu}; \ \ 
{\rm outgoing}\ Z: \  (i\epsilon_n^{\mu})^* = -i\epsilon_n^{\mu}; \ \  {\rm incoming/outgoing:}\
W^\pm: \  \pm\epsilon_n^{\mu},\ \qquad~~~~\\
&& {\rm with}\quad  
\epsilon_n^\mu = - \frac{m_V}{n\cdot k}n^{\mu} \simeq\,  \frac{m_V}{2E}\big(-1;\theta,0,1\big)\,.
\eeqa
The light-like gauge-fixing vector $n^\mu$ is defined in Eq.~(\ref{eq:nmu2}).
The corresponding propagators are given in Eq.~(\ref{app:prop}). 
Photons are subjected to the same gauge conditions, but this has little practical bearing on their showering behavior. As usual, they have purely transverse external polarization states, and only their transverse modes contribute to collinear-enhanced physics. As discussed in Section~\ref{sec:interference}, the transverse photon and $Z$ propagators should be treated coherently within a parton shower.
The $h$ and $\phi^0/Z_n$ propagators should also be treated coherently. We will see an example of this, including the $Z_n$ component, in Appendix~\ref{sec:examples}.

%-----------------------------------------------------
\subsection{Feynman rules for three-point couplings}
\label{sec:three-points}
%-----------------------------------------------------

Feynman rules in GEG are largely similar to those of standard gauges. We list below many of the relevant three-point vertex rules. For brevity, we omit four-point interactions, which do not play a role in $1\to2$ splittings at this order.

Wherever explicitly referenced, we reckon all four-momenta as flowing into the vertex. We use the small arrows next to a particle line to indicate the flow of the momenta as well as the electric charge, where relevant. When no arrows labelled for the charged particles, charge conservation is implied at each vertex for the particles involved.
%Similar logic applies to the flow of electric charge on the boson lines $W^\pm$ and $\phi^\pm$. 
%We indicate these charge flows with an arrow on the field line where relevant. 
%For example, a charged $W$ with arrow pointing away from a vertex is labeled here as $W^-$. (Absence of an explicit arrow means that we could have diagrams with charge flowing in either direction.)

Gauge field polarization vectors $\epsilon^\mu$ are kept explicit at the vertices here, and can take on three possible values associated with the propagating gauge degrees of freedom: the two spacelike transverse polarizations $\epsilon_\pm^\mu$ (or $\epsilon_{xy}^\mu$), and the lightlike polarization $\propto \epsilon_n^\mu$.\footnote{An off-shell photon does not have a physical pole associated with its $\epsilon_n$ polarization, and the phase of that polarization can be set arbitrarily since there is no associated phase with the creation/annihilation of asymptotic states. A simple default would be to follow the same convention as for the $Z$ boson.} The on-shell values of these polarizations and a convenient phase convention have been provided at the end of the preceding subsection as in Eq.~(\ref{eq:phase}). The extension to off-shell momenta follows immediately. However, some care should be taken with respect to how these polarizations are oriented relative to momentum flows, whether a boson is reckoned as ``incoming'' or ``outgoing.'' In particular, if the four-momentum $k$ is measured outgoing from a vertex, one should use $\epsilon(-k)$. (In many cases this is equivalent to $\epsilon(k)^*$, but an exception occurs for $W^\pm_n$.) Including the polarization vectors in the vertices as such, the vector boson propagators will not carry Lorentz indices, as given in Eq.~(\ref{app:prop}).

\beqa
%
% ffgamma
\vcenter{\begin{picture}(100,80)(0,0)
\SetColor{Black} \SetWidth{2}
\Text(0, 5)[]{$f$}
\ArrowLine(  10, 5)(50,25)
\Photon(50,25)(50,60){3}{3}
\ArrowLine( 50,25)(85,5)
\Text(90,5)[l]{$f$}
%\Text(70, 45)[]{\rotatebox{26}{$\longleftarrow$}}
\Text(57,60)[l]{$\gamma$}
\end{picture}}
\hspace{-4.0in}
& =\ \ &  i e Q^{EM}_f \slashed{\epsilon}  \nonumber \\
% ffW
\vcenter{\begin{picture}(100,80)(0,0)
\SetColor{Black} \SetWidth{2}
\Text(0,5)[]{$f$}
\ArrowLine(  10,5)(50,25)
\Photon(50,25)(50,60){3}{3}
\ArrowLine( 50,25)(85,5)
\Text(90,5)[l]{$f'$}
%\Text(70, 45)[]{\rotatebox{26}{$\longleftarrow$}}
\Text(57,60)[l]{$W^{\pm}$}
\end{picture}}
\hspace{-4.0in}
& =\ \ & i \frac{g_2}{\sqrt{2}} \slashed{\epsilon} P_L \nonumber \\
%
% ffZ
\vcenter{\begin{picture}(100,80)(0,0)
\SetColor{Black} \SetWidth{2}
\Text(0,5)[]{$f$}
\ArrowLine(  10,5)(50,25)
\Photon(50,25)(50,60){3}{3}
\ArrowLine( 50,25)(85,5)
\Text(90,5)[l]{$f$}
%\Text(70, 45)[]{\rotatebox{26}{$\longleftarrow$}}
\Text(57,60)[l]{$Z$}
\end{picture}}
\hspace{-4.0in}
& =\ \ &  i g_Z \slashed{\epsilon} \left((T^3_f - Q^{\rm EM}_f s_W^2)P_L  - Q^{\rm EM}_f s_W^2 P_R\right) \nonumber \\
%
% f -> f phi+
\vcenter{\begin{picture}(100,80)(0,0)
\SetColor{Black} \SetWidth{2}
\Text(0,5)[]{$u$}
\ArrowLine(  10,5)(50,25)
\DashLine(50,25)(50,60){5}
\ArrowLine( 50,25)(85,5)
\Text(57,60)[l]{$\phi^\pm$}
\Text(90, 5)[l]{$d$}
\end{picture}}
\hspace{-4.0in}
& =\ \ & i \left(-y_d P_L  +  y_u P_R\right) \nonumber \\
%
% f -> f phi-
\vcenter{\begin{picture}(100,80)(0,0)
\SetColor{Black} \SetWidth{2}
\Text(0,5)[]{$d$}
\ArrowLine(  10,5)(50,25)
\DashLine(50,60)(50,25){5}
\ArrowLine( 50,25)(85,5)
\Text(57,60)[l]{$\phi^\pm$}
\Text(90, 5)[l]{$u$}
\end{picture}}
\hspace{-4.0in}
& =\ \ & i \left(y_u P_L  -  y_d P_R\right) \nonumber \\
%
% f -> f phi0
\vcenter{\begin{picture}(100,80)(0,0)
\SetColor{Black} \SetWidth{2}
\Text(0,5)[]{$f$}
\ArrowLine(  10,5)(50,25)
\DashLine(50,25)(50,60){5}
\ArrowLine( 50,25)(85,5)
\Text(57,60)[l]{$\phi^0$}
\Text(90, 5)[l]{$f$}
\end{picture}}
\hspace{-4.0in}
& =\ \ & (\delta_{fu}-\delta_{fd})\frac{y_f}{\sqrt{2}} \gamma_5 \nonumber  \\
%
% f -> f h
\vcenter{\begin{picture}(100,80)(0,0)
\SetColor{Black} \SetWidth{2}
\Text(0,5)[]{$f$}
\ArrowLine(  10,5)(50,25)
\DashLine(50,25)(50,60){5}
\ArrowLine( 50,25)(85,5)
\Text(57,60)[l]{$h$}
\Text(90, 5)[l]{$f$}
\end{picture}}
\hspace{-4.0in}
& =\ \ &  -i \frac{y_f}{\sqrt{2}} \nonumber \\
%
%VVV-1
\vcenter{\begin{picture}(100,80)(0,0)
\SetColor{Black} \SetWidth{2}
\Photon(  10,5)(50,25){3}{3}
\Photon(50,25)(50,60){3}{3}% \ArrowLine(52,62)(52,63)
\Photon(50,25)(85, 5){3}{3} %\LongArrow(90,1)(80,6)
\Text(0,5)[]{$Z$}
\Text(30,5)[]{\rotatebox{30}{$\longrightarrow$}}
\Text(30, -3)[]{$k_1$}
\Text(57,60)[l]{$W^-$}
\Text(40, 45)[]{\rotatebox{90}{$\longleftarrow$}}
\Text(31, 47)[]{$k_2$}
\Text(90, 5)[l]{$W^+$}
\Text(75, 20)[]{\rotatebox{-30}{$\longleftarrow$}}
\Text(80, 27)[]{$k_3$}
\end{picture}}
\hspace{-4.0in}
& =\ \ &   i g_2 c_W \,\epsilon_{ijk}(\epsilon_i\cdot \epsilon_j )\big(\epsilon_k\cdot(k_i-k_j)\big)  \quad  [\epsilon_{123} \equiv 1] \nonumber \\
%VVV-2
\vcenter{\begin{picture}(100,80)(0,0)
\SetColor{Black} \SetWidth{2}
\Photon(  10,5)(50,25){3}{3}
\Photon(50,25)(50,60){3}{3} %\ArrowLine(52,62)(52,63)
\Photon(50,25)(85, 5){3}{3}
\Text(0,5)[]{$\gamma$}
\Text(30,5)[]{\rotatebox{30}{$\longrightarrow$}}
\Text(30, -3)[]{$k_1$}
\Text(57,60)[l]{$W^-$}
\Text(40, 45)[]{\rotatebox{90}{$\longleftarrow$}}
\Text(31, 47)[]{$k_2$}
\Text(90, 5)[l]{$W^+$}
\Text(75, 20)[]{\rotatebox{-30}{$\longleftarrow$}}
\Text(80, 27)[]{$k_3$}
\end{picture}}
\hspace{-4.0in}
& =\ \ &   i e \,\epsilon_{ijk}(\epsilon_i\cdot \epsilon_j )\big(\epsilon_k\cdot(k_i-k_j)\big)  \nonumber \\
% h\phiW
\vcenter{\begin{picture}(100,80)(0,0)
\SetColor{Black} \SetWidth{2}
\DashLine(  10,5)(50,25){6}
\Photon(50,25)(50,60){3}{3}
\DashLine(50,25)(85, 5){5}
\Text(0,5)[]{$h$}
\Text(30,5)[]{\rotatebox{30}{$\longrightarrow$}}
\Text(30, -3)[]{$q$}
\Text(57,60)[l]{$W^{\pm}$}
\Text(90, 5)[l]{$\phi^{\mp}$}
\Text(70, 5)[]{\rotatebox{-30}{$\longleftarrow$}}
\Text(70, -3)[]{$p$}
\end{picture}}
\hspace{-4.0in}
& =\ \ &  \pm i \frac{g_2}{2} (q-p)\cdot \epsilon \nonumber \\
% h\phiZ
\vcenter{\begin{picture}(100,80)(0,0)
\SetColor{Black} \SetWidth{2}
\DashLine(  10,5)(50,25){6}
\Photon(50,25)(50,60){3}{3}
\DashLine(50,25)(85, 5){5}
\Text(0,5)[]{$h$}
\Text(30,5)[]{\rotatebox{30}{$\longrightarrow$}}
\Text(30, -3)[]{$q$}
\Text(57,60)[l]{$Z$}
\Text(90, 5)[l]{$\phi^{0}$}
\Text(70, 5)[]{\rotatebox{-30}{$\longleftarrow$}}
\Text(70, -3)[]{$p$}
\end{picture}}
\hspace{-4.0in}
& =\ \ &  \frac{g_Z}{2}(q-p)\cdot \epsilon \nonumber \\
% \phi\phiW-1
\vcenter{\begin{picture}(100,80)(0,0)
\SetColor{Black} \SetWidth{2}
\DashLine(  10,5)(50,25){6}
\Photon(50,25)(50,60){3}{3}
\DashLine(50,25)(85, 5){5}
\Text(0,5)[]{$\phi^0$}
\Text(30,5)[]{\rotatebox{30}{$\longrightarrow$}}
\Text(30, -3)[]{$q$}
\Text(57,60)[l]{$W^{\pm}$}
\Text(90, 5)[l]{$\phi^{\mp}$}
\Text(70, 5)[]{\rotatebox{-30}{$\longleftarrow$}}
\Text(70, -3)[]{$p$}
\end{picture}}
\hspace{-4.0in}
& =\ \ &  \frac{g_2}{2}(q-p)\cdot \epsilon \nonumber \\
% \phi\phiZ-2
\vcenter{\begin{picture}(100,80)(0,0)
\SetColor{Black} \SetWidth{2}
%\DashArrowLine(  10,5)(50,25){6}
\DashLine(  10,5)(50,25){6}
\Photon(50,25)(50,60){3}{3}
%\DashArrowLine(50,25)(85, 5){5}
\DashLine(50,25)(85, 5){5}
\Text(0,5)[]{$\phi^+$}
\Text(30,5)[]{\rotatebox{30}{$\longrightarrow$}}
\Text(30, -3)[]{$q$}
\Text(57,60)[l]{$Z$}
\Text(90, 5)[l]{$\phi^-$}
\Text(70, 5)[]{\rotatebox{-30}{$\longleftarrow$}}
\Text(70, -3)[]{$p$}
\end{picture}}
\hspace{-4.0in}
& =\ \ &  ig_Z\frac{c_{2W}}{2}(q-p)\cdot \epsilon\nonumber \\
% \phi\phiA
\vcenter{\begin{picture}(100,80)(0,0)
\SetColor{Black} \SetWidth{2}
%\DashArrowLine(  10,5)(50,25){6}
\DashLine(  10,5)(50,25){6}
\Photon(50,25)(50,60){3}{3}
%\DashArrowLine(50,25)(85, 5){5}
\DashLine(50,25)(85, 5){5}
\Text(0,5)[]{$\phi^+$}
\Text(30,5)[]{\rotatebox{30}{$\longrightarrow$}}
\Text(30, -3)[]{$q$}
\Text(57,60)[l]{$\gamma$}
\Text(90, 5)[l]{$\phi^-$}
\Text(70, 5)[]{\rotatebox{-30}{$\longleftarrow$}}
\Text(70, -3)[]{$p$}
\end{picture}}
\hspace{-4.0in}
& =\ \ &  ie(q-p)\cdot \epsilon \nonumber \\
%
% hWW
\vcenter{\begin{picture}(100,80)(0,0)
\SetColor{Black} \SetWidth{2}
\DashLine(  10,5)(46,24){6}
\Text(50,25)[]{$\boldsymbol \otimes$}
\Photon(50,29)(50,60){3}{3} % \ArrowLine(52,62)(52,63)
\Photon(54,24)(85, 5){3}{3}
\Text(0,5)[]{$h$}
\Text(57,60)[l]{$W^-$}
\Text(90, 5)[l]{$W^+$}
\end{picture}}
\hspace{-4.0in}
& =\ \ &  i g_2 m_W \, \epsilon_{W^+}\!\!\cdot\epsilon_{W^-} \nonumber \\
%
% hZZ
\vcenter{\begin{picture}(100,80)(0,0)
\SetColor{Black} \SetWidth{2}
\DashLine(  10,5)(46,24){6}
\Text(50,25)[]{$\boldsymbol \otimes$}
\Photon(50,29)(50,60){3}{3}
\Photon(54,24)(85, 5){3}{3}
\Text(0,5)[]{$h$}
\Text(57,60)[l]{$Z$}
\Text(90, 5)[l]{$Z$}
\end{picture}}
\hspace{-4.0in}
& =\ \ &  i g_Z m_Z \, \epsilon_{Z_1}\!\!\cdot\epsilon_{Z_2} \nonumber \\
%
% h\phi+\phi-
\vcenter{\begin{picture}(100,80)(0,0)
\SetColor{Black} \SetWidth{2}
\DashLine(  10,5)(46,24){6}
\Text(50,25)[]{$\boldsymbol \otimes$}
%\DashArrowLine(50,29)(50,60){6}
%\DashArrowLine(85, 5)(54,24){6}
\DashLine(50,29)(50,60){6}
\DashLine(85, 5)(54,24){6}
\Text(0,5)[]{$h$}
\Text(57,60)[l]{$\phi^-$}
\Text(90, 5)[l]{$\phi^+$}
\end{picture}}
\hspace{-4.0in}
& =\ \ &  -i \frac{\lambda_h v}{2} \nonumber \\
%
% h\phi0\phi0
\vcenter{\begin{picture}(100,80)(0,0)
\SetColor{Black} \SetWidth{2}
\DashLine(  10,5)(46,24){6}
\Text(50,25)[]{$\boldsymbol \otimes$}
\DashLine(50,29)(50,60){6}
\DashLine(85, 5)(54,24){6}
\Text(0,5)[]{$h$}
\Text(57,60)[l]{$\phi^0$}
\Text(90, 5)[l]{$\phi^0$}
\end{picture}}
\hspace{-4.0in}
& =\ \ &  -i \frac{\lambda_h v}{2} \nonumber \\
%
% hhh
\vcenter{\begin{picture}(100,80)(0,0)
\SetColor{Black} \SetWidth{2}
\DashLine(  10,5)(46,24){6}
\Text(50,25)[]{$\boldsymbol \otimes$}
\DashLine(50,29)(50,60){6}
\DashLine(85, 5)(54,24){6}
\Text(0,5)[]{$h$}
\Text(57,60)[l]{$h$}
\Text(90, 5)[l]{$h$}
\end{picture}}
\hspace{-4.0in}
& =\ \ &   -i \frac{3\lambda_h v}{2} \nonumber \\
%
%VVs-1
\vcenter{\begin{picture}(100,80)(0,0)
\SetColor{Black} \SetWidth{2}
\Photon(  10,5)(46,24){3}{3}
\Text(50,25)[]{$\boldsymbol \otimes$}
\DashLine(50,29)(50,60){6}
\Photon(54,24)(85, 5){3}{3}
\Text(0,5)[]{$Z$}
\Text(57,60)[l]{$\phi^{\pm}$}
\Text(90, 5)[l]{$W^{\mp}$}
\end{picture}}
\hspace{-4.0in}
& =\ \ &   -i g_2 s^2_W m_Z\ \epsilon_Z\cdot \epsilon_W \nonumber \\
%
%VVs-2
\vcenter{\begin{picture}(100,80)(0,0)
\SetColor{Black} \SetWidth{2}
\Photon(  10,5)(46,24){3}{3}
\Text(50,25)[]{$\boldsymbol \otimes$}
\DashLine(50,29)(50,60){6}
\Photon(54,24)(85, 5){3}{3}
\Text(0,5)[]{$\gamma$}
\Text(57,60)[l]{$\phi^{\pm}$}
\Text(90, 5)[l]{$W^{\mp}$}
\end{picture}}
\hspace{-4.0in}
& =\ \ &   i e m_W\ \epsilon_\gamma\cdot \epsilon_W \nonumber 
\eeqa
\vskip 0.1cm 
The symbol $\otimes$ denotes the mass (or $v$) insertion from the EWSB.

\vskip 0.5cm

%----------------------------------------------------
\subsection{Example calculations with GEG}
\label{sec:examples}
%----------------------------------------------------

Calculations in high energy processes involving longitudinal vector bosons can be complicated in dealing with gauge artifacts, often exhibiting artificial ``bad high energy behavior'' containing factors of $E/v$. Here we show some explicit examples to demonstrate how to calculate ultra-collinear splitting amplitudes in GEG, where all such amplitudes are automatically free of such artifacts and are simply proportional to the VEV. We focus in detail on the specific massive fermion splitting $t_s \rightarrow W^+_{\longit} b_s$, where the fermion helicity $s=L,R$ is preserved. This calculation is also trivially adapted to cases where one or both fermion is a massless flavor, such as the usual $u_L \to W^+_\longit d_L$, and is straightforward to extend to $Z_\longit$ boson emission with appropriate replacements of couplings and remainder polarization phases. We also outline below the diagrammatic construction of a few other processes for illustration.

We first reemphasize that the longitudinal gauge boson $W_{\longit}^+$ in GEG should be interpolated by both the Goldstone field $\phi^+$ and  the remainder gauge field $W_n^+$, leading us to break up the splitting amplitude as
\be
i{\mathcal M}(t_s\rightarrow W_{\longit}^+ b_s) \,=\, i{\mathcal M}(t_s\rightarrow \phi^+ b_s) \,+\, i{\mathcal M}(t_s\rightarrow W^+_n b_s) .
\ee
Applying the three-point Feynman rules in Sec.~\ref{sec:three-points}, and taking the exact collinear limit ($\theta, k_T \to 0$) to extract the leading behavior, we have for the LH process\footnote{Note that for the charge-conjugate process, producing $W_n^-$, we would instead use the remainder polarization vector times $(-1)$: $-\epsilon_n$.}
\be
i{\mathcal M}(t_L\rightarrow \phi^+ b_L) 
   &\,=\,     & i\,\bar{u}(b_L)(y_t P_R - y_b P_L)u(t_L) \nonumber \\
   &\,\simeq\,& i\,y_t\sqrt{2E_b}\frac{m_t}{\sqrt{2E_t}}  \,-\, i\,y_b\frac{m_b}{\sqrt{2E_b}}\sqrt{2E_t}  \nonumber \\
   &\,\simeq\,& i\,v \left(\frac{y_t^2}{\sqrt{2}}\sqrt{\zb} \,-\, \frac{y_b^2}{\sqrt{2}} \frac{1}{\sqrt{\zb}}\right),  \nonumber  \\
i{\mathcal M}(t_L\rightarrow W_n^+ b_L) 
   &\,=\,     &  i\,\frac{g_2}{\sqrt{2}} \bar{u}(b_L)\big(\slashed{\epsilon}_n(W) P_L \big) u(t_L)  \nonumber \\
   &\,\simeq\,&  i\,\frac{g_2}{\sqrt{2}}\cdot 2\sqrt{2E_b} \left(-\frac{m_W}{2E_W}\right) \sqrt{2E_t}  \nonumber \\
   &\,=\,     & -i\,v \frac{g_2^2}{\sqrt{2}} \frac{\sqrt{\zb}}{z} \,.                                     
\ee
The full LH splitting amplitude is then
\be
i{\mathcal M}(t_L\rightarrow  W_{\longit}^+ b_L )
     \,=\, i\,v \frac{1}{z\sqrt{\zb}} \left(\frac{1}{\sqrt{2}}(y_t^2\zb-y_b^2)z- \frac{1}{\sqrt{2}}g_2^2\zb\right).
\ee
Plugging this in Eq.~(\ref{eq:split}), we have the splitting function 
\be
\frac{d{\mathcal P}_{t_L\rightarrow  W^+_\longit b_L}}{dz\,d\ktsq} = \frac{1}{16\pi^2} \frac{v^2}{\tktft} \left(\frac{1}{z}\right)  \left(\frac{1}{\sqrt{2}}(y_t^2\zb-y_b^2)z- \frac{1}{\sqrt{2}}g_2^2\zb\right)^2.
\label{eq:tL}                                    
\ee

As for the RH transition $t_R \rightarrow  W^+_\longit b_R$, there is no analogous amplitude for $W_n$ at $\order(v)$ due to the absence of RH charged-currents, so the amplitude is dominated by the Yukawa contribution,
\be
i{\mathcal M}(t_R\rightarrow W^+_\longit b_R) 
    & \,\simeq\, & i{\mathcal M}(t_R\rightarrow \phi^+ b_R) \nonumber \\
    & \,=\,      & i\,\bar{u}(b_R)(y_t P_R - y_b P_L)u(t_R) \nonumber  \\
    & \,\simeq\, & i\,y_t\frac{m_b}{\sqrt{2E_b}}\sqrt{2E_t} \,-\, i\,y_b\sqrt{2E_b}\frac{m_t}{\sqrt{2E_t}}  \nonumber \\
    & \,=\,      & i\,v \frac{y_t y_b}{\sqrt{2}}\left({1\over \sqrt{\zb}}-\sqrt{\zb}\right) \nonumber \\
    & \,=\,      & i\,v \frac{y_t y_b}{\sqrt{2}} \frac{z}{\sqrt{\zb}} ,
\ee
and the splitting function is
\beq
\frac{d{\mathcal P}_{t_R\rightarrow W^+_\longit b_R}}{dz\,d\ktsq} \,=\, \frac{1}{16\pi^2} \frac{v^2}{\tktft} z   \left(\frac{1}{\sqrt{2}}y_t y_b z \right)^2   \,=\,  \frac{1}{16\pi^2} \frac{v^2}{\tktft}  \left(\frac12 y_t^2 y_b^2 z^3 \right).     \label{eq:tR}                  
\eeq
Of course, given the small value of $y_b$, this process ends up becoming highly suppressed in practice. The results in Eqs.~(\ref{eq:tL}) and~(\ref{eq:tR}) lead to some of the formulas in Table \ref{tab:broken_fermion_splittings}.

\begin{figure}[t]
\begin{center}
\begin{subfigure}[t]{0.495\textwidth}
\includegraphics[width=200pt]{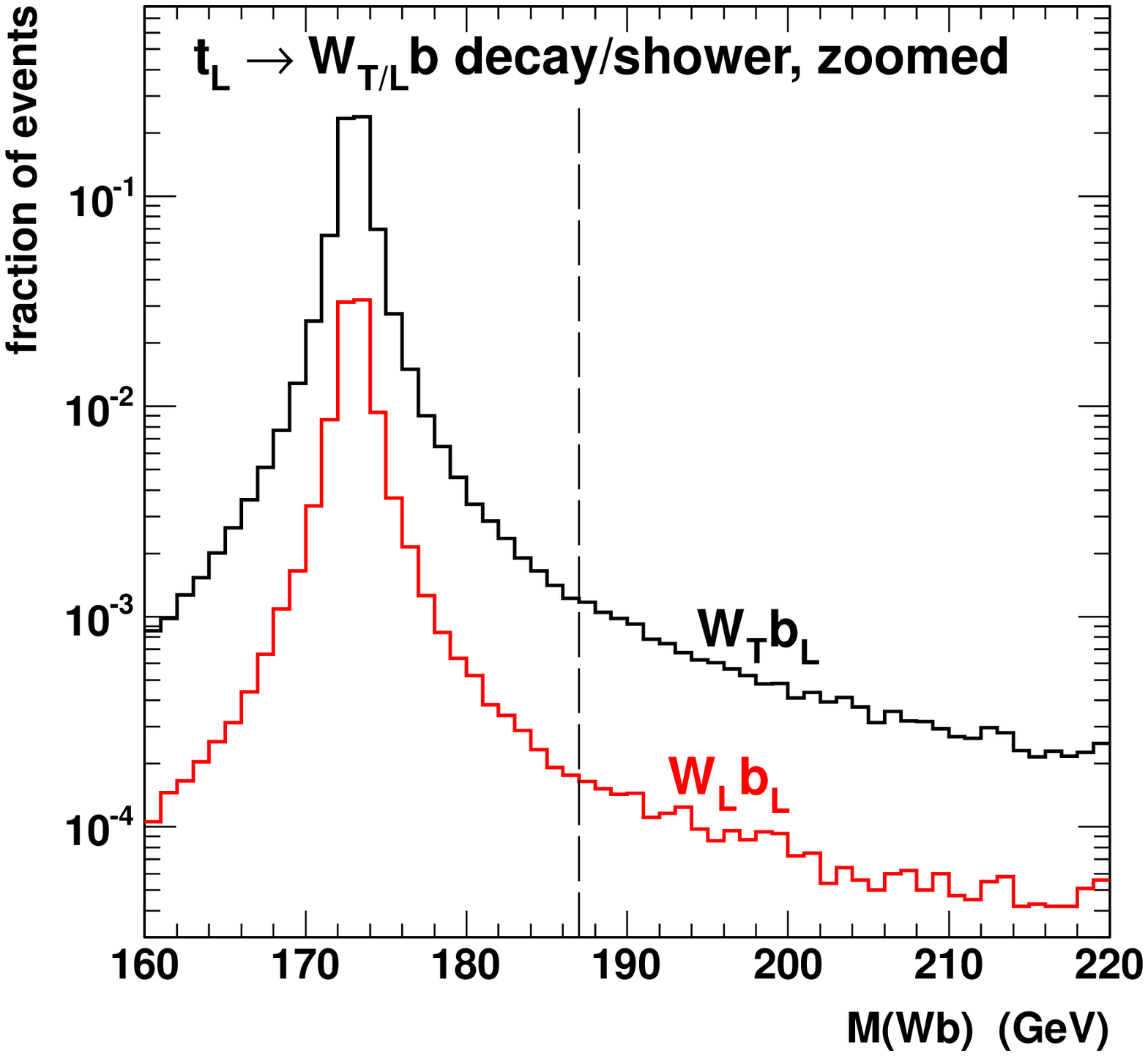}
\vspace{-0.5cm}\caption{}\end{subfigure}
\begin{subfigure}[t]{0.495\textwidth}
\includegraphics[width=200pt]{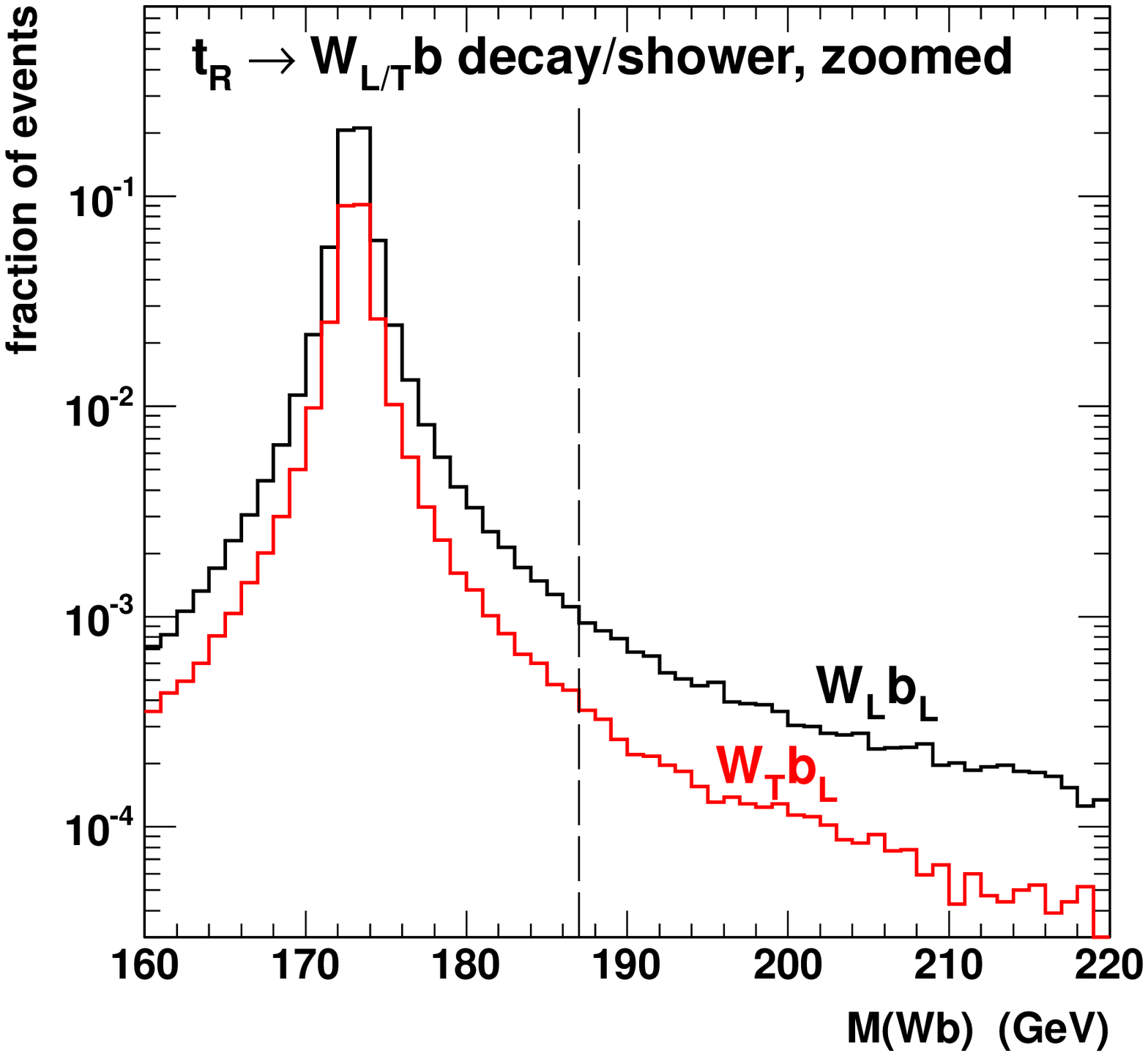}
\vspace{-0.5cm}\caption{}\end{subfigure}
\vspace{-0.6cm}
\end{center}
\caption[]{
Invariant mass distributions for EW decay/splitting of a 10~TeV polarized top quark for (a)~conventional-collinear $t_L \to W_T^+ b_L$ and ultra-collinear $t_L \to W_\longit^+ b_L$, and (b)~conventional-collinear $t_R \to W_\longit^+ b_L$ and ultra-collinear $t_R \to W_T^+ b_L$. Decay and shower are matched at 187~GeV (vertical dashed line). The conventional-collinear contributions correspond to the upper histograms, while the ultra-collinear contributions correspond to the lower histograms. 
}
\label{app:top}
\end{figure}

When combined with conventional collinear top quark splittings, the ultra-collinear splittings become important for modeling the approach to the top resonance peak. This includes as well the process $t_R \to W_T^+ b_L$. We show these individual shower contributions and their continuity with a simple Breit-Wigner model of top decay (weighted by $\Gamma_t(M(Wb))/\Gamma_t(m_t)$) in Fig.~\ref{app:top}. Here we have taken 10~TeV top quarks of either helicity, zooming into near the top quark pole, and set a decay/shower matching threshold of 187~GeV. All polarizations are measured in ``lab frame'' (as opposed to the top's rest frame). QCD and other electroweak showering effects are not incorporated.

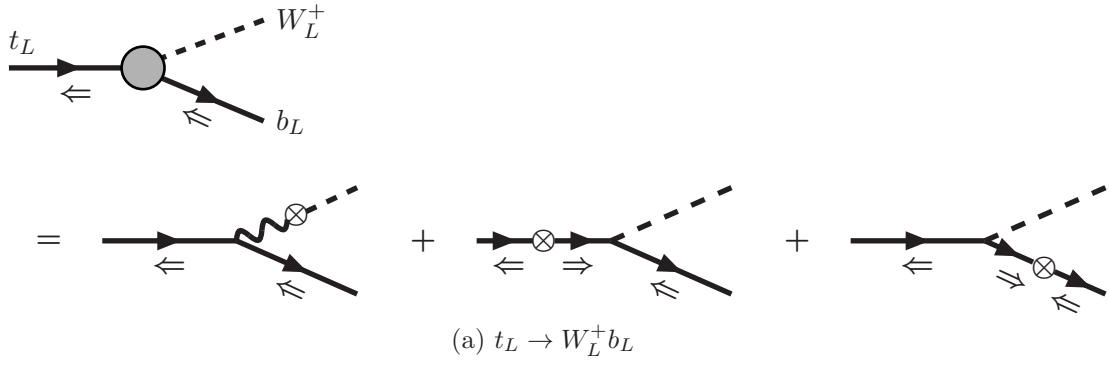
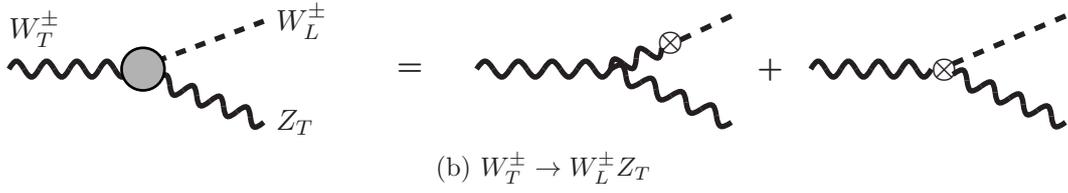
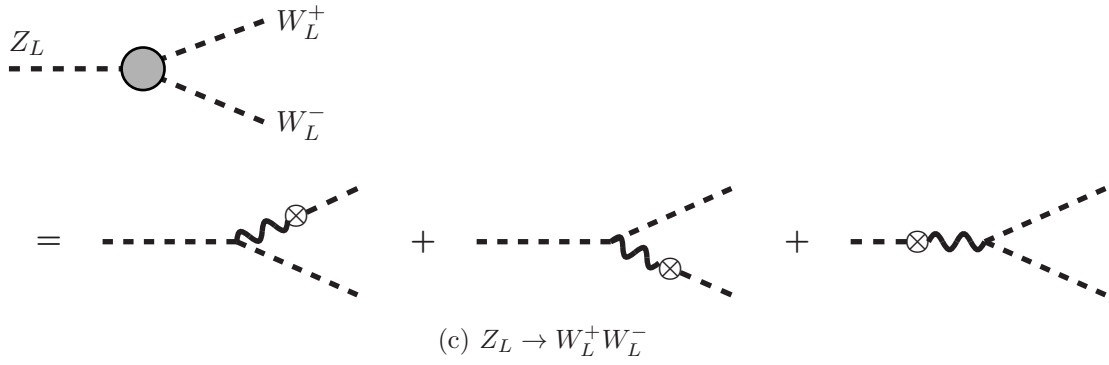
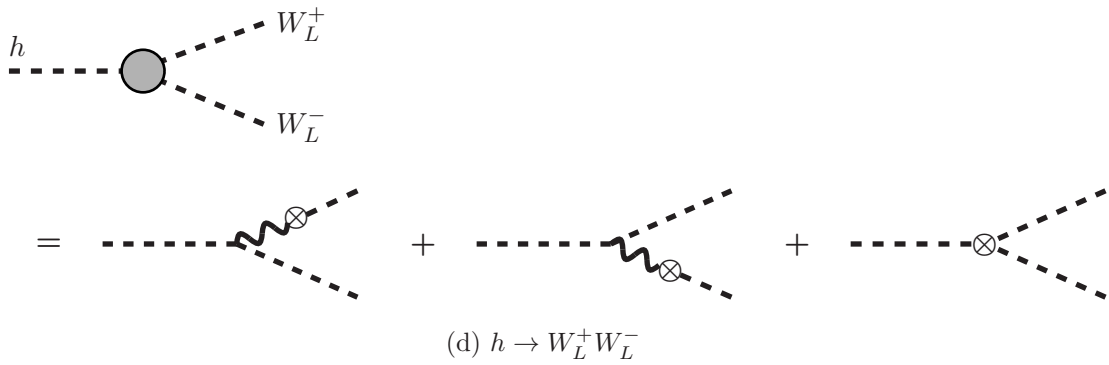
\begin{figure}[t!]
%ffL
\centering
\begin{subfigure}[t]{\textwidth}
\centering
%\begin{center}
\begin{picture}(400,120)(0,0)
\SetColor{Black} \SetWidth{2}
\ArrowLine(  0,90)(45,90)
%\Photon(50,75)(69,83){3}{2}
\DashLine(55,93)(95,108){4}
\ArrowLine( 55,87)(95,70)
\Text(25,80)[]{$\boldsymbol \Leftarrow$}
\Text(70,71)[]{\rotatebox{-25}{$\boldsymbol \Leftarrow$}}
 \Text(0,100)[l]{$t_L$}
 \Text(100,108)[l]{$W^+_{\longit}$}
 \Text(100,70)[l]{$b_L$}
 \SetWidth{1}
  \GCirc(50,90){8}{0.7}
\SetOffset(35,0)
\SetWidth{2}
\Text(-20,25)[]{$\boldsymbol =$}
\ArrowLine(  0,25)(50,25)
\Photon(50,25)(69,33){3}{2}
\DashLine(76,36)(95,45){4}
\ArrowLine( 50,25)(95,5)
\Text(72.5,35)[]{$\boldsymbol \otimes$}
\Text(25,15)[]{$\boldsymbol \Leftarrow$}
\Text(70,6)[]{\rotatebox{-25}{$\boldsymbol \Leftarrow$}}
\Text(120,25)[]{$\boldsymbol +$}
\SetOffset(175,0)
\ArrowLine(  0,25)(21,25)
\ArrowLine(  29,25)(50,25)
\DashLine(50,25)(95,45){6}
\ArrowLine( 50,25)(95,5)
\Text(25,25)[]{$\boldsymbol \otimes$}
\Text(12,15)[]{$\boldsymbol \Leftarrow$}
\Text(38,15)[]{$\boldsymbol \Rightarrow$}
\Text(70,6)[]{\rotatebox{-25}{$\boldsymbol \Leftarrow$}}
\Text(120,25)[]{$\boldsymbol +$}
\SetOffset(315,0)
\ArrowLine(  0,25)(50,25)
\DashLine(50,25)(95,45){6}
\ArrowLine( 50,25)(68,17)    % +45, -20 overall
\ArrowLine( 76,13.4)(95,5)
\Text(72.5,15)[]{$\boldsymbol \otimes$}
\Text(25,15)[]{$\boldsymbol \Leftarrow$}
\Text(60,10)[]{\rotatebox{-25}{$\boldsymbol \Rightarrow$}}
\Text(80,1)[]{\rotatebox{-25}{$\boldsymbol \Leftarrow$}}
\end{picture}
%\end{center}
\caption{$t_L\to W_{\longit}^+b_L$}
\label{fig:tWb}
\vspace{9mm}
\end{subfigure}

%%2\vspace{10pt}

%%%%
%TTL
\begin{subfigure}[t]{\textwidth}
\centering
%\begin{center}
\begin{picture}(400,60)(0,0)
\SetColor{Black} \SetWidth{2}
\Photon(  0,25)(45,25){3}{4}
%\Photon(50,75)(69,83){3}{2}
\DashLine(55,28)(95,43){4}
\Photon( 55,21)(95,5){3}{4}
 \Text(0,40)[l]{$W^{\pm}_T$}
 \Text(100,43)[l]{$W^{\pm}_{\longit}$}
 \Text(100,5)[l]{$Z_T$}
 \SetWidth{1}
  \GCirc(50,25){8}{0.7}
%\Text(72.5,85)[]{$\boldsymbol \otimes$}
%
\SetOffset(160,0)
\SetWidth{2}
\Text(-10,25)[]{$\boldsymbol =$}
\Photon(  15,25)(65,25){3}{4}
\Photon(65,25)(84,33){3}{2}
\DashLine(91,36)(110,45){4}
\Photon( 65,25)(110,5){3}{4}
\Text(87.5,35)[]{$\boldsymbol \otimes$}
\Text(125,25)[]{$\boldsymbol +$}
\SetOffset(300,0)
\Photon(  0,25)(45,25){3}{4}
\DashLine(53,27)(95,45){4}
%\Photon( 50,25)(68,17){3}{2}   % +45, -20 overall
\Photon( 53,23)(95,5){3}{4}
\Text(50,25)[]{$\boldsymbol \otimes$}
%\Text(120,25)[]{$\boldsymbol +$}
%
\end{picture}
%\end{center}
\caption{$W_T^{\pm}\to W_{\longit}^{\pm}Z_T$}
\label{fig:TTL}
\vspace{9mm}
\end{subfigure} 

%LLL
\begin{subfigure}[t]{\textwidth}
\centering
%\begin{center}
\begin{picture}(400,120)(0,0)
\SetColor{Black} \SetWidth{2}
\DashLine(  0,90)(45,90){4}
%\Photon(50,75)(69,83){3}{2}
\DashLine(55,93)(95,108){4}
\DashLine( 55,87)(95,70){4}
 \Text(0,100)[l]{$Z_{\longit}$}
 \Text(100,108)[l]{$W^+_{\longit}$}
 \Text(100,70)[l]{$W^-_{\longit}$}
 \SetWidth{1}
  \GCirc(50,90){8}{0.7}
%\Text(72.5,85)[]{$\boldsymbol \otimes$}
%
\SetOffset(20,0)
\SetWidth{2}
\Text(-5,25)[]{$\boldsymbol =$}
\DashLine(  15,25)(65,25){4}
\Photon(65,25)(84,33){3}{2}
\DashLine(91,36)(110,45){4}
\DashLine( 65,25)(110,5){4}
\Text(87.5,35)[]{$\boldsymbol \otimes$}
\Text(135,25)[]{$\boldsymbol +$}
\SetOffset(175,0)
\DashLine(  0,25)(50,25){4}
\DashLine(50,25)(95,45){4}
\Photon( 50,25)(68,17){3}{2}   % +45, -20 overall
\DashLine( 76,13.4)(95,5){4}
\Text(72.5,15)[]{$\boldsymbol \otimes$}
\Text(120,25)[]{$\boldsymbol +$}
\SetOffset(315,0)
\DashLine(  0,25)(21,25){4}
\Photon(  29,25)(50,25){3}{2}
\DashLine(50,25)(95,45){4}
\DashLine( 50,25)(95,5){4}
\Text(25,25)[]{$\boldsymbol \otimes$}
\end{picture}
%\end{center}
\caption{$Z_{\longit}\to W_{\longit}^+W^-_{\longit}$}
\label{fig:LLL}
\vspace{9mm}
\end{subfigure}

%%\vspace{10pt}

%%%%%%%
%hLL
\begin{subfigure}[t]{\textwidth}
\centering
%\begin{center}
\begin{picture}(400,120)(0,0)
\SetColor{Black} \SetWidth{2}
\DashLine(  0,90)(45,90){4}
%\Photon(50,75)(69,83){3}{2}
\DashLine(55,93)(95,108){4}
\DashLine( 55,87)(95,70){4}
 \Text(0,100)[l]{$h$}
 \Text(100,108)[l]{$W^+_{\longit}$}
 \Text(100,70)[l]{$W^-_{\longit}$}
 \SetWidth{1}
  \GCirc(50,90){8}{0.7}
%\Text(72.5,85)[]{$\boldsymbol \otimes$}
%
\SetOffset(20,0)
\SetWidth{2}
\Text(-5,25)[]{$\boldsymbol =$}
\DashLine(  15,25)(65,25){4}
\Photon(65,25)(84,33){3}{2}
\DashLine(91,36)(110,45){4}
\DashLine( 65,25)(110,5){4}
\Text(87.5,35)[]{$\boldsymbol \otimes$}
\Text(135,25)[]{$\boldsymbol +$}

\SetOffset(175,0)
\DashLine(  0,25)(50,25){4}
\DashLine(50,25)(95,45){4}
\Photon( 50,25)(68,17){3}{2}   % +45, -20 overall
\DashLine( 76,13.4)(95,5){4}
\Text(72.5,15)[]{$\boldsymbol \otimes$}
\Text(120,25)[]{$\boldsymbol +$}
\SetOffset(315,0)
\DashLine(  0,25)(45,25){4}
%\Photon(  29,25)(50,25){3}{2}
\DashLine(55,27)(95,45){4}
\DashLine( 55,23)(95,5){4}
\Text(50,25)[]{$\boldsymbol \otimes$}
\end{picture}
%\end{center}
\caption{$h\to W_{\longit}^+W^-_{\longit}$}
\label{fig:hLL}
\end{subfigure}

\caption{Representative ultra-collinear splittings with multiple contributing diagrams. The effects of the VEV are indicated schematically by the symbol $\otimes$. }
\label{fig:ultra_collinear}
\end{figure}

We have seen above how GEG allows us to organize the amplitude's dependence on EWSB by explicitly decomposing it into individual mass-insertion terms, or equivalently VEV-insertion terms. External-state fermion mass insertions are found by Taylor-expanding the fermion Dirac spinors, and external-state gauge boson mass insertions are found via the remainder polarization $\epsilon_n$. For more general processes, there may also be three-point interactions that function as VEV-insertions, such as interactions between the scalars or the $h V^\mu V_\mu$ vertices (listed in Sec.~\ref{sec:three-points}). Generally, we may rather straightforwardly construct any ultra-collinear amplitude at $\order(v)$ by adding together diagrams with exactly one mass-insertion or EWSB interaction. Besides helping to organize a calculation, this approach serves as a convenient tool for visualizing where different EWSB contributions arise in a given amplitude. Figures \ref{fig:ultra_collinear} provide several examples, including
\begin{itemize}
\item Fig.~\ref{fig:tWb}: $t_L\rightarrow  W^+_\longit b_L$, representative calculation for Table \ref{tab:broken_fermion_splittings};
\item Fig.~\ref{fig:TTL}: $W^{\pm}_T\rightarrow W^{\pm}_\longit Z_T$, representative calculation for Table \ref{tab:broken_vector_splittings};
\item Fig.~\ref{fig:LLL}: $Z_\longit \rightarrow W^+_\longit W^-_\longit$, representative calculation for Table \ref{tab:broken_scalar_splittings};
\item Fig.~\ref{fig:hLL}: $h \rightarrow W^+_\longit W^-_\longit$, representative calculation for Table \ref{tab:broken_scalar_splittings}.
\end{itemize}

%%%%%%%%%%%%%%%%%%%%%%%%%%%%%%%%%%%%%%%%%%%%%
\section{Coherent Showering}
\label{app:split}
%%%%%%%%%%%%%%%%%%%%%%%%%%%%%%%%%%%%%%%%%%%%%

Showering involving superpositions of different particle species can be described using density matrix formalism.
The initial value of the density matrix is proportional to the outer product of production amplitudes
$$\rho_{ij} \propto {\mathcal M}_i^{{\rm (prod)}*} {\mathcal M}_j^{\rm (prod)} ,$$ 
tracing out over other details of the rest of the event. Here, the indices run over the species. We nominally assign the state its smallest possible kinematic mass (zero for $\gamma/Z$, $m_Z$ for $h/Z_L$), and subsequently reweight/veto the splitting probability and adjust the global kinematics as necessary (see Section~\ref{sec:mass_effects}). 
This prescription specifically becomes relevant when evolving near kinematic thresholds.

The probability for an initial mixed quantum state to subsequently split into a specific exclusive final state, {\it e.g.}\ $\gamma/Z \to e_L^- e_R^+$ or $\nu_L\bar\nu_R$, must be computed by generalizing the splitting functions to Hermitian splitting {\it matrices} $d{\mathcal P}_{ij}$. The exclusive splitting rates are then computed by tracing against the normalized density matrix:
\beq
d{\cal P} \,=\, \frac{\rho_{ij}\ d{\mathcal P}_{ji}}{{\rm tr}[\rho]}.
\label{eq:rhoSplittingAppendix}
\eeq
If a boson is not split, the Sudakov evolution of $\rho$ proceeds analogous to mixed-state radioactive decay:
\beq
d\rho_{ij} \,=\, -\frac12\sum_{\rm channels}(\rho_{ik}d{\mathcal P}_{kj} + d{\mathcal P}_{ik}\rho_{kj}).  \label{eq:rhoEvolution}
\eeq
As usual, this just represents the wave-function running, now applied to multi-component states. The splitting matrices for an initial mixed quantum state are computed from outer products of splitting amplitudes, convolved with the mixed propagators. Representing the propagator matrix as ${\cal D}_{ij}$, and the amputated splitting amplitudes as ${\cal M}^{\rm (split)}_i$, the generalization from single-state evolution is
\beq
d{\cal P} \,\propto\, \frac{1}{q^4}|{\cal M^{\rm (split)}}|^2 \;\,\rightarrow\,\; d{\cal P}_{ij} \,\propto\, {\cal M}^{\rm (split)*}_k {\cal D}_{ki}^* {\cal D}_{jl} {\cal M}^{\rm (split)}_l. \label{eq:splittingMatrix}
\eeq
Using the relativistic approximation $q^2 \simeq (k_T^2 + \zb m_B^2 + z m_C^2)/z\zb$ for final-state splitting, this modifies Eq.~(\ref{eq:split}) to the more complicated form
\beq
\left[\frac{d{\mathcal P}_{A\rightarrow B+C}}{dz\,d\ktsq}\right]_{ij} \,\simeq\,  {1\over 16\pi^2} \ \frac{1}{z\zb} \  {\cal M}^{\rm (split)*}_k {\cal D}_{ki}^* {\cal D}_{jl} {\cal M}^{\rm (split)}_l   \ .
\eeq
In the massless limit with unmixed propagators, ${\cal D}_{ij} = i\delta_{ij}/q^2$, the form of the splitting matrix reduces to $d{\mathcal P}_{ij} \propto {\mathcal M}_i^{{\rm (split)}*}{\mathcal M}_j^{\rm (split)}/q^4$.

In more complete generality, a mixed state can split into another mixed state, leading to an enlarged set of indices for the splitting matrices. However, in most cases, the final-state density matrices are fully determined by the initial-state density matrices, such that in practice a single pair of indices suffices.

While the formalism is basis-independent, we default to some standard bases in our EW shower approach. Within the unbroken phase (Section~\ref{sec:unbroken}), we present neutral gauge and scalar splitting functions in the interaction basis $(B^0,W^0)$, $(H^0,H^{0*})$. In the broken phase (Section~\ref{sec:broken}), we present them in the mass basis $(\gamma,Z)$, $(h,Z_{\longit})$. The corresponding propagator matrices in the unbroken-phase basis, including the effects of EWSB, are\footnote{The shower formalism automatically accounts for logarithmic running effects in the wavefunction factors for these propagators. We do not attempt to account for mass renormalization effects, as the masses are anyway of power-suppressed importance at very high virtualities. Additional perturbative corrections near the weak scale are also neglected.}
\beq
{\mathcal D}_{B^0B^0} = \frac{i \cos^2\theta_W}{q^2} + \frac{i \sin^2\theta_W}{q^2-m_Z^2}\,,\qquad
{\mathcal D}_{W^0W^0} = \frac{i \sin^2\theta_W}{q^2} + \frac{i \cos^2\theta_W}{q^2-m_Z^2}\,,
\nonumber
\eeq
\beq
{\mathcal D}_{B^0W^0} = {\mathcal D}_{W^0B^0} =  \frac{i\cos\theta_W\sin\theta_W(-m_Z^2)}{q^2(q^2-m_Z^2)}  
\label{eq:BWpropagators}
\eeq
for the gauge bosons ($\theta_W$ is the weak mixing angle), and
\beq
{\mathcal D}_{H^{0}H^{0*}} = {\mathcal D}_{H^{0*}H^{0}} = \frac{i/2}{q^2-m_h^2} + \frac{i/2}{q^2-m_Z^2}\,, 
\nonumber
\eeq
\beq
{\mathcal D}_{H^{0}H^{0}} = {\mathcal D}_{H^{0*}H^{0*}} = \frac{i/2}{q^2-m_h^2} - \frac{i/2}{q^2-m_Z^2} ,
\label{eq:hZpropagators}
\eeq
for the neutral scalars. In the mass basis, the matrices are diagonal and have entries corresponding to the usual poles:
\beq
 {\cal D}_{\gamma\gamma} = \frac{i}{q^2}\,, \quad {\cal D}_{ZZ} = \frac{i}{q^2-m_Z^2}\,, \quad {\cal D}_{\gamma Z} = {\cal D}_{Z\gamma} = 0
\eeq
\beq
 {\cal D}_{hh} = \frac{i}{q^2-m_h^2}\,, \quad {\cal D}_{Z_{\longit}Z_{\longit}} = \frac{i}{q^2-m_Z^2}\,, \quad {\cal D}_{h Z_{\longit}} = {\cal D}_{Z_{\longit}h} = 0.
\eeq

Similar considerations apply in the application and generation of PDFs~\cite{Ciafaloni:2005fm}. The $\gamma/Z$ and (in principle) $h/Z_{\longit}$ PDFs should each properly be treated as $2\times2$ matrices, and hard process cross sections sourced by these PDFs computed by tracing against the hard matrix elements. The PDF evolution equations involve matrix-valued splitting functions. In the high-$k_T$/high-virtuality limit, these follow straightforwardly from the splitting functions presented in the Section~\ref{sec:unbroken}. However, unless working well above the TeV-scale, mass effects can still be important. The above propagator modifications must then be applied at the (spacelike) virtual leg emerging from a splitting.

%%%%%%%%%%%%%%%%%%%%%%%%%%%%%%%%%%%%%%%%%%%
\section{Final-State Shower Simulation}
\label{sec:FSR}
%%%%%%%%%%%%%%%%%%%%%%%%%%%%%%%%%%%%%%%%%%%

In order to facilitate studies of final-state weak showering at the level of exclusive rates, we have programmed a variation of the {\tt PYTHIA6}~\cite{Sjostrand:2006za} timelike virtuality-ordered parton shower. Basic collinear QCD is included by default, extended to the massive showering formalism outlined in Section~\ref{sec:split}, and including purely ultra-collinear processes. In addition, the full set of weak showering processes described in this paper has been added, with a number of novel features compared to standard showering programs, outlined in the main text. In particular, see Section~\ref{sec:novel_features}. Here we describe a few additional technicalities of the implementation.

Splitting functions in the virtuality-ordered shower are simple to relate to those in the $k_T$-ordered shower, which we have used by default for most of the presentation. Using the relativistic/collinear approximation for a splitting $A \to B+C$, we get
\beq
Q^2 \,\simeq\, \frac{1}{z\zb} (k_T^2 + \zb m_B^2 + z m_C^2) \, .  \label{eq:kTtoQ}
\eeq
Working in $\log Q$, we can build the translation
\beq
\frac{d{\cal P}}{dz \, d\log Q^2} \,\simeq\,  \frac{1}{z\zb} \frac{Q^2}{(Q^2 - m_A^2)^2} \left( \tilde k_T^4 \frac{d{\cal P}}{dz \, d\ktsq}  \right) \, .
\eeq
The function in parentheses goes either as $k_T^2$ or as $v^2$. For given $Q$, $z$, and daughter masses, the former is simple to derive either by inverting the approximate 
Eq.~(\ref{eq:kTtoQ}) or by using exact kinematics. For the energy-sharing variable $z$, we use CM-frame three-momentum fraction $|\vec k_B|/(|\vec k_B|+|\vec k_C|)$. To approximately model the phase space effects in the nonrelativistic limit, we further weight the splitting probabilities by a velocity factor $|\vec k_B| |\vec k_C|/E_B E_C$. We also suppress splittings at angles larger than $\theta \approx \pi/2$, where the collinear shower would be highly untrustworthy.

As in {\tt PYTHIA6}, the input to the shower is a ``hard'' partonic configuration with some characteristic virtuality scale, assumed here to be large compared to the weak scale. Evolution is based on a simple recoiler method, whereby particles are showered in pairs. (At the current level, no dipole coherence effects or color/isospin flows are incorporated, nor are they strictly necessary at leading-log level, but they would be possible to include in more advanced approaches.) Each particle in a pair undergoes a trial QCD/EW Sudakov evolution, defined in the hard event's rest frame, and ignoring the possible evolution of its sister. In general, each particle may undergo a $1\to2$ splitting and acquire an off-shell mass. Kinematics are then adjusted within the pair's rest frame, by boosting each showered system along the pair's axis to preserve momentum and energy. If the summed masses from the trial evolutions exceeds the original pair's mass, the more off-shell splitting is vetoed, and that particle's evolution restarted. The procedure is easily recursed to build up completely showered events, with the two daughters from a given splitting serving as paired sisters in subsequent evolution.

Kinematic back-reaction effects are also incorporated, as discussed in Section~\ref{sec:mass_effects} and parametrized in Eq.~\ref{eq:weight}. The kinematic re-arrangments required by setting a daughter off-shell through its secondary showering can have a sizable effect on the mother's splitting rate. We introduce this back-reaction factor as an additional weight multiplying the daughter's splitting probability. In our virtuality-ordered implementation, the virtuality of the mother (invariant mass of the daughter pair) remains unchanged, so $Q^*=Q$. The Jacobian for the transformation is then simply $|dz^*/dz|$, and its explicit form is tied to our kinematic prescription above.  Within the mother splitting $A \to B+C$, assume that particle $B$ with momentum-fraction $z$ is the one to be set off-shell: $B \to B^*$. Within the $A$ rest-frame, the direction of $B$ ($C$) is held at a fixed angle $\Theta$ ($\pi-\Theta$) relative to $A$'s boost axis from the CM-frame. The angle $\Theta$ has a one-to-one mapping to both the old $z$ and the new $z^*$, and is a useful intermediate variable. Another useful intermediate variable is the ratio $Y \equiv z^2/\zb^2$, and the analogous $Y^*$. The Jacobian can then be built up in pieces as
\beq 
 \left| \frac{dz^*}{dz} \right|  \,=\,  \left| \frac{dY^*}{dz^*} \right|^{-1} \, \left| \frac{dY^*}{d\Theta} \right| \, \left| \frac{dY}{d\Theta} \right|^{-1} \, \left| \frac{dY}{dz} \right|  \, ,
\eeq
where, 
\beq
\frac{dY}{dz} \,=\, \frac{2z}{\zb^3}
\eeq
and
\beq
\frac{dY}{d\cos\Theta} \,=\, \frac{\calA(\bar\calB-\calB)\cos^2\Theta + 2\calA(\bar\calC - \calC)\cos\Theta + (\calB\bar\calC - \bar\calB\calC)}{(\calA\cos^2\Theta + \bar\calB\cos\Theta + \bar\calC)^2}  \, .
\eeq
The symbols $\calA$, etc, here are shorthand for various quantities built out of $A$'s velocity $\beta_A$, and daughter kinematics in its rest-frame: the $A$-frame three-momentum magnitude of either of the daughters $P$, and their individual $A$-frame energies and kinematic masses $E_B$, $E_C$, $m_B$, $m_C$. We have
\beq
\begin{matrix}
\calA \,\equiv\,  \beta_A^2 P^2 \, , \quad \calB \,\equiv\, 2\beta_A P E_B \, , \quad \bar\calB \,\equiv\, -2\beta_A P E_C \, , \\ \\
\calC \,\equiv\, P^2 + \beta_A^2 m_B^2 \, , \quad \bar\calC \,\equiv\, P^2 + \beta_A^2 m_C^2 \, .
\end{matrix}
\eeq
Analogous formulas hold with $Y^*$ and $z^*$, defining the coefficients $\calA$, etc, using the $A$-frame kinematic quantities redefined with $B$ set off-shell. (Prescriptions yielding simpler analytic formulas than ours almost certainly exist.) The differential splitting function of the mother must also be re-evaluated using the off-shell daughter kinematics. This is much simpler, as there the main effect is just the change in $z$. Explicit EWSB mass factors for the daughters, which appear in the numerators of the ultra-collinear splitting functions, are not adjusted from their on-shell values.

Angular-ordering may also be invoked. If the showering pair was itself produced from a splitting, the event-frame angles of each daughter splitting and mother splitting can be compared, and the former splitting(s) vetoed if it has a larger angle. This veto may be applied selectively depending on the nature of the splitting and its parent splitting.

In our approach, parton shower evolution is automatically matched onto decay for $W^\pm$, $Z$, Higgs, and top. This matching is particularly simple in the virtuality-ordered shower. Particles that survive down to their decay/shower matching scale are assigned masses drawn from a Breit-Wigner distribution and final-state flavors assigned according to known branching fractions. In practice, we also weight the Breit-Wigner distribution accounting for the different available decay phase space at different off-shell virtualities. Similar to a shower splitting, the decays are then further weighted with back-reaction factors, if the decaying particle was itself produced in a splitting. The back-reaction factor here is applied as a simple probabilistic veto.

Finally, we re-emphasize that the neutral bosons $\gamma/Z_T$ and $h/Z_{\longit}$ are produced and evolved as general quantum mixed states. They are assigned initial kinematic masses of zero and $m_Z$, respectively, and given nontrivial $2\times 2$ density matrices that evolve via matrix-valued Sudakov factors. There is one major practical difference in implementing these Sudakovs relative to simple number-valued Sudakovs. In the latter case, a given particle's wavefunction decreases in magnitude as its evolution proceeds, but the surviving probability is an automatic outcome of the differential splitting rates integrated via monte carlo. In practice, these splitting rates are integrated over $z$ with the expedient of over-estimator functions, and vetoed-down to the true rates. In the matrix-valued case, however, the wavefunction can also {\it rotate}, and capturing this effect using over-estimator functions and a veto algorithm does not appear to be as straightforward. Instead, we use explicit formulas for the $z$-integrated splitting matrices at each virtuality step. These formulas are necessarily approximate, but we have verified that they yield results similar to what would be obtained by costly brute-force numerical integration.

%\input{extra}

%%%%%%%%%%%%%%
% References
%%%%%%%%%%%%%%

\bibliography{lit}

\begin{thebibliography}{103}
\expandafter\ifx\csname natexlab\endcsname\relax\def\natexlab#1{#1}\fi
\expandafter\ifx\csname bibnamefont\endcsname\relax
  \def\bibnamefont#1{#1}\fi
\expandafter\ifx\csname bibfnamefont\endcsname\relax
  \def\bibfnamefont#1{#1}\fi
\expandafter\ifx\csname citenamefont\endcsname\relax
  \def\citenamefont#1{#1}\fi
\expandafter\ifx\csname url\endcsname\relax
  \def\url#1{\texttt{#1}}\fi
\expandafter\ifx\csname urlprefix\endcsname\relax\def\urlprefix{URL }\fi
\providecommand{\bibinfo}[2]{#2}
\providecommand{\eprint}[2][]{\url{#2}}

\bibitem[{\citenamefont{Collins et~al.}(1985)\citenamefont{Collins, Soper, and
  Sterman}}]{Collins:1984kg}
\bibinfo{author}{\bibfnamefont{J.~C.} \bibnamefont{Collins}},
  \bibinfo{author}{\bibfnamefont{D.~E.} \bibnamefont{Soper}}, \bibnamefont{and}
  \bibinfo{author}{\bibfnamefont{G.~F.} \bibnamefont{Sterman}},
  \emph{\bibinfo{title}{{Transverse Momentum Distribution in Drell-Yan Pair and
  W and Z Boson Production}}}, \bibinfo{journal}{Nucl. Phys.}
  \textbf{\bibinfo{volume}{B250}}, \bibinfo{pages}{199} (\bibinfo{year}{1985}).

\bibitem[{\citenamefont{Collins et~al.}(1989)\citenamefont{Collins, Soper, and
  Sterman}}]{Collins:1989gx}
\bibinfo{author}{\bibfnamefont{J.~C.} \bibnamefont{Collins}},
  \bibinfo{author}{\bibfnamefont{D.~E.} \bibnamefont{Soper}}, \bibnamefont{and}
  \bibinfo{author}{\bibfnamefont{G.~F.} \bibnamefont{Sterman}},
  \emph{\bibinfo{title}{{Factorization of Hard Processes in QCD}}},
  \bibinfo{journal}{Adv. Ser. Direct. High Energy Phys.}
  \textbf{\bibinfo{volume}{5}}, \bibinfo{pages}{1} (\bibinfo{year}{1989}),
  \eprint{hep-ph/0409313}.

\bibitem[{\citenamefont{Bengtsson and Sjostrand}(1987)}]{Bengtsson:1986et}
\bibinfo{author}{\bibfnamefont{M.}~\bibnamefont{Bengtsson}} \bibnamefont{and}
  \bibinfo{author}{\bibfnamefont{T.}~\bibnamefont{Sjostrand}},
  \emph{\bibinfo{title}{{A Comparative Study of Coherent and Noncoherent Parton
  Shower Evolution}}}, \bibinfo{journal}{Nucl. Phys.}
  \textbf{\bibinfo{volume}{B289}}, \bibinfo{pages}{810} (\bibinfo{year}{1987}).

\bibitem[{\citenamefont{Sjostrand et~al.}(2008)\citenamefont{Sjostrand, Mrenna,
  and Skands}}]{Sjostrand:2007gs}
\bibinfo{author}{\bibfnamefont{T.}~\bibnamefont{Sjostrand}},
  \bibinfo{author}{\bibfnamefont{S.}~\bibnamefont{Mrenna}}, \bibnamefont{and}
  \bibinfo{author}{\bibfnamefont{P.~Z.} \bibnamefont{Skands}},
  \emph{\bibinfo{title}{{A Brief Introduction to PYTHIA 8.1}}},
  \bibinfo{journal}{Comput.Phys.Commun.} \textbf{\bibinfo{volume}{178}},
  \bibinfo{pages}{852} (\bibinfo{year}{2008}), \eprint{0710.3820}.

\bibitem[{\citenamefont{Bahr et~al.}(2008)}]{Bahr:2008pv}
\bibinfo{author}{\bibfnamefont{M.}~\bibnamefont{Bahr}} \bibnamefont{et~al.},
  \emph{\bibinfo{title}{{Herwig++ Physics and Manual}}}, \bibinfo{journal}{Eur.
  Phys. J.} \textbf{\bibinfo{volume}{C58}}, \bibinfo{pages}{639}
  (\bibinfo{year}{2008}), \eprint{0803.0883}.

\bibitem[{\citenamefont{Gleisberg et~al.}(2009)\citenamefont{Gleisberg, Hoeche,
  Krauss, Schonherr, Schumann, Siegert, and Winter}}]{Gleisberg:2008ta}
\bibinfo{author}{\bibfnamefont{T.}~\bibnamefont{Gleisberg}},
  \bibinfo{author}{\bibfnamefont{S.}~\bibnamefont{Hoeche}},
  \bibinfo{author}{\bibfnamefont{F.}~\bibnamefont{Krauss}},
  \bibinfo{author}{\bibfnamefont{M.}~\bibnamefont{Schonherr}},
  \bibinfo{author}{\bibfnamefont{S.}~\bibnamefont{Schumann}},
  \bibinfo{author}{\bibfnamefont{F.}~\bibnamefont{Siegert}}, \bibnamefont{and}
  \bibinfo{author}{\bibfnamefont{J.}~\bibnamefont{Winter}},
  \emph{\bibinfo{title}{{Event Generation with SHERPA 1.1}}},
  \bibinfo{journal}{JHEP} \textbf{\bibinfo{volume}{02}}, \bibinfo{pages}{007}
  (\bibinfo{year}{2009}), \eprint{0811.4622}.

\bibitem[{\citenamefont{Knapp et~al.}(2003)\citenamefont{Knapp, Heck, Sciutto,
  Dova, and Risse}}]{Knapp:2002vs}
\bibinfo{author}{\bibfnamefont{J.}~\bibnamefont{Knapp}},
  \bibinfo{author}{\bibfnamefont{D.}~\bibnamefont{Heck}},
  \bibinfo{author}{\bibfnamefont{S.~J.} \bibnamefont{Sciutto}},
  \bibinfo{author}{\bibfnamefont{M.~T.} \bibnamefont{Dova}}, \bibnamefont{and}
  \bibinfo{author}{\bibfnamefont{M.}~\bibnamefont{Risse}},
  \emph{\bibinfo{title}{{Extensive Air Shower Simulations at the Highest
  Energies}}}, \bibinfo{journal}{Astropart. Phys.}
  \textbf{\bibinfo{volume}{19}}, \bibinfo{pages}{77} (\bibinfo{year}{2003}),
  \eprint{astro-ph/0206414}.

\bibitem[{\citenamefont{Cirelli et~al.}(2009)\citenamefont{Cirelli, Kadastik,
  Raidal, and Strumia}}]{Cirelli:2008pk}
\bibinfo{author}{\bibfnamefont{M.}~\bibnamefont{Cirelli}},
  \bibinfo{author}{\bibfnamefont{M.}~\bibnamefont{Kadastik}},
  \bibinfo{author}{\bibfnamefont{M.}~\bibnamefont{Raidal}}, \bibnamefont{and}
  \bibinfo{author}{\bibfnamefont{A.}~\bibnamefont{Strumia}},
  \emph{\bibinfo{title}{{Model-Independent Implications of the $e^\pm$,
  Anti-Proton Cosmic Ray Spectra on Properties of Dark Matter}}},
  \bibinfo{journal}{Nucl. Phys.} \textbf{\bibinfo{volume}{B813}},
  \bibinfo{pages}{1} (\bibinfo{year}{2009}), \bibinfo{note}{[Addendum: Nucl.
  Phys.B873,530(2013)]}, \eprint{0809.2409}.

\bibitem[{\citenamefont{Cirelli et~al.}(2011)\citenamefont{Cirelli, Corcella,
  Hektor, Hutsi, Kadastik, Panci, Raidal, Sala, and Strumia}}]{Cirelli:2010xx}
\bibinfo{author}{\bibfnamefont{M.}~\bibnamefont{Cirelli}},
  \bibinfo{author}{\bibfnamefont{G.}~\bibnamefont{Corcella}},
  \bibinfo{author}{\bibfnamefont{A.}~\bibnamefont{Hektor}},
  \bibinfo{author}{\bibfnamefont{G.}~\bibnamefont{Hutsi}},
  \bibinfo{author}{\bibfnamefont{M.}~\bibnamefont{Kadastik}},
  \bibinfo{author}{\bibfnamefont{P.}~\bibnamefont{Panci}},
  \bibinfo{author}{\bibfnamefont{M.}~\bibnamefont{Raidal}},
  \bibinfo{author}{\bibfnamefont{F.}~\bibnamefont{Sala}}, \bibnamefont{and}
  \bibinfo{author}{\bibfnamefont{A.}~\bibnamefont{Strumia}},
  \emph{\bibinfo{title}{{PPPC 4 DM ID: A Poor Particle Physicist Cookbook for
  Dark Matter Indirect Detection}}}, \bibinfo{journal}{JCAP}
  \textbf{\bibinfo{volume}{1103}}, \bibinfo{pages}{051} (\bibinfo{year}{2011}),
  \bibinfo{note}{[Erratum: JCAP1210,E01(2012)]}, \eprint{1012.4515}.

\bibitem[{\citenamefont{Arkani-Hamed et~al.}(2015)\citenamefont{Arkani-Hamed,
  Han, Mangano, and Wang}}]{Arkani-Hamed:2015vfh}
\bibinfo{author}{\bibfnamefont{N.}~\bibnamefont{Arkani-Hamed}},
  \bibinfo{author}{\bibfnamefont{T.}~\bibnamefont{Han}},
  \bibinfo{author}{\bibfnamefont{M.}~\bibnamefont{Mangano}}, \bibnamefont{and}
  \bibinfo{author}{\bibfnamefont{L.-T.} \bibnamefont{Wang}},
  \emph{\bibinfo{title}{{Physics Opportunities of a 100 TeV Proton-Proton
  Collider}}} (\bibinfo{year}{2015}), \eprint{1511.06495}.

\bibitem[{\citenamefont{Mangano et~al.}(2016{\natexlab{a}})}]{Mangano:2016jyj}
\bibinfo{author}{\bibfnamefont{M.~L.} \bibnamefont{Mangano}}
  \bibnamefont{et~al.}, \emph{\bibinfo{title}{{Physics at a 100 TeV $pp$
  Collider: Standard Model processes}}} (\bibinfo{year}{2016}{\natexlab{a}}),
  \eprint{1607.01831}.

\bibitem[{\citenamefont{Golling et~al.}(2016)}]{Golling:2016gvc}
\bibinfo{author}{\bibfnamefont{T.}~\bibnamefont{Golling}} \bibnamefont{et~al.},
  \emph{\bibinfo{title}{{Physics at a 100 TeV $pp$ Collider: Beyond the
  Standard Model phenomena}}} (\bibinfo{year}{2016}), \eprint{1606.00947}.

\bibitem[{\citenamefont{Abramowski et~al.}(2013)}]{Abramowski:2013ax}
\bibinfo{author}{\bibfnamefont{A.}~\bibnamefont{Abramowski}}
  \bibnamefont{et~al.} (\bibinfo{collaboration}{HESS}),
  \emph{\bibinfo{title}{{Search for Photon-Linelike Signatures from Dark Matter
  Annihilations with H.E.S.S.}}}, \bibinfo{journal}{Phys. Rev. Lett.}
  \textbf{\bibinfo{volume}{110}}, \bibinfo{pages}{041301}
  (\bibinfo{year}{2013}), \eprint{1301.1173}.

\bibitem[{\citenamefont{Lefranc and Moulin}(2015)}]{Lefranc:2015vza}
\bibinfo{author}{\bibfnamefont{V.}~\bibnamefont{Lefranc}} \bibnamefont{and}
  \bibinfo{author}{\bibfnamefont{E.}~\bibnamefont{Moulin}},
  \emph{\bibinfo{title}{{Dark Matter Search in the Inner Galactic halo with
  H.E.S.S. I and H.E.S.S. II}}} (\bibinfo{year}{2015}), \eprint{1509.04123}.

\bibitem[{\citenamefont{Carr et~al.}(2015)}]{Carr:2015hta}
\bibinfo{author}{\bibfnamefont{J.}~\bibnamefont{Carr}} \bibnamefont{et~al.}
  (\bibinfo{collaboration}{CTA Consortium}), \emph{\bibinfo{title}{{Prospects
  for Indirect Dark Matter Searches with the Cherenkov Telescope Array (CTA)}}}
  (\bibinfo{year}{2015}), \eprint{1508.06128}.

\bibitem[{\citenamefont{Dawson et~al.}(2014)\citenamefont{Dawson, Ismail, and
  Low}}]{Dawson:2014pea}
\bibinfo{author}{\bibfnamefont{S.}~\bibnamefont{Dawson}},
  \bibinfo{author}{\bibfnamefont{A.}~\bibnamefont{Ismail}}, \bibnamefont{and}
  \bibinfo{author}{\bibfnamefont{I.}~\bibnamefont{Low}},
  \emph{\bibinfo{title}{{Redux on ``When is the Top Quark a Parton"}}},
  \bibinfo{journal}{Phys. Rev.} \textbf{\bibinfo{volume}{D90}},
  \bibinfo{pages}{014005} (\bibinfo{year}{2014}), \eprint{1405.6211}.

\bibitem[{\citenamefont{Han et~al.}(2015)\citenamefont{Han, Sayre, and
  Westhoff}}]{Han:2014nja}
\bibinfo{author}{\bibfnamefont{T.}~\bibnamefont{Han}},
  \bibinfo{author}{\bibfnamefont{J.}~\bibnamefont{Sayre}}, \bibnamefont{and}
  \bibinfo{author}{\bibfnamefont{S.}~\bibnamefont{Westhoff}},
  \emph{\bibinfo{title}{{Top-Quark Initiated Processes at High-Energy Hadron
  Colliders}}}, \bibinfo{journal}{JHEP} \textbf{\bibinfo{volume}{04}},
  \bibinfo{pages}{145} (\bibinfo{year}{2015}), \eprint{1411.2588}.

\bibitem[{\citenamefont{Almeida et~al.}(2009)\citenamefont{Almeida, Lee, Perez,
  Sung, and Virzi}}]{Almeida:2008tp}
\bibinfo{author}{\bibfnamefont{L.~G.} \bibnamefont{Almeida}},
  \bibinfo{author}{\bibfnamefont{S.~J.} \bibnamefont{Lee}},
  \bibinfo{author}{\bibfnamefont{G.}~\bibnamefont{Perez}},
  \bibinfo{author}{\bibfnamefont{I.}~\bibnamefont{Sung}}, \bibnamefont{and}
  \bibinfo{author}{\bibfnamefont{J.}~\bibnamefont{Virzi}},
  \emph{\bibinfo{title}{{Top Jets at the LHC}}}, \bibinfo{journal}{Phys. Rev.}
  \textbf{\bibinfo{volume}{D79}}, \bibinfo{pages}{074012}
  (\bibinfo{year}{2009}), \eprint{0810.0934}.

\bibitem[{\citenamefont{Kane et~al.}(1984)\citenamefont{Kane, Repko, and
  Rolnick}}]{Kane:1984bb}
\bibinfo{author}{\bibfnamefont{G.~L.} \bibnamefont{Kane}},
  \bibinfo{author}{\bibfnamefont{W.}~\bibnamefont{Repko}}, \bibnamefont{and}
  \bibinfo{author}{\bibfnamefont{W.}~\bibnamefont{Rolnick}},
  \emph{\bibinfo{title}{{The Effective $W^\pm$, $Z^0$ Approximation for
  High-Energy Collisions}}}, \bibinfo{journal}{Phys.Lett.}
  \textbf{\bibinfo{volume}{B148}}, \bibinfo{pages}{367} (\bibinfo{year}{1984}).

\bibitem[{\citenamefont{Dawson}(1985)}]{Dawson:1984gx}
\bibinfo{author}{\bibfnamefont{S.}~\bibnamefont{Dawson}},
  \emph{\bibinfo{title}{{The Effective W Approximation}}},
  \bibinfo{journal}{Nucl.Phys.} \textbf{\bibinfo{volume}{B249}},
  \bibinfo{pages}{42} (\bibinfo{year}{1985}).

\bibitem[{\citenamefont{Chanowitz and Gaillard}(1985)}]{Chanowitz:1985hj}
\bibinfo{author}{\bibfnamefont{M.~S.} \bibnamefont{Chanowitz}}
  \bibnamefont{and} \bibinfo{author}{\bibfnamefont{M.~K.}
  \bibnamefont{Gaillard}}, \emph{\bibinfo{title}{{The TeV Physics of Strongly
  Interacting W's and Z's}}}, \bibinfo{journal}{Nucl. Phys.}
  \textbf{\bibinfo{volume}{B261}}, \bibinfo{pages}{379} (\bibinfo{year}{1985}).

\bibitem[{\citenamefont{{ATLAS and CMS collaborations}}()}]{LHCVBF}
\bibinfo{author}{\bibnamefont{{ATLAS and CMS collaborations}}},
  \emph{\bibinfo{title}{{Measurements of the Higgs Boson Production and Decay
  Rates and Constraints on its Couplings from a Combined ATLAS and CMS Analysis
  of the LHC $pp$ Collision Data at $\sqrt{s} = 7$ and~8 TeV}}},
  \eprint{{ATLAS-CONF-2015-044,CMS-PAS-HIG-15-002}}.

\bibitem[{\citenamefont{Melles}(2001)}]{Melles:2000gw}
\bibinfo{author}{\bibfnamefont{M.}~\bibnamefont{Melles}},
  \emph{\bibinfo{title}{{Subleading Sudakov Logarithms in Electroweak
  High-Energy Processes to All Orders}}}, \bibinfo{journal}{Phys. Rev.}
  \textbf{\bibinfo{volume}{D63}}, \bibinfo{pages}{034003}
  (\bibinfo{year}{2001}), \eprint{hep-ph/0004056}.

\bibitem[{\citenamefont{Beenakker and Werthenbach}(2000)}]{Beenakker:2000kb}
\bibinfo{author}{\bibfnamefont{W.}~\bibnamefont{Beenakker}} \bibnamefont{and}
  \bibinfo{author}{\bibfnamefont{A.}~\bibnamefont{Werthenbach}},
  \emph{\bibinfo{title}{{New Insights into the Perturbative Structure of
  Electroweak Sudakov Logarithms}}}, \bibinfo{journal}{Phys. Lett.}
  \textbf{\bibinfo{volume}{B489}}, \bibinfo{pages}{148} (\bibinfo{year}{2000}),
  \eprint{hep-ph/0005316}.

\bibitem[{\citenamefont{Moretti et~al.}(2006)\citenamefont{Moretti, Nolten, and
  Ross}}]{Moretti:2006ea}
\bibinfo{author}{\bibfnamefont{S.}~\bibnamefont{Moretti}},
  \bibinfo{author}{\bibfnamefont{M.}~\bibnamefont{Nolten}}, \bibnamefont{and}
  \bibinfo{author}{\bibfnamefont{D.}~\bibnamefont{Ross}},
  \emph{\bibinfo{title}{{Weak Corrections to Four-Parton Processes}}},
  \bibinfo{journal}{Nucl.Phys.} \textbf{\bibinfo{volume}{B759}},
  \bibinfo{pages}{50} (\bibinfo{year}{2006}), \eprint{hep-ph/0606201}.

\bibitem[{\citenamefont{Dittmaier et~al.}(2012)\citenamefont{Dittmaier, Huss,
  and Speckner}}]{Dittmaier:2012kx}
\bibinfo{author}{\bibfnamefont{S.}~\bibnamefont{Dittmaier}},
  \bibinfo{author}{\bibfnamefont{A.}~\bibnamefont{Huss}}, \bibnamefont{and}
  \bibinfo{author}{\bibfnamefont{C.}~\bibnamefont{Speckner}},
  \emph{\bibinfo{title}{{Weak Radiative Corrections to Dijet Production at
  Hadron Colliders}}}, \bibinfo{journal}{JHEP} \textbf{\bibinfo{volume}{1211}},
  \bibinfo{pages}{095} (\bibinfo{year}{2012}), \eprint{1210.0438}.

\bibitem[{\citenamefont{Kuhn et~al.}(2005{\natexlab{a}})\citenamefont{Kuhn,
  Kulesza, Pozzorini, and Schulze}}]{Kuhn:2004em}
\bibinfo{author}{\bibfnamefont{J.~H.} \bibnamefont{Kuhn}},
  \bibinfo{author}{\bibfnamefont{A.}~\bibnamefont{Kulesza}},
  \bibinfo{author}{\bibfnamefont{S.}~\bibnamefont{Pozzorini}},
  \bibnamefont{and} \bibinfo{author}{\bibfnamefont{M.}~\bibnamefont{Schulze}},
  \emph{\bibinfo{title}{{Logarithmic Electroweak Corrections to Hadronic Z+1
  Jet Production at Large Transverse Momentum}}}, \bibinfo{journal}{Phys.
  Lett.} \textbf{\bibinfo{volume}{B609}}, \bibinfo{pages}{277}
  (\bibinfo{year}{2005}{\natexlab{a}}), \eprint{hep-ph/0408308}.

\bibitem[{\citenamefont{Kuhn et~al.}(2005{\natexlab{b}})\citenamefont{Kuhn,
  Kulesza, Pozzorini, and Schulze}}]{Kuhn:2005az}
\bibinfo{author}{\bibfnamefont{J.~H.} \bibnamefont{Kuhn}},
  \bibinfo{author}{\bibfnamefont{A.}~\bibnamefont{Kulesza}},
  \bibinfo{author}{\bibfnamefont{S.}~\bibnamefont{Pozzorini}},
  \bibnamefont{and} \bibinfo{author}{\bibfnamefont{M.}~\bibnamefont{Schulze}},
  \emph{\bibinfo{title}{{One-Loop Weak Corrections to Hadronic Production of
  $Z$ Bosons at Large Transverse Momenta}}}, \bibinfo{journal}{Nucl. Phys.}
  \textbf{\bibinfo{volume}{B727}}, \bibinfo{pages}{368}
  (\bibinfo{year}{2005}{\natexlab{b}}), \eprint{hep-ph/0507178}.

\bibitem[{\citenamefont{Kuhn et~al.}(2006)\citenamefont{Kuhn, Kulesza,
  Pozzorini, and Schulze}}]{Kuhn:2005gv}
\bibinfo{author}{\bibfnamefont{J.~H.} \bibnamefont{Kuhn}},
  \bibinfo{author}{\bibfnamefont{A.}~\bibnamefont{Kulesza}},
  \bibinfo{author}{\bibfnamefont{S.}~\bibnamefont{Pozzorini}},
  \bibnamefont{and} \bibinfo{author}{\bibfnamefont{M.}~\bibnamefont{Schulze}},
  \emph{\bibinfo{title}{{Electroweak Corrections to Hadronic Photon Production
  at Large Transverse Momenta}}}, \bibinfo{journal}{JHEP}
  \textbf{\bibinfo{volume}{03}}, \bibinfo{pages}{059} (\bibinfo{year}{2006}),
  \eprint{hep-ph/0508253}.

\bibitem[{\citenamefont{Kuhn et~al.}(2007)\citenamefont{Kuhn, Kulesza,
  Pozzorini, and Schulze}}]{Kuhn:2007qc}
\bibinfo{author}{\bibfnamefont{J.~H.} \bibnamefont{Kuhn}},
  \bibinfo{author}{\bibfnamefont{A.}~\bibnamefont{Kulesza}},
  \bibinfo{author}{\bibfnamefont{S.}~\bibnamefont{Pozzorini}},
  \bibnamefont{and} \bibinfo{author}{\bibfnamefont{M.}~\bibnamefont{Schulze}},
  \emph{\bibinfo{title}{{Electroweak Corrections to Large Transverse Momentum
  Production of $W$ Bosons at the LHC}}}, \bibinfo{journal}{Phys. Lett.}
  \textbf{\bibinfo{volume}{B651}}, \bibinfo{pages}{160} (\bibinfo{year}{2007}),
  \eprint{hep-ph/0703283}.

\bibitem[{\citenamefont{Kuhn et~al.}(2008)\citenamefont{Kuhn, Kulesza,
  Pozzorini, and Schulze}}]{Kuhn:2007cv}
\bibinfo{author}{\bibfnamefont{J.~H.} \bibnamefont{Kuhn}},
  \bibinfo{author}{\bibfnamefont{A.}~\bibnamefont{Kulesza}},
  \bibinfo{author}{\bibfnamefont{S.}~\bibnamefont{Pozzorini}},
  \bibnamefont{and} \bibinfo{author}{\bibfnamefont{M.}~\bibnamefont{Schulze}},
  \emph{\bibinfo{title}{{Electroweak Corrections to Hadronic Production of $W$
  Bosons at Large Transverse Momenta}}}, \bibinfo{journal}{Nucl. Phys.}
  \textbf{\bibinfo{volume}{B797}}, \bibinfo{pages}{27} (\bibinfo{year}{2008}),
  \eprint{0708.0476}.

\bibitem[{\citenamefont{Hollik et~al.}(2008)\citenamefont{Hollik, Kasprzik, and
  Kniehl}}]{Hollik:2007sq}
\bibinfo{author}{\bibfnamefont{W.}~\bibnamefont{Hollik}},
  \bibinfo{author}{\bibfnamefont{T.}~\bibnamefont{Kasprzik}}, \bibnamefont{and}
  \bibinfo{author}{\bibfnamefont{B.~A.} \bibnamefont{Kniehl}},
  \emph{\bibinfo{title}{{Electroweak Corrections to $W$-Boson Hadroproduction
  at Finite Transverse Momentum}}}, \bibinfo{journal}{Nucl. Phys.}
  \textbf{\bibinfo{volume}{B790}}, \bibinfo{pages}{138} (\bibinfo{year}{2008}),
  \eprint{0707.2553}.

\bibitem[{\citenamefont{Becher and Garcia~i Tormo}(2013)}]{Becher:2013zua}
\bibinfo{author}{\bibfnamefont{T.}~\bibnamefont{Becher}} \bibnamefont{and}
  \bibinfo{author}{\bibfnamefont{X.}~\bibnamefont{Garcia~i Tormo}},
  \emph{\bibinfo{title}{{Electroweak Sudakov Effects in $W, Z$ and $\gamma$
  Production at Large Transverse Momentum}}}, \bibinfo{journal}{Phys. Rev.}
  \textbf{\bibinfo{volume}{D88}}, \bibinfo{pages}{013009}
  (\bibinfo{year}{2013}), \eprint{1305.4202}.

\bibitem[{\citenamefont{Denner and
  Pozzorini}(2001{\natexlab{a}})}]{Denner:2000jv}
\bibinfo{author}{\bibfnamefont{A.}~\bibnamefont{Denner}} \bibnamefont{and}
  \bibinfo{author}{\bibfnamefont{S.}~\bibnamefont{Pozzorini}},
  \emph{\bibinfo{title}{{One Loop Leading Logarithms in Electroweak Radiative
  Corrections. 1. Results}}}, \bibinfo{journal}{Eur. Phys. J.}
  \textbf{\bibinfo{volume}{C18}}, \bibinfo{pages}{461}
  (\bibinfo{year}{2001}{\natexlab{a}}), \eprint{hep-ph/0010201}.

\bibitem[{\citenamefont{Denner and
  Pozzorini}(2001{\natexlab{b}})}]{Denner:2001gw}
\bibinfo{author}{\bibfnamefont{A.}~\bibnamefont{Denner}} \bibnamefont{and}
  \bibinfo{author}{\bibfnamefont{S.}~\bibnamefont{Pozzorini}},
  \emph{\bibinfo{title}{{One Loop Leading Logarithms in Electroweak Radiative
  Corrections. 2. Factorization of Collinear Singularities}}},
  \bibinfo{journal}{Eur. Phys. J.} \textbf{\bibinfo{volume}{C21}},
  \bibinfo{pages}{63} (\bibinfo{year}{2001}{\natexlab{b}}),
  \eprint{hep-ph/0104127}.

\bibitem[{\citenamefont{Chiu et~al.}(2008{\natexlab{a}})\citenamefont{Chiu,
  Golf, Kelley, and Manohar}}]{Chiu:2007yn}
\bibinfo{author}{\bibfnamefont{J.-y.} \bibnamefont{Chiu}},
  \bibinfo{author}{\bibfnamefont{F.}~\bibnamefont{Golf}},
  \bibinfo{author}{\bibfnamefont{R.}~\bibnamefont{Kelley}}, \bibnamefont{and}
  \bibinfo{author}{\bibfnamefont{A.~V.} \bibnamefont{Manohar}},
  \emph{\bibinfo{title}{{Electroweak Sudakov Corrections Using Effective Field
  Theory}}}, \bibinfo{journal}{Phys. Rev. Lett.}
  \textbf{\bibinfo{volume}{100}}, \bibinfo{pages}{021802}
  (\bibinfo{year}{2008}{\natexlab{a}}), \eprint{0709.2377}.

\bibitem[{\citenamefont{Chiu et~al.}(2008{\natexlab{b}})\citenamefont{Chiu,
  Golf, Kelley, and Manohar}}]{Chiu:2007dg}
\bibinfo{author}{\bibfnamefont{J.-y.} \bibnamefont{Chiu}},
  \bibinfo{author}{\bibfnamefont{F.}~\bibnamefont{Golf}},
  \bibinfo{author}{\bibfnamefont{R.}~\bibnamefont{Kelley}}, \bibnamefont{and}
  \bibinfo{author}{\bibfnamefont{A.~V.} \bibnamefont{Manohar}},
  \emph{\bibinfo{title}{{Electroweak Corrections in High Energy Processes using
  Effective Field Theory}}}, \bibinfo{journal}{Phys. Rev.}
  \textbf{\bibinfo{volume}{D77}}, \bibinfo{pages}{053004}
  (\bibinfo{year}{2008}{\natexlab{b}}), \eprint{0712.0396}.

\bibitem[{\citenamefont{Chiu et~al.}(2008{\natexlab{c}})\citenamefont{Chiu,
  Kelley, and Manohar}}]{Chiu:2008vv}
\bibinfo{author}{\bibfnamefont{J.-y.} \bibnamefont{Chiu}},
  \bibinfo{author}{\bibfnamefont{R.}~\bibnamefont{Kelley}}, \bibnamefont{and}
  \bibinfo{author}{\bibfnamefont{A.~V.} \bibnamefont{Manohar}},
  \emph{\bibinfo{title}{{Electroweak Corrections Using Effective Field Theory:
  Applications to the LHC}}}, \bibinfo{journal}{Phys. Rev.}
  \textbf{\bibinfo{volume}{D78}}, \bibinfo{pages}{073006}
  (\bibinfo{year}{2008}{\natexlab{c}}), \eprint{0806.1240}.

\bibitem[{\citenamefont{Chiu et~al.}(2009)\citenamefont{Chiu, Fuhrer, Kelley,
  and Manohar}}]{Chiu:2009mg}
\bibinfo{author}{\bibfnamefont{J.-y.} \bibnamefont{Chiu}},
  \bibinfo{author}{\bibfnamefont{A.}~\bibnamefont{Fuhrer}},
  \bibinfo{author}{\bibfnamefont{R.}~\bibnamefont{Kelley}}, \bibnamefont{and}
  \bibinfo{author}{\bibfnamefont{A.~V.} \bibnamefont{Manohar}},
  \emph{\bibinfo{title}{{Factorization Structure of Gauge Theory Amplitudes and
  Application to Hard Scattering Processes at the LHC}}},
  \bibinfo{journal}{Phys. Rev.} \textbf{\bibinfo{volume}{D80}},
  \bibinfo{pages}{094013} (\bibinfo{year}{2009}), \eprint{0909.0012}.

\bibitem[{\citenamefont{Chiu et~al.}(2010)\citenamefont{Chiu, Fuhrer, Kelley,
  and Manohar}}]{Chiu:2009ft}
\bibinfo{author}{\bibfnamefont{J.-y.} \bibnamefont{Chiu}},
  \bibinfo{author}{\bibfnamefont{A.}~\bibnamefont{Fuhrer}},
  \bibinfo{author}{\bibfnamefont{R.}~\bibnamefont{Kelley}}, \bibnamefont{and}
  \bibinfo{author}{\bibfnamefont{A.~V.} \bibnamefont{Manohar}},
  \emph{\bibinfo{title}{{Soft and Collinear Functions for the Standard
  Model}}}, \bibinfo{journal}{Phys. Rev.} \textbf{\bibinfo{volume}{D81}},
  \bibinfo{pages}{014023} (\bibinfo{year}{2010}), \eprint{0909.0947}.

\bibitem[{\citenamefont{Ciafaloni et~al.}(2000)\citenamefont{Ciafaloni,
  Ciafaloni, and Comelli}}]{Ciafaloni:2000rp}
\bibinfo{author}{\bibfnamefont{M.}~\bibnamefont{Ciafaloni}},
  \bibinfo{author}{\bibfnamefont{P.}~\bibnamefont{Ciafaloni}},
  \bibnamefont{and} \bibinfo{author}{\bibfnamefont{D.}~\bibnamefont{Comelli}},
  \emph{\bibinfo{title}{{Electroweak Bloch-Nordsieck Violation at the TeV
  Scale: 'Strong' Weak Interactions?}}}, \bibinfo{journal}{Nucl. Phys.}
  \textbf{\bibinfo{volume}{B589}}, \bibinfo{pages}{359} (\bibinfo{year}{2000}),
  \eprint{hep-ph/0004071}.

\bibitem[{\citenamefont{Ciafaloni et~al.}(2001)\citenamefont{Ciafaloni,
  Ciafaloni, and Comelli}}]{Ciafaloni:2000gm}
\bibinfo{author}{\bibfnamefont{M.}~\bibnamefont{Ciafaloni}},
  \bibinfo{author}{\bibfnamefont{P.}~\bibnamefont{Ciafaloni}},
  \bibnamefont{and} \bibinfo{author}{\bibfnamefont{D.}~\bibnamefont{Comelli}},
  \emph{\bibinfo{title}{{Electroweak Double Logarithms in Inclusive Observables
  for a Generic Initial State}}}, \bibinfo{journal}{Phys. Lett.}
  \textbf{\bibinfo{volume}{B501}}, \bibinfo{pages}{216} (\bibinfo{year}{2001}),
  \eprint{hep-ph/0007096}.

\bibitem[{\citenamefont{Ciafaloni and Comelli}(2005)}]{Ciafaloni:2005fm}
\bibinfo{author}{\bibfnamefont{P.}~\bibnamefont{Ciafaloni}} \bibnamefont{and}
  \bibinfo{author}{\bibfnamefont{D.}~\bibnamefont{Comelli}},
  \emph{\bibinfo{title}{{Electroweak Evolution Equations}}},
  \bibinfo{journal}{JHEP} \textbf{\bibinfo{volume}{11}}, \bibinfo{pages}{022}
  (\bibinfo{year}{2005}), \eprint{hep-ph/0505047}.

\bibitem[{\citenamefont{Baur}(2007)}]{Baur:2006sn}
\bibinfo{author}{\bibfnamefont{U.}~\bibnamefont{Baur}},
  \emph{\bibinfo{title}{{Weak Boson Emission in Hadron Collider Processes}}},
  \bibinfo{journal}{Phys. Rev.} \textbf{\bibinfo{volume}{D75}},
  \bibinfo{pages}{013005} (\bibinfo{year}{2007}), \eprint{hep-ph/0611241}.

\bibitem[{\citenamefont{Bell et~al.}(2010)\citenamefont{Bell, Kuhn, and
  Rittinger}}]{Bell:2010gi}
\bibinfo{author}{\bibfnamefont{G.}~\bibnamefont{Bell}},
  \bibinfo{author}{\bibfnamefont{J.}~\bibnamefont{Kuhn}}, \bibnamefont{and}
  \bibinfo{author}{\bibfnamefont{J.}~\bibnamefont{Rittinger}},
  \emph{\bibinfo{title}{{Electroweak Sudakov Logarithms and Real Gauge-Boson
  Radiation in the TeV Region}}}, \bibinfo{journal}{Eur.Phys.J.}
  \textbf{\bibinfo{volume}{C70}}, \bibinfo{pages}{659} (\bibinfo{year}{2010}),
  \eprint{1004.4117}.

\bibitem[{\citenamefont{Chiesa et~al.}(2013)\citenamefont{Chiesa, Montagna,
  Barz{\`e}, Moretti, Nicrosini, Piccinini, and Tramontano}}]{Chiesa:2013yma}
\bibinfo{author}{\bibfnamefont{M.}~\bibnamefont{Chiesa}},
  \bibinfo{author}{\bibfnamefont{G.}~\bibnamefont{Montagna}},
  \bibinfo{author}{\bibfnamefont{L.}~\bibnamefont{Barz{\`e}}},
  \bibinfo{author}{\bibfnamefont{M.}~\bibnamefont{Moretti}},
  \bibinfo{author}{\bibfnamefont{O.}~\bibnamefont{Nicrosini}},
  \bibinfo{author}{\bibfnamefont{F.}~\bibnamefont{Piccinini}},
  \bibnamefont{and}
  \bibinfo{author}{\bibfnamefont{F.}~\bibnamefont{Tramontano}},
  \emph{\bibinfo{title}{{Electroweak Sudakov Corrections to New Physics
  Searches at the LHC}}}, \bibinfo{journal}{Phys. Rev. Lett.}
  \textbf{\bibinfo{volume}{111}}, \bibinfo{pages}{121801}
  (\bibinfo{year}{2013}), \eprint{1305.6837}.

\bibitem[{\citenamefont{Christiansen and
  Sj{\"o}strand}(2014)}]{Christiansen:2014kba}
\bibinfo{author}{\bibfnamefont{J.~R.} \bibnamefont{Christiansen}}
  \bibnamefont{and}
  \bibinfo{author}{\bibfnamefont{T.}~\bibnamefont{Sj{\"o}strand}},
  \emph{\bibinfo{title}{{Weak Gauge Boson Radiation in Parton Showers}}},
  \bibinfo{journal}{JHEP} \textbf{\bibinfo{volume}{1404}}, \bibinfo{pages}{115}
  (\bibinfo{year}{2014}), \eprint{1401.5238}.

\bibitem[{\citenamefont{Krauss et~al.}(2014)\citenamefont{Krauss, Petrov,
  Schoenherr, and Spannowsky}}]{Krauss:2014yaa}
\bibinfo{author}{\bibfnamefont{F.}~\bibnamefont{Krauss}},
  \bibinfo{author}{\bibfnamefont{P.}~\bibnamefont{Petrov}},
  \bibinfo{author}{\bibfnamefont{M.}~\bibnamefont{Schoenherr}},
  \bibnamefont{and}
  \bibinfo{author}{\bibfnamefont{M.}~\bibnamefont{Spannowsky}},
  \emph{\bibinfo{title}{{Measuring Collinear W Emissions Inside Jets}}},
  \bibinfo{journal}{Phys. Rev.} \textbf{\bibinfo{volume}{D89}},
  \bibinfo{pages}{114006} (\bibinfo{year}{2014}), \eprint{1403.4788}.

\bibitem[{\citenamefont{Bauer and Ferland}(2016)}]{Bauer:2016kkv}
\bibinfo{author}{\bibfnamefont{C.~W.} \bibnamefont{Bauer}} \bibnamefont{and}
  \bibinfo{author}{\bibfnamefont{N.}~\bibnamefont{Ferland}},
  \emph{\bibinfo{title}{{Resummation of Electroweak Sudakov Logarithms for Real
  Radiation}}} (\bibinfo{year}{2016}), \eprint{1601.07190}.

\bibitem[{\citenamefont{Aaboud et~al.}(2016)}]{Aaboud:2016ylh}
\bibinfo{author}{\bibfnamefont{M.}~\bibnamefont{Aaboud}} \bibnamefont{et~al.}
  (\bibinfo{collaboration}{ATLAS}), \emph{\bibinfo{title}{{Measurement of $W$
  Boson Angular Distributions in Events with High Transverse Momentum Jets at
  $\sqrt{s}=$ 8~TeV Using the ATLAS Detector}}} (\bibinfo{year}{2016}),
  \eprint{1609.07045}.

\bibitem[{\citenamefont{Ciafaloni et~al.}(2011)\citenamefont{Ciafaloni,
  Comelli, Riotto, Sala, Strumia, and Urbano}}]{Ciafaloni:2010ti}
\bibinfo{author}{\bibfnamefont{P.}~\bibnamefont{Ciafaloni}},
  \bibinfo{author}{\bibfnamefont{D.}~\bibnamefont{Comelli}},
  \bibinfo{author}{\bibfnamefont{A.}~\bibnamefont{Riotto}},
  \bibinfo{author}{\bibfnamefont{F.}~\bibnamefont{Sala}},
  \bibinfo{author}{\bibfnamefont{A.}~\bibnamefont{Strumia}}, \bibnamefont{and}
  \bibinfo{author}{\bibfnamefont{A.}~\bibnamefont{Urbano}},
  \emph{\bibinfo{title}{{Weak Corrections are Relevant for Dark Matter Indirect
  Detection}}}, \bibinfo{journal}{JCAP} \textbf{\bibinfo{volume}{1103}},
  \bibinfo{pages}{019} (\bibinfo{year}{2011}), \eprint{1009.0224}.

\bibitem[{\citenamefont{Cavasonza et~al.}(2015)\citenamefont{Cavasonza,
  Krämer, and Pellen}}]{Cavasonza:2014xra}
\bibinfo{author}{\bibfnamefont{L.~A.} \bibnamefont{Cavasonza}},
  \bibinfo{author}{\bibfnamefont{M.}~\bibnamefont{Krämer}}, \bibnamefont{and}
  \bibinfo{author}{\bibfnamefont{M.}~\bibnamefont{Pellen}},
  \emph{\bibinfo{title}{{Electroweak Fragmentation Functions for Dark Matter
  Annihilation}}}, \bibinfo{journal}{JCAP} \textbf{\bibinfo{volume}{1502}},
  \bibinfo{pages}{021} (\bibinfo{year}{2015}), \eprint{1409.8226}.

\bibitem[{\citenamefont{Bauer et~al.}(2015)\citenamefont{Bauer, Cohen, Hill,
  and Solon}}]{Bauer:2014ula}
\bibinfo{author}{\bibfnamefont{M.}~\bibnamefont{Bauer}},
  \bibinfo{author}{\bibfnamefont{T.}~\bibnamefont{Cohen}},
  \bibinfo{author}{\bibfnamefont{R.~J.} \bibnamefont{Hill}}, \bibnamefont{and}
  \bibinfo{author}{\bibfnamefont{M.~P.} \bibnamefont{Solon}},
  \emph{\bibinfo{title}{{Soft Collinear Effective Theory for Heavy WIMP
  Annihilation}}}, \bibinfo{journal}{JHEP} \textbf{\bibinfo{volume}{01}},
  \bibinfo{pages}{099} (\bibinfo{year}{2015}), \eprint{1409.7392}.

\bibitem[{\citenamefont{Ovanesyan et~al.}(2015)\citenamefont{Ovanesyan,
  Slatyer, and Stewart}}]{Ovanesyan:2014fwa}
\bibinfo{author}{\bibfnamefont{G.}~\bibnamefont{Ovanesyan}},
  \bibinfo{author}{\bibfnamefont{T.~R.} \bibnamefont{Slatyer}},
  \bibnamefont{and} \bibinfo{author}{\bibfnamefont{I.~W.}
  \bibnamefont{Stewart}}, \emph{\bibinfo{title}{{Heavy Dark Matter Annihilation
  from Effective Field Theory}}}, \bibinfo{journal}{Phys. Rev. Lett.}
  \textbf{\bibinfo{volume}{114}}, \bibinfo{pages}{211302}
  (\bibinfo{year}{2015}), \eprint{1409.8294}.

\bibitem[{\citenamefont{Baumgart
  et~al.}(2015{\natexlab{a}})\citenamefont{Baumgart, Rothstein, and
  Vaidya}}]{Baumgart:2014vma}
\bibinfo{author}{\bibfnamefont{M.}~\bibnamefont{Baumgart}},
  \bibinfo{author}{\bibfnamefont{I.~Z.} \bibnamefont{Rothstein}},
  \bibnamefont{and} \bibinfo{author}{\bibfnamefont{V.}~\bibnamefont{Vaidya}},
  \emph{\bibinfo{title}{{Calculating the Annihilation Rate of Weakly
  Interacting Massive Particles}}}, \bibinfo{journal}{Phys. Rev. Lett.}
  \textbf{\bibinfo{volume}{114}}, \bibinfo{pages}{211301}
  (\bibinfo{year}{2015}{\natexlab{a}}), \eprint{1409.4415}.

\bibitem[{\citenamefont{Baumgart
  et~al.}(2015{\natexlab{b}})\citenamefont{Baumgart, Rothstein, and
  Vaidya}}]{Baumgart:2014saa}
\bibinfo{author}{\bibfnamefont{M.}~\bibnamefont{Baumgart}},
  \bibinfo{author}{\bibfnamefont{I.~Z.} \bibnamefont{Rothstein}},
  \bibnamefont{and} \bibinfo{author}{\bibfnamefont{V.}~\bibnamefont{Vaidya}},
  \emph{\bibinfo{title}{{Constraints on Galactic Wino Densities from Gamma Ray
  Lines}}}, \bibinfo{journal}{JHEP} \textbf{\bibinfo{volume}{04}},
  \bibinfo{pages}{106} (\bibinfo{year}{2015}{\natexlab{b}}),
  \eprint{1412.8698}.

\bibitem[{\citenamefont{Baumgart and Vaidya}(2015)}]{Baumgart:2015bpa}
\bibinfo{author}{\bibfnamefont{M.}~\bibnamefont{Baumgart}} \bibnamefont{and}
  \bibinfo{author}{\bibfnamefont{V.}~\bibnamefont{Vaidya}},
  \emph{\bibinfo{title}{{Semi-Inclusive Wino and Higgsino Annihilation to
  LL'}}} (\bibinfo{year}{2015}), \eprint{1510.02470}.

\bibitem[{\citenamefont{Lee et~al.}(1977)\citenamefont{Lee, Quigg, and
  Thacker}}]{Lee:1977eg}
\bibinfo{author}{\bibfnamefont{B.~W.} \bibnamefont{Lee}},
  \bibinfo{author}{\bibfnamefont{C.}~\bibnamefont{Quigg}}, \bibnamefont{and}
  \bibinfo{author}{\bibfnamefont{H.}~\bibnamefont{Thacker}},
  \emph{\bibinfo{title}{{Weak Interactions at Very High-Energies: The Role of
  the Higgs Boson Mass}}}, \bibinfo{journal}{Phys.Rev.}
  \textbf{\bibinfo{volume}{D16}}, \bibinfo{pages}{1519} (\bibinfo{year}{1977}).

\bibitem[{\citenamefont{Mangano et~al.}(2003)\citenamefont{Mangano, Moretti,
  Piccinini, Pittau, and Polosa}}]{Mangano:2002ea}
\bibinfo{author}{\bibfnamefont{M.~L.} \bibnamefont{Mangano}},
  \bibinfo{author}{\bibfnamefont{M.}~\bibnamefont{Moretti}},
  \bibinfo{author}{\bibfnamefont{F.}~\bibnamefont{Piccinini}},
  \bibinfo{author}{\bibfnamefont{R.}~\bibnamefont{Pittau}}, \bibnamefont{and}
  \bibinfo{author}{\bibfnamefont{A.~D.} \bibnamefont{Polosa}},
  \emph{\bibinfo{title}{{ALPGEN, A Generator for Hard Multiparton Processes in
  Hadronic Collisions}}}, \bibinfo{journal}{JHEP}
  \textbf{\bibinfo{volume}{0307}}, \bibinfo{pages}{001} (\bibinfo{year}{2003}),
  \eprint{hep-ph/0206293}.

\bibitem[{\citenamefont{Hook and Katz}(2014)}]{Hook:2014rka}
\bibinfo{author}{\bibfnamefont{A.}~\bibnamefont{Hook}} \bibnamefont{and}
  \bibinfo{author}{\bibfnamefont{A.}~\bibnamefont{Katz}},
  \emph{\bibinfo{title}{{Unbroken $SU(2)$ at a 100 TeV Collider}}},
  \bibinfo{journal}{JHEP} \textbf{\bibinfo{volume}{1409}}, \bibinfo{pages}{175}
  (\bibinfo{year}{2014}), \eprint{1407.2607}.

\bibitem[{\citenamefont{Rizzo}(2014)}]{Rizzo:2014xma}
\bibinfo{author}{\bibfnamefont{T.~G.} \bibnamefont{Rizzo}},
  \emph{\bibinfo{title}{{Exploring New Gauge Bosons at a 100 TeV Collider}}},
  \bibinfo{journal}{Phys. Rev.} \textbf{\bibinfo{volume}{D89}},
  \bibinfo{pages}{095022} (\bibinfo{year}{2014}), \eprint{1403.5465}.

\bibitem[{\citenamefont{Manohar et~al.}(2015)\citenamefont{Manohar, Shotwell,
  Bauer, and Turczyk}}]{Manohar:2014vxa}
\bibinfo{author}{\bibfnamefont{A.}~\bibnamefont{Manohar}},
  \bibinfo{author}{\bibfnamefont{B.}~\bibnamefont{Shotwell}},
  \bibinfo{author}{\bibfnamefont{C.}~\bibnamefont{Bauer}}, \bibnamefont{and}
  \bibinfo{author}{\bibfnamefont{S.}~\bibnamefont{Turczyk}},
  \emph{\bibinfo{title}{{Non-Cancellation of Electroweak Logarithms in
  High-Energy Scattering}}}, \bibinfo{journal}{Phys. Lett.}
  \textbf{\bibinfo{volume}{B740}}, \bibinfo{pages}{179} (\bibinfo{year}{2015}),
  \eprint{1409.1918}.

\bibitem[{\citenamefont{Chen et~al.}()\citenamefont{Chen, Han, and
  Tweedie}}]{EWshower}
\bibinfo{author}{\bibfnamefont{J.}~\bibnamefont{Chen}},
  \bibinfo{author}{\bibfnamefont{T.}~\bibnamefont{Han}}, \bibnamefont{and}
  \bibinfo{author}{\bibfnamefont{B.}~\bibnamefont{Tweedie}},
  \emph{\bibinfo{title}{{In preparation}}}.

\bibitem[{\citenamefont{Dokshitzer}(1977)}]{Dokshitzer:1977sg}
\bibinfo{author}{\bibfnamefont{Y.~L.} \bibnamefont{Dokshitzer}},
  \emph{\bibinfo{title}{{Calculation of the Structure Functions for Deep
  Inelastic Scattering and e+ e- Annihilation by Perturbation Theory in Quantum
  Chromodynamics.}}}, \bibinfo{journal}{Sov. Phys. JETP}
  \textbf{\bibinfo{volume}{46}}, \bibinfo{pages}{641} (\bibinfo{year}{1977}),
  \bibinfo{note}{[Zh. Eksp. Teor. Fiz.73,1216(1977)]}.

\bibitem[{\citenamefont{Gribov and Lipatov}(1972)}]{Gribov:1972ri}
\bibinfo{author}{\bibfnamefont{V.~N.} \bibnamefont{Gribov}} \bibnamefont{and}
  \bibinfo{author}{\bibfnamefont{L.~N.} \bibnamefont{Lipatov}},
  \emph{\bibinfo{title}{{Deep Inelastic e p Scattering in Perturbation
  Theory}}}, \bibinfo{journal}{Sov. J. Nucl. Phys.}
  \textbf{\bibinfo{volume}{15}}, \bibinfo{pages}{438} (\bibinfo{year}{1972}),
  \bibinfo{note}{[Yad. Fiz.15,781(1972)]}.

\bibitem[{\citenamefont{Altarelli and Parisi}(1977)}]{Altarelli:1977zs}
\bibinfo{author}{\bibfnamefont{G.}~\bibnamefont{Altarelli}} \bibnamefont{and}
  \bibinfo{author}{\bibfnamefont{G.}~\bibnamefont{Parisi}},
  \emph{\bibinfo{title}{{Asymptotic Freedom in Parton Language}}},
  \bibinfo{journal}{Nucl. Phys.} \textbf{\bibinfo{volume}{B126}},
  \bibinfo{pages}{298} (\bibinfo{year}{1977}).

\bibitem[{\citenamefont{Bjorken}(1969)}]{Bj}
\bibinfo{author}{\bibfnamefont{J.~D.} \bibnamefont{Bjorken}},
  \emph{\bibinfo{title}{{Asymptotic Sum Rules at Infinite Momentum}}},
  \bibinfo{journal}{Phys. Rev.} \textbf{\bibinfo{volume}{179}},
  \bibinfo{pages}{1547} (\bibinfo{year}{1969}).

\bibitem[{\citenamefont{Sjostrand et~al.}(2006)\citenamefont{Sjostrand, Mrenna,
  and Skands}}]{Sjostrand:2006za}
\bibinfo{author}{\bibfnamefont{T.}~\bibnamefont{Sjostrand}},
  \bibinfo{author}{\bibfnamefont{S.}~\bibnamefont{Mrenna}}, \bibnamefont{and}
  \bibinfo{author}{\bibfnamefont{P.}~\bibnamefont{Skands}},
  \emph{\bibinfo{title}{{PYTHIA 6.4 Physics and Manual}}},
  \bibinfo{journal}{JHEP} \textbf{\bibinfo{volume}{05}}, \bibinfo{pages}{026}
  (\bibinfo{year}{2006}), \eprint{hep-ph/0603175}.

\bibitem[{\citenamefont{Nagy and Soper}(2007)}]{Nagy:2007ty}
\bibinfo{author}{\bibfnamefont{Z.}~\bibnamefont{Nagy}} \bibnamefont{and}
  \bibinfo{author}{\bibfnamefont{D.~E.} \bibnamefont{Soper}},
  \emph{\bibinfo{title}{{Parton Showers with Quantum Interference}}},
  \bibinfo{journal}{JHEP} \textbf{\bibinfo{volume}{0709}}, \bibinfo{pages}{114}
  (\bibinfo{year}{2007}), \eprint{0706.0017}.

\bibitem[{\citenamefont{Cahn et~al.}(1987)\citenamefont{Cahn, Ellis, Kleiss,
  and Stirling}}]{Cahn:1986zv}
\bibinfo{author}{\bibfnamefont{R.~N.} \bibnamefont{Cahn}},
  \bibinfo{author}{\bibfnamefont{S.~D.} \bibnamefont{Ellis}},
  \bibinfo{author}{\bibfnamefont{R.}~\bibnamefont{Kleiss}}, \bibnamefont{and}
  \bibinfo{author}{\bibfnamefont{W.~J.} \bibnamefont{Stirling}},
  \emph{\bibinfo{title}{{Transverse Momentum Signatures for Heavy Higgs
  Bosons}}}, \bibinfo{journal}{Phys. Rev.} \textbf{\bibinfo{volume}{D35}},
  \bibinfo{pages}{1626} (\bibinfo{year}{1987}).

\bibitem[{\citenamefont{Barger et~al.}(1988{\natexlab{a}})\citenamefont{Barger,
  Han, and Phillips}}]{Barger:1988mr}
\bibinfo{author}{\bibfnamefont{V.~D.} \bibnamefont{Barger}},
  \bibinfo{author}{\bibfnamefont{T.}~\bibnamefont{Han}}, \bibnamefont{and}
  \bibinfo{author}{\bibfnamefont{R.~J.~N.} \bibnamefont{Phillips}},
  \emph{\bibinfo{title}{{Improving the Heavy Higgs Boson Two Charged Lepton -
  Two Neutrino Signal}}}, \bibinfo{journal}{Phys. Rev.}
  \textbf{\bibinfo{volume}{D37}}, \bibinfo{pages}{2005}
  (\bibinfo{year}{1988}{\natexlab{a}}).

\bibitem[{\citenamefont{Kleiss and Stirling}(1988)}]{Kleiss:1987cj}
\bibinfo{author}{\bibfnamefont{R.}~\bibnamefont{Kleiss}} \bibnamefont{and}
  \bibinfo{author}{\bibfnamefont{W.~J.} \bibnamefont{Stirling}},
  \emph{\bibinfo{title}{{Tagging the Higgs}}}, \bibinfo{journal}{Phys. Lett.}
  \textbf{\bibinfo{volume}{B200}}, \bibinfo{pages}{193} (\bibinfo{year}{1988}).

\bibitem[{\citenamefont{Barger et~al.}(1990)\citenamefont{Barger, Cheung, Han,
  and Phillips}}]{Barger:1990py}
\bibinfo{author}{\bibfnamefont{V.~D.} \bibnamefont{Barger}},
  \bibinfo{author}{\bibfnamefont{K.-m.} \bibnamefont{Cheung}},
  \bibinfo{author}{\bibfnamefont{T.}~\bibnamefont{Han}}, \bibnamefont{and}
  \bibinfo{author}{\bibfnamefont{R.~J.~N.} \bibnamefont{Phillips}},
  \emph{\bibinfo{title}{{Strong $W^{+} W^{+}$ Scattering Signals at $p p$
  Supercolliders}}}, \bibinfo{journal}{Phys. Rev.}
  \textbf{\bibinfo{volume}{D42}}, \bibinfo{pages}{3052} (\bibinfo{year}{1990}).

\bibitem[{\citenamefont{Beenakker and Werthenbach}(2002)}]{Beenakker:2001kf}
\bibinfo{author}{\bibfnamefont{W.}~\bibnamefont{Beenakker}} \bibnamefont{and}
  \bibinfo{author}{\bibfnamefont{A.}~\bibnamefont{Werthenbach}},
  \emph{\bibinfo{title}{{Electroweak Two Loop Sudakov Logarithms for On-Shell
  Fermions and Bosons}}}, \bibinfo{journal}{Nucl. Phys.}
  \textbf{\bibinfo{volume}{B630}}, \bibinfo{pages}{3} (\bibinfo{year}{2002}),
  \eprint{hep-ph/0112030}.

\bibitem[{\citenamefont{Srivastava and Brodsky}(2002)}]{Srivastava:2002mw}
\bibinfo{author}{\bibfnamefont{P.~P.} \bibnamefont{Srivastava}}
  \bibnamefont{and} \bibinfo{author}{\bibfnamefont{S.~J.}
  \bibnamefont{Brodsky}}, \emph{\bibinfo{title}{{A Unitary and Renormalizable
  Theory of the Standard Model in Ghost Free Light Cone Gauge}}},
  \bibinfo{journal}{Phys. Rev.} \textbf{\bibinfo{volume}{D66}},
  \bibinfo{pages}{045019} (\bibinfo{year}{2002}), \eprint{hep-ph/0202141}.

\bibitem[{\citenamefont{Dams and Kleiss}(2004)}]{Dams:2004vi}
\bibinfo{author}{\bibfnamefont{C.}~\bibnamefont{Dams}} \bibnamefont{and}
  \bibinfo{author}{\bibfnamefont{R.}~\bibnamefont{Kleiss}},
  \emph{\bibinfo{title}{{The Electroweak Standard Model in the Axial Gauge}}},
  \bibinfo{journal}{Eur. Phys. J.} \textbf{\bibinfo{volume}{C34}},
  \bibinfo{pages}{419} (\bibinfo{year}{2004}), \eprint{hep-ph/0401136}.

\bibitem[{\citenamefont{Wulzer}(2014)}]{Wulzer:2013mza}
\bibinfo{author}{\bibfnamefont{A.}~\bibnamefont{Wulzer}},
  \emph{\bibinfo{title}{{An Equivalent Gauge and the Equivalence Theorem}}},
  \bibinfo{journal}{Nucl. Phys.} \textbf{\bibinfo{volume}{B885}},
  \bibinfo{pages}{97} (\bibinfo{year}{2014}), \eprint{1309.6055}.

\bibitem[{\citenamefont{Yao and Yuan}(1988)}]{Yao:1988aj}
\bibinfo{author}{\bibfnamefont{Y.-P.} \bibnamefont{Yao}} \bibnamefont{and}
  \bibinfo{author}{\bibfnamefont{C.~P.} \bibnamefont{Yuan}},
  \emph{\bibinfo{title}{{Modification of the Equivalence Theorem Due to Loop
  Corrections}}}, \bibinfo{journal}{Phys. Rev.} \textbf{\bibinfo{volume}{D38}},
  \bibinfo{pages}{2237} (\bibinfo{year}{1988}).

\bibitem[{\citenamefont{Bagger and Schmidt}(1990)}]{Bagger:1989fc}
\bibinfo{author}{\bibfnamefont{J.}~\bibnamefont{Bagger}} \bibnamefont{and}
  \bibinfo{author}{\bibfnamefont{C.}~\bibnamefont{Schmidt}},
  \emph{\bibinfo{title}{{Equivalence Theorem Redux}}}, \bibinfo{journal}{Phys.
  Rev.} \textbf{\bibinfo{volume}{D41}}, \bibinfo{pages}{264}
  (\bibinfo{year}{1990}).

\bibitem[{\citenamefont{He et~al.}(1992)\citenamefont{He, Kuang, and
  Li}}]{He:1992nga}
\bibinfo{author}{\bibfnamefont{H.-J.} \bibnamefont{He}},
  \bibinfo{author}{\bibfnamefont{Y.-P.} \bibnamefont{Kuang}}, \bibnamefont{and}
  \bibinfo{author}{\bibfnamefont{X.-y.} \bibnamefont{Li}},
  \emph{\bibinfo{title}{{On the Precise Formulation of Equivalence Theorem}}},
  \bibinfo{journal}{Phys. Rev. Lett.} \textbf{\bibinfo{volume}{69}},
  \bibinfo{pages}{2619} (\bibinfo{year}{1992}).

\bibitem[{\citenamefont{Kunszt and Soper}(1988)}]{Kunszt:1987tk}
\bibinfo{author}{\bibfnamefont{Z.}~\bibnamefont{Kunszt}} \bibnamefont{and}
  \bibinfo{author}{\bibfnamefont{D.~E.} \bibnamefont{Soper}},
  \emph{\bibinfo{title}{{On the Validity of the Effective $W$ Approximation}}},
  \bibinfo{journal}{Nucl. Phys.} \textbf{\bibinfo{volume}{B296}},
  \bibinfo{pages}{253} (\bibinfo{year}{1988}).

\bibitem[{\citenamefont{Borel et~al.}(2012)\citenamefont{Borel, Franceschini,
  Rattazzi, and Wulzer}}]{Borel:2012by}
\bibinfo{author}{\bibfnamefont{P.}~\bibnamefont{Borel}},
  \bibinfo{author}{\bibfnamefont{R.}~\bibnamefont{Franceschini}},
  \bibinfo{author}{\bibfnamefont{R.}~\bibnamefont{Rattazzi}}, \bibnamefont{and}
  \bibinfo{author}{\bibfnamefont{A.}~\bibnamefont{Wulzer}},
  \emph{\bibinfo{title}{{Probing the Scattering of Equivalent Electroweak
  Bosons}}}, \bibinfo{journal}{JHEP} \textbf{\bibinfo{volume}{06}},
  \bibinfo{pages}{122} (\bibinfo{year}{2012}), \eprint{1202.1904}.

\bibitem[{\citenamefont{Han et~al.}(1992)\citenamefont{Han, Valencia, and
  Willenbrock}}]{Han:1992hr}
\bibinfo{author}{\bibfnamefont{T.}~\bibnamefont{Han}},
  \bibinfo{author}{\bibfnamefont{G.}~\bibnamefont{Valencia}}, \bibnamefont{and}
  \bibinfo{author}{\bibfnamefont{S.}~\bibnamefont{Willenbrock}},
  \emph{\bibinfo{title}{{Structure Function Approach to Vector Boson Scattering
  in $pp$ Collisions}}}, \bibinfo{journal}{Phys. Rev. Lett.}
  \textbf{\bibinfo{volume}{69}}, \bibinfo{pages}{3274} (\bibinfo{year}{1992}),
  \eprint{hep-ph/9206246}.

\bibitem[{\citenamefont{Ababekri et~al.}(2016)\citenamefont{Ababekri, Dulat,
  Isaacson, Schmidt, and Yuan}}]{Ababekri:2016kkj}
\bibinfo{author}{\bibfnamefont{M.}~\bibnamefont{Ababekri}},
  \bibinfo{author}{\bibfnamefont{S.}~\bibnamefont{Dulat}},
  \bibinfo{author}{\bibfnamefont{J.}~\bibnamefont{Isaacson}},
  \bibinfo{author}{\bibfnamefont{C.}~\bibnamefont{Schmidt}}, \bibnamefont{and}
  \bibinfo{author}{\bibfnamefont{C.~P.} \bibnamefont{Yuan}},
  \emph{\bibinfo{title}{{Implication of CMS Data on Photon PDFs}}}
  (\bibinfo{year}{2016}), \eprint{1603.04874}.

\bibitem[{\citenamefont{Alva et~al.}(2015)\citenamefont{Alva, Han, and
  Ruiz}}]{Alva:2014gxa}
\bibinfo{author}{\bibfnamefont{D.}~\bibnamefont{Alva}},
  \bibinfo{author}{\bibfnamefont{T.}~\bibnamefont{Han}}, \bibnamefont{and}
  \bibinfo{author}{\bibfnamefont{R.}~\bibnamefont{Ruiz}},
  \emph{\bibinfo{title}{{Heavy Majorana Neutrinos from $W\gamma$ Fusion at
  Hadron colliders}}}, \bibinfo{journal}{JHEP} \textbf{\bibinfo{volume}{02}},
  \bibinfo{pages}{072} (\bibinfo{year}{2015}), \eprint{1411.7305}.

\bibitem[{\citenamefont{Bagger et~al.}(1995)\citenamefont{Bagger, Barger,
  Cheung, Gunion, Han, Ladinsky, Rosenfeld, and Yuan}}]{Bagger:1995mk}
\bibinfo{author}{\bibfnamefont{J.}~\bibnamefont{Bagger}},
  \bibinfo{author}{\bibfnamefont{V.~D.} \bibnamefont{Barger}},
  \bibinfo{author}{\bibfnamefont{K.-m.} \bibnamefont{Cheung}},
  \bibinfo{author}{\bibfnamefont{J.~F.} \bibnamefont{Gunion}},
  \bibinfo{author}{\bibfnamefont{T.}~\bibnamefont{Han}},
  \bibinfo{author}{\bibfnamefont{G.~A.} \bibnamefont{Ladinsky}},
  \bibinfo{author}{\bibfnamefont{R.}~\bibnamefont{Rosenfeld}},
  \bibnamefont{and} \bibinfo{author}{\bibfnamefont{C.~P.} \bibnamefont{Yuan}},
  \emph{\bibinfo{title}{{CERN LHC Analysis of the Strongly Interacting W W
  System: Gold Plated Modes}}}, \bibinfo{journal}{Phys. Rev.}
  \textbf{\bibinfo{volume}{D52}}, \bibinfo{pages}{3878} (\bibinfo{year}{1995}),
  \eprint{hep-ph/9504426}.

\bibitem[{\citenamefont{Agashe et~al.}(2005)\citenamefont{Agashe, Contino, and
  Pomarol}}]{Agashe:2004rs}
\bibinfo{author}{\bibfnamefont{K.}~\bibnamefont{Agashe}},
  \bibinfo{author}{\bibfnamefont{R.}~\bibnamefont{Contino}}, \bibnamefont{and}
  \bibinfo{author}{\bibfnamefont{A.}~\bibnamefont{Pomarol}},
  \emph{\bibinfo{title}{{The Minimal Composite Higgs Model}}},
  \bibinfo{journal}{Nucl. Phys.} \textbf{\bibinfo{volume}{B719}},
  \bibinfo{pages}{165} (\bibinfo{year}{2005}), \eprint{hep-ph/0412089}.

\bibitem[{\citenamefont{Giudice et~al.}(2007)\citenamefont{Giudice, Grojean,
  Pomarol, and Rattazzi}}]{Giudice:2007fh}
\bibinfo{author}{\bibfnamefont{G.~F.} \bibnamefont{Giudice}},
  \bibinfo{author}{\bibfnamefont{C.}~\bibnamefont{Grojean}},
  \bibinfo{author}{\bibfnamefont{A.}~\bibnamefont{Pomarol}}, \bibnamefont{and}
  \bibinfo{author}{\bibfnamefont{R.}~\bibnamefont{Rattazzi}},
  \emph{\bibinfo{title}{{The Strongly-Interacting Light Higgs}}},
  \bibinfo{journal}{JHEP} \textbf{\bibinfo{volume}{06}}, \bibinfo{pages}{045}
  (\bibinfo{year}{2007}), \eprint{hep-ph/0703164}.

\bibitem[{\citenamefont{Alwall et~al.}(2014)\citenamefont{Alwall, Frederix,
  Frixione, Hirschi, Maltoni, Mattelaer, Shao, Stelzer, Torrielli, and
  Zaro}}]{Alwall:2011uj}
\bibinfo{author}{\bibfnamefont{J.}~\bibnamefont{Alwall}},
  \bibinfo{author}{\bibfnamefont{R.}~\bibnamefont{Frederix}},
  \bibinfo{author}{\bibfnamefont{S.}~\bibnamefont{Frixione}},
  \bibinfo{author}{\bibfnamefont{V.}~\bibnamefont{Hirschi}},
  \bibinfo{author}{\bibfnamefont{F.}~\bibnamefont{Maltoni}},
  \bibinfo{author}{\bibfnamefont{O.}~\bibnamefont{Mattelaer}},
  \bibinfo{author}{\bibfnamefont{H.~S.} \bibnamefont{Shao}},
  \bibinfo{author}{\bibfnamefont{T.}~\bibnamefont{Stelzer}},
  \bibinfo{author}{\bibfnamefont{P.}~\bibnamefont{Torrielli}},
  \bibnamefont{and} \bibinfo{author}{\bibfnamefont{M.}~\bibnamefont{Zaro}},
  \emph{\bibinfo{title}{{The Automated Computation of Tree-Level and
  Next-to-Leading Order Differential Cross Sections, and Their Matching to
  Parton Shower Simulations}}}, \bibinfo{journal}{JHEP}
  \textbf{\bibinfo{volume}{07}}, \bibinfo{pages}{079} (\bibinfo{year}{2014}),
  \eprint{1405.0301}.

\bibitem[{\citenamefont{Cacciari et~al.}(2008)\citenamefont{Cacciari, Salam,
  and Soyez}}]{Cacciari:2008gp}
\bibinfo{author}{\bibfnamefont{M.}~\bibnamefont{Cacciari}},
  \bibinfo{author}{\bibfnamefont{G.~P.} \bibnamefont{Salam}}, \bibnamefont{and}
  \bibinfo{author}{\bibfnamefont{G.}~\bibnamefont{Soyez}},
  \emph{\bibinfo{title}{{The Anti-$k_t$ Jet Clustering Algorithm}}},
  \bibinfo{journal}{JHEP} \textbf{\bibinfo{volume}{04}}, \bibinfo{pages}{063}
  (\bibinfo{year}{2008}), \eprint{0802.1189}.

\bibitem[{\citenamefont{Agashe et~al.}(2007)\citenamefont{Agashe, Davoudiasl,
  Gopalakrishna, Han, Huang, Perez, Si, and Soni}}]{Agashe:2007ki}
\bibinfo{author}{\bibfnamefont{K.}~\bibnamefont{Agashe}},
  \bibinfo{author}{\bibfnamefont{H.}~\bibnamefont{Davoudiasl}},
  \bibinfo{author}{\bibfnamefont{S.}~\bibnamefont{Gopalakrishna}},
  \bibinfo{author}{\bibfnamefont{T.}~\bibnamefont{Han}},
  \bibinfo{author}{\bibfnamefont{G.-Y.} \bibnamefont{Huang}},
  \bibinfo{author}{\bibfnamefont{G.}~\bibnamefont{Perez}},
  \bibinfo{author}{\bibfnamefont{Z.-G.} \bibnamefont{Si}}, \bibnamefont{and}
  \bibinfo{author}{\bibfnamefont{A.}~\bibnamefont{Soni}},
  \emph{\bibinfo{title}{{LHC Signals for Warped Electroweak Neutral Gauge
  Bosons}}}, \bibinfo{journal}{Phys. Rev.} \textbf{\bibinfo{volume}{D76}},
  \bibinfo{pages}{115015} (\bibinfo{year}{2007}), \eprint{0709.0007}.

\bibitem[{\citenamefont{Dennis et~al.}(2007)\citenamefont{Dennis, Karagoz,
  Servant, and Tseng}}]{Dennis:2007tv}
\bibinfo{author}{\bibfnamefont{C.}~\bibnamefont{Dennis}},
  \bibinfo{author}{\bibfnamefont{M.}~\bibnamefont{Karagoz}},
  \bibinfo{author}{\bibfnamefont{G.}~\bibnamefont{Servant}}, \bibnamefont{and}
  \bibinfo{author}{\bibfnamefont{J.}~\bibnamefont{Tseng}},
  \emph{\bibinfo{title}{{Multi-W Events at LHC from a Warped Extra Dimension
  with Custodial Symmetry}}} (\bibinfo{year}{2007}), \eprint{hep-ph/0701158}.

\bibitem[{\citenamefont{Agashe et~al.}(2013)}]{Agashe:2013hma}
\bibinfo{author}{\bibfnamefont{K.}~\bibnamefont{Agashe}} \bibnamefont{et~al.}
  (\bibinfo{collaboration}{Top Quark Working Group}), in
  \emph{\bibinfo{booktitle}{{Proceedings, Community Summer Study 2013: Snowmass
  on the Mississippi (CSS2013): Minneapolis, MN, USA, July 29-August 6, 2013}}}
  (\bibinfo{year}{2013}), \eprint{1311.2028},
  \urlprefix\url{http://inspirehep.net/record/1263763/files/arXiv:1311.2028.pdf}.

\bibitem[{\citenamefont{Agashe et~al.}(2008)\citenamefont{Agashe, Belyaev,
  Krupovnickas, Perez, and Virzi}}]{Agashe:2006hk}
\bibinfo{author}{\bibfnamefont{K.}~\bibnamefont{Agashe}},
  \bibinfo{author}{\bibfnamefont{A.}~\bibnamefont{Belyaev}},
  \bibinfo{author}{\bibfnamefont{T.}~\bibnamefont{Krupovnickas}},
  \bibinfo{author}{\bibfnamefont{G.}~\bibnamefont{Perez}}, \bibnamefont{and}
  \bibinfo{author}{\bibfnamefont{J.}~\bibnamefont{Virzi}},
  \emph{\bibinfo{title}{{LHC Signals from Warped Extra Dimensions}}},
  \bibinfo{journal}{Phys. Rev.} \textbf{\bibinfo{volume}{D77}},
  \bibinfo{pages}{015003} (\bibinfo{year}{2008}), \eprint{hep-ph/0612015}.

\bibitem[{\citenamefont{Lillie et~al.}(2007)\citenamefont{Lillie, Randall, and
  Wang}}]{Lillie:2007yh}
\bibinfo{author}{\bibfnamefont{B.}~\bibnamefont{Lillie}},
  \bibinfo{author}{\bibfnamefont{L.}~\bibnamefont{Randall}}, \bibnamefont{and}
  \bibinfo{author}{\bibfnamefont{L.-T.} \bibnamefont{Wang}},
  \emph{\bibinfo{title}{{The Bulk RS KK-Gluon at the LHC}}},
  \bibinfo{journal}{JHEP} \textbf{\bibinfo{volume}{09}}, \bibinfo{pages}{074}
  (\bibinfo{year}{2007}), \eprint{hep-ph/0701166}.

\bibitem[{\citenamefont{Frederix and Maltoni}(2009)}]{Frederix:2007gi}
\bibinfo{author}{\bibfnamefont{R.}~\bibnamefont{Frederix}} \bibnamefont{and}
  \bibinfo{author}{\bibfnamefont{F.}~\bibnamefont{Maltoni}},
  \emph{\bibinfo{title}{{Top Pair Invariant Mass Distribution: A Window on New
  Physics}}}, \bibinfo{journal}{JHEP} \textbf{\bibinfo{volume}{01}},
  \bibinfo{pages}{047} (\bibinfo{year}{2009}), \eprint{0712.2355}.

\bibitem[{\citenamefont{Han}(2008)}]{Han:2008xb}
\bibinfo{author}{\bibfnamefont{T.}~\bibnamefont{Han}},
  \emph{\bibinfo{title}{{The 'Top Priority' at the LHC}}},
  \bibinfo{journal}{Int. J. Mod. Phys.} \textbf{\bibinfo{volume}{A23}},
  \bibinfo{pages}{4107} (\bibinfo{year}{2008}), \eprint{0804.3178}.

\bibitem[{\citenamefont{Degrande et~al.}(2011)\citenamefont{Degrande, Gerard,
  Grojean, Maltoni, and Servant}}]{Degrande:2010kt}
\bibinfo{author}{\bibfnamefont{C.}~\bibnamefont{Degrande}},
  \bibinfo{author}{\bibfnamefont{J.-M.} \bibnamefont{Gerard}},
  \bibinfo{author}{\bibfnamefont{C.}~\bibnamefont{Grojean}},
  \bibinfo{author}{\bibfnamefont{F.}~\bibnamefont{Maltoni}}, \bibnamefont{and}
  \bibinfo{author}{\bibfnamefont{G.}~\bibnamefont{Servant}},
  \emph{\bibinfo{title}{{Non-Resonant New Physics in Top Pair Production at
  Hadron Colliders}}}, \bibinfo{journal}{JHEP} \textbf{\bibinfo{volume}{03}},
  \bibinfo{pages}{125} (\bibinfo{year}{2011}), \eprint{1010.6304}.

\bibitem[{\citenamefont{Han and Ruiz}(2014)}]{Han:2013sea}
\bibinfo{author}{\bibfnamefont{T.}~\bibnamefont{Han}} \bibnamefont{and}
  \bibinfo{author}{\bibfnamefont{R.}~\bibnamefont{Ruiz}},
  \emph{\bibinfo{title}{{Higgs Bosons from Top Quark Decays}}},
  \bibinfo{journal}{Phys. Rev.} \textbf{\bibinfo{volume}{D89}},
  \bibinfo{pages}{074045} (\bibinfo{year}{2014}), \eprint{1312.3324}.

\bibitem[{\citenamefont{Mangano
  et~al.}(2016{\natexlab{b}})\citenamefont{Mangano, Plehn, Reimitz, Schell, and
  Shao}}]{Plehn:2015cta}
\bibinfo{author}{\bibfnamefont{M.~L.} \bibnamefont{Mangano}},
  \bibinfo{author}{\bibfnamefont{T.}~\bibnamefont{Plehn}},
  \bibinfo{author}{\bibfnamefont{P.}~\bibnamefont{Reimitz}},
  \bibinfo{author}{\bibfnamefont{T.}~\bibnamefont{Schell}}, \bibnamefont{and}
  \bibinfo{author}{\bibfnamefont{H.-S.} \bibnamefont{Shao}},
  \emph{\bibinfo{title}{{Measuring the Top Yukawa Coupling at 100 TeV}}},
  \bibinfo{journal}{J. Phys.} \textbf{\bibinfo{volume}{G43}},
  \bibinfo{pages}{035001} (\bibinfo{year}{2016}{\natexlab{b}}),
  \eprint{1507.08169}.

\bibitem[{\citenamefont{de~Florian et~al.}(2016)}]{deFlorian:2016spz}
\bibinfo{author}{\bibfnamefont{D.}~\bibnamefont{de~Florian}}
  \bibnamefont{et~al.} (\bibinfo{collaboration}{LHC Higgs Cross Section Working
  Group}), \emph{\bibinfo{title}{{Handbook of LHC Higgs Cross Sections: 4.
  Deciphering the Nature of the Higgs Sector}}} (\bibinfo{year}{2016}),
  \eprint{1610.07922}.

\bibitem[{\citenamefont{Morrissey et~al.}(2012)\citenamefont{Morrissey, Plehn,
  and Tait}}]{Morrissey:2009tf}
\bibinfo{author}{\bibfnamefont{D.~E.} \bibnamefont{Morrissey}},
  \bibinfo{author}{\bibfnamefont{T.}~\bibnamefont{Plehn}}, \bibnamefont{and}
  \bibinfo{author}{\bibfnamefont{T.~M.~P.} \bibnamefont{Tait}},
  \emph{\bibinfo{title}{{Physics Searches at the LHC}}},
  \bibinfo{journal}{Phys. Rept.} \textbf{\bibinfo{volume}{515}},
  \bibinfo{pages}{1} (\bibinfo{year}{2012}), \eprint{0912.3259}.

\bibitem[{\citenamefont{Barger et~al.}(1988{\natexlab{b}})\citenamefont{Barger,
  Han, and Ohnemus}}]{Barger:1987re}
\bibinfo{author}{\bibfnamefont{V.~D.} \bibnamefont{Barger}},
  \bibinfo{author}{\bibfnamefont{T.}~\bibnamefont{Han}}, \bibnamefont{and}
  \bibinfo{author}{\bibfnamefont{J.}~\bibnamefont{Ohnemus}},
  \emph{\bibinfo{title}{{Heavy Leptons at Hadron Supercolliders}}},
  \bibinfo{journal}{Phys. Rev.} \textbf{\bibinfo{volume}{D37}},
  \bibinfo{pages}{1174} (\bibinfo{year}{1988}{\natexlab{b}}).

\end{thebibliography}
\bibliographystyle{apsper}

\end{document}